\documentstyle[12pt,amssymb,amscd,righttag]{amsart}

\newtheorem{thm}{Theorem}[section]
\newtheorem{lem}{Lemma}[section]
\newtheorem{pro}{Proposition}[section]
\newtheorem{cor}{Corollary}[section]

\theoremstyle{definition}

\newtheorem{exmp}{Example}[section]

\theoremstyle{remark}

\newtheorem{rem}{Remark}[section]

\numberwithin{equation}{section}
\renewcommand{\epsilon}{\varepsilon}
\renewcommand{\phi}{\varphi}

\newcommand{\romm}{r^{\bullet(m)}}
\newcommand{\dnb}{\Delta_{\nu,\bullet}}
\newcommand{\Kerb}{\Ker^{\bullet}}
\newcommand{\Bb}{\overline{B}}
\newcommand{\web}{\wedge^{\bullet}}
\newcommand{\cPe}{\cP^{\bullet}}
\newcommand{\rk}{\operatorname{rk}}
\newcommand{\cP}{\cal{P}}
\newcommand{\cD}{\cal{D}}
\newcommand{\cH}{\cal{H}}
\newcommand{\cL}{\cal{L}}
\newcommand{\cF}{\cal{F}}
\newcommand{\ya}{y_1}
\newcommand{\yb}{y_2}
\newcommand{\sig}{\sigma}
\newcommand{\sio}{\sig_0}
\newcommand{\sia}{\sig_1}
\newcommand{\dromb}{\left(\DR(M)\right)_2}
\newcommand{\tno}{\theta_{\nu,\bullet}}
\newcommand{\lrs}{\widetilde{\phantom{abab}}}
\newcommand{\llrs}{\widetilde{\phantom{abababab}}}
\newcommand{\es}{\underset{\llrs}{\hrulefill}}
\newcommand{\sst}{\scriptstyle}
\newcommand{\df}{\partial}
\newcommand{\den}{\Delta_\nu}

\newcommand{\DR}{DR^{\bullet}}
\newcommand{\oma}{\omega_1}
\newcommand{\omb}{\omega_2}
\newcommand{\Ce}{C^{\bullet}}
\newcommand{\Ge}{G^{\bullet}}
\newcommand{\Hh}{H^{\bullet}}
\newcommand{\cHh}{\cal{H}^{\bullet}}
\newcommand{\kab}{_{k=1,2}}
\newcommand{\Me}{\widetilde{M}}
\newcommand{\Mm}{\overline{M}}
\newcommand{\Qq}{\overline{Q}}
\newcommand{\Wan}{W^{\bullet}_a(\nu)}
\newcommand{\dew}{\det W^{\bullet}_a(\nu)}
\newcommand{\deh}{\det H^{\bullet}(M_\nu)}
\newcommand{\mg}{\left(M_\nu,g_M\right)}

\newcommand{\din}{\Delta_{\nu,i}}
\newcommand{\djn}{\Delta_{\nu,j}}
\newcommand{\dtn}{\Delta_{\nu,t}}
\newcommand{\djon}{\Delta_{\nu_0,j}}
\newcommand{\djot}{\Delta_{\nu_0,t}}
\newcommand{\djna}{\Delta_{\nu_1,j}}
\newcommand{\dtna}{\Delta_{\nu_1,t}}
\newcommand{\doo}{\Delta^{\bullet}}
\newcommand{\don}{\Delta^{\bullet}_{\nu}}
\newcommand{\donoo}{\doo_{\nu_0}}
\newcommand{\dono}{\Delta^{\bullet}_{\nu;0}}
\newcommand{\doaa}{\Delta^{\bullet}_{1,1}}
\newcommand{\domb}{\Delta^{\bullet}_{M_2}}
\newcommand{\dombn}{\Delta^{\bullet}_{M_2,N}}
\newcommand{\znj}{\zeta_{\nu,j}}
\newcommand{\znb}{\zeta_{\nu,\bullet}}
\newcommand{\Op}{\operatorname{Op}}
\newcommand{\ind}{\operatorname{ind}}
\newcommand{\supp}{\operatorname{supp}}
\newcommand{\Dom}{\operatorname{Dom}}
\newcommand{\res}{\operatorname{res}}
\newcommand{\Ker}{\operatorname{Ker}}
\newcommand{\Ree}{\operatorname{Re}}
\newcommand{\Det}{\bold{D}\bold{e}\bold{t}}
\newcommand{\Tr}{\operatorname{Tr}}
\newcommand{\Spp}{\operatorname{Sp}}
\newcommand{\tr}{\operatorname{tr}}
\newcommand{\id}{\operatorname{id}}
\newcommand{\Aut}{\operatorname{Aut}}
\newcommand{\mno}{M_{\nu_0}}
\newcommand{\Pii}{\Pi^{\bullet}}
\newcommand{\Wano}{W_a^{\bullet}(\nu_0)}
\newcommand{\dewj}{\det W_a^j(\nu)}
\newcommand{\dewo}{\det W_a^{\bullet}(\nu_0)}
\newcommand{\tonga}{T_0(M_{\nu(\gamma)},Z;a)}
\newcommand{\fan}{\phi^{an}_\nu}
\newcommand{\fano}{\phi^{an}_{\nu_0}}
\newcommand{\We}{W^{\bullet}}
\newcommand{\Ve}{V^{\bullet}}
\newcommand{\Aa}{A^{\bullet}}
\newcommand{\Spec}{\operatorname{Spec}}
\newcommand{\Cone}{\operatorname{Cone}}
\newcommand{\Co}{\Cone^{\bullet}}
\newcommand{\uno}{U_{\nu_0}}
\newcommand{\Rn}{R_\nu}
\newcommand{\Ron}{R^{\bullet}_\nu}
\newcommand{\rs}{\stackrel{\rightarrow}{_{\sim}}}

\newcommand{\wC}{\Bbb C}
\newcommand{\wR}{\Bbb R}

\newcommand{\wZ}{\Bbb Z}
\newcommand{\rbo}{\wR^2\setminus(0,0)}
\newcommand{\Fe}{F^{\bullet}}
\newcommand{\Ke}{K^{\bullet}}
\newcommand{\Le}{L^{\bullet}}
\newcommand{\Image}{\operatorname{Im}}
\newcommand{\ronce}{{\rho}_{\nu,c}}
\newcommand{\renc}{r_{\nu,c}}
\newcommand{\vcna}{v^c_{\nu *}}
\newcommand{\vna}{v_{\nu *}}
\newcommand{\ao}{\alpha_0}
\newcommand{\bo}{\beta_0}
\newcommand{\kn}{k_\nu}
\newcommand{\kno}{k_{\nu_0}}
\newcommand{\Ejn}{E^{j,Neu}_{t,x,x}}
\newcommand{\Ejd}{E^{j,Dir}_{t,x,x}}
\newcommand{\ha}{h_1}
\newcommand{\hb}{h_2}
\newcommand{\ua}{u_1}
\newcommand{\ub}{u_2}
\newcommand{\xa}{x_1}
\newcommand{\xb}{x_2}
\newcommand{\dfg}{\df_{\gamma}}
\newcommand{\bgo}{\big|_{\gamma=0}}
\newcommand{\gjln}{G^j_{\lambda}(\nu)}
\newcommand{\eps}{\epsilon}
\newcommand{\ianma}{i^*_{N,M_1}}
\newcommand{\mab}{M_{[-1/2,1/2]}}
\newcommand{\Eo}{e^{\bullet}}
\newcommand{\Eotxaxb}{e^{\bullet}_{t,x_1,x_2}}
\newcommand{\eo}{E^{\bullet}}
\newcommand{\eotxy}{E^{\bullet}_{t,x,y}}
\newcommand{\ettxa}{E_{t-\tau,x,*}}
\newcommand{\eottxa}{\ettxa^{\bullet}}

\newcommand{\Eotxy}{e^{\bullet}_{t,x,y}}
\newcommand{\eotxaxb}{E^{\bullet}_{t,x_1,x_2}}

\newcommand{\eotaxaa}{E^{\bullet}_{t_1,x_1,*}}
\newcommand{\eotuxaa}{\eo_{\tau,x_1,*}}
\newcommand{\eottaxb}{\eo_{t-\tau,*,x_2}}

\newcommand{\eotbay}{E^{\bullet}_{t_2,*,y}}

\newcommand{\eotaxaz}{E^{\bullet}_{t_1,x_1,z}}

\newcommand{\eotbzxb}{E^{\bullet}_{t_2,z,x_2}}

\newcommand{\eotuxa}{E^{\bullet}_{\tau,x,*}}
\newcommand{\eottay}{E^{\bullet}_{t-\tau,*,y}}
\newcommand{\etzazb}{E_{t,z_1,z_2}}
\newcommand{\eotzazb}{E^{\bullet}_{t,z_1,z_2}}
\newcommand{\etxaxb}{E_{t,x_1,x_2}}
\newcommand{\etaxza}{E_{t_1,x,z_1}}
\newcommand{\etbzbx}{E_{t_2,z_2,x}}
\newcommand{\etzba}{E_{t_1+t_2,z_2,z_1}}
\newcommand{\qaza}{Q_{1,z_1}}
\newcommand{\qbzb}{Q_{2,z_2}}
\newcommand{\dza}{d_{z_1}}
\newcommand{\dzb}{d_{z_2}}
\newcommand{\rzb}{r_{z_2}}
\newcommand{\aza}{A_{z_1}}

\newcommand{\deza}{\delta_{z_1}}
\newcommand{\dezb}{\delta_{z_2}}
\newcommand{\iama}{i^*_{M_1}}
\newcommand{\Pe}{P^{\bullet}}
\newcommand{\pom}{P^{\bullet(m)}}

\begin{document}

\pagestyle{style}
\vspace{5mm}
\title[Generalized Ray-Singer conjecture.\,I.\,A\;manifold\;with\;boundary]
{Generalized Ray-Singer conjecture.~I.\\A manifold with a smooth boundary }
\author{S.M. Vishik}
\dedicatory{For my parents}
\address{Max-Planck Institut f\"ur Mathematik,
Gottfried-Claren-Stra{\ss}e~26, 53225 Bonn, Germany}
\email{senia@@mpim-bonn.mpg.de}
\date{ }
\maketitle
\vskip .5cm

\begin{abstract}
This paper is devoted to a proof of a generalized Ray-Singer
conjecture for a manifold with boundary (the Dirichlet and
the Neumann boundary conditions are independently given on each connected
component of the boundary and the transmission boundary condition
is given on the interior boundary).
The Ray-Singer conjecture \cite{RS} claims that for a closed manifold
the combinatorial and the analytic torsion norms on the determinant
of the cohomology are equal. For a manifold with boundary the ratio between
the analytic torsion and the combinatorial torsion is computed.
Some new general properties of the Ray-Singer
analytic torsion are found. The proof does~not use any computation of
eigenvalues and its asymptotic expansions or explicit expressions
for the analytic torsions of any special classes of manifolds.
\end{abstract}

\tableofcontents
Torsion invariants for manifolds which are not simply connected were
introduced by K. Reidemeister in \cite{Re1}, \cite{Re2}, where he obtained
with the help of such invariants a full $PL$-classification
of three-dimensional lens spaces. These invariants were generalized by
W. Franz to multi-dimensional $PL$-manifolds in \cite{Fr}. As the result
of this generalization he obtained a $PL$-classification of lens
spaces of any dimension. (These torsions were the first invariants
of manifolds which are not homotopy invariants.) J.H.C. Whitehead
in \cite{Wh} and G. de Rham in \cite{dR3} introduced torsion invariants
for smooth manifolds. G. de Rham in \cite{dR3} proved that a spherical
Clifford-Klein manifold (i.e., the quotient of a sphere under
the fixed-point free action of a finite group of rotations) is determined
up to an isometry by its fundamental group and by its Reidemeister torsions.
The Whitehead torsion for a homotopy equivalence between finite cell
complexes was introduced in \cite{Wh} as a generalization of the Reidemeister
torsion invariants defined in \cite{Re1}, \cite{Fr}, and \cite{dR3}.
(Its values are in the Whitehead group
$\operatorname{Wh}(\pi_1)$ of the fundamental group $\pi_1$.)
The Whitehead torsion is connected with Whitehead's theory of simple
homotopy types (\cite{Wh}, \cite{dRMK}, \cite{Mi}, Section~7).
Some modifications of Reidemeister torsions were considered
by J. Milnor in \cite{Mi}, Sections~8, 12, and by V. Turaev
in \cite{T}, Section~3. The scalar Reidemeister torsion is a global invariant
of a cell decomposition of a manifold and of an acyclic representation
of its fundamental group. It is an invariant of the $PL$-structure of
a manifold. The Reidemeister torsion for an arbitrary finite-dimensional
unimodular representation of the fundamental group can be defined
as a canonical norm
on the determinant line of the cohomology of a manifold (with the coefficients
in the local system defined by this representation). It is some kind
of multiplicative analog of the Euler characteristic in the case
of odd-dimensional manifolds. (The Euler characteristic of a closed manifold
is trivial in the odd-dimensional case.) Formulas for the Reidemeister
torsions of a direct product of manifolds (\cite{KwS}) are analogous
to the multiplicative property of the Euler characteristic.

The Ray-Singer analytic torsion was introduced in \cite{RS}
for a closed Riemannian manifold $\left(M,g_M\right)$ with an acyclic
orthogonal representation of the fundamental group $\pi_1(M)$.
It is equal to a product of the corresponding powers of the determinants
of the Laplacians on differential forms $\DR(M)$. These determinants
are regularized with the help of the zeta-functions of the Laplacians.
(The scalar Reidemeister torsion also can be written by the analogous
formula, where Riemannian Laplacians are replaced by the combinatorial ones.)
The Ray-Singer analytic torsion is defined with the help of a Riemannian
metric $g_M$ but it is independent of $g_M$ in the acyclic case.
(This assertion was proved in \cite{RS}, Theorem~2.1.)
So it is an invariant of a smooth structure on $M$. It has the properties
analogous to the properties of the Reidemeister torsion (\cite{RS},
Sections~2, 7). The Ray-Singer conjecture (\cite{RS}) claims that for
an acyclic representation $\rho$ of the fundamental group of a closed
manifold $M$ the Reidemeister torsion of $\left(M,\rho\right)$ (which
is defined for any smooth triangulation of $M$) is equal to the Ray-Singer
analytic torsion of $\left(M,\rho\right)$. This conjecture was independently
proved by W. M\"uller in \cite{Mu1} and by J. Cheeger in \cite{Ch} for
closed manifolds. The Ray-Singer analytic torsion can also be defined
for any finite-dimensional unitary representation $\rho$ of $\pi_1(M)$.
In this case the Ray-Singer torsion is the norm on the determinant
line $\det\Hh\left(M,\rho\right)$. For instance, it is defined
for a trivial one-dimensional representation. So the analytic
torsion norm provides us with a canonical norm on the determinant line
of the de Rham complex of a manifold.
(The Ray-Singer formula for an arbitrary finite-dimensional unitary
representation $\rho$ of $\pi_1(M)$ in the case, when $M$ is a smooth
closed manifold, claims that the Ray-Singer norm
on $\det\Hh\left(M,\rho\right)$ is equal to the Reidemeister norm
on $\det\Hh\left(M,\rho\right)$.)

Let $\left(M,g_M\right)$ be a manifold with a smooth boundary $\df M$
and with the Dirichlet and the Neumann boundary conditions independently
given on the connected components of $\df M$. Let $Z\subset\df M$ be a union
of the components of $\df M$ where the Dirichlet boundary conditions
are given. Let $F_{\rho}$ be a local system with a fiber $\wC^m$ defined
by a unitary representation $\rho\colon\pi_1(M)\to U(m)$. Then the Ray-Singer
torsion norm $T_0\left(M,Z;F_{\rho}\right)$ is defined on
$\det\Hh\left(M,Z;F_{\rho}\right)$. It is independent of $g_M$ (if $g_M$
is a direct product metric near $\df M$) and it depends on a flat
Hermitian metric on the fibers $F_{\rho}$ (for a general $(M,Z)$).
A flat Hermitian structure on $F_{\rho}$ defines a norm on the line
$\det\left(F_x,M,Z\right):=\otimes_k\left(\det F_{x_k}\right)^{\chi\left(M_k,
Z\cap\df M_k\right)}$, where the product is over the full set
of representatives $F_{x_k}$ of fibers of $F_{\rho}$ over the connected
components $M_k$ of $M$ (with one such a fiber $F_{x_k}$ for each $M_k$,
$x_k\in M_k$, $\det F_x:=\wedge^{\max}F_x$). The tensor product of this
norm and of $T_0\left(M,Z;F_{\rho}\right)$ is a modified Ray-Singer norm on
$\det\Hh\left(M,Z;F_{\rho}\right)\otimes\det\left(F_x,M,Z\right)$ and
it does~not depend on $g_M$ and on a flat Hermitian metric on $F_{\rho}$
(\cite{V1}). The Ray-Singer torsion norm for the de Rham complex
of $(M,Z)$ with the coefficients in the direct sum of any finite-dimensional
local system $F_{\rho}$ and of the dual one $F_{\rho}^{\vee}$ is defined
in \cite{V2}.
In this case the Reidemeister torsion
$\tau_0\left(M,Z;F_{\rho}\oplus F_{\rho}^{\vee}\right)$ (i.e., the one
for $(M,Z)$ with the coefficients in $F_{\rho}\oplus F_{\rho}^{\vee}$)
is well-defined, because the fibers of the line bundle
$\det\left(F_{\rho}\oplus F_{\rho}^{\vee}\right)$ have the canonical norm
in accordance with the local system structure. In this case,
the Ray-Singer torsion differs from the Reidemeister torsion by an explicit
factor (which is computed in \cite{V2}) but this torsion does~not depend
on $g_M$ (if $g_M$ is a direct product metric near $\df M$). This definition
of the Ray-Singer torsion norm
does~not use a Hermitian structure in the fibers of $F_{\rho}$.
In \cite{Mu2} another Ray-Singer torsion was introduced for the de Rham
complex of a closed $\left(M,g_M\right)$ with the coefficients in a local
system $F_{\rho}$, defined by a unimodular finite-dimensional representation
$\rho$ of $\pi_1(M)$. This torsion is defined with the help of an arbitrary
Hermitian metric $h_{\rho}$ in the fibers of $F_{\rho}$ and it depends
in general on this metric.
(For a non-unitary representation $\rho$ there are no Hermitian metrics
on $F_{\rho}$, which are flat with respect to the canonical flat structure.)
It was proved in \cite{Mu2} that in the case of an odd-dimensional $M$
the Ray-Singer torsion, defined with the help of a Hermitian metric
$h_{\rho}$, is independent of $\left(h_{\rho},g_M\right)$
and is equal to the Reidemeister torsion.
(The Reidemeister torsion is canonically defined for any unimodular
finite-dimensional representation of $\pi_1$. In the case
of an odd-dimensional closed $M$ it is independent of a flat Hermitian
metric on $\det F_{\rho}$, since the Euler characteristic
in this case is equal to zero for each connected component of $M$.)
The Ray-Singer torsion, defined with the help of $h_{\rho}$, depends
on $\left(h_{\rho},g_M\right)$ for a general even-dimensional $M$.
The definition  of the Ray-Singer torsion for any finite-dimensional
representation $\rho$ of $\pi_1(M)$ for a closed $\left(M,g_M\right)$
equiped with a Hermitian metric $h_{\rho}$ (on the fibers
of the corresponding vector bundle) is given
in \cite{BZ1}, \cite{BZ2}. In \cite{BZ2} the Ray-Singer metric
on the determinant line, corresponding to a finite flat exact sequence
$\left(\Fe,d_F\right)$ of finite-dimensional flat vector bundles over $M$
is computed (in terms of $g_M$ and of Hermitian metrics on $F^j$).

The Gaussian integral of $\exp\left(-(Sx,x)\right)$, where $S$ is a positive
self-adjoint operator in a finite-dimensional Hilbert space $H$, $\dim H=n$,
is equal to $\left(2\pi\right)^{n/2}\left(\det S\right)^{-1/2}$.
The Ray-Singer torsion appears naturally in the computations of asymptotic
expansions for analogous infinite-dimensional integrals
of $\exp\left(-ikI(A)\right)$, where $I(A)$ possesses
an infinite-dimensional symmetry group $G$ (\cite{Sc}, \cite{Wi1},
\cite{Wi2}). For instance, the Chern-Simons action
$$
I(A):=\left(4\pi\right)^{-1}\int_M\Tr\left(A\wedge dA+2/3A\wedge A\wedge A
\right)
$$
on a trivialized principal
$G$-bundle $P_G$ over a closed orientable three-dimensional manifold
(where $G=SU_N$ and $\Tr$ is the trace in the $N$-dimensional geometrical
representation of $G$, and where $A$ is a connection form) is invariant
under the gauge transformations $A\to gAg^{-1}-dg\cdot g^{-1}=:A_g$
for a smooth $g\colon M\to G$ (where $A_g$ is the same connection but
with respect to another trivialization of $P_G$, i.e., with respect
to another smooth section $G\to P_G$). Stationary points of $I(A)$
are the flat connections $A_\alpha$ (i.e., such that the curvature
$F\left(A_\alpha\right)$ is equal to zero). The asymptotic
of an integral of $\exp\left(-ikI(A)\right)$ as $k\to+\infty$,
$k\in\wZ_+$, is computed by the stationary phase method.
The principal term of the contribution of a point $A_\alpha$ into this
integral (in the case when the flat connection $A_\alpha$ is an isolated
one) has as its absolute value the square root of the Ray-Singer
torsion of $M$ with coefficients in the local system, defined by a flat
connection $A_\alpha$, with the Lie algebra $\frak{g}$ of $G$ as its fibers
(see \cite{Wi1}; \cite{Wi2}, 2.2; \cite{BW}, 2).

The Reidemeister torsion was essentially used in \cite{Wi2}, 4, for
the computation of the volume of a moduli space $\cal{M}$ of the fundamental
group representations for a closed two-dimensional surface. In this
case the Reidemeister torsion is a section of $\left|\det\right|T^*\cal{M}$,
i.e., it is a density on $\cal{M}$.

\medskip
This paper is devoted to a proof of a generalized Ray-Singer conjecture
for manifolds with a smooth boundary (and also for transmission
boundary conditions given on the interior boundaries). We suppose that
the local system is trivial. The proof of the Ray-Singer conjecture
for non-unitary local systems and for manifolds with
corners will be the subject of a subsequent paper.

Let $\left(M,g_M\right)$ be a Riemannian manifold with a smooth boundary
$\df M$ and let the Dirichlet and the Neumann boundary conditions be
independently given on the connected components of $\df M$. Let $g_M$ be
a direct product metric near $\df M$. Then the Ray-Singer torsion of
$\left(M,g_M\right)$ is defined as a norm on the determinant line
$\det\Hh\left(M,Z\right)$. (Here $Z$ is the union of the connected
components of $\df M$ where the Dirichlet boundary conditions are given.)
This norm is independent of $g_M$ (for direct product metrics $g_M$
near $\df M$).
The Reidemeister torsion of $(M,Z)$ is an invariant of the $PL$-structure
of $(M,Z)$ and it is a norm on the same determinant line. The torsion norms
are defined in Section~\ref{Sec1}. The Ray-Singer norm differs from
the Reidemeister norm on $\det\Hh\left(M,Z\right)$ for a general
$\df M\ne\emptyset$. Their ratio is computed in Theorem~\ref{T4} below.

Let $\left(M,g_M\right)$ be obtained by gluing
two Riemannian manifolds $\left(M_j,g_{M_j}\right)$ along the common
component $N$ of their boundaries, $M:=M_1\cup_N M_2$ (where $N$
is a closed smooth manifold of codimension one in $M$). Let $g_M$
be a direct product metric near $N$. Then, as it is proved
in Theorem~\ref{T1}, the Ray-Singer torsion norm $T_0(M,Z)$
on $\det\Hh\left(M,Z\right)$ is equal to the tensor product
of the Ray-Singer norms
$T_0\left(M_1,Z_1\cup N\right)\otimes T_0\left(M_2,Z_2\cup N\right)\otimes T_0
(N)$ ($Z_k:=Z\cap\df\Mm_k$), where the line $\det\Hh(M,Z)$ is identified with
the tensor product of the lines
$\det\Hh\left(M_1,Z_1\cup N\right)\otimes\det\Hh\left(M_2,Z_2\cup N\right)
\otimes\det\Hh(N)$ by the short exact sequence of the de Rham
complexes
\begin{equation}
0\!\to\!\DR\!\left(\!M_1,\!Z_1\!\cup\! N\!\right)\!\oplus\!\DR\!\left(\!M_2,
\!Z_2\!\cup\! N\!\right)\!\to\!\DR\!(\!M,\!Z)\!\to\!\DR\!(N)\!\to\! 0,
\label{X3010}
\end{equation}
where $\DR(M,Z)$ is the relative de Rham complex of smooth forms
with the zero  geometrical restrictions to $Z$, the left arrow
is the natural inclusion, and the right arrow is $\sqrt 2$ times
a geometrical restriction. For the Reidemeister norm this assertion
is also true and the identification of the determinant lines
is given by the analogous exact sequence of cochain complexes. However
in this case the right arrow is the geometrical restriction of cochains
(without additional factor $\sqrt 2$). Let $(M,Z)$ be obtained
by gluing two manifolds $\left(M_1,Z_1\right)$ and $\left(M_2,Z_2\right)$
along the common component $N$ of their boundaries, $M:=M_1\cup_N M_2$.
Then the ratio of the square of the Ray-Singer norm and the square
of the Reidemeister norm for $(M,Z)$ is equal to the product
of the same ratios for $\left(M_1,Z_1\cup N\right)$,
$\left(M_2,Z_2\cup N\right)$, and for $N$
with an additional factor $2^{-\chi(N)}$. So the assertion
of Theorem~\ref{T1} claims that it is possible to calculate
the Ray-Singer norm by cutting of a manifold into pieces
which are manifolds with smooth boundaries. The main theorems
of this paper are consequences of Theorem~\ref{T1}. This theorem
provides us with the gluing formula for the Ray-Singer torsion norms.
Such a gluing  formula is a new one.

In the case of a manifold with a smooth boundary, the Ray-Singer torsion
$T_0(M,Z)$ is a function not only of $(M,Z)$ but also of the phase $\theta$
of a cut of the spectral plane $\wC$ (because the zeta-functions
$\zeta_j(s)$ for the Laplacians $\Delta_j$ on $DR^j(M,Z)$ are defined
for $\Ree\,s>(\dim M)/2$ as the sums $\sum\lambda^{-s}$
over the nonzero eigenvalues, and $\lambda^{-s}$ is defined as
$\lambda^{-s}_{(\theta)}:=\exp\left(-s\log_{(\theta)}\lambda\right)$, where
$\theta-2\pi<\Image\log_{(\theta)}\lambda<\theta$, $\theta\notin2\pi\wZ$).
In fact, $T_0(M,Z;\theta)$ (as well as $\zeta_j(s)$) depends only
on $[\theta/2\pi]$. The zeta-function regularization of the
$\det'\left(\Delta_j\right)$ (i.e., of the product of all the nonzero
eigenvalues of $\Delta_j$, including their multiplicities) is defined as
$\exp\left(-\df_s\zeta_j(s)|_{s=0}\right)$. The analytic continuation
of $\zeta_j(s)$ is regular at zero. The zeta-function $\zeta_j(s;m)$
depends on $m:=[\theta/2\pi]$, $\theta\notin 2\pi\wZ$, as follows:
\begin{align*}
\zeta_j(s;m+1)                               & =\exp(-2\pi is)\zeta(s;m),\\
\operatorname{det}'\left(\Delta_j;m+1\right) & =\exp\left(2\pi i\zeta_j(0)
\right)\operatorname{det}'\left(\Delta_j;m\right).
\end{align*}
The number $\zeta_j(0)$ is independent of $m$, and the number
$\zeta_j(0)+\dim\Ker\Delta_j$ can be interpreted as the regularized
dimension of the space $DR^j(M)$. This regularized dimension depends
not only on the space $DR^j(M)$ but it also depends on a positive
definite self-adjoint elliptic differential operator of a positive order,
which acts in $DR^j(M)$. This dimension is a real number but it is not
an integer in the case of the Laplacians on $\DR(M)$
for a general closed even-dimensional $\left(M,g_M\right)$. Hence,
$\det\left(\Delta_j;m\right)$ depends on $m$ for such $\left(M,g_M\right)$.
The number $\zeta_j(0)$ is an integer for a generalized Laplacian
on a closed odd-dimensional $\left(M,g_M\right)$, according to \cite{BGV},
Theorem~2.30, or to \cite{Gr}, Theorem~1.6.1. It is equal to zero when
$M$ is closed, $\dim M$ is odd, and $\dim\Ker\Delta_j=0$.

Even in such a simple case as for an interval $(I,\df I)$ with the Dirichlet
boundary conditions, the dependence of $T_0(M,Z;m)$ on $m$ is nontrivial.
The ratio of the torsion $T_0\left(M,Z;[\theta/2\pi]\right)$
and the Reidemeister torsion norm is computed in Theorem~\ref{TA424}.

\medskip
The paper is organized as follows. In Section~\ref{Sec1} we deduce
a generalization of the Ray-Singer conjecture from the gluing formula
for Ray-Singer torsion norms. This formula is proved in Theorem~\ref{T1}.
The proof uses $\nu$-transmission interior boundary conditions on $N$,
where $\nu=(\alpha,\beta)\in\rbo$. These interior boundary problems
give us a smooth in $\nu$ family of spectral problems on $M$. Such a problem
for $\nu=(1,1)$ coincides (in a spectral sense) with the spectral problem
for a glued $M$. For $\nu=(0,1)$ or for $\nu=(1,0)$ it is a direct sum
of spectral problems on $M_1$ and on $M_2$, i.e., the two pieces of $M$
are completely disconnected.
So this family provides us with a smooth process of cutting (in a spectral
sense) of $M$ in two pieces $M_1$ and $M_2$.
Let $M=M_1\cup_N M_2$ be obtained by gluing $M_1$ and $M_2$
along the common component $N$ of their boundaries.
Then the Ray-Singer norm $T_0\left(M_{\nu},Z\right)$ on the determinant
line $\det\Hh\left(M_{\nu},Z\right)$ for the de Rham complex
$\DR\left(M_{\nu},Z\right)$ with $\nu$-transmission conditions on $N$
is defined. The short exact sequence for $\DR\left(M_{\nu},Z\right)$,
similar to (\ref{X3010}), has the same the first and third terms
as (\ref{X3010}).
The homomorphisms $r_{\nu}\colon\DR\left(M_{\nu},Z\right)\to\DR(N)$ are
of the form $r_{\nu}=\left(\alpha i^*_1+\beta i^*_2\right)/|\nu|$, where
$i^*_j\omega_j$ are the geometrical restrictions to $N$ for the components
$\omega_j$ of
$\omega=\left(\omega_1,\omega_2\right)\in\DR\left(M_{\nu},Z\right)$.
Note that $r_{(1,1)}=\sqrt 2\,i^*$. (This is the reason of the appearance
of $\sqrt 2\,i^*$ in the exact sequence (\ref{X3010}), connected with
the gluing formula.) In Lemma~\ref{L2} we prove that the gluing property
for analytic torsion norms (Theorem~\ref{T1}) is equivalent
to the independence of $\nu$ of the norms on
$\det\Hh\left(M_1,Z_1\cup N\right)\otimes\det\Hh\left(M_2,Z_2\cup N\right)
\otimes\det\Hh(N)$ induced by $T_0\left(M_{\nu},Z\right)$. (Here
the identification of the determinant lines is defined by the short exact
sequence for $\DR\left(M_{\nu},Z\right)$.) The latter assertion is proved
in Section~\ref{Sec2}. First we prove that the norm induced by
the Ray-Singer torsion $T_0\left(M_{\nu},Z\right)$ is locally independent
of $\nu$ in the case when $\alpha\beta\ne 0$ (where $\nu=(\alpha,\beta)$).
We do this in Sections~2.3, 2.5, and 2.6 with the help of explicit
variation formulas for the scalar Ray-Singer torsion
$T\left(M_{\nu},Z\right)$ (if $\nu$ depends smoothly on a parameter).
We define a family (in $\nu$) of homomorphisms to identify
finite-dimensional subcomplexes $\Wan$ of $\DR\left(M_{\nu},Z\right)$.
(The complexes $\Wan$ are spanned by the eigenforms
of the Laplacians with eigenvalues less than a fixed number $a>0$.
We suppose that $a$ is~not an eigenvalue
of $\Delta_j\left(M_{\nu},Z\right)$ for $0\le j\le n$.) Then we
compute the actions of these homomorphisms on the determinant lines.
These identifications are not canonical; we choose some particular
(quite natural) identifications for $\nu$ sufficiently close
to $\nu_0$ such that $\ao\bo\ne 0$.

Then it is enough to prove the continuity in $\nu\in\rbo$ of the norm on
$\det\Hh\left(M_1,Z_1\cup N\right)\otimes\det\Hh\left(M_2,Z_2\cup N\right)
\det\Hh(N)$, which is induced by the Ray-Singer norm
$T_0\left(M_{\nu},Z\right)$. We prove in Section~2.7 that the truncated
scalar analytic torsion $T\left(M_{\nu},Z;a\right)$, corresponding to
the eigenvalues $\lambda$ of $\Delta_j\left(M_{\nu},Z\right)$ which are
greater than $a$, is locally continuous in $\nu$. Then we prove that
the norm, induced by the analytic torsion norm $T_0\left(\Wan\right)$
of a finite-dimensional complex $\Wan$, is locally continuous in $\nu$.
The latter assertion is proved in Sections~2.2, 2.4, and 2.7 with the use
of the cone of the homomorphism
$\Ron(a)\colon\Wan\to\Ce\left(X_{\nu},Z\cap X\right)$ (where
$\Ron(a)$ is the integration of differential forms from $\We_a$
over the simplexes of a given smooth triangulation $X$ of $M$, and
$\Ce\left(X_{\nu},Z\cap X\right)$ is the corresponding cochain complex).
This homomorphism is a quasi-isomorphism for any $\nu\in\rbo$
(Proposition~\ref{PA87}). We can conclude that the analytic torsion
norm on $\det\Hh\left(\Cone\Rn(a)\right)=\wC$ (for a fixed $\nu$)
corresponds to an acyclic finite-dimensional complex and is defined
by the derivatives at zero  of the zeta-functions for self-adjoint
finite-dimensional invertible operators. So these norms are locally
continuous in $\nu$. (This is proved in Section~2.7.) Then the local
continuity of the norm induced
by $T_0\left(\Wan\right)$ on $\det\Hh\left(M_1,Z_1\cup N\right)\otimes\det
\Hh\left(M_2,Z_2\cup N\right)\otimes\det\Hh(N)$ follows from the continuity
of the norm (on the same determinant line) induced
by $T_0\left(\Ce\left(X_{\nu},Z\cap X\right)\right)$ and from the identity:
$$
T_0\left(\We_a\right)=T_0\left(\Ce\left(X_{\nu},Z\cap X\right)\right)\big/
\left\|1\right\|^2_{T_0(\Co\Rn(a))}.
$$
This identity is proved in Lemma~\ref{L2.4}.

The use of the cone of $\Ron(a)$ allows us to avoid difficulties, connected
with the fact that some positive eigenvalues of the Laplacians
$\doo\left(M_{\nu},Z\right)$ tend to $0$ as $\nu=(\alpha,\beta)$
tends to $\nu_0=(1,0)$ (or to $\nu_0=(0,1)$). The dimensions of
$\Hh\left(M_{\nu},Z\right)$ essentially change when $\nu$, $\alpha\beta\ne 0$,
is replaced by $\nu_0$. (Only the Euler characteristic $\chi\left(\Hh\left(M
_{\nu},Z\right)\right)$ does~not change when $\nu$ is replaced by $\nu_0$.)
It is impossible to find for a general $N$ the precise asymptotic expressions
for the eigenvalues $\lambda$, which tend to zero as $\nu\to\nu_0$,
and especially to find the asymptotics of the corresponding
eigenforms $\omega_{\lambda}$ of $\doo\left(M_{\nu},Z\right)$.
So the continuity of the norm induced by $T_0\left(M_{\nu},Z\right)$
(viewed as a function of $\nu$) at the point $\nu_0$
cannot be proved for a general $M$ (obtained by gluing two pieces $M_1$
and $M_2$ along $N$) with the help of separate computations
of the asymptotic expressions for the scalar torsion
$T\left(M_{\nu},Z\right)$ and for the measure
on $\det\Hh\left(M_{\nu},Z\right)$ defined by harmonic forms.
The proof of the classical Ray-Singer conjecture in \cite{Ch}
and the proof in \cite{Mu2} (in the case of unimodular representations
of $\pi_1(M)$)
are based on asymptotic computations of such quantities for a manifold
with boundary $M_u:=M\setminus S_u$, where $S_u$ is a tubular neighborhood
of an embedded sphere $S^k\hookrightarrow M^n$ as the radius $u$
of the tubular neighborhood (in the normal to $S^k$ direction) tends
to zero. (It is also supposed in \cite{Ch} that $S_u$ is a direct product
on $S^k\times D^{n-k}$ and that $g_M|_{S_u}$ is a direct product metric
on $S^k\times D^{n-k}$.)

To give a rigorous proof of the assertions above used in the proof
of the gluing formula, it is necessary to prove a lot of analytic
propositions. We do it in Sections~2.2, 2.6, 2.7, and in Section~3.
The theory of $\zeta$- and $\theta$-functions
in the case  of $\nu$-transmission interior boundary conditions is elaborated
in Section~3. The precise estimates of the corresponding $\zeta$-functions
in vertical strips are obtained in Section~3.4. These estimates
allow us using the inverse Mellin transform to derive the information
about the densities on $M$, $N$, and $\df M$
for the asymptotic expansions as $t\to+0$ of $\theta$-functions
from the properties of the densities for appropriate $\zeta$-functions.

\section{Analytic torsion and the Ray-Singer conjecture}
\label{Sec1}
\subsection{Analytic and combinatorial torsions norms}
The analytic torsion norm appears in the following finite-dimensional
algebraic situation.
Let $(\Aa,d)$ be a finite complex of finite-dimensional
Hilbert spaces.
{\em The determinant of $(\Aa,d)$} is the
tensor product
$$
{\otimes}_j(\Lambda^{\max}A^j)^{(-1)^{j+1}} =:
{\det}(\Aa),
$$
where $\Lambda^{\max}A^j=:{\det}A^j$ is the top exterior
power of the linear space $A^j$ and where $L^{-1}$ is the dual space
$L^{\vee}$ for a one-dimensional vector space $L$ over $\wC$.
The natural Hilbert norm $\left\|\cdot\right\|^2_{\det A}$ is defined
by the Hilbert norms on $A^j$.

The determinant of the cohomology ${\det}\Hh(A)$ of $(\Aa,d)$
is also defined and there is a natural norm on it (since $H^j(A)$
is the subquotient of $A^j$). The differential $d$ provides us
with the identification
\begin{equation*}
f(d):{\det}(\Aa) \simeq {\det}\Hh(A).
\end{equation*}
However in the general case this identification is not an isometry
of the norms
$\left\|\cdot\right\|^2_{\det A}$ and $\left\|\cdot\right\|^2_{\det\Hh(A)}$.
For $f(d)$ to be an isometry it is necessary to multiply
$\left\|\cdot\right\|^2_{\det\Hh(A)}$
by the {\em scalar~analytic~torsion\/} of a complex $(\Aa,d)$,
which is defined as
\begin{equation}
T(\Aa,d) = \exp\bigl(\Sigma(-1)^jj{\df}_s{\zeta}_j(s)|_{s=0}\bigr).
\label{A2}
\end{equation}
Here $\zeta_j(s)=\mathop{\sum}'\lambda^{-s}$ is the sum%
\footnote{The function $\lambda^{-s}$ is defined
as $\exp\left(-s\log\lambda\right)$ where $\log\lambda\in\wR$
for $\lambda\in\wR_+$.}
over all the nonzero eigenvalues $\lambda\ne 0$ (including their
multiplicities)
of the nonnegative (i.e., if $\lambda\ne 0$ then $\lambda>0$) self-adjoint
operator $(d^*d+dd^*)|A^j$. The derivative $\df_s\zeta_j(s)|_{s=0}$
is equal to $-\log\det'\left(\left(d^*d+dd^*\right)|_{A^j}\right)$ (i.e.,
it is equal to the sum of $(-\log\lambda)\in\wR$ over all the nonzero
eigenvalues $\lambda$).

It is enough to prove the assertion (\ref{A2}) in the case
of a two-terms complex
$d\colon F_0\rs F_1$, where $\dim F_j=1$, $e_j\in F_j$, $de_0=\mu e_1$,
$\mu\ne 0$, and where $\left\|e_0\right\|^2=1=\left\|e_1\right\|^2$.
In this case
the element $e_1\otimes e_0^{-1}\in\det\left(\Fe\right)$ is of the unit
norm and the square of the norm of the corresponding element
$\mu^{-1}\in\wC$ from $\wC=\det 0=\det\Hh(F)$ is equal
to $\left|\mu^{-2}\right|^2$. If the norm $\left|\mu^{-1}\right|^2$
is multiplied by the scalar analytic torsion for $\Fe$, namely by
$\exp\left(\log\det\left(d^*d\right)\right)=\exp\left(\log\det\left(dd^*\right)
\right)=\left|\mu\right|^2$ then the isomorphism between $\det(\Fe)$
and $\wC=\det 0$ (defined by $d$) becomes an isometry.

This finite-dimensional definition make sense also for the
infinite-dimensional de Rham complex of a closed smooth manifold.
In this case the analytic torsion is the norm on the determinant
of the cohomology of this manifold. Let $\left(\DR(M),d\right)$ be
the de Rham complex of smooth differential forms
(with the values in $\wC$) on a closed manifold $M$.
The scalar analytic torsion for a closed Riemannian
manifold $(M,g_M)$ is defined by the same formula (\ref{A2}),
where $d^*=\delta$ (relative to $g_M$) and $(d^*d+dd^*)|DR^j(M)$
is the Laplace-Beltrami operator $\Delta_j$. In this case the series,
which defines $\zeta_j(s)$, converges for $\Ree\,s>(\dim M)/2$.
The analytic function $\zeta_j(s)$ can be analytically (meromorphically)
continued to the whole complex plane.
It is known that $\zeta_j(s)$ has
simple poles and that it is regular at zero (\cite{Se2}).

The cohomology $\Hh\left(DR(M)\right)$ are canonically identified
(by the integration of the forms over the simplexes) with the cohomology
$\Hh(M)$ of $M$. This follows from the de Rham theorem.
The Hodge theorem claims that each element of $H^j\left(DR(M)\right)$
has one and only one representative in the space of harmonic
forms $\Ker{\Delta}_j$. The natural norm on $\Ker{\Delta}_j$
(defined by the Riemannian metric $g_M$) provides us
with the norm $\left\|\cdot\right\|^2_{\det\Hh(M)}$ on $\det\Hh(M)$.
For an odd-dimensional $M$ this norm depends on $g_M$.

\noindent{\bf Definition}. The {\em analytic torsion norm} $T_0(M)$ on
${\det}\Hh(M)$  is the norm
\begin{equation}
T_0(M) :=\left\|\cdot\right\|^2_{\det \Hh(M)}\cdot\exp\left(\Sigma(-1)^j
j\df_s\zeta_j(s)|_{s=0}\right).
\label{A3}
\end{equation}

The main property of this norm is its independence of a Riemannian
metric $g_M$. So it is an invariant of a smooth structure on $M$.
Let us suppose that $g_M=g_M(\gamma)$ depends smoothly
on a parameter $\gamma\in\wR^1$. Then the variation formulas in \cite{RS},
Theorems~2.1, 7.3 (or in \cite{Ch}, Theorem~3.10, (3.22)), claim that
\begin{equation}
\dfg\!\sum_j\!\left(-1\right)^j\!j\df_s\zeta_{j,\gamma}(s)\big|_{s=0}\!=
\!\sum_j\!\left(-1\right)^j\!\left(-\Tr\!\left(\exp\left(-t\Delta_{j,\gamma}
\right)\alpha\right)^0\!+\!\Tr\!\left(\cH_{j,\gamma}\alpha\right)\right).
\label{X597}
\end{equation}
Here $\cH_{j,\gamma}$ is the kernel of the orthogonal projection operator
from $DR^j(M)$ onto $\Ker\Delta_j\!\left(M,g_M(\gamma)\right)$,
$\alpha:=*^{-1}_\gamma\dfg\left(*_\gamma\right)$ ($*_\gamma$ corresponds
to $g_M(\gamma)$) and
$\Tr\!\left(\exp\!\left(\!-t\Delta_{j,\gamma}\right)\!\alpha\right)^0$ is
the constant coefficient in the asymptotic expansion as $t\to+0$
($n:=\dim M$):
\begin{equation}
\Tr\left(\exp\left(-t\Delta_{j,\gamma}\right)\alpha\right)=
\sum^l_{k=0}m_{j,k}t^{-n/2+k}+o\left(t^l\right).
\label{X595}
\end{equation}

The existence of the asymptotic expansion (\ref{X595}) follows from \cite{Gr},
Theorem~1.6.1, or from \cite{BGV}, Theorem~2.30. For a family of norms
$\left\|\cdot\right\|^2(\gamma)$ on $\det\Hh(M)$ defined by the harmonic
forms $\Ker\left(\Delta_j\left(M,g_M(\gamma)\right)\right)$
the following equality holds for any fixed $\mu\in\det\Hh(M)$,
$\mu\ne 0$ (\cite{RS}, Section~7):
$$
\dfg\log\left\|\mu\right\|^2_{\det\Hh(M)}(\gamma)=-\sum\left(-1\right)^j
\Tr\left(\cH_{j,\gamma}\alpha\right).
$$
Hence, (\ref{X597}) involves the equality
\begin{equation}
\dfg\log T_0\left(M,g_M\right)=\sum\left(-1\right)^{j+1}m_{j,n/2}.
\label{X596}
\end{equation}

Since $k$ in (\ref{X595}) are integers, we see that the right
side of (\ref{X596}) is zero for odd $n$. For even $n$, $n=2l$, the right
side of (\ref{X596}) is also equal to zero, since $m_{j,l}=-m_{2l-j,l}$.
This fact follows from the equalities
\begin{gather*}
\dfg\left(*^{-1}_\gamma *_\gamma\right)=0,\qquad\qquad\alpha=-*\alpha*^{-1},\\
\Tr\left(\exp\left(-t\Delta_j\right)\alpha\right)=\Tr\left(\left(*\exp
\left(-t\Delta_j\right)*^{-1}\right)(-\alpha)\right)=-\Tr\left(\exp\left(-t
\Delta_{n-j}\right)\alpha\right),
\end{gather*}
(since they involve the equalities $m_{j,k}=-m_{n-j,k}$, where $n$ is even
and $k\in\wZ_+\cup 0$).

The analytic torsion norm can be interpreted (in an intuitional sense)
as the norm, corresponding to an element $v\in\det\DR(M)$ ($v$ is defined
up to a multiplicative constant $c\in{\wC}$, $|c|=1$, and its ``torsion norm''
is equal to one). The space $\det\DR(M)$ and $L_2$-norm on it are not
defined but the space $\det\Hh(M)$ and the analytic torsion norm $T_0(M)$
on it are rigorously defined. For a finite-dimensional complex
the analytic torsion norm on the determinant of its cohomology corresponds
to the norm on the determinant of the complex defined by the Hilbert
structures on the terms of this complex. The analytic torsion
norm is (in some sense) a multiplicative Euler characteristic
useful for odd-dimensional manifolds.

The same definition of $T_0(M)$ make sense also in the case  when $M$
is a compact Riemannian manifold with a smooth boundary
$\df M=\cup{N_i}$ and with the Dirichlet or the Neumann boundary
conditions given independently on each connected component
$N_i$ of $\df M$. Let the metric $g_M$
be a direct product metric near $\df M$.
Then $T_0(M)$ is independent of $g_M$ as in the case of a closed
manifold (this is proved below).

Let $X$ be a smooth triangulation of $M$ and let $\left(\Ce(X),d_c\right)$
be a cochain complex of $X$ ( with complex coefficients). Then
each $C^j(X)$ has the Hilbert structure defined by the orthonormal basis
of basic cochains $\{\delta_e\}$, where $\delta_e(e_1)$ is 1 for
$e_1=e$ and $0$ for $e_1\ne e$. Hence the  scalar torsion
$T\left(\Ce(X),d_c\right)$ is also defined.

The {\em combinatorial torsion} $\tau_0(X)$ is defined as the following norm
on the determinant of the cohomology $\Hh\left(C(X),d_c\right)=\Hh(M)$ :
\begin{equation}
\tau_0(X):=\left\|\cdot\right\|^2_{\det\Hh(C(X))}\cdot T\left(\Ce(X),d_c\right)
\label{A4}
\end{equation}
(where $H^j\left(C(X)\right)$ is the subquotient of $C^j(X)$ and so
it has the natural Hilbert structure induced from $C^j(X)$).
The norm (\ref{A4}) is invariant under
any regular subdivisions of $X$. So this norm is an invariant of the
combinatorial structure of $M$ (which is completely defined by a smooth
structure on $M$). This norm corresponds to the Hilbert
norm on $\det\Ce(X)$, defined by the basic cochains.

Let $M$ be a manifold with a smooth boundary $\df M=\cup N_i$, where $N_i$
are the connected components of $\df M$. Let $Z$ be the union of
$N_i$ where the Dirichlet boundary conditions are given. Set $V:=X\cap Z$.
Then (\ref{A4}) (where $\Hh(C(X))$ and $T\left(\Ce(X),d_c\right)$
are replaced by $\Hh(C(X,V))$ and by $T\left(\Ce(X,V),d_c\right)$)
provides us with the definition of the norm $\tau_0(X,V)$.
This norm is an invariant of the combinatorial structure on $(M,Z)$
(\cite{Mi}, Sections~7, 8, 9).

\subsection{Gluing formulas}

The Ray-Singer conjecture claims that for a closed smooth manifold
$M$ the norms $\tau_0(M)$ and $T_0(M)$ on the same one-dimensional space
$\det\!\Hh\!(\!M\!)$ are equal%
\footnote{The cohomology $\Hh\left(DR(M)\right)$ and
$\Hh\left(C(X)\right)$ are identified (according to the de Rham theorem)
by the homomorphism of the integration of forms from $\DR(M)$
over the simplexes of a smooth triangulation $X$ of $M$.}
\begin{equation}
\tau_0(M)=T_0(M).
\label{A5}
\end{equation}

How to prove such a formula in a natural way? It is necessary to find
a general property of the analytic torsion which involves
the equality (\ref{A5}). Such a property can be formulated
as follows.
Let $(M,\df M)$ be a Riemannian manifold with a smooth boundary
and with the Dirichlet or the Newmann boundary conditions given
independently on the connected components of $\df M$.
Let a closed codimension one submanifold $N$ of $M$, $N\cap\df M=\emptyset$,
divides $M$ in two pieces $M_1$ and $M_2$ (glued along $N$),
$M=M_1\cup_N M_2$,
and let a metric $g_M$ be a direct product metric near $N$ and
near $\df M$. Let $T_0(M_k,N)$ be the analytic torsion norm for $M_k$
(with the Dirichlet boundary conditions on $N$), and let the boundary
conditions on the connected components of $\df M$ belonging to $\df M_k$
be the same as for $T_0(M)$. The following assertion central in this paper.
\begin{thm}[Gluing property]
The analytic torsion norm $T_0(M,Z)$ is the tensor product of
the analytic torsion norms for $(M_1,Z_1\cup N)$, $(M_2,Z_2N)$, and for $N$
\begin{equation}
\phi_{an}T_0(M,Z)=T_0(M_1,Z_1\cup N)\otimes T_0(M_2,Z_2\cup N)\otimes T_0(N),
\label{A6}
\end{equation}
where $Z_k:=Z\cap\df\Mm_k$.
\label{T1}
\end{thm}

The identification $\phi_{an}$ (in (\ref{A6})) of $\det\Hh(M,Z)$ with
the tensor product of the three one-dimensional spaces:
\begin{multline}
\phi_{an}\colon\det\Hh(M,Z)\rightarrow \det\Hh
(M_1,Z_1\cup N)\otimes\det\Hh(M_2,Z_2\cup N)\otimes\det\Hh(N)=\\
=:\Det(M,N,Z)
\label{B1}
\end{multline}
is defined by the long cohomology exact sequence corresponding
to the following short exact sequence of the de Rham complexes:
\begin{equation}
0\!\rightarrow\!\DR\!(M_1,\!Z_1\cup N)\!\oplus\!\DR\!(M_2,\!Z_2\cup N)
\!\rightarrow\!\DR\!(M_{1,1},\!Z)\!\stackrel{r}{\rightarrow}\!\DR\!(N)
\!\rightarrow\! 0.
\label{A7}
\end{equation}

The relative de Rham complex $\left(\DR(M_k,Z_k\cup N),d\right)$ (where $d$
is the exterior derivative of differential forms) consists of the smooth
forms $\omega$ on $\Mm_k$, having the zero geometrical restriction
to $N:i^*_k \omega=0$ (where $i_k\colon N\subset \df M_k\hookrightarrow M_k$)
and also having the zero restrictions to the components of
$\df M\cap M_k$, where the Dirichlet boundary conditions are given
(i.e., to $Z_k$). The complex $\left(DR\left(M_{1,1}\right),d\right)$ consists
of the pairs $(\oma,\omb)$ of smooth differential forms
$\omega_k\in\DR\left(M_k,Z_k\right)$
(i.e., $\omega_k$ have the zero geometrical restrictions to the
corresponding components of $\df M\cap \Mm_k$), which have the
same geometrical restrictions to $N$:
\begin{equation*}
i_1^*\,\oma=i_2^*\,\omb.
\end{equation*}
The differential $d(\oma,\omb)$ in $\DR\left(M_{1,1}\right)$ is defined
as $(d\oma, d\omb)$. The left arrow in (\ref{A7})
is the natural inclusion of $\oplus_k\DR\left(M_k,Z_k\cup N\right)$
into $\DR\left(M_{1,1},Z\right)$.
The right arrow $r$ in (\ref{A7}) is not a usual geometrical
restriction but is the one multiplied by $\sqrt 2$ :
\begin{equation}
r(\oma,\omb)=\sqrt 2\,i_k^*\omega_k\in \DR(N).
\label{A9}
\end{equation}

To define $\phi_{an}$ it is necessary to introduce a natural
identification of $\Hh\left(DR(M,Z)\right)$ with
$\Hh\left(DR\left(M_{1,1},Z\right)\right)$.
(The short exact sequence (\ref{A7}) provides us with the identification
$$
\phi_{an}: \det \Hh\left(DR\left(M_{1,1},Z\right)\right)\rs \Det (M,N,Z),
$$
but not with the identification of $\det \Hh\left(DR(M,Z)\right)$ with
$\Det(M,N,Z)$.)
We show in Proposition \ref{P1}
(for any given metric $g_M$) that not only
all the eigenvalues with their multiplicities but also all the
eigenforms of the natural Laplacian $\Delta_{1,1}$ on
$\DR\left(M_{1,1},Z\right)$ are the same as for the Laplacian on $\DR(M,Z)$.
Thus, the operator $\Delta_{1,1}(g_M)$ in a very strict spectral sense
is the same as $\Delta(g_M)$.

The homotopy operator between the identity operator
on $\DR\left(M_{1,1},Z\right)$ and the projection operator
from $\DR\left(M_{1,1},Z\right)$
onto $\Ker^{\bullet}\Delta_{1,1}=\Ker^{\bullet}\Delta$ is obtained with
the help of the Green function $G_{1,1}$ for the operator $\Delta_{1,1}$
(Lemma \ref{L1}). This homotopy operator provides us with the canonical
identification of  $\Hh\left(DR\left(M_{1,1},Z\right)\right)$
with $\Ker\doaa$. So it defines the identification
of $\Hh\left(DR\left(M_{1,1},Z\right)\right)$
with $\Ker\doo=\Hh\left(DR(M,Z)\right)$ (since $\Ker\doo$ is canonically
identified with $\Ker\doo_{1,1}$).

To prove Theorem~\ref{T1} we introduce a family of interior
boundary conditions on $N$ and show that the induced norm
$\phi^{\nu}_{an}T_0\left(M_\nu,Z\right)$ on $\Det (M,N,Z)$ is independent
of $\nu$ (where $\nu=(\alpha,\beta)\in\rbo$ are the parameters of interior
boundary conditions on $N$). Namely
\begin{equation}
\fan T_0\left(M_\nu,Z\right)=c_0 T_0\left(M_1,Z_1\cup N\right)\otimes T_0
\left(M_2,Z_2\cup N\right)\otimes T_0(N)
\label{A10}
\end{equation}
with some positive $c_0$ which may depend on $\left(M,g_M,\df M\right)$
and on the boundary conditions on $\df M$ but {\em does~not depend}
on the parameters $(\alpha,\beta)=\nu$. Suppose that the formula (\ref{A10})
holds for any gluing two pieces $M_1$ and $M_2$ along a closed $N$,
$M=M_1\cup_N M_2$, where the factor $c_0$ is independent of $\nu$.
Then it is easy to conclude that $c_0=1$ (Lemma~\ref{L2}).
In (\ref{A10}) $T_0(M_\nu,Z)$ is the analytic torsion norm
for the de Rham complex $\left(\DR(M_\nu,Z),d\right)$.
This complex consists of the pairs of smooth forms $(\oma,\omb)$
such that $\omega_k\in\DR\left(\Mm_k,Z_k\right)$ has the zero geometrical
restrictions to $Z_k:=Z\cap\df M_k$%
\footnote{$Z$ is the union of the components of $\df M$ where the Dirichlet
boundary conditions are given.}
and that the following transmission condition holds for the geometrical
restrictions $i^*_k\omega_k$ of $\omega_k$ to $N$
\begin{equation}
\alpha i^*_1\,\oma=\beta i^*_2\,\omb.
\label{A11}
\end{equation}
The analytic
torsion norm $T_0\left(M_{\nu},Z\right)$ is defined for an arbitrary
$\nu=(\alpha,\beta)\in\rbo$. There is a canonical identification
of $H^j\left(\DR\left(M_{\nu},Z\right)\right)$ with the space
of the corresponding harmonic forms
$\Ker\left(\den|DR^j\left(M_{\nu},Z\right)\right)$ (Lemma~\ref{L1}).
This identification (similarly to the case
of $\DR\left(M_{1,1},Z\right)$) is obtained by the homotopy operator,
which is defined using the Green function for the Laplacian
$\Delta_{\nu}$. (This Laplacian is an elliptic self-adjoint operator
by Theorem~\ref{TAT705}.) The boundary conditions for $\Delta_{\nu}$ on $N$
and $\df M$ are elliptic (and differential).
The Green function $G_\nu$ for $\Delta_{\nu}$ exists (and depends smoothly
on $\nu\ne(0,0)$) according to Theorem~\ref{TAT705} and
to Proposition~\ref{PA3002}.
This identification provides us with the natural norms on
$H^j\left(\DR\left(M_\nu,Z\right)\right)=: H^j\left(M_\nu,Z\right)$
and on $\det\Hh\left(M_\nu,Z\right)$. The scalar analytic torsion
$T\left(M_\nu,Z\right)$ is defined
by $\zeta_{\nu,j}(s):={\sum}'\lambda^{-s}_i$ for $\Ree\,s>(\dim M)/2$
(where the sum is over all the nonzero eigenvalues $\lambda_i$
of the Laplacian $\djn:=\den|DR^j\left(M_\nu,Z\right)$ with their
multiplicities). These functions $\znj$ can be continued
to meromorfic functions on the whole complex plane with simple poles
and regular at zero. (This statement is proved
in Theorem~\ref{TAT705} and in Proposition~\ref{PA3002} below.)

The {\em analytic torsion norm} on $\det \Hh\left(M_\nu,Z\right)$ is the norm
$$
T_0(M_\nu,Z)=\left\|\cdot\right\|^2_{\det\Hh(M_\nu,Z)}\exp\left(\sum (-1)^jj
{\df}_s\zeta_{\nu,j}(s)\big|_{s=0}\right).
$$

The identification $\phi_{\nu}^{an}$ in (\ref{A10} ) is defined
by the short exact sequence of the de Rham complexes (where
$Z_k:=Z\cap\df\Mm_k$):
\begin{equation}
0\to\DR\left(M_1,Z_1\cup N\right)\oplus\DR\left(M_2,Z_2\cup N
\right)\to\DR\left(M_{\alpha,\beta},\!Z\right) @>{r_{\alpha,\beta}}>>
\DR(N)\to 0.
\label{A12}
\end{equation}
The left arrow in (\ref{A12}) is the natural inclusion and the right
arrow $r_{\alpha,\beta}$ is
\begin{equation}
r_{\alpha,\beta}(\oma,\omb):=(\alpha^2+\beta^2)^{-1/2}
(\beta i^*_1\oma+\alpha i^*_2\omb).
\label{A13}
\end{equation}
For $(\alpha,\beta)=(1,1)$ we have $r_{1,1}=\sqrt 2\,i^*_k\omega_k$.
This corresponds to (\ref{A9}). Hence, $\phi_{an}$ is equal
to $\phi^{an}_{\alpha,\beta}$ for $(\alpha,\beta)=(1,1)$.

The complex $\DR\left(M_\nu,Z\right)$ for the values $(0,1)$ and $(1,0)$
of $\nu$ is the direct sum of the de Rham complexes
of all the smooth forms (with the zero geometrical restriction to $Z_k$)
on one of the manifolds $M_k$ and of all the smooth forms with the zero
geometrical restriction to $Z_j\cup N$ on another piece $M_j$
of the manifold $M$. Thus, the two pieces of $M$ are completely
disconnected with respect to $\DR\left(M_\nu,Z\right)$ for these special
values of~$\nu$.
The family of spectral problems on $\DR\left(M_\nu,Z\right)$
for $\nu\in\rbo$ provides us with a smooth deformation between
a spectral problem on $M$ (without any interior boundary conditions) and
the direct sum of spectral problems on $\left(M_1,Z_1\right)$ and
on $\left(M_2,Z_2\cup N\right)$. So this family of interior boundary
problems is (in a spectral sense) a kind of a smooth cutting of $M$
in two disconnected pieces.

Let $\left(M_1,N\right)$ be a compact smooth Riemannian manifold
$\left(M_1,g_{M_1}\right)$ with a smooth boundary  $\df M_1$
and let $N$ be a union
of some connected components of $\df M_1$. Let a metric $g_{M_1}$
be a direct product metric near the boundary. Then (as it follows
from the equality (\ref{A6})) the analytic torsion norm
$T_0\left(M_{1},N\right)$
on $\det\Hh\left(DR\left(M_1,N\right)\right)$ does~not depend on~$g_{M_1}$.
To prove this it is enough to take as $\left(M,g_M\right)$ a closed
manifold $M=M_1\cup_N M_1$ with a mirror symmetric
(with respect to $N$) Riemannian metric $g_M$ which coincides with $g_{M_1}$
on each piece $M_1$ of $M$ ($g_{M_1}$ is a direct product
metric near $N$ and so $g_M$ is smooth on $M$).
Since the torsions $T_0(M)$ and $T_0(N)$ are independent of $g_{M_1}$
and of $g_N=g_{M_1}|_TN$ we see that $T_0\left(M_1,N\right)$ does~not depend
on $g_{M_1}$.

It follows from the equality (\ref{A10}) with $c_0=1$ that
$T_0\left(M_\nu,Z\right)$ does~not depend on $g_M$.
Indeed, $T_0\left(M_j,Z_j\cup N\right)$ and $T_0(N)$ are independent
of $g_M$, and the identification $\fan$ is also independent of $g_M$.
(Here $M$ is a manifold with a smooth boundary $\df M$,
$N\cap\df M=\emptyset$, the Dirichlet boundary conditions are given
on a union $Z$ of some components of $\df M$, the Neumann boundary
conditions are given on $\df M\setminus Z$, and $g_M$ is a direct product
metric near $\df M$ and near $N$, $Z_k:=Z\cap\df\Mm_k$.)

Since $\DR\left(M_{0,1},Z\right)$ is the direct sum
$\DR\left(M_1,Z_1\right)\oplus\DR\left(M_2,Z_2\cup N\right)$
of the de Rham complexes ($Z_k:=Z\cap\df\Mm_k$), we see that
the analytic torsion norm $T_0\left(M_{0,1}\right)$ is canonically equal
to the tensor product of norms:
\begin{equation}
T_0\left(M_{0,1},Z\right)=T_0\left(M_1,Z_1\right)\otimes T_0\left(M_2,Z_2
\cup N\right).
\label{X1}
\end{equation}
The determinant line in (\ref{X1}) is the tensor product
\begin{equation*}
\det\Hh\left(M_{0,1},Z\right)=\det\Hh\left(M_1,Z_1\right)\otimes
\det\Hh\left(M_2,Z_2\cup N\right)
\end{equation*}
(where $\Hh\left(M_1,Z_1\right)$ and $\Hh\left(M_2,Z_2\cup N\right)$ are
the relative cohomology).

The formula (\ref{A6}) claims for $\nu=(0,1)$ that
\begin{equation}
\phi^{an}_{0,1}T_0\left(M_{0,1},Z\right)=T_0\left(M_1,Z_1\cup N\right)
\otimes T_0\left(M_2,Z_2\cup N\right)\otimes T_0(N).
\label{X2}
\end{equation}
It follows from the definition of the exact sequence (\ref{A12}) that
$\phi^{an}_{0,1}$ is the identity on the component
$\det\Hh\left(M_2,Z_2\cup N\right)$ of
$\det\Hh\left(M_{0,1},Z\right)$. The following theorem is an immediate
consequence of (\ref{X1}) and (\ref{X2}). Let $N$ be a union of some
connected components of $\df M_1$, let $M_1$ be a compact Riemannian
manifold with a smooth boundary $\df M_1$ and let $Z_1$ be a union
of some connected components  of $\df M_1$ not belonging to $N$.
Suppose that the metric $g_{M_1}$ is a direct product metric near $\df M_1$.

\begin{thm}[Gluing of boundary components]
The equality holds
\begin{equation}
\phi_{an}T_0\left(M_1,Z_1\right)=T_0\left(M_1,Z_1\cup N\right)\otimes T_0(N).
\label{X3}
\end{equation}
\label{T2}
\end{thm}

The identification of the determinant lines in (\ref{X3})
\begin{equation}
\phi_{an}\colon\det\Hh\left(M_1,Z_1\right)\rs\det\Hh\left(M_1,Z_1\cup N\right)
\otimes\det\Hh(N)
\label{X31}
\end{equation}
is defined by the short exact sequence of the de Rham complexes:
\begin{equation}
0\rightarrow\DR\left(M_{1},Z_1\cup N\right)\rightarrow\DR\left(M_{1},Z_1
\right)\rightarrow\DR(N)\rightarrow 0,
\label{B2}
\end{equation}
where the left arrow is the natural inclusion, and the right arrow is the
geometrical restriction.

\begin{exmp}
Formula (\ref{X3}) contains the Lerch formula (\cite{WW}, 13.21, 12.32)
for the derivative at zero of the zeta-function of Riemann
$\zeta(s)$ (defined for $\Ree\,s>1$ as $\sum_{n\ge 1}n^{-s}$):
\begin{equation*}
\df _s\zeta(s)|_{s=0}=-2^{-1}\log 2\pi.
\end{equation*}

Indeed, let $M$ be an interval $(0,b]\subset\wR$ with the Dirichlet
boundary conditions at $0$ and the Neumann conditions at $b$.
Set $N$ be a point $b$. Then the formula (\ref{X3}) claims in this case that
\begin{equation}
T_0\left((0,b]\right)=T_0\left((0,b)\right)\otimes T_0(b).
\label{X5}
\end{equation}
The cohomology $\Hh\left((0,b]\right)=\Hh\left([0,b],0\right)$ are trivial.
The scalar analytic torsion $T((0,b])$ is equal to
$\exp\left(-\df _s\zeta_1(s;M)|_{s=0}\right)$,
where $\zeta_1(s;M)$ is the zeta-function for the Laplacian on
$DR^1\left((0,b]\right)$. This zeta-function for $\Ree\,s>1/2$
is defined by the series
\begin{equation*}
\zeta_1(s;M)=\sum _{n\ge 0}\left(\left((\pi/{2b})(2n+1)\right)^2\right)
^{-s}.
\end{equation*}
So $\zeta_1(s;M)=(\pi/2b)^{-2s}\left(1-2^{-2s}\right)\zeta(2s)$
for $\Ree\,s>1/2$, where $\zeta(s)$ is the zeta-function of Riemann.
Hence, the latter equality between the analytic continuations
of $\zeta_1(s;M)$ and of $\zeta(2s)$ holds for all $s\in\wC$, and
$\df_s\zeta_1(s)|_{s=0}=2\zeta(0)\log 2$.

The determinant line $\det\Hh(M)$ on the left in (\ref{X5}) is
canonically isomorphic to $\wC$, and the $T_0(M)$-norm of the element
$1\in\wC$ is equal to
\begin{equation*}
\left\|1\right\|^2_{T_0(M)}=\exp\left(-\zeta'_1(0;M)\right)
=\exp\left(-2\,\zeta(0)\log 2\right)=2.
\end{equation*}
(Note, that the function $2\,\zeta(2s)$ is the zeta-function
for the Laplacian $\Delta=\left(-\df^2/\df x^2\right)$ on functions
on the circle of the length $2\pi$. As the circle is odd-dimensional,
then the value of $\,2\,\zeta(2s)$ at zero is equal to $-\dim\Ker\Delta=-1$.
Hence, $2\,\zeta(0)=-1$.)

The scalar analytic torsion $T\left((0,b)\right)$ is equal to
$\exp\!\left(\!-\df_s\zeta_1\!\left(s;\!M,\!N\right)\right)$, where
$\zeta_1\!(s;\!M,\!N)$ for $\Ree\,s>1/2$ is defined by the series
$$
\zeta_1(s;M,N)=\sum_{n\ge 1}\left(\left((\pi/b)n\right)^2\right)^{-s}=
(\pi/b)^{-2s}\zeta(2s).
$$
Hence, this equality holds for all $s\in\wC$, and the scalar analytic
torsion is equal to
\begin{equation*}
T\left((0,b)\right)=\exp\left(-2\,\zeta'(0)+2\,\zeta(0)\log(\pi/b)\right)=
\exp\left(-\log(\pi/b)-2\zeta'(0)\right).
\end{equation*}

The identification of the determinant lines on the right and on the left
in (\ref{X5}) is defined by the cohomology exact sequence
\begin{equation}
0\rightarrow H^0(b)\rightarrow H^1\left((0,b)\right)\rightarrow 0.
\label{X9}
\end{equation}
The element $1\in H^0(b)$ (of the norm $1$) is mapped by (\ref{X9})
to the element $(dx/b)$ of the norm $\|dx/b\|^2=b^{-1}$.
The element $h=1^{-1}\otimes (dx/b)$, corresponding to the element
$1\in{\wC}=\det\Hh\left((0,b]\right)$, has the norm $b^{-1}$.
So the equality (\ref{X5}) claims that
$$
\log 2=-\log b -\log(\pi/b)-2\,\zeta'(0).
$$
Thus the equality $\zeta'(0)=-2^{-1}\log(2\pi)$ is a particular case%
\footnote{In this paper the proofs of the equality (\ref{X3}), of Theorem
\ref{T1}, and of the equality (\ref{A10}) with $c_0=1$ do~not use the
Lerch formula. So we have obtained (by the way) a new proof of the
Lerch formula.}
of Theorem~\ref{T2}.
\end{exmp}

\medskip
The natural $L_2$-norm on $\oplus_j\DR\left(\Mm_j\right)$ is defined by
\begin{equation}
\left(v_1,v_1\right):=\int_M\left(v_1\wedge *\bar{v}_1\right),
\label{Y2}
\end{equation}
where $\left(v_1\wedge*\bar{v}_1\right)$ is a real density on $M$,
corresponding to $v_1\wedge *\bar{v}_1$.
\begin{lem}
The Green functions $G_\nu$ for the Laplacians $\don$ provide us
with the homotopy operator in the complex $\DR\left(M_\nu,Z\right)$
\begin{equation}
K_\nu:=\delta G_\nu
\label{Y0}
\end{equation}
between the orthogonal projection operator
$p_{\cH}\colon\DR\left(M_\nu,Z\right)\rightarrow\Ker\left(\don\right)$ and
the identity operator on $\DR\left(M_\nu,Z\right)$. The following equality
holds in $\DR\left(M_\nu,Z\right)$ :
\begin{equation*}
dK_\nu+K_\nu d=\id- p_{\cH}.
\end{equation*}
\label{L1}
\end{lem}

\noindent{\bf Proof}. The Green function for $\don$ maps
the $L_2$-completion $\dromb$ of $\DR(M)%
$\footnote{$\dromb$ coincides with the $L_2$-completion
of $\DR\left(M_\nu,Z\right)$ and with the $L_2$-completion
of $\oplus_j\DR\left(\Mm_j\right)$.}
into the $\Dom\left(\don\right)$ (Theorem~\ref{TAT705}).
The $\Dom\left(\don\right)$ is defined as the domain
of definition $D\left(\don\right)$ for $\don$
in $\DR\left(M_\nu,Z\right)$ completed with respect to the graph topology
norm $\left\|\omega\right\|^2_{graph}\!:=\left\|\omega\right\|^2_2+
\left\|\don\omega\right\|^2_2$ for $\omega\in D\left(\don\right)$
(where $\left\|\omega\right\|^2_2:=(\omega,\omega)$ is the $L_2$-norm
(\ref{Y2})). The Green function $G_\nu$ maps $\DR\left(M_\nu,Z\right)$
into $D\left(\don\right)$ (since, by Theorem~\ref{TAT705}, $\don$ is
a nonnegative elliptic differential operator with elliptic boundary
conditions). The definition of the Green function claims that
\begin{equation}
\don G_\nu=\id-p_{\cH},
\label{B3}
\end{equation}
on $\dromb$ (where $\don\omega$ for $\omega\in\Dom\left(\don\right)$
is defined as $\lim_i\don\omega_i$ for $\omega_i\in D\left(\don\right)$,
$\left\|\omega-\omega_i\right\|^2_{graph}\to 0$). In particular, this
equality holds on $\DR\left(M_\nu,Z\right)\subset\dromb$.

The $D\left(\don\right)\subset\dromb$ is defined as follows. The adjoint
to $d_\nu$ operator $\delta_\nu$ in $\oplus_j\DR\left(\Mm_j\right)$
is defined on elements $v_2=(\oma,\omb)$, where $\omega_k$ are smooth
differential forms on $\Mm_k$ and the linear functional
$$
l_{v_2}(v_1)=<dv_1,v_2>=\int_M(dv_1\wedge*\bar{v}_2)
$$
is continuous in $\DR\left(M_\nu,Z\right)$ with respect to the $L_2$-norm
(\ref{Y2}) of $v_1\in\DR\left(M_\nu,Z\right)$. For such an element
$v_2=(\oma,\omb)$ the form $*v_2=(*\oma,*\omb)$ has the zero geometrical
restriction to $\df M\setminus Z$, and the following transmission
condition has to hold on $N$ for $v_2$:
\begin{equation}
\beta\,i^*_{N,1}(*\oma)=\alpha\,i^*_{N,2}(*\omb),
\label{Y3}
\end{equation}
where $i_{N,k}\colon\ N\subset\df M_k\hookrightarrow M_k$. (These boundary
conditions for $v_2$ are consequences of Stokes' formula.)

The domain $D\left(\don\right)\subset\DR\left(M_\nu,Z\right)$
is defined as the set of $\omega\in\DR\left(M_\nu,Z\right)$ such that
\begin{equation}
\omega\in D(\delta_\nu),\qquad d\omega\in D(\delta_\nu),\qquad \delta
\omega\in\DR\left(M_\nu,Z\right).
\label{Y4}
\end{equation}

Note that $dG_\nu\omega=G_\nu d\omega$
for $\omega\in\DR\left(M_\nu,Z\right)$ (this equality follows
from Stokes' formula). Hence the identity (\ref{B3}) can be represented
on $\DR\left(M_\nu,Z\right)$ as
$$
Kd+dK=id-p_{\cH}.
$$
Thus the lemma is proved.\ \ \ $\Box$

\begin{cor}
The homotopy operator (\ref{Y0}) defines a canonical identification
between the cohomology $\Hh\left(DR\left(M_\nu,Z\right)\right)$ and
the space of harmonic forms $\Kerb\left(\den\right)$.
\label{C1}
\end{cor}

Let for simplicity $g_M$ be a direct product metric near
$N$. Let the Dirichlet boundary conditions be given on a union $Z$
of some connected components of $\df M$ and the Neumann conditions
be given on $\df M\setminus Z$. Then the following holds.
\begin{pro}
The eigenforms of $\Delta\left(M,Z;g_M\right)$ are the same
as the eigenforms of $\Delta_{1,1}$.
\label{P1}
\end{pro}

\noindent{\bf Proof.} Let $\nu$ be equal to $(1,1)$. The conditions (\ref{Y4})
for $\omega\!=\!(\oma,\omb)\!\in\!\DR\left(M_{1,1},\!Z\right)$ are equivalent
on $N$ to the following ones:
\begin{alignat}{2}
i^*_{N,1}\oma & =i^*_{N,2}\omb, & \qquad i^*_{N,1}(*\oma)
& =i^*_{N,2}(*\omb),
\label{Y5}\\
i^*_{N,1}(*d\oma) & =i^*_{N,2}
(*d\omb), & \qquad i^*_{N,1}(*d*\oma) & =i^*_{N,2}(*d*\omb),
\label{Y6}
\end{alignat}
where $*$ is the star-operator for the Riemannian metric $g_M$.

The equalities (\ref{Y5}) claim that the restrictions to $N$ of the forms
$\oma$ and $\omb$ are the same (i.e., they are the same smooth sections
of $\wedge^{\bullet}T^*M|_N$). The equalities (\ref{Y5}) and (\ref{Y6})
are equivalent to the assertion that the following pairs of forms have
the same restrictions to $N$ (as the smooth sections of
$\wedge^{\bullet}T^*M|_N$):
\begin{equation}
\left\{d\oma,d\omb\right\}, \qquad \left\{\delta\oma,\delta\omb\right\},
\qquad \left\{\oma,\omb\right\}.
\label{Y7}
\end{equation}
Any eigenform for $\doo\left(M,Z;g_M\right)$ belongs
to $D\left(\doo_{1,1}\right)$. So it is an eigenform
for $\doo_{1,1}$. Let $\omega=(\oma,\omb)\in D\left(\doo_{1,1}\right)$ be
an eigenform for $\Delta_{1,1}$:
\begin{equation}
\doo_{1,1}\omega=\left(\doo\oma,\doo\omb\right)=\lambda(\oma,\omb).
\label{Y8}
\end{equation}
Then%
\footnote{All the eigenforms of $\don$ (for $\nu\in\rbo$) are
$C^{\infty}$-smooth on $\Mm_k$, as it follows from Theorem~\ref{TAT705}.}
$\omega_k$ are $C^\infty$-forms on $\Mm_k$ and (as it follows
from (\ref{Y7}), (\ref{Y8})) the restrictions of the following
pairs of forms are the same as the sections of $\wedge^{\bullet}T^*M|_N$
(for $k=0,1,2\dots$):
\begin{equation}
\left\{\Delta^k\oma,\Delta^k\omb\right\},\qquad \left\{d\Delta^k\oma,d\Delta^k
\omb\right\},\qquad \left\{\delta\Delta^k\oma,\delta\Delta^k\omb\right\}.
\label{Y9}
\end{equation}
So $\omega=(\oma,\omb)$ is a $C^\infty$-form on $M=M_1\cup_N M_2$.
In fact, it follows from (\ref{Y5}) and from the identity
of the restrictions to $N$ of $\Delta\oma$ and $\Delta\omb$ that
$\left(\Delta_I\otimes\id\right)\omega_k$ have (for $k=1,2$) the same
restrictions from $M_k$ to $N$. (The Laplacian $\Delta$ is equal to
$\id_I\otimes\Delta_N+\Delta_I\otimes\id_N$ with respect to the direct
product structure $I\times N$ in the neighborhood of
$N=0\times N\hookrightarrow I\times N\hookrightarrow M$,
$0\in I\setminus\df I$.) Hence, according to (\ref{Y5}) and (\ref{Y6}),
the $2$-jets of $\oma$ and of $\omb$ are the same on $N$. The identity
between the $(2k+1)$-jets of $\omega_k$ on $N$ follows (by induction)
from (\ref{Y9}). Thus, $\omega$ is an eigenform for $\Delta_M$:
$\Delta_M\omega=\lambda\omega$. The proposition is proved.\ \ \ $\Box$

\subsection{Properties of analytic and combinatorial torsion norms}
One of the main properties of the analytic torsion norm is as follows.
Let $M$ be a manifold $M_1\times M_2$ with a direct product metric.
One of these Riemannian manifolds, for instance $M_1$,
can have a nonempty boundary $\df M_1$.
In this case let $g_{M_1}$ be a direct product metric near
$\df M_1$, and let the Dirichlet boundary conditions be given
on the components $Z=Z_1\times M_2$
of $\left(\df M_1\right)\times M_2=\df\left(M_1\times M_2\right)$.
Let the Neumann boundary conditions be given on
$\df\left(M_{1}\times M_{2}\right)\setminus\left(Z_{1}\times M_{2}\right)=
\left(\df M_1\setminus Z_1\right)\times M_2$.

The K\"unneth formula for the cohomology claims that
\begin{equation}
H^j(M,Z)=\oplus_{i+k=j}H^i\left(M_1,Z_1\right)\otimes H^k\left(M_2\right).
\label{B4}
\end{equation}

So the determinant of the cohomology of $\DR(M,Z)$ is the tensor product
\begin{equation}
\det\Hh(M,Z)=\left(\det\Hh\left(M_1,Z_1\right)\right)^{\chi(M_2)}\otimes
\det\Hh\left(M_2\right)^{\chi(M_1,Z_1)}.
\label{V4}
\end{equation}

\begin{pro}
The identification (\ref{V4}) induces the isomorphism of the analytic torsion
norm $T_0(M,Z)$ with the tensor product
\begin{equation}
T_0(M,Z)=T_0\left(M_1,Z_1\right)^{\otimes\chi(M_2)}\otimes T_0\left(M_2\right)
^{\otimes \chi(M_1,Z_1)},
\label{V1}
\end{equation}
where $\chi(M_1,Z_1)$, $\chi\left(M_2\right)$ are the Euler characteristics.
\label{P2}
\end{pro}

\begin{rem}
It is shown above that the analytic torsion norms $T_0(M,Z)$,
$T_0\left(M_1,Z_1\right)$, and $T_0\left(M_2\right)$ are independent
of Riemannian metrics $g_M$,\ $g_{M_k}$ which are supposed to be direct
product metrics near $\df M$, $\df M_1$. So, if the equality (\ref{V1})
holds for a direct product metric on $\left(M_1,\df M_1\right)\times M_2$,
(where $g_{M_1}$ is a direct product metric near $\df M_1$) then this
equality holds for any metric $g_M$ (which is supposed to be a direct
product metric near $\df M$).
\end{rem}

\noindent{\bf Proof.} The scalar analytic torsion $T(M)$ for a direct
product metric on $M=M_1\times M_2$ is equal to
\begin{equation}
T(M,Z)=T\left(M_1,Z_1\right)^{\chi(M_2)}T\left(M_2\right)^{\chi(M_1,Z_1)}.
\label{V2}
\end{equation}
This statement is proved in \cite{RS}, Theorem~2.5, in the case
of an acyclic local system over $M_1$. In the general case, (\ref{V2})
follows from
the proof of Theorem~2.5 in \cite{RS} and from the following equality
(where $\lambda\ne 0$, $m(i,\lambda,M_2)$ is the dimension of the
$\lambda$-eigenspace for $\Delta_{M_2,i}$,
$m\left(j,0,M_1\right):=\dim\Ker\Delta_{M_1,Z_1;j}$):
$$
\sum\left(-1\right)^{i+j}(i+j)m\left(i,\lambda,M_2\right)m\left(j,0,M_1\right)=
\left(\sum\left(-1\right)^iim\left(i,\lambda,M_2\right)\right)\chi\left(M_1,Z
_1\right),
$$
which holds, since the alternating sum over $i$ of $m(i,\lambda,M_2)$
is equal to zero (for any nonzero $\lambda$).

For such a metric on $M$ the following canonical identifications are
the isometries between the natural Hilbert structure on the space
of harmonic forms $\Ker\Delta_j(M,Z)$ and the tensor products (and
the direct sums) of the Hilbert structures on harmonic forms for
$\Delta_{\bullet}\left(M_1,Z_1\right)$ and $\Delta_{\bullet}\left(M_2\right)$:
\begin{equation}
\Ker\Delta_j(M,Z)=\oplus_{i+k=j}\Ker\Delta_i\left(M_1,Z_1\right)\otimes
\Ker\Delta_k\left(M_2\right).
\label{V3}
\end{equation}
These Hilbert structures induce the norms on
$$
\det\Hh(M,Z)=\det\Ker\Delta_{\bullet}\left(M,Z;g_M\right),
\ \det\Hh\left(M_1,Z_1\right)=\det\Ker\Delta_{\bullet}\left(M_1,Z_1;g_{M_1}
\right)
$$
and on $\det\Hh\left(M_2\right)=\det\Ker\Delta_{\bullet}\left(M_2,
g_{M_2}\right)$ such that the identification (\ref{V4}) of the determinant
lines is an isometry:
\begin{equation}
\left\|\cdot\right\|^2_{\det\Ker\Delta_M}=\left(\left\|\cdot\right\|^2
_{\det\Ker\Delta_{M_1,Z_1}}\right)^{\chi(M_2)}\left(\left\|\cdot\right\|^2
_{\det\Ker\Delta_{M_2}}\right)^{\chi(M_1,Z_1)}.
\label{V5}
\end{equation}
The equality (\ref{V1}) follows from (\ref{V2}),(\ref{V4}),
and (\ref{V5}).\ \ \ $\Box$

The following lemma makes it possible to use the variations on $\nu$
in the proof of Theorem~\ref{T1}. Let $\left(M,g_M\right)$ be a compact
Riemannian manifold with a smooth boundary $\df M$ and let
$N\hookrightarrow M\setminus\df M$ be a smooth closed codimension one
submanifold of $M$ with a trivial normal bundle $\left(TM|_N\right)/TN$
such that $M=M_1\cup_N M_2$ is obtained by gluing two its pieces
$M_1$ and $M_2$ along $N$.
Let $g_M$ be a direct product metric near $\df M$ and near $N$.
Let $Z$ be a union of some connected components of $\df M$
where the Dirichlet boundary conditions are given and let the Neumann
boundary conditions be given on $\df M\setminus N$.
\begin{lem}
Let us suppose that the norm $\fan T_0\left(M_{\nu},Z\right)$ is independent
of $\nu\in\rbo$ for any such $\left(M,g_M,N,Z\right)$%
\footnote{The equivalent formulation is as follows. Let $M$ be obtained
by gluing along $N$, i.e., $M=M_1\cup_N M_2$, and let it be equiped
with a Riemannian metric $g_M$, which is a direct product metric near
$\df M$ and near $N$. Then it is supposed that the norm
$\fan T_0\left(M_\nu,Z\right)$ is independent of $\nu\in\rbo$.}
(where
the identification $\fan$ is defined by the exact sequence (\ref{A12})
of the de Rham complexes and by Lemma~\ref{L1}). Then the factor $c_0$
in the gluing formula (\ref{A10}) for $\fan T_0\left(M_{\nu},Z\right)$
is equal to one.
\label{L2}
\end{lem}

\begin{rem}
Theorem~\ref{T1} is a direct consequence of Lemma~\ref{L2} and
of the assertion that $\fan T_0\left(M_{\nu},Z\right)$ is independent
of $\nu\in\rbo$. Indeed, $T_0\left(M_{1,1},Z\right)$ coincides with
$T_0(M,Z)$ (according to Proposition~\ref{P1}) and the identifications
$\fan$ and $\phi_{an}$ are the same. Hence the formula (\ref{A10}),
where $c_0$ is equal to one and $\nu=(1,1)$, is the gluing formula
of Theorem~\ref{T1}.
\end{rem}
\begin{rem}
The assertion that the norm $\fan T_0\left(M_{\nu},Z\right)$ does not
depend on $\nu$ is equivalent to the independence of $\nu$
of the factor $c_0$ in (\ref{A10}).
\end{rem}

\medskip
\noindent{\bf Proof.} The factor $c_0$ in (\ref{A10}) lies in $\wR_+$.
If $c_0$ is independent of $\nu$ for $\left(M,g_M,N,Z\right)$ then
\begin{equation}
\phi^{an}_{1,0}\,T_0\left(M_{1,0},Z\right)=\phi^{an}_{0,1}\,T_0\left(M_{0,1},
Z\right).
\label{Z0}
\end{equation}
It follows from (\ref{Z0}) and from (\ref{X1}), (\ref{X31}), and (\ref{A12})
that there are the equalities with the same positive constant $c_0$ as
in (\ref{A10}) for $\left(M,g_M,N,Z\right)$ (where $Z_k:=Z\cap\df M_k$):
\begin{align}
\phi_{an}T_0\left(M_1,Z_1\right) & =c_0 T_0\left(M_1,Z_1\cup N\right)\otimes T
_0(N), \label{Z1}\\
\phi_{an}T_0\left(M_2,Z_2\right) & =c_0 T_0\left(M_2,Z_2\cup N\right)\otimes T
_0(N).
\label{Z2}
\end{align}
We can conclude from (\ref{Z1}) and (\ref{Z2}) that the factor
$c_0=c_0(N,g_N)$ is defined by $N$, $g_N$ and that it {\em does not~depend}
on $M_1$, $M_2$, $M$, and $g_M$ (it is independent also of $\nu$).

Let $M_1$ in (\ref{Z1}) be a manifold $M_1=N\times I$ with a direct
product metric. Then $\df M_1=N\cup N$ and (\ref{Z1}) claims in this case
that
\begin{equation}
\phi_{an}T_0(N\times I)=c_0(N,g_N)^2\,T_0(N\times I,N\times\df I)
\otimes T_0(N)^{\otimes 2},
\label{Z3}
\end{equation}
where the identification $\phi_{an}$ is defined by the exact sequence
(\ref{B2}). It follows from (\ref{Z3}) and from the multiplicative
property (\ref{V1}) that
\begin{equation}
T_0(N)^{\chi(I)}\otimes T_0(I)^{\chi(N)}=c_0^2\,T_0(N)^
{\chi(I,\df I)}\otimes T_0(I,\df I)^{\chi(N)}\otimes T_0(N)^2,
\label{TT1}
\end{equation}
where $c_0:=c_0(N,g_N)$ depends on $N$ and on $g_N$ only. Then the following
equality is a consequence of (\ref{TT1}) and of the identification
(\ref{X31}) (defined by the exact sequence (\ref{B2})):
\begin{equation}
T_0(I)^{\chi(N)}=c_0(N,g_N)^2\,T_0(I,\df I)^{\chi(N)}\otimes
T_0(\df I)^{\chi(N)}.
\label{TT2}
\end{equation}

Note that $T_0(\df I)$ is the standart norm on $\det\Hh(\df I)$ which is
canonically identified with $\wC$ (up to a possible factor $(-1)$ in the
identification). Namely $\left\|1\right\|^2_{T_0}=1$ for
$1\in\wC$. An immediate consequence of the equality (\ref{Z2}) for $M_1=I$,
$N=\df I$ and of (\ref{TT2}) is the following:
\begin{equation}
c_0\left(N,g_N\right)^2=c_0(\df I)^{\chi(N)}.
\label{TT3}
\end{equation}
Hence, it is enough to prove that $c_0(\df I)=1$, and it will be done now.

Let $\,I\,$ be an interval $[0,a]$. The scalar analytic torsions for $I$
and for $(I,\df I)$ are equal: $T(I)=T(I,\df I)$, since
\begin{align}
\zeta_1(s;I)=\zeta_0(s;I,\df I),
\label{TT5}\\
\zeta_1(s;I,\df I)=\zeta_0(s;I,\df I),
\label{TT6}
\end{align}
(where $\zeta_j(s;M,Z)$ is the $\zeta$-function of the Laplacian on
$\left(DR^j(M,Z),g_M\right)$). The equality (\ref{TT6}) follows from
the identification of the eigenforms, defined by the exterior derivative
$d$, and the equality (\ref{TT5}) follows from the identification
of the eigenforms defined by the Riemannian $*$ on $I$.

The cohomology exact sequence for the pair $(I,\df I)$ is
\begin{equation}
0\rightarrow H^0(I)\rightarrow H^0(\df I)\rightarrow H^1(I,\df I)
\rightarrow 0.
\label{TT4}
\end{equation}

The complex (\ref{TT4}) is acyclic and so the determinant $D$
of its cohomology is canonically isomorphic to $\wC$.
The components of (\ref{TT4}) are equiped with the natural Hilbert
structures (because they are the spaces of harmonic forms on the interval
$I\subset\wR$ with the standart metric). Hence, there is the induced norm
$\left\|\cdot\right\|_{D}$ on $D=\wC$. We have to prove that
$\|1\|^2_D=1$
for $1\in\wC=D$. This equality is equavalent to the assertion that
$c_0(\df I)$ is equal to one.

The norm of the element $a^{-1/2}\cdot 1\in DR^0(I)$ is equal to $1$.
(It is a harmonic form and it represents an element from $H^0(I)$).
Its image in $H^0(\df I)$ is as follows:
$$
-a^{-1/2}\cdot[0]+a^{-1/2}\cdot[a]\in H^0(\df I)=H^0(0\cup a).
$$
The norm of the element
$a^{-1/2}\cdot dx\in DR^1(I,\df I)$ is equal to $1$ in $H^1(I,\df I)$ and
an element $-a^{1/2}\cdot[0]$ is mapped by the differential
of the exact sequence (\ref{TT4}) to the harmonic form
$a^{-1/2}dx\in H^1(I,\df I)$. (The arrows in (\ref{TT4})
are of the topological nature. So the latter statement is obtained using
\begin{equation*}
a^{1/2}=\int_{[0,a]}a^{-1/2}\ dx=\left(a^{-1/2}dx,(I,\df I)\right),
\end{equation*}
where $(I,\df I)$ is the fundamental class of $H_1(I,\df I)$.)

The corresponding volume element $\left(-a^{-1/2}[0]+a^{-1/2}[a]\right)
\wedge\left(-a^{1/2}[0]\right)=[0]\wedge[a]$ in  $\det H^0(\df I)$
is an element with the norm one. Hence, $c_0(\df I)=1$.
The equality $c_0\left(N,g_N\right)=1$ (for a union $N$ of some connected
components of $\df M_1$) follows from the equality $c_0(\df I)=1$ and from
(\ref{TT3}). The lemma is proved.\ \ \ $\Box$

Let $M=M_1\cup_N M_2$ be obtained by gluing $M_1$ and $M_2$ along $N$
(as in Theorem \ref{T1}), and let $X$  be a smooth triangulation
of $M$ such that $M_k$ and $N$ are invariant under $X$. Namely
$X=X_1\cup_W X_2$, where $X_k$ is a smooth triangulation of a manifold
$M_k$ with a smooth boundary $\df M_k=N\cup\left(\df M\cap \Mm_k\right)$.
(Here $N\subset{M}$ is a smooth closed manifold of codimension one
in $M$ such that $N$ divides $M$ in two pieces $M_1$ and $M_2$
as in Theorem~\ref{T1}, $N\cap\df M=\emptyset$, and $W:=X\cap N=X_k\cap N$.)

Let $Z$ be a union of the connected components of $\df M$, where
the Dirichlet boundary conditions are given. Set $V:=X\cap Z$,
$Z_k:=\df M_k\cap Z$, $V_k:=X_k\cap Z_k$. The exact sequence of cochain
complexes
\begin{equation}
0\rightarrow\oplus_{k=1,2}\Ce(X_k,W\cup V_k)\rightarrow\Ce(X,V)
\stackrel{i^*_N}{\longrightarrow}\Ce(W)\rightarrow 0
\label{K1}
\end{equation}
(where the left arrow is the natural inclusion and the right arrow is the
geometrical restriction of cochains) provides us with the identification
\begin{equation*}
\phi_c\colon \det \Hh(X,V)\rs\det \Hh
(X_1,W\cup V_1)\otimes\det \Hh(X_2,W\cup V_2)\otimes\det
\Hh(W).
\end{equation*}

By the definition of the combinatorial torsion norm on the determinant line
(determined by the prefered basises of the basic cochains)
the following statement holds.
\begin{pro}
Under the conditions above, the combinatorial torsion norms are equal:
\begin{equation}
\phi_c\tau_0(X,V)=\tau_0(X_1,W\cup V_1)\otimes\tau_0(X_2,
W\cup V_2)\otimes\tau_0(W).
\label{K3}
\end{equation}
\label{P3}
\end{pro}
\noindent This combinatorial equality is analogous to the gluing formula
of Theorem \ref{T1}. But it is necessary to note as follows.

\begin{rem}
\label{Rem1}
The formulas (\ref{K3}) and (\ref{A6}) correspond to the
{\em different identifications} $\phi_c$ and $\phi_{an}=\phi^{an}_{1,1}$
between one pair of the canonically identified%
\footnote{The cohomology are identified according to the de Rham theorem
by the integration over the simplexes of $X$ of the corresponding
differential forms. The spaces of harmonic forms
$\Ker\Delta_{\bullet}(M,Z)$ and $\Ker\Delta_{\bullet}\left(M_{1,1},Z\right)$
are canonically identified by Proposition~\ref{P1}.}
one-dimensional spaces
$$
\det \Hh(X,V)=\det \Hh(M,Z),
$$
and the triple tensor products of three other pairs of the canonically
identified spaces
$$
\det\Hh(X_k,W\cup V_k)=\det\Hh(M_k,N\cup Z_k),\quad
\det\Hh(W)=\det\Hh(N).
$$
(Note that $\phi_c$ is defined by the exact sequence (\ref{K1}), where
the right arrow $i^*_N$ is the restriction of the cochains.
However, in the exact sequence (\ref{A7}), which defines $\phi_{an}$,
the right arrow is equal
to $\sqrt 2\,i^*_N$ for the common geometrical restriction $i^*_N$ to $N$
of pairs $\omega=(\oma,\omb)$ of smooth differential forms $\omega_k$ on
$\Mm_k$ such that $i^*_{N,1}\oma=i^*_{N,2}\omb$.)
\end{rem}

Let $X$ be a smooth triangulation of a compact manifold with boundary
$(M,\df M)$. Let $Z$ and $Y$ be disjoint unions of some connected
components of $\df M$ such that $Z\cap Y=\emptyset$. Let $V=X\cap Z$,
$F=X\cap Y$. Then the exact sequence
\begin{equation*}
0\rightarrow \Ce(X,V\cup F)\rightarrow \Ce(X,V)\rightarrow
\Ce(F)\rightarrow 0
\end{equation*}
(where the left arrow is the natural inclusion of cochains and the right
arrow is the restriction of cochains) defines the identification
\begin{equation*}
\phi^F_c\colon \det \Hh(X,V)\rs\det\Hh(X,V\cup F)\otimes\det\Hh(F).
\end{equation*}

The following assertion is an immediate consequence
of the definition of the combinatorial torsion norm.
\begin{pro}
The combinatorial torsion norm of $(X,V)$ is equal to the tensor product
of the following combinatorial torsion norms:
\begin{equation*}
\phi_c\tau_0(X,V)=\tau_0(X,V\cup F)\otimes \tau_0(F).
\end{equation*}
\label{P4}
\end{pro}

\noindent This combinatorial equality is similar to the gluing formula
of Theorem~\ref{T2}.

\vspace{3mm}
Let $e(M,Z)$ be the logarithm of the ratio between the analytic and
the combinatorial torsion norms:
\begin{equation*}
e(M,Z):=\log_2\left(T_0(M,Z)/\tau_0(X,V)\right)
\end{equation*}
(where $T_0(M,Z)/\tau_0(X,V):=\left\|l\right\|^2_{T_0(M,Z)}/\|l\|^2_{\tau_0
(X,V)}$ for an arbitrary nonzero element $l$ of the determinant line
$\det\Hh(M,Z)=\det\Hh(X,V)$).

\begin{rem}
\label{Rem2}
It is proved above that $e(M,\df M)$ {\em does~not depend}
on a metric $g_M$, if $g_M$ is a direct product metric near $\df M$.
\end{rem}

\begin{lem}
1. Let $\left(S,g_S\right)$ be a closed Riemannian manifold.
Then the following identity holds, if $g_{M\times S}$ is a direct
product metric near $\df(M\times S)=\df M\times S$:
\begin{equation}
e(M\times S,Z\times S)=\chi(M,Z)e(S)+e(M,Z)\chi(S)
\label{K5.1}
\end{equation}
($\chi(M,Z)$ is the relative Euler characteristic of $M$ modulo
$Z\subset \df M$).

2. Let $Y$ be a union of some connected components of $\,\df M\setminus Z$.
Then
\begin{equation}
e(M,Z)=e(M,Y\cup Z)+e(Y).
\label{K5.2}
\end{equation}
\label{L3}
\end{lem}

\noindent{\bf Proof.} The equality (\ref{K5.2}) follows from Theorem
\ref{T2} and from Proposition \ref{P4}. (In this case, $\phi_c=\phi_{an}$.)
The equality (\ref{K5.1}) follows from Proposition~\ref{P2} and
from the multiplicative property of the combinatorial torsion norms.
Namely let $K$ be a smooth
triangulation of $S$ and let $V=X\cap Z$. Then the identification
of the determinants of the cohomology defined by (\ref{B4}) and (\ref{V4})
is an isometry of the combinatorial torsion norms:
$$
\tau_0(X\times K,V\times K)=\tau_0(X,V)^{\chi(K)}\otimes \tau_0(K)^{\chi(X,V)}.
$$
The same identification of the cohomology is the isometry (\ref{V1})
of the analytic torsion norms, if the metric $g_{M\times S}$ is a direct
product metric near $\df(M\times S)$. Hence, the identity (\ref{K5.1})
holds for such metrics $g_{M\times S}$.\ \ \ $\Box$

\begin{rem}
It follows from (\ref{K5.2}) and from Remark~\ref{Rem2} that $e(M,Z)$
does~not depend on $g_M$ for any union $Z$ of the connected components
of $\df M$ (in particular for $Z=\emptyset$).
\end{rem}

\subsection{Generalized Ray-Singer conjecture}
\subsubsection{Properties of the ratio of the analytic and the combinatorial
torsion norms}
Lemma~\ref{L2} claims that Theorem \ref{T1} follows from (\ref{A10})
{\em with $c_0$ independent of $\nu$}. So it is enough to prove that
the norm $\fan T_0\left(M_\nu,Z\right)$ is independent of $\nu\in\rbo$
(under the same conditions on $M$, $g_M$, $N$, and $Z$ as in (\ref{A10})
and in Lemma~\ref{L2}). The latter assertion is proved
in Section~\ref{Sec2}.  In the remaining part of Section~\ref{Sec1}
we prove a generalization of the Ray-Singer conjecture for manifolds
with boundary (and with the transmission condition (\ref{A11})
on the interior  boundary) using the gluing formula of Theorem~\ref{T1}.
This formula has the following consequence.

Let $M=M_1\cup_N M_2$ be obtained by gluing $M_1$ and $M_2$ along $N$.
\begin{lem}
Under the conditions of Lemma~\ref{L2}, on $\left(g_M,N,Z\right)$
the following holds:
\begin{equation*}
e(M,Z)=e(M_1,Z_1\cup N)+e(M_2,Z_2\cup N)+e(N)-\chi(N).
\end{equation*}
\label{L4}
\end{lem}

\noindent{\bf Proof.} This identity is an immediate consequence
of Theorem~\ref{T1} and of the following commutative diagram:
\begin{gather}
\begin{array}{ccccc}
\det\Hh(M,Z)  & & @>{\phi_{an}=\phi^{an}_{1,1}}>\llrs> & & {\Det}(M,Z,N)\\
\Vert \sst{R} & & & & \Vert\big\downarrow\sst{R} \\
\det\Hh(X,V) & @>{\phi_c}>\lrs> & {\Det}(X,V,W) & @>{A^H_W}>\lrs>
& {\Det}(X,V,W)\\
\ \ \ \ \big\uparrow\,\wr\,\sst{d_c} & & \ \ \ \big\uparrow\,\wr\,\sst{d_c} &
& \ \ \ \big\uparrow\,\wr\,
\sst{d_c} \\
\det\Ce(X,V) & @>{(\ref{K1})}>\lrs> & {\Det}\Ce(X,V,W) & @>{A_W}>\lrs>
& {\Det}\Ce(X,V,W)
\end{array}
\label{K7}
\end{gather}

Here
\begin{gather}
\begin{split}
{\Det}(M,Z,N) & :=\left(\otimes\kab\det \Hh\left(M_k,N\cup Z_k\right)\right)
\otimes\det\Hh(N),\\
{\Det}(X,V,W) & :=\left(\otimes\kab\det\Hh\left(X_k,W\cup V_k\right)\right)
\otimes\det\Hh(W),\\
{\Det}\Ce(X,V,W) & :=\left(\otimes\kab\det\Ce\left(X_k,W\cup V_k\right)\right)
\otimes\det\Ce(W),
\end{split}
\label{K71}
\end{gather}
$$
A_W:=\id_X\oplus\sqrt 2\id_W\in\Aut\left(\oplus\kab
\Ce\left(X_k,W\cup V_k\right)\right)\oplus\Aut\Ce(W),
$$
$A^H_W$ is the induced by $A_W$ operator on the determinant of the
cohomology, \\
$R$ is the identification induced by the integration of differential
forms over the simplexes of $X$ (by the de Rham theorem),\\
$\phi_c$ and $\phi_{an}$ are the identifications induced by (\ref{K1})
and by (\ref{A7}) in a view of Proposition~\ref{P1}.

The commutativity of (\ref{K7}) follows from the commutativity of the
diagram \\
\vspace{3mm}
$$
\begin{array}{ccccccc}
0\rightarrow\oplus\kab\! & \!\DR\left(M_k,N\cup Z_k\right)\! &
\!\rightarrow\! & \!\DR\left(M_{1,1},Z\right)\! & @>{\sqrt 2\,i^*_N}>> &
\!\DR(N)\! &\!\rightarrow 0 \\
 & \big\downarrow\sst{R} & & \big\downarrow\sst{R} & &
\big\downarrow\sst{R} & \\
0\rightarrow\oplus\kab\! & \!\Ce\left(X_k,W\cup V_k\right)\! &
\!\rightarrow\! & \!\Ce(X,V)\! &
\!@>{i^*_N}>> \Ce(W) @>{\sqrt 2\id}>>\! & \!\Ce(W)\! & \rightarrow 0
\end{array}
$$

\medskip
The induced action of $\sqrt 2\,\id$ on $l_W\in\det\Ce(W)$ is
$l_W\rightarrow 2^{-\chi(W)/2}l_W$ (where $\chi(W)=\chi(N)$ is
the Euler characteristic). So the induced action of $A_W$ and of
$A^H_W$ on $l\in{\Det}(M,Z,N)=\left(\otimes\kab\det\Ce\left(X_k,W
\cup V_k\right)\right)\otimes\det\Ce(W)$ is
\begin{equation}
l\rightarrow 2^{-\chi(N)/2}l.
\label{K9}
\end{equation}
(The identification of the determinant lines is defined by $R$ and
by $d_c$ in the right column of (\ref{K7}).)

For an arbitrary nonzero $m\in\Det(M,Z,N)$ the following equality
is deduced from (\ref{K9}) and from the commutativity of (\ref{K7}):
\begin{equation}
\left(\phi_{an}T_0(M,Z)\right)(m)=2^{-\chi(N)}\left(\phi_c T_0(M,Z)
\right)(m).
\label{K10}
\end{equation}
Theorem~\ref{T1} and Proposition~\ref{P3} claim that
\begin{gather}
\begin{split}
\phi_{an}T_0(M,Z) & =T_0\left(M_1,N\cup Z_1\right)\otimes T_0\left(M_2,N
\cup Z_2\right)\otimes T_0(N),\\
\phi_c\tau_0(X,V) & =\tau_0\left(X_1,W\cup V_1\right)\otimes\tau_0
\left(X_2,W\cup V_2\right)\otimes\tau_0(W).
\end{split}
\label{K11}
\end{gather}

The isometries (\ref{K10}) and (\ref{K11}) involve the equality
\begin{multline*}
e(M,Z)=\log_2\left({T_0(M,Z)/\tau_0(X,V)}\right)=\\
=-\chi(N)+\left(\sum\kab\log_2\left(T_0\left(M_k,N\cup Z_k\right)/\tau_0
\left(X_k,W\cup V_k\right)\right)\right)+\log_2\left(T_0(N)/\tau_0(W)\right).
\end{multline*}
Thus the lemma is proved.\ \ \ $\Box$

\medskip
Let $\nu=(\alpha,\beta)\in\rbo$ and let $\left(\Ce\left(X_\nu,V\right),d_c
\right)$ be the complex of pairs of cochains
$\left(c_1,c_2\right)$, $c_k\in\Ce\left(X_k,V_k\right)$,
with the $\nu$-transmission boundary condition (similar to (\ref{A11}))
on $W\subset\df X_k$ between their geometrical restrictions
\begin{equation}
\alpha i^*_{W,1}c_{1}=\beta i^*_{W,2}c_{2}.
\label{K12}
\end{equation}

The integration over the simplexes provides us with a quasi-isomorphism
of the complexes:
$$
\Rn :\left(\DR\left(M_\nu,Z\right),d\right)\rightarrow \left(\Ce\left(X_\nu,V
\right),d_c\right)
$$
(i.e., $\Rn$ induces an isomorphism between the corresponding cohomology).

The morphism of complexes $r_{\nu,c}\colon\left(\Ce\left(X_\nu,V\right),d_c
\right)\to\Ce\left(W,d_c\right)$ is defined by analogy with the definition
of $r_{\nu}$. Its value on each element
$\left(c_{1},c_{2}\right)\in\Ce\left(X_\nu,V\right)$ is
\begin{equation*}
r_{\nu,c}\left(c_{1},c_{2}\right)=(\alpha^2 +\beta^2)^{-1/2}\left(\beta i^*
_{W,1}c_{1}+\alpha i^*_{W,2}c_2\right).
\end{equation*}

The vertical arrows in the following diagram of complexes are
quasi-isomorphisms%
\footnote{$R_\nu$ is a quasi-isomorphism according
to Proposition~\ref{PA87}.}
\begin{gather}
\begin{array}{ccccccc}
\ & & & & & & \\
0\rightarrow\oplus\kab & \!\DR\left(M_k,N\cup Z_k\right)\! & \rightarrow &\!\DR
\left(M_\nu,Z\right)\!
& \stackrel{r_\nu}{\longrightarrow} & \!\DR(N)\! & \rightarrow 0  \\
 & \big\downarrow\sst{R} & & \big\downarrow\sst{R_\nu} & & \big\downarrow
\sst{R} & \\
0\rightarrow\oplus\kab & \Ce\left(X_k,W\cup V_k\right) & \stackrel{j}
{\rightarrow} &
\Ce\left(X_\nu,V\right) & \stackrel{r_{\nu,c}}{\longrightarrow} & \Ce(W) &
\rightarrow 0\\
\ & & & & & &
\end{array}
\label{LL1}
\end{gather}
This diagram is commutative.
The left horisontal arrows in it are the natural inclusions.
Let $\phi^c_\nu$ be the identification
\begin{equation}
\phi^c_\nu\colon \,\det\Hh\left(C(X_\nu,V)\right)\stackrel{R_\nu}{=}
\det\Hh\left(M_\nu,Z\right)\to\Det(M,Z,N),
\label{LL2}
\end{equation}
defined by the bottom row of this diagram.

\begin{rem}
The equality
\begin{equation}
\phi^c_\nu=\phi^{an}_\nu
\label{LL3}
\end{equation}
follows from the commutativity of (\ref{LL1}). But $\phi^c_{1,1}\ne\phi_c$
(in contrast with the identity $\phi^{an}_{1,1}=\phi_{an}$).
According to (\ref{K10}) it holds that
\begin{equation*}
\phi^c_{1,1}=\phi_{an}=2^{-\chi(N)}\phi_c.
\end{equation*}
\label{R4}
\end{rem}

The space $C^j(X_\nu,V)$ is a subspace
of $\oplus\kab C^j\left(X_k,V_k\right)$. The Hilbert structure
on $C^j\left(X_k,V_k\right)$ is defined by the orthonormal basises
of cochains $\{\delta_e\}$ (parametrized by $j$-dimensional simplexes
$e$ of $X_k\backslash V_k$). So the Hilbert structures
on $\Ce\left(X_\nu,V\right)$ and on $\det\Ce\left(X_\nu,V\right)$
are defined. The {\em scalar combinatorial torsion} is defined as in
(\ref{A2}):
\begin{equation*}
T\left(\Ce\left(X_\nu,V\right),d_c\right):=\exp\left(\sum\left(-1\right)^jj
\df_s\zeta^c_{j,\nu}(s)\big|_{s=0}\right),
\end{equation*}
where
$\zeta^c_{j,\nu}(s):=\Tr'\left(\left(\Delta^c_{j,\nu}\right)^{-s}\right)$
is the
sum ${\sum}'\lambda^{-s}$ over all the nonzero eigenvalues $\lambda$
of the finite-dimensional operator
$\Delta^c_{j,\nu}\!=\!\left(d^*_cd_c+d_cd^*_c|C^j\left(X_\nu,V\right)\right)$
(with their multiplicities), $d^*_c$ is adjoint to $d_c$
in $C^j\left(X_\nu,V\right)$ with respect to the Hilbert structure
in $\Ce\left(X_\nu,V\right)$.

{\em The combinatorial torsion} is the following norm
on $\det\Hh\left(C\left(X_\nu,V\right)\right)$%
\footnote{It is isomorphic to $\det\Hh\left(M_\nu,Z\right)$ under
the quasi-isomorphism $\Rn$.}:
\begin{equation}
\tau_{0}\left(X_\nu,Z\right):=\left\|\cdot\right\|^2_{\det\Hh\left(C\left(
X_\nu,V\right)\right)}\cdot T\left(\Ce\left(X_\nu,V\right),d_c\right),
\label{LL6}
\end{equation}
where the norm on $\det\Hh\left(C\left(X_\nu,V\right)\right)$ is defined
by the Hilbert structures on the subquotions
$H^j\left(C\left(X_\nu,V\right)\right)$ of the Hilbert spaces
$C^j\left(X_\nu,V\right)$.

\begin{rem}
For each $\nu=(\alpha,\beta)\in\rbo$ the combinatorial torsion
$\tau_{0}\left(X_\nu,V\right)$ is an invariant of the combinatorial
structure defined by a smooth triangulation of the triplet $[(M,\df M);Z;N]$,
where $M$ is a manifold with a smooth boundary $\df M$,\ $Z$ is a union
of some connected components of $\df M$, and $N$ is a smooth codimension
one closed submanifold of~$M$ with a trivial normal bundle
$\left(TM|_{N}\right)/TN$.
\end{rem}

\begin{pro}
The combinatorial torsion norm $\tau_{0}\left(X_\nu,Z\right)$ is isometric
under the identification (\ref{LL2}) to the tensor product
of the combinatorial torsion norms:
\begin{equation*}
\phi^c_\nu\tau_{0}\left(X_\nu,V\right)=\tau_{0}\left(X_{1},W\cup V_{1}\right)
\otimes\tau_{0}\left(X_{2},W\cup V_{2}\right)\otimes\tau_{0}(W).
\end{equation*}
\label{P5}
\end{pro}

\noindent{\bf Proof.} Under the identification (\ref{LL2}),
the Hilbert space $\Ce\left(X_\nu,V\right)$ is isometric
to the tensor product of the Hilbert spaces
on $\det\Ce\left(X_k,W\cup V_k\right)$ (for $k=1,2$) and on $\det\Ce(W)$.
(The Hilbert structures on on $\Ce\left(X_\nu,V\right)$,
$\Ce\left(X_k,W\cup V_k\right)$, and on $\Ce(W)$ are defined above.)
Indeed, let $\rho_{\nu,c}\colon C^j(W)\rightarrow C^j\left(X_\nu,V\right)$
be linear maps defined for $w\in C^j(W)$ by
\begin{equation}
\rho_{\nu,c}(w)=\left(\alpha^2+\beta^2\right)^{-1/2}\left(\beta w,
\alpha w\right)\in C^j\left(X_\nu,V\right)
\label{LL8}
\end{equation}
Then $r_{\nu,c}\rho_{\nu,c}=\id$ on $\Ce(W)$,\ $\rho_{\nu,c}$
is an isometry between $C^j(W)$ and ${\Image}\,\rho_{\nu,c}$,
and $\Image\,\rho_{\nu,c}$ is the orthogonal complement
in $\Ce\left(X_\nu,V\right)$ to the image of the natural inclusion
$j\colon\oplus\kab\Ce\left(X_k,W\cup V_k\right)\hookrightarrow\Ce\left(X_\nu,
V\right)$ (where $j$ is an isometry onto ${\Image}\,j$).
So the identification $\phi^c_\nu$ is the isometry of the combinatorial
torsion norms.\ \ \ $\Box$

The number $e\left(M_\nu,Z\right)\in\wR$ is defined as the logarithm
of the ratio between the analytic and the combinatorial torsion norms:
\begin{equation*}
e\left(M_\nu,Z\right):=\log_{2}\left(T_{0}\left(M_\nu,Z\right)/\tau_{0}
\left(X_\nu,V\right)\right).
\end{equation*}

\begin{cor}
Under the conditions of Lemma~\ref{L2}, the equality holds:
\begin{equation}
e\left(M_\nu,Z\right)=e\left(M_{1},Z_{1}\cup N\right)+e\left(M_{2},Z_{2}
\cup N\right)+e(N),
\label{LL10}
\end{equation}
where $Z$ is a union of some connected components of $\,\df M$ and
$Z_k=Z\cap\df M_k$.
\label{C2}
\end{cor}

\begin{cor}
$e\left(M_\nu,Z\right)$ is independent of $\nu\in\rbo$.
\label{C3}
\end{cor}

\begin{cor}
For an arbitrary $\nu\in\rbo$ the equality holds:
\begin{equation}
e\left(M_\nu,Z\right)-e(M,Z)=\chi(N).
\label{LL11}
\end{equation}
\label{C4}
\end{cor}
This equality follows from Lemma~\ref{L4} and from (\ref{LL10}).

\begin{rem}
Even for $\nu=(1,1)$ the number $e\left(M_\nu,Z\right)$ differs
from $e(M,Z)$ in the case $\chi(N)\ne 0$.
\label{R6}
\end{rem}

\subsubsection{Ratio of the analytic torsion norm and the combinatorial
torsion norm for spheres and disks. Spherical Morse surgeries}

The values of $e(M)$ and $e(M,\df M)$, where $M$ is a sphere $S^n$ or
a disk $D^n$ (with a direct product metric near $\df D^n=S^{n-1}$)
are deduced now from Lemma~\ref{L4}.

\begin{lem}
1. For all the spheres, $e\left(S^n\right)$ is zero.\\
2. For even-dimensional disks, $e\left(D^{2n}\right)$ and
$e\left(D^{2n},\df D^{2n}\right)$ are
zero.\\
3. For all odd-dimensional disks, $e\left(D^{2n+1}\right)$ and
$e\left(D^{2n+1},\df D^{2n+1}\right)$ are equal to one.
\label{L5}
\end{lem}

\noindent{\bf Proof.} A closed interval $D^1$ is obtained by gluing
two intervals $D^1=D^1\cup_{pt}D^1$ in their common boundary point.
Lemma~\ref{L4} claims in this case that
\begin{equation}
e\left(D^1\right)=2e\left(D^1,pt\right)+e(pt)-\chi(pt).
\label{M1}
\end{equation}
Since $e(pt)=0$, we see that (\ref{K5.2}) involves the equalities
\begin{equation}
e\left(D^1\right)=e\left(D^1,pt\right)=e\left(D^1,\df D^1\right).
\label{M1.1}
\end{equation}
Hence (\ref{M1}) involves $e\left(D^1\right)=\chi(pt)=1$.

A circle $S^1$ is obtained by gluing two intervals, namely
$S^1=D^1\cup_{\df D^1}D^1$. So, according to Lemma~\ref{L4} and
to (\ref{M1.1}), we have
\begin{equation}
e\left(S^1\right)=2e\left(D^1\right)+e\left(\df D^1\right)-\chi\left(\df D^1
\right)=0.
\label{M3}
\end{equation}

Suppose (by the induction hypothesis) that $e(S^m)=0$ for $m\le n-1$.
The sphere $S^n$ (for $n\ge 2$) is the union
$\left(D^{n-1}\times S^1\right)\bigcup_{S^{n-2}\times S^1}
\left(D^2\times S^{n-2}\right)=S^n$
Indeed, $S^n=\left\{(x_{1},\ldots,x_{n+1})\in{\wR}^{n+1}\colon\sum x^2_j=1
\right\}$, the disk $D^2$ in $D^2\times S^{n-2}$ in the decomposition
above corresponds
to $\left\{(x_{1},x_{2})\colon x^2_{1}+x^2_{2}\le\epsilon\right\}$ and
$S^{n-2}=\left\{(x_j)\in S^n,\ x_{1}=x_{2}=0\right\}$.
Lemma~\ref{L4} claims in this case that
(since $\chi\left(S^{n-2}\times S^1\right)=0$)
\begin{equation*}
e\left(S^n\right)=e\left(D^{n-1}\times S^1,S^{n-2}\times S^1\right)+e\left(D
^2\times S^{n-2},S^1\times S^{n-2}\right)+e\left(S^{n-2}\times S^1\right).
\end{equation*}

The equalities below are deduced from the induction hypothesis, from
Lemma~\ref{L3} ((\ref{K5.2}), (\ref{K5.1})), and from (\ref{M3}):
\begin{gather}
\begin{split}
e\left(D^{n-1}\times S^1,S^{n-2}\times S^1\right) & =e\left(D^{n-1}\times S^1
\right)-e\left(S^{n-2}\times S^1\right), \\
e\left(D^2\times S^{n-2},S^1\times S^{n-2}\right) & =e\left(D^2\times S^{n-2}
\right)-e\left(S^1\times S^{n-2}\right),\\
e\left(S^1\times S^{n-2}\right)=0, &\qquad e\left(D^{n-1}\times S^1\right)=0,\\
e\left(D^2\times S^{n-2}\right) & =\chi\left(S^{n-2}\right)e(D^2).
\end{split}
\label{M6}
\end{gather}

Hence the combinatorial torsion norm is equal to the analytic torsion norm
for all odd-dimensional spheres $S^{2m+1}$:
\begin{equation}
e\left(S^{2m+1}\right)=0, \qquad T_{0}\left(S^{2m+1}\right)=\tau_{0}\left(S
^{2m+1}\right).
\label{M7}
\end{equation}

It follows from Lemma~\ref{L3} and from (\ref{M3}) that
$e\left(D^2\right)=e\left(D^2,\df D^2\right)$. It is deduced from
Lemma~\ref{L4} and from (\ref{M3}) that
$e\left(S^2\right)=2e\left(D^2\right)$. According to (\ref{M6}) the equality
$e\left(S^{2m}\right)=0$
for all even-dimensional spheres is a consequence of the equality
$e\left(S^2\right)=0$.

Let $\left(M,g_{M}\right)$ be any closed Riemannian manifold
of even dimension $2n$. Then the scalar analytic torsion
$T\left(M,g_{M}\right)$ is equal to $1$.
(This equality was proved in \cite{RS}, Theorem~2.1, with the help
of the equality
\begin{equation*}
\sum(-1)^jj\,m(\lambda,j)=0,
\end{equation*}
where $\lambda$ is an arbitrary nonzero eigenvalue of $\Delta_j$
on $DR^j(M)$ and $m(\lambda,j)$ is its multiplicity. The latter assertion
follows from the symmetry relation $m(\lambda,j)=m(\lambda,2n-j)$,
which is obtained applying the operator $*$ for a Riemannian
metric $g_M$ to the $\lambda$-eigenforms for $\Delta_j$.)
So (in particular) the torsion norm $T_{0}\left(S^2\right)$ is equal to
$\left\|\cdot\right\|^2_{\det\Hh\left(S^2\right)}$, where the norm
on $\Hh\left(S^2\right)$ is the norm defined by $g_M$ on the harmonic
forms $\Ker\doo$.
(The unduced norm $\left\|\cdot\right\|^2_{\det\Hh\left(S^2\right)}$
does~not depend on the metric $g_{S^2}$, as it follows from the invariance
of $T_0\left(M,g_M\right)$ with respect to $g_M$, proved above.)

Let $v$ be a volume of $S^2$ relative to a Riemannian metric $g_{S^2}$.
Then the element $h\in\det\Hh\left(S^2\right)$ defined below
is of the norm $1$:
\begin{equation*}
h=\left(v^{-1/2}\cdot 1_{S^2}\right)^{-1}\otimes\left(v^{-1/2}(*1_{S^2})
\right)^{-1},\qquad \left\|h\right\|^2_{\det\Hh\left(S^2\right)}=1
\end{equation*}
(here $1_{S^2}$ is the constant $1\in DR^0\left(S^2\right)$ and $*1_{S^2}$ is
the $g_{S^2}$-volume form).

The sphere $S^2$ has a cell decomposition%
\footnote{This $CW$-complex (cell stratification) has a subdivision
which is a $C^1$-triangulation of $S^2$. So as the combinatorial torsion
is defined also for $CW$-complexes and as it is invariant under subdivisions,
$\tau\left(S^2\right)$ can be computed from this cell stratification
(\cite{Mi}, Sections~7, 8, 12.3).}
$X_{S^2}\colon X:=D^2\cup_{\df D^2}pt$. Hence the element
$h_c\in\det\Ce\left(X_{S^2}\right)$ defined below  is of the norm $1$:
\begin{equation*}
h_c=\left(\delta_{pt}\right)^{-1}\otimes\left(\delta_{D^2}\right)^{-1},
\ \|h_c\|^2_{\det\Ce\left(X_{S^2}\right)}=1.
\end{equation*}
(For this cell-decomposition $d_c=0$, and so
$\det\Ce\left(X_{S^2}\right)$ is the same as $\det\Hh\left(S^2\right)$
without the $d_c$-identification. The cochains $\delta_{pt}$,\ $\delta_{D^2}$
are the basic elements  in $H^0\left(S^2\right)$, $H^2\left(S^2\right)$.)

The integration homomorphism $R\colon\DR\left(S^2\right)\rightarrow\Ce\left(X
_{S^2}\right)$ maps $1_{S^2}$ to $\delta_{pt}$ and $*1_{S^2}$
to $v\cdot\delta_{D^2}$. So $R(h)=h_c$ and we have
\begin{equation}
e\left(S^2\right)=0, \qquad e\left(S^{2m}\right)=0.
\label{N1}
\end{equation}
The equalities below follow from Lemmas~\ref{L3}, \ref{L4},
and from (\ref{M7}),
(\ref{N1}):
\begin{align*}
0=e\left(S^n\right) & =2e\left(D^n,\df D^n\right)+e\left(S^{n-1}\right)-
\chi\left(S^{n-1}\right),\\
e\left(D^n,\df D^n\right) & =e\left(D^n\right)-e\left(S^{n-1}\right)=
e\left(D^n\right),\\
e\left(D^n,\df D^n\right) & =2^{-1}\chi\left(S^{n-1}\right)=e\left(D^n\right).
\end{align*}
Lemma~\ref{L5} is proved.\ \ \ $\Box$

The equality
$e\left(D^{m+1}\times S^n\right)=e\left(D^{m+1}\times S^n,\df\left(D^{m+1}
\times S^n\right)\right)$ holds by Lemmas~\ref{L3} and \ref{L5}.

\begin{cor}
For arbitrary $n$, $m\ge 0$ the equality holds:
\begin{equation}
e\left(D^{m+1}\times S^n\right)=e\left(S^m\times D^{n+1}\right).
\label{N3}
\end{equation}
\label{C5}
\end{cor}
(According to Lemma~\ref{L5}, each side of (\ref{N3}) is equal to $2$
in the case of a pair of even numbers $(m,n)$
and it is equal to zero for other pairs $(m,n)$.)

Let $M$ be a compact manifold with a smooth boundary $\df M$ and let
$Z$ be a union of some connected components of $\df M$. Let $\Me$
be obtained by some spherical Morse surgery (with a trivial normal bundle)
of $M$ (i.e., there exists a manifold
$\left(M_{1},\df M_{1}\right)\subset M\setminus\df M$,\
$M_{1}\simeq D^{m+1}\times S^n$, $m+n+1=\dim M$, with
$\df M_{1}=S^m\times S^n$,\ $M=M_{1}\cup_{\df M}M_{2}$, such that
$\Me=\Me_{1}\cup_{\df\Me_{1}}M_{2}$ is obtained by gluing
$\Me_{1}=S^{m}\times D^{n+1}$ and $M_2$ by a diffeomorphism
$f\colon\df\Me_{1}\rs\df M_{2}$).

Let the metrics $g_{M}$ and $g_{\Me}$ be direct product metrics near
$\df M$ and $\df \Me$.
(It is proved above that the numbers $e(M,Z)$ and $e\left(\Me,Z\right)$
do~not depend on the metrics $g_{M}$, $g_{\Me}$ supposed to be
direct product metrics near $\df M$ and near $\df\Me$.)
\begin{lem}
The number $e(M,Z)$ is invariant under the spherical Morse surgeries
with a trivial normal bundle, i.e., the equality holds
\begin{equation}
e(M,Z)=e\left(\Me,Z\right).
\label{S1}
\end{equation}
\label{L6}
\end{lem}

\noindent{\bf Proof.} The metrics $g_{M}$ and $g_{\Me}$ can be replaced
by Riemannian metrics on $M$ and $\Me$ which are direct product metrics
on $\df M_{1}\times I$ and $\df\Me_{1}\times I$ near $\df M_{1}\subset M$
and near $\df\Me_{1}\subset\Me$ (and which are direct product metrics
near $\df M$ and $\df\Me$). Lemma~\ref{L4} claims in this case that
\begin{gather}
\begin{split}
e(M,Z) & =e\left(M_{1},\df M_{1}\right)+e\left(M_{2},\df M_{1}\cup Z\right)+
e\left(\df M_{1}\right)-\chi\left(\df M_{1}\right),\\
e(\Me,Z) & =e\left(\Me_{1},\df\Me_{1}\right)+e\left(M_{2},\df\Me_{1}\cup Z
\right)+e\left(\df\Me_{1}\right)-\chi\left(\df\Me_{1}\right).
\end{split}
\label{S2}
\end{gather}

The smooth closed manifolds $\df\Me_{1}$ and
$\df M_{1}$ are diffeomorphic. Hence
$$
e\left(\df M_{1}\right)=e\left(\df\Me_{1}\right),\quad\chi\left(\df M_{1}
\right)=\chi\left(\df\Me_{1}\right),\quad
e\left(M_{2},\df M_{1}\cup Z\right)=e\left(M_{2},\df\Me_1\cup Z\right).
$$
Corollary~\ref{C5} and Lemmas~\ref{L3} and \ref{L5} claim that
$e\left(M_{1},\df M_{1}\right)=e\left(\Me_{1},\df\Me_{1}\right)$.
So the equality (\ref{S1}) follows from (\ref{S2}).\ \ \ $\Box$

\subsubsection{Proof of the generalized Ray-Singer conjecture}

\begin{thm}[Classical Ray-Singer conjecture]
For any closed Riemannian \\ manifold $\left(M,g_{M}\right)$ its analytic
torsion norm is equal to the combinatorial torsion norm
\begin{equation*}
T_{0}(M)=\tau_{0}(M).
\end{equation*}
\label{T3}
\end{thm}
\noindent{\bf Proof.} There is a smooth Morse function $f$ on a direct
product $M\times I$ (i.e., a function with the nondegenerate isolated
critical points with different critical values) such that the following
holds. Its minimum value is equal to zero, $f^{-1}(0)=M\times\df I$,
and the zero is~not a critical value of $f$. Its maximum value
$\max_{M\times I}f$ equals $1$ and the maximum value level is the only
one point. Namely $f^{-1}(1)$ is an interior point of $M\times(I,\df I)$.

As $f^{-1}(1-\epsilon)$ (where $\epsilon>0$ is very small) is
a~sphere $S^n$ ($n=\dim M$), there exists a sequence of spherical
Morse surgeries (given by transformations of levels $f^{-1}(x)$,
$x\in(0,1-\eps)$ for $x$ divided by critical values) such that their
composition is a transformation of a manifold%
\footnote{The manifold $M$ is~not supposed to be orientable.}
$M\cup M=M\times\df I=f^{-1}(0)$ into $S^n=f^{-1}(1-\eps)$.

Aa a consequence of Lemma~\ref{L6} in this case we get
\begin{equation*}
2e(M)=e(M\cup M)=e\left(S^n\right).
\end{equation*}

Lemma \ref{L5}\ claims that $0=e\left(S^n\right)=e(M)$.
Thus, the Ray-Singer conjecture is proved.\ \ \ $\Box$

\medskip
Let $\left(M,g_{M}\right)$ be a compact Riemannian manifold with a smooth
boundary $\df M$. Let $Z$ be a union of some connected components of $\df M$
and let $g_{M}$ be a direct product metric near $\df M$. The following
two theorems are generalizations of the Ray-Singer conjecture.
\begin{thm}
Under the conditions above, the following equality holds for a manifold
with a smooth boundary:
\begin{equation}
T_{0}(M,Z)=2^{\chi(\df M)/2}\tau_{0}(M,Z).
\label{N8}
\end{equation}
\label{T4}
\end{thm}

\noindent{\bf Proof.} Lemma~\ref{L3} claims that
$e(M,Z)=e(M,\df M)+e(\df M\setminus Z)$.
According to Theorem~\ref{T3}, $e(\df M\setminus Z)$ is equal to zero.
Hence $e(M,Z)=e(M,\df M)$. In the case of $\df M\ne\emptyset$ there is
a mirror-symmetric closed Riemannian manifold $P=M\cup_{\df M}M$
obtained by gluing two copies of $\left(M,g_M\right)$ along $\df M$.
According to Lemma~\ref{L4}, we have
\begin{equation*}
e(P)=2e(M,\df M)+e(\df M)-\chi(\df M).
\end{equation*}
Theorem~\ref{T3} claims that $e(\df M)=0=e(P)$. Thus, we get
\begin{equation*}
e(M,Z)=e(M,\df M)=2^{-1}\chi(\df M),
\end{equation*}
which is equivalent to (\ref{N8}).\ \ \ $\Box$

Let $\left(M,Z,g_{M}\right)$ be as in Theorem \ref{T4}.
Let $N$ be a codimension one in $M$ two-sided in $M$ closed submanifold
$N\subset M\setminus\df M$. Let $M$ be obtained by gluing $M_1$ and $M_2$
along $N$. Let $g_{M}$ be a direct product metric
near $N$ and let the $\nu$-transmission boundary conditions (\ref{A11})
be given on $N$ (where $\nu=(\alpha,\beta)\in\rbo$).
\begin{thm}
The analytic torsion norm is expressed by the combinatorial torsion
norm (in the case of the $\,\nu$-transmission
interior boundary condition on $N$) as follows
\begin{equation*}
T_{0}(M_\nu,Z)=2^{\chi(\df M)/2+\chi(N)}\tau_{0}(M_\nu,Z).
\end{equation*}
\label{T5}
\end{thm}

\noindent{\bf Proof.} The equality (\ref{LL11}) claims that
$e(M_\nu,Z)=e(M,Z)+\chi(N)$.
So the assertion of the theorem follows from Theorem~\ref{T4} and from
the equality (\ref{LL11}).\ \ \ $\Box$

\begin{rem}
This proof of the generalizations of the Ray-Singer conjecture does~not
use any explicit expressions for the scalar analytic torsions
of any special classes of manifolds. The proofs in \cite{Mu1}, \cite{Ch}
of the classical Ray-Singer conjecture essentially used the explicit
expressions for the scalar analytic torsions for spheres and lens spaces.
(The latter expressions were obtained by D.B.~Ray in \cite{Ra}.
He computed there the scalar analytic torsion for lens
spaces and spheres  with homogeneous metrics by explicit
calculations of the $\zeta$-functions for the corresponding
Laplacians using Gegenbauer's polynomials.) The proof in \cite{Mu1} used
the precise estimates of \cite{DP} for the eigenvalues of the corresponding
combinatorial Laplacians. The proof of \cite{Ch} used the Lerch formula
for the derivative at zero of the zeta-function of Riemann (\cite{WW},
13.21, 12.32). We don't use this formula. (Its new proof is obtained
here.) Our proof of the generalized Ray-Singer formula is based
on a gluing property for the  analytic torsion norms. This property
is proved here for a general gluing two Riemannian manifolds
by a diffeomorphism of some connected components of their boundaries.
It is proved without any computations of asymptotics of eigenvalues
and eigenforms for the corresponding Laplacians.
\end{rem}

\section{Gluing formula for analytic torsion norms.
Proof of Theorem~1.1}
\label{Sec2}
\subsection{Strategy of the proof}

In Section~\ref{Sec1} the generalized Ray-Singer conjecture for a manifold
with a smooth boundary is deduced from Theorem \ref{T1}.
Namely it is deduced from the gluing formula
\begin{equation}
\phi_{an}T_{0}(M,Z)=T_{0}\left(M_{1},Z_1\cup N\right)\otimes T_{0}\left(M_{2},
Z_2\cup N\right)\otimes T_{0}(N),
\label{ST1}
\end{equation}
which holds under the conditions of Theorem \ref{T1} (where
$Z_k:=Z\cap\df M_k$). The identification $\phi_{an}$ in (\ref{ST1})
is defined in (\ref{B1}) with the help of the exact sequence (\ref{A7})
of the de Rham complexes. It is proved in Lemma~\ref{L2} that
the equality (\ref{ST1}) follows from the assertion that under
the conditions of Theorem~\ref{T1}, the induced analytic torsion norm%
\footnote{The identification $\fan$ is defined by the short exact
sequence (\ref{A12}).}
$\fan T_0\left(M_{\nu}\right)$ does~not depend on a parameter $\nu$
of the interior boundary conditions. The latter statement means that
the equality
\begin{equation}
\fan T_{0}\left(M_\nu,Z\right)=c_{0}T_{0}\left(M_1,Z_1\cup N\right)
\otimes T_{0}\left(M_{2},Z_2\cup N\right)\otimes T_{0}(N),
\label{ST2}
\end{equation}
holds with a positive constant $c_0$ which is independent of $\nu\in\rbo$.
(However, it {\em is~not supposed} in Lemma~\ref{L2} that $c_0$
{\em is independent of} $M,N,g_{M}$, and $Z$.)

The strategy of the proof of the equality (\ref{ST2}) is as follows.
First we prove that $c_{0}$ is constant on each of four
connected components
\begin{equation}
U_j\subset U:=\{(\alpha,\beta)\in{\wR}
^2\colon\alpha\beta\ne 0\}.
\label{ST3}
\end{equation}

Then it is enough to prove that $c_{0}(\nu)$ is continuous
as a function of $\nu$ for $\nu\in\rbo$. These two
assertions provide us with a proof of the equality (\ref{ST2}).

Let $\nu_{0}\in U$ and let $a>0$ be a number not belonging to the spectrum
$S(\nu_{0}):=\bigcup_i\Spec\Delta_i(\mno,g_{M})\subset\wR_+$
of the Laplacians on $\DR\left(M_\nu,Z\right)$. This spectrum is discrete
according to Theorem~\ref{TAT705}. In particular, each its eigenvalue
is of a finite multiplicity. Let $W^i_a(\nu)$
be a subspace of $DR^i\left(M_\nu,Z\right)$, spanned by all
the eigenforms $\omega_\lambda$ for $\din :=\Delta_i(M_\nu,g_{M})$ with
their eigenvalues $\lambda\le a$. Then $dW^i_a(\nu)\subset W_a^{i+1}(\nu)$.
So $\left(\Wan,d\right)$ is a finite-dimensional subcomplex
of $\left(\DR\left(M_\nu,Z\right),d\right)$ equiped with the natural
Hilbert structures on $\Wan\hookrightarrow\DR\left(M_{\nu},Z\right)$
(defined by $g_M$).

Let $\left\|\cdot\right\|^2_{\dew}$ be the induced norm on
$\dew$. For $\nu$ very close to $\nu_{0}$ it holds also that $a\notin S(\nu)$
(Proposition~\ref{PA3002}).
By the definition of $\Wan$, its cohomology $H^j\left(\Wan\right)$
are canonically identified with the space of harmonic forms
${\Ker}\,\Delta_j\mg$.
The differential $d$ in $\Wan$ induces the identification
\begin{equation}
d_{W}\colon\dew\rs\det\Ker\Delta_{\bullet}\mg.
\label{S5}
\end{equation}

According to Lemma~\ref{L1} there is a canonical identification between
the harmonic forms and the cohomology of the de Rham complex
(the latter one is independent of $g_M$):
\begin{equation}
\Ker\,\Delta_i\mg=H^i\left(DR\left(M_\nu,Z\right)\right).
\label{S6}
\end{equation}

So there is the induced canonical identification of the determinant lines:
\begin{equation}
\det{\Ker}\Delta_{\bullet}\mg=\det\Hh\left(DR\left(M_\nu,Z\right),d\right).
\label{S7}
\end{equation}

Let $\left\|\cdot\right\|^2_{\deh}$ be a norm on
$\det\Hh\left(M_\nu,Z\right):=\det\Hh\left(DR\left(M_{\nu},Z\right),d\right)$
induced by the identifications (\ref{S6}) and (\ref{S7})
from the Hilbert structure on the harmonic forms $\Ker\Delta_{\bullet}\mg$.
(This structure is defined by the Riemannian metric $g_M$.)

The identification (\ref{S5}) is~not an isometry of the norms
$\left\|\cdot\right\|^2_{\deh}$ and $\left\|\cdot\right\|^2_{\dew}$
in general. The norm $\left\|\cdot\right\|^2_{\deh}$ has to be multiplied
by an additional factor for the identification (\ref{S5}) to become
an isometry. This factor is the {\em scalar analytic torsion}
of a complex $\left(\Wan,d\right)$, defined by the general
formula (\ref{A2}). We can conclude that the {\em analytic torsion norm}
$T_{0}\left(M_\nu,Z\right)$ on the determinant
of $\Hh\left(DR\left(M_\nu,Z\right)\right)$ is isometric (under
the identifications (\ref{S5}) and (\ref{S7})) to the norm
\begin{equation}
T_{0}\left(M_\nu,Z;a\right):=\left\|\cdot\right\|^2_{\dew}\exp\left(\sum(-1)^j
j\df_s\znj(s;a)|_{s=0}\right).
\label{S8}
\end{equation}
The zeta-function $\znj(s;a)$ is defined for $\Ree \,s>(\dim \,M)/2$
by the series $\sum_{\lambda>0}\lambda^{-s}$, where the sum is over
all the eigenvalues $\lambda$ of $\Delta_j\left(\mg\right)$ (including
their multiplicities), such that $\lambda>a$. This $\zeta$-function
can be continued meromorphically to the whole complex plane
$\wC$ and it is regular at zero. The latter assertion follows from
Theorem~\ref{TAT705} and from the equality (which is obvious
for $\Ree\,s>(\dim M)/2$):
\begin{equation}
\znj(s;a)=\znj(s)-\sum_{0<\lambda\le a}\lambda^{-s}.
\label{S9}
\end{equation}
(The series for $\znj(s)$, $\Ree\,s>(\dim M)/2$, is the sum over
all the nonzero eigenvalues of $\Delta_j\mg$ with their multiplicities,
where $\lambda^{-s}:=\exp(-s\log\lambda)$ and $\log\lambda\in{\wR}$
for $\lambda>0$).

The identifications $d_{W}$ (\ref{S5}) and $\phi^{\nu}_{an}$
(the latter one is defined  with the help of (\ref{A12})) provide us
(under the conditions of Lemma~\ref{L2}) with the identification:
\begin{equation}
\fan(a)\colon\dew\rs\Det(M,N,Z)
\label{S10}
\end{equation}
($\Det (M,N,Z)$%
\footnote{To remind, $Z$ is the union of the connected components
of $\df M$, where the Dirichlet boundary conditions are given.
The Neumann boundary conditions are given on $\df M\setminus Z$.}
is defined in (\ref{B1})). The assertion that $c_{0}(\nu)$ is independent
of $\nu$ on each connected component $U_j$ of $U$ (\ref{ST3}) is equivalent
to the following one. The analytic torsion norm $T_0\left(M_\nu,Z;a\right)$
is transformed (under the identification (\ref{S10})) into the norm
on $\Det(M,N,Z)$:
\begin{equation}
\fan(a)\!\circ\! T_{0}\!\left(M_\nu,Z;a\right)\!=\!c_{0}(\nu)
T_{0}\!\left(M_{1},Z_1\cup N\right)\!\otimes\! T_{0}\!\left(M_{2},Z_2
\cup N\right)\!\otimes\! T_{0}(N),
\label{S11}
\end{equation}
where $c_{0}(\nu)$ is {\em constant on each connected component $U_j$}.

The action of $\phi^{\nu}_{an}(a)$ is as follows (by its definition):
\begin{equation*}
\phi^{\nu}_{an}(a)T_{0}\left(M_\nu,Z;a\right)=T\left(M_\nu,Z;a\right)\phi
^{\nu}_{an}(a)\circ\left\|\cdot\right\|^2_{\dew},
\end{equation*}
where the scalar analytic torsion $T\left(M_\nu,Z;a\right)$ is defined
as the scalar factor in (\ref{S8}):
\begin{equation}
T\left(M_\nu,Z;a\right):=\exp\left(\sum(-1)^jj\df_s\znj(s;a)|_{s=0}\right).
\label{S14}
\end{equation}

Let $\nu(\gamma)$, $\gamma\in(\epsilon,\epsilon)\subset{\wR}$, be a smooth
curve in $U$ (\ref{ST3}) and let $\nu(0)=\nu_{0}$.
Let $\Pi_j(\nu_{0};a)$ be an ortogonal projection operator from
$\left(DR^j(M)\right)_{2}$ onto $W^j_a(\nu_{0})$ (relative to the natural
Hilbert  structure (\ref{Y2}) in $\left(DR^j(M)\right)_2$). Let $p_{1}$
be a linear operator in $\dromb$, mapping $(\oma,\omb)\in\dromb$
to $(\oma,0)$. (Respectively $p_{2}$ maps $(\oma,\omb)$ to $(0,\omb)$.)

Let $\nu$ and $\nu_{0}$ be arbitrary points from $U$. Then the following
isomorphism of the de Rham complexes is defined (where $\kn:=\alpha/\beta$
for $\nu=(\alpha,\beta)\in U$):
\begin{equation}
\!v_\nu\!=\!v_{\nu\nu_{0}}\colon\DR\!\left(\mno,Z\right)\rs\DR\!\left(M_\nu,
Z\right),\;
v_\nu(\oma,\omb)\!:=\!\left(\oma,\!(k_\nu/k_{\nu_0})\!\cdot\!\omb\right).
\label{ST20}
\end{equation}

Thus the induced isomorphism is defined:
\begin{equation*}
v_{\nu *}\colon\Hh\left(DR\left(\mno,Z\right)\right)\to\Hh\left(DR\left(M_\nu,
Z\right)\right).
\end{equation*}

Let $a$ be a positive number from ${\wR}_+\setminus S(\nu_{0})$.
Then for $\nu$ very close to $\nu_{0}$ the number $a$ is also from
${\wR}_+\setminus S(\nu)$ (Proposition~\ref{PA3002}). The complexes
$\Wan$ and $\Wano$ are isomorphic as abstract finite-dimensional
complexes (and (\ref{ST20}) provides us with a natural
but not canonical isomorphism of these complexes). We have to compute
the action of $\phi^{\nu}_{an}$ on the norms $\left\|\cdot\right\|^2_{\dew}$
for $\nu$ very close to $\nu_{0}$. However $\left\|\cdot\right\|^2_{\Wan}$
are the norms on different complexes. So it is necessary to define
some isomorphism between $\Wano$ and $\Wan$ and then to compute its
action on $\left\|\cdot\right\|^2_{\dewo}$ and on the space $\Det(M,N,Z)$.
The choice (\ref{ST22}) of such an identification is done below.

For $\nu$ very close to $\nu_{0}$ the subspaces $\Wan$ and $\Wano$
are very close in the $L_{2}$-completion $\dromb$
of $\DR\left(M_\nu,Z\right)=:\DR(\nu)$, according
to Proposition~\ref{PA3002}. So the following isomorphism of these
finite-dimensional complexes is well-defined:
\begin{multline}
g_\nu=\Pii(\nu;a)\cdot v_{\nu\nu_{0}}\cdot j_{\nu_{0}}\colon\\
\Wano\hookrightarrow\left(\DR(\nu_{0}),d\right)\stackrel{v_{\nu\nu_{0}}}
{@>>\lrs>}\left(\DR(\nu),d\right)\ @>>{\Pii(\nu;a)}>\ \left(\Wan,d\right),
\label{ST22}
\end{multline}
where $j_{\nu_{0}}$ is the natural inclusion of $\Wano$ and $\Pii(\nu;a)$
is the orthogonal projection operator onto $\Wan$. Its action on the norm
$\left\|\cdot\right\|^2_{\dewo}$ is computed by the following lemma.

\begin{lem}
Let $l$ be an arbitrary nonzero element of $\dewo$. Then the equality
holds for any smooth variation $\nu(\gamma)$ of $\nu_{0}=\nu(0)$:
\begin{equation}
\dfg\log\|g_\nu l\|^2_{\dew}\bgo=-2\,\dfg\log(k_\nu)\bgo
\left(\sum(-1)^j\Tr\left(p_{2}\Pi^j(\nu_{0};a)\right)\right).
\label{ST23}
\end{equation}
\label{L2.2}
\end{lem}

In (\ref{ST23}) the rank (i.e., the dimension of the image) of the operator
$p_{2}\Pi^j(\nu_{0};a)$ is less or equal to $\dim\Wano$. This operator acts
in $\dromb$.

\medskip
Then the following lemma provides us with the variation formula
for $T\left(M_\nu,Z;a\right)$.

\begin{lem}
For $\gamma=0$ the equality holds:
\begin{equation*}
\dfg\log T\left(M_\nu,Z;a\right)=2\,\dfg\log(\kn)\bgo\sum(-1)^j b_{1,j}
\left(\mno,Z;a\right),
\end{equation*}
where $k_\nu:=\alpha/\beta$ for $\nu=(\alpha,\beta)\in U$. Here
$b_{1,j}(\mno,Z;a)$ is a constant coefficient (i.e., $t^0$-coefficient
$q_{0}$) in the asymptotic expansion as $t\to+0$ of the trace
of the operator below (acting in $\left(DR^j(M)\right)_2$):
\begin{equation}
\Tr\left(p_{1}\left\{\exp\left(-t\djon\right)\left(1-\Pi^j(\nu_{0};a)
\right)\right\}\right)\sim q_{-n}t^{-n/2}+q_{-n+1}t^{-(n-1)/2}
+\ldots+q_{0}t^0+\ldots
\label{S16}
\end{equation}
\label{L2.1}
\end{lem}

\begin{rem}
\label{Rem2.1}
The operators $\exp\left(-t\djon\right)$ and $\Pi^j(\nu_{0};a)$
acting in the $L_{2}$-completion $\left(DR^j(M)\right)_{2}$
of $DR^j\left(\mno,Z\right)$ (which coincides with the $L_{2}$-completion
of $DR^j(M)$) have their images in the domain of definition of the Laplacian
$D\!\left(\djon\right)\!\subset\! DR^j\!\left(\!\mno,Z\right)$. The existence
of the asymptotic expansion (\ref{S16}) follows from Theorem~\ref{TAT706}.
The coefficients $q_m$ with $m\le-1$ in  (\ref{S16}) are independent
of $a$. The coefficients $\tilde{q}_m$ of the asymptotic expansion
for $\Tr\bigl(p_{1}\exp(-t\djon)\bigr)$ are equal to the sums
of the integrals over $M_{1}$ and over $\df M_1\supset N$
of the locally defined densities on $M_{1}$ and  on $\df M_1$
(by Theorem~\ref{TAT706}). However, in the general case we cannot
represent $\Tr\left(p_{1}\Pi^j(\nu_{0};a)\right)$ as an integral
of a locally defined density (because there are no universal local
formulas for the eigenforms $\omega_{\lambda}$ of $\djon$).
Hence there is no universal local formula for a coefficient $q_{0}$
in (\ref{S16}) but there are such formulas for $q_m=\tilde{q}_m$ with $m<0$.
\end{rem}

\begin{cor}
For an arbitrary nonzero $l\in\dewo$ the equality holds
\begin{equation}
\dfg\!\log\|g_{\nu}l\|^2_{T_{0}\left(M_\nu,Z\right)}\!=\!2\dfg\!\left(
\log\kn\!\right)\bgo\!\left(\sum(-1)^j\!\left(b_{1,j}\!\left(\mno,Z\right)
\!-\!\dim W^j_a\right)\!\right),
\label{ST30}
\end{equation}
where $b_{1,j}\left(\mno,Z\right)$ is a constant coefficient
(i.e., the $t^0$-coefficient) of the asymptotic
expansion of $\Tr\left(p_{1}\exp\left(-t\djon\right)\right)$ relative
to $t\to +0$ and $p_{1}\exp\left(-t\djon\right)$ is the operator acting
in $\left(DR^j(M)\right)_2$.
\label{C2.1}
\end{cor}

\begin{rem}
Note that in the right side of (\ref{ST30}) there are the Euler
characteristic $\chi\left(M_\nu,Z\right):=\sum(-1)^j\dim W^j_a$ and
the alternating sum of the integrals $b_{1,j}\left(\mno,Z\right)$
(over $M_1$ and over $\df M_1$) of the locally defined densities
(Remark~\ref{Rem2.1}). (Here $Z$ is the union of the connected components
of $\df M$, where the Dirichlet boundary conditions are given). The number
$\chi\left(M_\nu,Z\right)$ is also equal to the sum of the integrals
over $M$, $N$, and over $\df M$ of the locally defined densities.
\end{rem}

Let $\nu(\gamma)$ be a smooth variation of a point $\nu_{0}\in U$
(\ref{ST3}). Let $l(\gamma)\in\det W_a^{\bullet}\left(\nu(\gamma)\right)$
be a variation of an arbitrary nonzero element $l\in\dewo$ such that
$\phi^{an}_{\nu(\gamma)}(a)\circ l(\gamma)$ is a fixed (nonzero) element
of $\Det(M,N,Z)$. Then the equality (\ref{S11}) (where the factor
$c_{0}(\nu)$ is constant on each connected component of $U$ (\ref{ST3}))
is equivalent to the assertion that for any such a variation $l(\gamma)$
its analytic torsion norm is independent of $\gamma$:
\begin{equation}
\dfg\left\|l(\gamma)\right\|^2_{\tonga}\bgo=0.
\label{ST31}
\end{equation}

Corollary~\ref{C2.1} provides us with the formula (\ref{ST30}) for
a variation of the analytic torsion norm $\|g_{\nu(\gamma)}l\|^2_{\tonga}$
(where $l\in\dewo$). The assertion (\ref{ST31}) is equivalent
to the following identity:
\begin{equation}
\dfg\log\left\|g_{\nu(\gamma)}l\right\|^2_{\tonga}\bgo=\dfg\log
\left\|g_{\nu *}f\right\|^2\bgo,
\label{ST32}
\end{equation}
where $f$ is an arbitrary nonzero element of $\Det(M,N,Z)$ (for instance,
$f=\fano(a)\circ l$) and
$g_{\nu *}=\fan(a)\circ g_\nu\circ\left(\fano(a)\right)^{-1}$ is defined
by the following commutative diagram, where $\nu\in U$ is very close
to $\nu_{0}$:
\begin{gather*}
\begin{CD}
\Det(M,N,Z)             @>>{g_{\nu *}}>              \Det(M,N,Z)\\
{\sst{\fano(a)}}\,\big\uparrow\wr\wr @.  \wr\wr\big\uparrow\sst{\fan(a)} @.\\
\dewo                   @>>{g_\nu=g_{\nu\nu_{0}}}>  \dew
\end{CD}
\end{gather*}
The norm on the right in the equality (\ref{ST32}) is an arbitrary
Hilbert norm in one-dimensional space $\Det(M,N,Z)$. The value
of the expression on the right in (\ref{ST32}) is independent
of such a norm.

The action of the isomorphism
$g_\nu=g_{\nu\nu_{0}}\colon\Wano\to\Wan$ on $\Det(M,N,Z)$ is described
by the following lemma.
\begin{lem}
For an arbitrary element $f\in\Det(M,N,Z)$ the equality holds:
\begin{equation}
\dfg\log\left\|g_{\nu *}f\right\|^2\bgo=-2\dfg\left(\log\kn\right)
\bgo\sum(-1)^jb_{2,j}\left(\mno,Z\right),
\label{ST34}
\end{equation}
where $b_{2,j}\left(\mno,Z\right)$ is the constant coefficient (i.e.,
the $t^0$-coefficient) in the asymptotic expansion (relative to $t\to +0$)
for the trace of the operator $p_{2}\exp\left(-t\djon\right)$ acting
in $\left(DR^j(M)\right)_2$.
\label{L2.3}
\end{lem}

Here $p_{2}$ is the operator $p_{2}\colon(\oma,\omb)\to(0,\omb)$
for $\omega_k\in\left(DR\left(\Mm_k\right)\right)_2$.

\begin{rem}
Note that $\Tr\exp\left(-t\djon\right)=\sum_j\Tr\left(p_j\exp\left(-t\djon
\right)\right)$. So we have
\begin{equation*}
-\sum(-1)^jb_{2,j}\left(\mno,Z\right)=\sum(-1)^jb_{1,j}\left(\mno,Z\right)
-\chi\left(\mno,Z\right).
\end{equation*}
Hence the equality (\ref{ST32}) follows from (\ref{ST30}) and (\ref{ST34}).
\end{rem}

\medskip
Thus Lemmas~\ref{L2.2}--\ref{L2.3} provide us with a proof of the assertion
that the factor $c_{0}(\nu)$ is independent of $\nu$ on each connected
component $U_j$ of $U$ (\ref{ST3}).

\subsection{Continuity of the analytic torsion norms}
To prove that $c_0(\nu)$ is independent of $\nu\in\rbo$, it is enough%
\footnote{The factor $c_0(\nu)$ is constant on each connected component
$U_j$ of $U$ (\ref{ST3}), and $U$ is dense in $\rbo$.}
to show that the norm $\phi^{an}_\nu\circ T_{0}(M_\nu)$ on $\Det(M,N,Z)$
is continuous in $\nu\in\rbo$. The following norms on $\Det(M,N,Z)$ are
the same for an arbitrary $a\ge 0$:
\begin{equation}
\fan(a)\circ T_{0}\left(M_\nu,Z;a\right)=\fan\circ T_{0}\left(M_\nu,Z\right).
\label{S301}
\end{equation}

Let us prove the continuity of $\,\fan T_{0}\left(M_\nu,Z\right)$
as a function of $\nu$ at a point $\nu_{0}\in\rbo$. (The series
of lemmas above provides us with the proof of this assertion
in the case when $\nu_{0}\in U$ (\ref{ST3}). But now this will be proved
at an arbitrary $\nu_{0}\in\rbo$, for instance,
at  $\nu_{0}\in{\wR}^2\setminus\left(U\cup(0,0)\right)$.)
By (\ref{S301}), it is enough to obtain the continuity in $\nu$
at $\nu=\nu_{0}$ of the norm $\fan(a)\circ T_{0}(M_\nu;a)$
on $\Det(M,N,Z)$ for a fixed $a>0$ such that
$a\notin S(\nu_{0}):=\cup_j\Spec\left(\djon\right)$.
Since $a\notin S(\nu_{0})$, we see that $a\notin S(\nu)$ for $\nu$ very close
to $\nu_{0}$. (The latter assertion follows from Proposition~\ref{PA3002}.
It claims that the resolvents
$\Ge_\lambda(\nu):=\left(\don-\lambda\right)^{-1}$ for
$\lambda\notin\Spec(\dnb)$ form a smooth in $(\lambda,\nu)$
family of bounded operators in $\dromb$ and that $\Spec\left(\dnb\right)$
is discrete. As $\Ge_a(\nu_{0})$ is bounded in $\dromb$, the operator
$\Ge_a(\nu)$ is also bounded for $\nu$, close to $\nu_{0}$, and so
$a\notin\Spec(\dnb)$ for such $\nu$.) The assertion below claims that
the truncated scalar analytic torsion (\ref{S14}) is a locally
continuous function.%
\footnote{This truncated scalar analytic torsion is a continuous function
on the set of $\nu\in\rbo$ such that $a\notin\cup_i\Spec\left(\din\right)$.}

\begin{pro}
The scalar analytic torsion $T\left(M_\nu,Z;a\right)$ is continuous
in $\nu$ at $\nu_{0}$.
\label{PA300}
\end{pro}

Thus, the continuity of $\fan T_{0}\left(M_\nu,Z\right)$ (as a function
of $\nu$) at $\nu_0$ is equivalent to the condition
that the norm on $\Det(M,N,Z)$
\begin{equation}
\fan(a)\circ\left\|\cdot\right\|^2_{\dew}
\label{S302}
\end{equation}
is continuous in $\nu$ at $\nu_{0}$.
The continuity of the norm (\ref{S302}) is deduced from the following
finite-dimensional algebraic lemma.
Let
\begin{equation}
f\colon (\Aa,d_{A})\to (\Ve,d_{V})
\label{S303}
\end{equation}
be a quasi-isomorphism of finite complexes of finite-dimensional Hilbert
spaces. Let $f_*\colon \det\Hh(A)\rs\det\Hh(W)$ be the induced
identification of the determinant lines. Let $T_{0}(\Aa)$ and $T_{0}(\Ve)$
be the analytic torsion norms (\ref{A3}) on the determinant lines
identified by $f_*:\det\Hh(A)=\det\Hh(V)$. Let $(\Co f,d)$,
$\Cone^j\!f\!=\!A^{j-1}\oplus V^j$,  be a simple
complex, associated with the bicomplex (\ref{S303}):
\begin{equation}
d_{\Cone}\colon {\Cone}^j\!\to\! {\Cone}^{j+1},\quad
d_{\Cone}(x,y)\!=\!\left(-d_{A}x,fx+d_{V}y\right)
\label{X100}
\end{equation}
(for $(x,y)\in A^{j+1}\oplus V^j$). Then $\Co\!f$ is an acyclic finite
complex of finite-dimensional Hilbert spaces ($\Cone^j\! f$ is the direct
sum of Hilbert spaces $A^{j+1}$ and $V^j$), $\Hh(\Cone\! f)\!=0$.
Hence $\det\Hh(\Cone f)$ is canonically identified with $\wC$ and
the analytic torsion norm for $\Co f$ is a norm on ${\wC}\,$.
The ratio $T_{0}(V)/T_{0}(A)\in{\wR}_+$ is defined as the ratio between
the two norms on the one-dimensional spaces $\det\Hh(V)$ and $\det\Hh(A)$
identified by $f_*$.

\begin{lem}
Under the conditions above, the equality holds:
\begin{equation}
\left\|1\right\|^2_{T_{0}(\Co f)}=T_{0}(\Ve)/T_{0}(\Aa),
\label{S304}
\end{equation}
where the left side is the analytic torsion norm
of $1\in\wC=\det\Hh(\Cone f)$.
\label{L2.4}
\end{lem}

Let $a>0$ be a number from ${\wR}_+\setminus S(\nu_{0})$. Then there
exists an open neighborhood $\uno(a)$ of $\nu_{0}\in\uno(a)\subset\rbo$
such that $a\notin S(\nu)$ for $\nu\in\uno(a)$ (Proposition~\ref{PA3002}).
The family of complexes $\left(\Wan,d\right)$ of Hilbert spaces
is continuous on $\uno(a)$ in the following sense.

The operator $\Pi^j_a(\nu):=\Pi^j(\nu;a)$ is a finite rank projection
operator in $\left(DR^j(M)\right)_{2}$ with its image $W^j_a(\nu)$:
\begin{equation*}
\Pi^j_a(\nu)\colon\left(DR^j(M)\right)_{2}\to W^j_a(\nu)\subset DR^j\left(
M_\nu,Z\right)\subset\left(DR^j(M)\right)_{2}.
\end{equation*}

\begin{pro}
The family of operators $\Pii_a(\nu)$ is continuous in $\nu$
for $\nu\in\uno(a)$ with respect to the operator norm in $\dromb$.
The same is true for the families
\begin{gather*}
\begin{split}
d\Pii_a(\nu) & \colon\dromb\to DR^{\bullet+1}(M_\nu)\subset
\left(DR^{\bullet+1}(M)\right)_{2}, \\
\delta\Pii_a(\nu) & \colon\dromb\to DR^{\bullet-1}(M_\nu)
\subset\left(DR^{\bullet-1}(M)\right)_{2}.
\end{split}
\end{gather*}
These are the families of finite rank operators.
\label{PA85}
\end{pro}

\noindent{\bf Proof.} It follows from Proposition~\ref{PA3002} that
if $a\notin S(\nu_{0})$ then there exists an $\eps>0$ such that
$(a-\eps,a+\eps)\cap S(\nu)=\emptyset$ for $\nu$ sufficiently close
to $\nu_{0}$. Hence
$\{\lambda\colon a-\eps<|\lambda|<a+\eps\}\cap S(\nu)=\emptyset$
for such $\nu$ (since $S(\nu)\subset\wR_+\cup 0$ by Theorem~\ref{TAT705}).
Thus, according to Proposition~\ref{PA3002}, the operators
$$
\Pii_a(\nu)={i\over 2\pi}\int_{\Gamma_a}\Ge_\lambda(\nu)d\lambda
$$
form a smooth in $\nu$ (for such $\nu$) family of finite rank operators
in $\dromb$ (where the circle $\Gamma=\{\lambda\colon |\lambda|=a\}$
is oriented opposite to the clockwise). The operators
$d\Pii_a(\nu)\colon\dromb\to\left(DR^{\bullet+1}(M)\right)_{2}$ form
(for such $\nu$) a smooth in $\nu$ family of finite rank operators
(according to Proposition~\ref{PA3002}.\ \ \ $\Box$

\begin{cor}
For $\nu$ sufficiently close to $\nu_{0}$ the family of operators
$\Pii_a(\nu)$ identifies the graded linear spaces
$\Wano:=\Image\Pii_a(\nu_{0})$ and $\Wan$. Such an identification
nearly commutes with $d$ in the following sense:
\begin{equation}
\left\|d\Pii_a(\nu)w-\Pi^{\bullet+1}_a(\nu)dw\right\|_{2}\le c(\nu,\nu_{0})
\|w\|_{2}
\label{S1100}
\end{equation}
(for any $w\in\Wano$), where $c(\nu,\nu_{0})\to +0$ as $\nu\to\nu_{0}$.
This identification also nearly commutes with $\delta$:
\begin{equation*}
\left\|\delta\Pii_a(\nu)w-\Pi^{\bullet-1}_a(\nu)\delta w\right\|_{2}\le
c(\nu,\nu_{0})\left\|w\right\|_2
\end{equation*}
for $w\in\Wano$ ($\left\|\cdot\right\|_2$ is the $L_2$-norm in $\dromb$).
\label{CA86}
\end{cor}

The estimate (\ref{S1100}) follows from the continuity (in $\nu$)
of the families
$d\Pii_a(\nu)$ and $\Pi^{\bullet+1}_a(\nu)$ since
the following operator norms tend to zero as $\nu\to\nu_{0}$:
$$
\left\|d\Pii_a(\nu)-d\Pii_a(\nu_{0})\right\|_{2}\to +0,\quad
\left\|\Pi^{\bullet+1}_a(\nu)-\Pi^{\bullet+1}_a(\nu_{0})\right\|_{2}\to +0.
$$
Indeed, for an arbitrary $w\in\Wano$ we have $d\Pii_a w=dw$. Hence
the estimates
\begin{multline*}
\left\|\Pi^{\bullet+1}_a(\nu)dw-dw\right\|_{2}\le\left\|\Pi^{\bullet+1}_a
(\nu)-\Pi^{\bullet+1}_a(\nu_{0})\right\|_{2}\cdot\|dw\|_{2}\le \\
\le C\cdot\left\|\Pi^{\bullet+1}_a(\nu)-\Pi^{\bullet+1}_a
(\nu_{0})\right\|_{2}\cdot\|w\|_{2}
\end{multline*}
are true because the differential $d\colon\Wano\to W_a^{\bullet+1}(\nu_{0})$
of a finite complex of finite-dimensional spaces is bounded
(with respect to the Hilbert norm induced from $\dromb$).

For each $\nu\in\rbo$ the combinatorial cochain complex
$\left(\Ce\left(X_{\nu},V\right),d\right)$ (with $V:=X\cap Z$) is defined
by the $\nu$-transmission condition (\ref{K12}). A homomorphism
of the integration of forms from $\Wan$ over the simplexes of $X$
\begin{equation}
R_\nu(a)\colon\left(\Wan,d\right)\to\left(\Ce(X_\nu,V),d\right)
\label{S305}
\end{equation}
is also defined for all $\nu\in\rbo$. For every such $\nu$ the following
variant of the de Rham theorem holds.

\begin{pro}
 $R_\nu(a)$ is a quasi-isomorphism.
\label{PA87}
\end{pro}

\noindent{\bf Proof.}
1. Let $R_\nu\colon\left(\DR(M_\nu,Z),d\right)\to\left(\Ce(X_\nu,V),d\right)$
be the integration homomorphism of pairs of forms
$(\oma,\omb)\in\DR(M_\nu,Z)$ over the simplexes of $X_j\setminus V_j$.
Then $R_\nu$ is a {\em quasi-isomorphism}.%
\footnote{This assertion claims that the analogy of the classical de Rham
theorem is true in the case of the $\nu$-transmission interior boundary
conditions. The classical de Rham theorem for smooth closed manifolds
was proved in \cite{dR1} (see also \cite{dR4}, Ch.~IV, \cite{W}, Ch.~IV,
\S~29). The explicit isomorphism between the \v{C}ech cohomology for a good
cover of a smooth closed $M$ and the de Rham cohomology of $M$ is defined
with the help of the de Rham-\v{C}ech complex (\cite{BT}, Ch.~II, \S~9).}

Indeed, in the commutative diagram (\ref{LL1}) the left and
the right vertical arrows are quasi-isomorphisms according to the de Rham
theorem for a closed manifold $N$ and for manifolds $M_{1}$ and $M_{2}$
with smooth boundaries. (The proof of the latter one is given in \cite{RS},
Proposition~4.2.) The cohomology exact sequences provide us with
the commutative diagram
$$
\begin{array}{ccccccc}
 @>\df_{D}>>\! &\!H^*\left(\oplus\kab\DR\left(M_k,N\cup Z_k\right)\right)\!
& \!\to\! & \!H^*\left(\DR\left(M_\nu,Z\right)\right)\! & \!@>>\left(r_\nu
\right)_*>\! & \!H^*\left(\DR(N)\right)\! & \!\to \\
 & \downarrow{\sst{R_*}} & & \downarrow{\sst{(R_\nu)_*}} & & \downarrow
{\sst{R_*}} & \\
 @>\df_c>>\!    &\!H^*\left(\oplus\kab\Ce\left(X_k,W\cup V_k\right)\right)\!
& \!@>>j_*>\! &\!H^*\left(\Ce\left(X_\nu,V\right)\right)\! & \!@>{\left(r
_{\nu,c}\right)_*}>>\! & \!H^*\left(\Ce(W)\right)\! & \!\to
\end{array}
$$
\begin{equation}
\label{X2040}
\end{equation}
with the exact rows, where the vertical arrows $R_*$ on the left and on the
right are isomorphisms (according to the de Rham theorem) and where
$\df_{D}=\df_c$ under the identifications $R_*$. Hence $\left(R_\nu\right)_*$
is also an isomorphism.

\medskip
The exactness of the top row in (\ref{X2040}) can be interpreted and
proved as follows.
The sheaf $\Fe_{\nu}:=\DR_{\nu}$ ($\nu=(\alpha,\beta)\in\rbo$) of germs
$(\oma,\omb)$ of pairs of $C^{\infty}$-forms $\omega_j$ on $M_j$ such
that%
\footnote{It is supposed that $\omega_j$ has the zero geometrical
restriction to $Z_k$ (at the points $x\in Z_k\subset\df M_k$).}
$\alpha i^*_1\oma=\beta i^*_2\omb$ (here $i^*_j$ are the geometrical
restrictions from $M_j$ to $N\hookrightarrow\df M_j$) is a $c$-soft sheaf.
(The latter notion means that the restriction
$\Gamma\left(M,F^j_{\nu}\right)\to\Gamma\left(K,i^{-1}_K F^j_{\nu}\right)$
is surjective for any compact $i_K\colon K\hookrightarrow M$, \cite{KS},
Definition~2.5.5.) The sheaf $F_{\nu}$ is $c$-soft since appropriate
smooth partitions of unity exist on $M$. The sequence of complexes of global
sections
$$
0\to\Gamma_c\left(M\setminus N,\Fe_{\nu}\right)\to\Gamma\left(M,\Fe_{\nu}
\right)\to\Gamma\left(M,i_{N,*}i^{-1}_N\Fe_{\nu}\right)\to 0
$$
(here $i_N\colon N\hookrightarrow M$) has the terms which possess
the following properties: \\
1) $\Gamma\left(M,\Fe_{\nu}\right)=\DR\left(M_{\nu},Z\right)$,\\
2) $\Gamma_c\left(M\setminus N,\Fe_{\nu}\right)$ is a subcomplex of
$\oplus\kab\DR\left(M_k,N\cup Z_k\right)$ and its natural inclusion
is a quasi-isomorphism. Indeed, if $\omega\in\DR\left(M_k,N\cup Z_k\right)$
is a closed form then $\omega=dv$ in a neighborhood of $N$ in $M_k$
(where $v$ is a smooth form with the zero geometrical restriction to $N$).
So $\omega-d(\phi v)=0$ in some neighborhood of $N$ in $M_k$ ($\phi$
is an appropriate cutting function). We have
$\Gamma_c\left(M\setminus N,\Fe_{\nu}\right)=\Gamma\left(M,j_{!}j^{-1}
\Fe_{\nu}\right)$, where $j\colon M\setminus N\hookrightarrow M$ and
$j_{!}$ is the direct image with proper supports,
$j^{-1}\Fe_{\nu}\simeq\DR|_{M\setminus N}$. The sheaf $j_{!}j^{-1}\Fe_{\nu}$
is $c$-soft according to \cite{KS}, Proposition~2.5.7. \\
3) $\Gamma\left(M,i_{N,*}i^{-1}_N\Fe_{\nu}\right)$ has a natural
homomorphism $q_{\nu}:=r_{\nu}\circ\left(i^*_1,i^*_2\right)$ onto
$\DR(N)$ (where $r_\nu$ is defined in (\ref{A13})) and $q_{\nu}$
is a quasi-isomorphism. In fact, if the form
$u=dt\wedge\omega_N(t)$ on $I\times N$ is closed then
it is exact, because then $d_N\omega_N(t)=0$ and so
$u=d\int^t_0\omega_N(\tau)d\tau$. (Here $t$ is the coordinate on $I$
and $t=0$ is the equation of $N=0\times N\hookrightarrow I\times N$,
$0\in I\setminus\df I$.) Hence $q_{\nu}$ is a quasi-isomorphism.
(This assertion follows also from the Poincar\'e lemma.) The sheaf
$i_{N,*}i^{-1}_N\Fe_{\nu}$ is $c$-soft by \cite{KS}, Proposition~2.5.7.

For a compact manifold $M$ the category of $c$-soft sheaves on $M$
is injective with respect to the functor of global sections
$\Gamma(M;\centerdot)$ (\cite{KS}, Proposition~2.5.10).
The complex $\Fe_{\nu}$
is a $c$-coft resolvent of a constructible sheaf (\cite{KS}, Chapter~VIII)
$\wC_{\nu}$ on $M$, which is isomorphic to $\wC_{M\setminus N}$ on
$M\setminus N$ and to $\wC_N$ on $N$ (where $\wC_X$ is a constant sheaf
on $X$), and the gluing map for $\wC_{\nu}$ is $\left|\nu\right|^{-1/2}
(\alpha,\beta)\colon\wC_N\to i^{-1}_N j_*\wC_{M\setminus N}=\wC_N\oplus\wC_N$
(i.e., $c\to\left|\nu\right|^{-1}(\beta c,\alpha c)$). The complexes
$j_{!}j^{-1}\Fe_{\nu}$ and $i_{N,*}i^{-1}_N\Fe_{\nu}$ are $c$-soft
resolvents of constructible sheaves
$j_{!}j^{-1}\wC_{\nu}=j_{!}\wC_{M\setminus N}$ and of
$i_{N,*}i^{-1}_N\wC_{\nu}$. (The latter one is isomorphic to $i_{N,*}\wC_N$
under $r_{\nu}$.) So the exactness of the cohomology sequence in the top
row of (\ref{X2040}) follows from \cite{KS}, (2.6.33), Remark~2.6.10.

\medskip
2. The projection operator
$p_{\cH}\colon\DR\left(M_\nu,Z\right)\hookrightarrow\dromb\to\Ker\left(\don
\right)$ provides us with the isomorphism
$p_{\cH*}\colon\Hh\left(DR\left(M_\nu,Z\right)\right)\to\Ker\left(\don\right)$
(by Lemma~\ref{L1}). So the inclusion
$i_a\colon\left(\Wan,d\right)\hookrightarrow\left(\DR\left(M_\nu,Z\right),d
\right)$ is a quasi-isomorphism and
$\left(i_a\right)_*\colon\Ker\left(\don\right)\rs\Hh\left(DR\left(M_\nu,
Z\right)\right)$
is equal to $\left(p_{\cH*}\right)^{-1}$ (since $p_{\cH}i_a=\id$ on
$\Ker\left(\don\right)$). From an obvious equality $R_\nu(a)=R_\nu i_a$
it follows that $R_\nu(a)$ is a quasi-isomorphism.\ \ \ $\Box$

Thus the assertion of Lemma~\ref{L2.4} can be applied to the bicomplex
(\ref{S305}). The result is as follows.
\begin{cor}
The equality holds:
\begin{equation}
T_{0}(\Ce(X_\nu,V))/\left\|1\right\|^2_{T_{0}(\Co\Rn(a))}=T_{0}\left(\We_a
\right).
\label{S306}
\end{equation}
\label{C2.3}
\end{cor}

The identifications $\fan(a)$ (for an arbitrary $a>0$) and $\fan$
are defined such that the following norms on ${\Det}(M,N,Z)$
are equal:
\begin{equation}
\fan(a)\circ\left\|\cdot\right\|^2_{\dew}=\fan\circ T_{0}(W_a).
\label{S307}
\end{equation}

Hence, as it follows from (\ref{S306}), (\ref{S307}), we have
\begin{equation}
\fan(a)\circ\|\cdot\|^2_{\dew}=\left(\fan\circ T_{0}\left(\Ce\left(X_\nu,V
\right)\right)\right)\circ\left(\left\|1\right\|^2\right)^{-1}_{T_{0}
(\Co(R_{\nu}(a)))}\ \ \ .
\label{S308}
\end{equation}

\begin{pro}
The factor $\left(\left\|1\right\|^2_{T_{0}(\Co(R_\nu(a)))}\right)^{-1}$
in (\ref{S308}) is a continuous function of $\nu\in\uno(a)$.
\label{PA88}
\end{pro}

\noindent{\bf Proof.} The complex $\Co\left(\Rn(a)\right)$ is acyclic
according to Proposition \ref{PA87}. Its scalar analytic torsion
\begin{equation}
\left\|1\right\|^2_{T_{0}(\Co(R_\nu(a)))}:=\exp\left(\sum_{j\ge -1}(-1)^j
j\zeta'_j(0)\right)
\label{S1093}
\end{equation}
is defined as in (\ref{A2}) by the $\zeta$-functions
of the ``Laplacians'' $L_\nu:=d^*_\nu d_\nu+d_\nu d^*_\nu$ of the complex
$\left(\Co\left(\Rn(a)\right),d_\nu\right)$.%
\footnote{The spaces $\Wan$ are equiped with the Hilbert structure from
$\left(\dromb,g_M\right)$. The spaces
$\Ce\left(X_\nu,V\right)\subset\Ce(X_{1})\oplus\Ce(X_{2})$
are equiped with the Hilbert structure defined by the basic cochains
in $\oplus\Ce(X_k)$ and
$\Co(R_\nu)=W_a^{\bullet-1}(\nu)\oplus\Ce\left(X_\nu,V\right)$
is the orthogonal direct sum of Hilbert spaces.}
Since the complex $\Co\left(\Rn(a)\right)$ is acyclic we see that these
Laplacians are positive definite. (So they have the zero kernels.)
Their determinants
$\det(\Delta^{\bullet}_\nu)$ are continuous positive functions of $\nu$
on $\uno(a)$ (and so the expression on the right in (\ref{S1093})
is a continuous function of $\nu\in\uno(a)$). The latter statement
is derived as follows.

\begin{pro}
Let $m\in\wZ_+$ and $m\ge m_{0}:=1+\min\{k\in\wZ_+:4k\ge\dim M\}$.
Then there exists a positive constant $C=C(M,N,Z,g_{M})$ independent of
$\nu\in\rbo$ (and of $m$ also) such that the following estimate holds
uniformly with respect to $x\in\Mm_{1}\cup\Mm_{2}$:
\begin{equation}
|\omega(x)|^2<C\sum^m_{i=0}\left\|\Delta^i_\nu\omega\right\|^2_{2}
\label{S1094}
\end{equation}
for all $\omega$ such that%
\footnote{The domain of definition of $D(\Delta^{\bullet}_\nu)$
for $\Delta^{\bullet}_\nu$ is defined by (\ref{Y4}) and (\ref{Y3}).}
\begin{equation}
\omega\in\DR(M_\nu,Z),\quad \omega\in D(\Delta^{\bullet}_\nu),\quad
\Delta_\nu\omega\in D(\Delta^{\bullet}_\nu),\dots,\Delta^m_\nu\omega\in
D(\Delta^{\bullet}_\nu).
\label{S10941}
\end{equation}
(Here $|\omega(x)|^2$ is the norm at $\wedge^{\bullet}T_x M$ defined
by $g_{M}$ and $\left\|\cdot\right\|^2_{2}$ is the $L_{2}$-norm in $\dromb$.)
\label{PA89}
\end{pro}

\begin{cor}
If $w\in\Wan$ then $w\in D(\Delta^m_\nu)$ for an arbitrary $m\in\wZ_+$.
So the following estimate holds uniformly with respect
to $x\in\Mm_{1}\cup\Mm_{2}$ and to $\nu\in\rbo$
\begin{equation}
\left|w(x)\right|^2<C_{1}\left\|w\right\|^2_{2},
\label{S1095}
\end{equation}
where $C_{1}=C_{1}\left(M,N,Z,g_{M}\right)> 0$.
\label{CA90}
\end{cor}

The graded Hilbert space $\Ce(X_\nu,V)$ is isomorphic to the direct sum
\begin{equation}
\Ce(X_\nu,V)=\Ce\left(X,N_X\cup V\right)\oplus\Ce(N_X),
\label{S1097}
\end{equation}
where $V:=X\cap Z$, $N_X:=X\cap N$, $\Ce\left(X,N_X\cap V\right)$
is a graded linear subspace of $\Ce\left(X_\nu,V\right)$ (with respect
to the natural inclusion), and the inclusion
$j_\nu\colon\Ce(N_X)\hookrightarrow\Ce\left(X_\nu,V\right)\subset\oplus_i
\Ce(X_i)$ is defined as
$j_\nu:=\left(\alpha^2+\beta^2\right)^{-1}(\beta\id,\alpha\id)$.
The space on the right in (\ref{S1097}) is independent of $\nu$.
Hence (\ref{S1097}) provides us with the isometric identification of the
graded Hilbert spaces
\begin{equation}
p_\nu\colon\Ce(X_{\nu_{0}},V)\rs\Ce(X_\nu,V).
\label{S1098}
\end{equation}

\begin{cor}
Let $\nu\in\uno(a)$ be sufficiently close to $\nu_0$.
Let $\Wan$ be identified with $\Wano$ by $\;\Pii_a(\nu)\colon\Wano\to\Wan$.
Let $\Ce(X_{\nu_{0}},V)$ be identified with $\Ce(X_\nu,V)$
by $p_\nu$ (\ref{S1098}). Then the estimate (\ref{S1095}) involves
that for such $\nu$ the family of homomorphisms of the integration over
the simplexes of $X$
\begin{equation*}
\Ron(a)\colon\left(\Wan,d\right)\to\left(\Ce(X_\nu,V),d_c\right)
\end{equation*}
is a continuous in $\nu$ family of quasi-isomorphisms between finite
complexes of finite-dimensional Hilbert spaces.
\label{CA91}
\end{cor}

Let $f_\nu\colon\left(\Fe(\nu),d_{F}(\nu)\right)\to\left(\Ke(\nu),d_{K}(\nu)
\right)$ be a family of homomorphisms between finite complexes of
finite-dimensional Hilbert spaces. Let the trivialization
of these two families of complexes be defined by the identifications
of the graded linear spaces
\begin{equation*}
\Pi_\nu\colon\Fe(\nu_{0})\to\Fe(\nu),\qquad
p_\nu\colon\Ke(\nu_{0})\to\Ke(\nu).
\end{equation*}
Let these idetifications be chosen such that $f_\nu$ becomes a continuous
family of the  homomorphisms
$$
f_\nu\colon\left(\Fe,d_{F}(\nu)\right)\to\left(\Ke,d_{K}(\nu)\right)
$$
between the continuous families of complexes with the fixed underlying
graded linear spaces $\Fe:=\Fe(\nu_{0})$ and $\Ke:=\Ke(\nu_{0})$.
Let the Hilbert structures on $F^j$ and $K^j$
are continuous functions of $\nu$ for all $j$. In this case, $f_\nu$
is called {\em a continuous family}. Then the following assertion is true.

\begin{pro}
Let $f_\nu$ be a continuous family. Then the determinants $\det(\Le_\nu)$ of
the Laplacians $\Le_\nu=d^*_\nu d_\nu+d_\nu d^*_\nu$ on
$\left(\Co f_\nu,d_\nu\right)$ are continuous functions of $\nu$.
\label{PA95}
\end{pro}

\noindent{\bf Proof.} The operator $d^*_\nu$ adjoint to the differential
$d_\nu$ of $\Co f_\nu$ (relative to the Hilbert structure on
$\Co f_\nu=F^{\bullet-1}\oplus\Ke$)%
\footnote{$\Co f$ is the direct sum of Hilbert spaces
$F^{\bullet+1}\oplus\Ke$ (with the Hilbert structures on $F^{\bullet+1}$
and $K^j$ depending continuously on $\nu$).}
is defined on the whole finite-dimensional space $\Co f_\nu$.
Since $d_\nu$ tends to $d_{\nu_{0}}$ (for instance, in the operator norm%
\footnote{As $\Co f_\nu$ is a finite-dimensional space, the weak
convergence of the operators acting in it is equivalent to the convergence
with respect to the operator norm.})
as $\nu\to\nu_{0}$ we see that $d^*_\nu$ also tends to $d^*_{\nu_{0}}$.
Thus $\Le_\nu\to\Le_{\nu_{0}}$ as $\nu\to\nu_{0}$ and
$\det\Le_\nu\to\det\Le_{\nu_{0}}$ (since the space $\Co f_\nu$ is
finite-dimensional).\ \ \ $\Box$

\begin{cor}
The functions $\det\left(\Le_\nu\right)$ of $\nu$ for $f_\nu=\Ron(a)$
are continuous and positive.
\label{CA96}
\end{cor}

\noindent The positivity of $\det\left(\Le_\nu\right)$ is equivalent
to the acyclicity of $\left(\Co\Rn(a),d_\nu\right)$
(where $d_\nu:=d_{\Cone(\Rn(a))}$).

Proposition \ref{PA88} is proved.\ \ \ $\Box$

\begin{rem}
Propositions \ref{PA85}, \ref{PA89}, and Corollary \ref{CA91} claim that under
the identifications (\ref{S1098}) and $\Pii_a(\nu)$, the Hilbert structures
on $\det\Co\left(\Rn(a)\right)$ and the differentials $d_\nu$ in
$\Co\left(\Rn(a)\right)$ are continuous in $\nu$ at $\nu_{0}$. Hence the
analytic torsion norms $\left\|\cdot\right\|^2_{T_{0}(\Co(\Rn(a)))}$ on
$\wC=\det\Hh\left(\Cone\left(\Rn(a)\right)\right)=\det 0$ are also
continuous in $\nu$ at $\nu_0$.
\end{rem}

According to (\ref{LL3}) we have $\fan=\phi^c_\nu$, where $\phi^c_\nu$
is defined by the bottom row of the commutative diagram (\ref{LL1}).
So the continuity of the norm $\fan(a)\circ\left\|\cdot\right\|^2_{\dew}$
on ${\Det}(M,N,Z)$ can be deduced from (\ref{S308}) and from the following
lemma.
\begin{lem}
The norm $\phi^c_\nu T_{0}\left(C\left(X_\nu,V\right)\right)$
on ${\Det}(M,N,Z)$ does~not depend on $\nu\in\uno(a)$.
\label{L2.5}
\end{lem}

The continuity in $\nu$ of the norm $\fan T_{0}\left(M_\nu,Z\right)$
on $\Det(M,N,Z)$ follows from (\ref{S301}), (\ref{S302}), and from
the continuity of the norm $\fan(a)\circ\left\|\cdot\right\|^2_{\dew}$.
(The latter assertion is proved above.) The equality (\ref{A10}) holds
with $c_{0}(\nu)$ which is constant and positive on each
connected component $U_j$ of $U$ (\ref{ST3}). Because the norm
$\fan T_{0}(M_\nu)$ on ${\Det}(M,N,Z)$ is continuous in $\nu\in\rbo$,
the equality (\ref{A10}) holds for all such $\nu$ with $c_{0}$
{\em independent of} $\nu$. Theorem~\ref{T1} follows from (\ref{A10})
and from the assertion of Lemma \ref{L2}.\ \ \ $\Box$

\begin{rem}
It is not important for the proofs of Theorem~\ref{T1} and
of (\ref{A10}) that the family of finite-dimensional complexes
$\left(\Ce\left(X_\nu,V\right),d_c\right)$ in (\ref{S308}) is
of a combinatorial nature.
It is enough for the proof to have a family of finite-dimensional complexes
$(\Fe_\nu,d_{F})$ which are defined locally in $\nu$ (i.e., for $\nu$
in a neighborhood of an arbitrary $\nu_{0}\in\rbo$) together with the data
as follows. Continuous families of quasi-isomorphisms
$f_\nu(a)\colon\left(\Wan,d\right)\to\left(\Fe_\nu,d_{F}\right)$ and
of Hilbert structures $h_\nu$ on $\Fe_\nu$ are defined.
A family $(\Fe_\nu,h_\nu)$ may depend on $a$ and on $\nu_{0}$
but it has to possess the property as follows. The norm
$\fan\circ\left(f_{\nu}(a)_*\right)^{-1}\circ T_{0}(\Fe_\nu,h_{\nu})$ on
$\Det(M,N,Z)$ is continuous in $\nu$ at $\nu_{0}$. (Here
$f_{\nu}(a)_*\colon\det\Hh(M_\nu,Z)\rs\det\Hh(F_\nu,d_{F})$ is the induced
identification.)
\end{rem}

\noindent{\bf Proof of Lemma~\ref{L2.5}.} Let $\psi_\nu$ be the
identification of the determinant lines defined by the bottom row
of the commutative diagram (\ref{LL1}):
\begin{equation*}
\psi_\nu\colon\det\Ce(X_\nu,V)\rs\left(\otimes\kab\det\Ce(X_k,W\cup V_k)
\right)\otimes\det\Ce(W)=:\Det\Ce(X,V,W)
\end{equation*}
(where $V$ is the induced smooth triangulation $Z\cap X$
of $Z\subset\df M$, $V_k:=V\cap\df M_k$ and $W:=X\cap N=N_X$).
The following diagram is commutative:
\begin{gather}
\begin{CD}
\det\Ce(X_\nu,V) @>{\psi_\nu}>\lrs>    \Det\Ce(X,V,W) \\
{\sst{d_c}}\,\big\downarrow\,\wr @.      {\sst{d_c}}\,\big\downarrow\,\wr @. \\
\det\Hh(X_\nu,V) @>{\phi^c_\nu}>\lrs>  \Det(X,V,W) \\
\ \ \ \| {\sst R}@.                    \ \ \ \| \sst{R} @.  \\
\det\Hh(M_\nu,Z) @>{\fan}>\lrs>        \Det(M,Z,N)
\end{CD}
\label{S309}
\end{gather}
(The determinant lines on the right in (\ref{S309}) are defined by
(\ref{K71}). The identification $d_c$ on the right in (\ref{S309})
is a triple tensor product of the  identifications
induced by $d_c$ on $\Ce(X_k,W\cup V_k)$ and on $\Ce(W)$. The identification
$R$ is defined by the integration over the simplexes of $X$.)
The commutativity of the diagram (\ref{S309}) is equivalent
to the definition (\ref{LL2}) of $\phi^c_\nu$. Since the identification
$d_c$ on the right in (\ref{S309}) is independent of $\nu$ we see that
the statement of Lemma~\ref{L2.5} is a consequence of the following
proposition.
\begin{pro}
The identification $\psi_\nu$ in (\ref{S309}) is an isometry between
the combinatorial norm $\left\|\cdot\right\|^2_{\det\Ce(X_\nu,V)}$
and the triple tensor product of the combinatorial norms on
$\det\Ce\left(X_k,W\cup V_k\right)$ ($k=1,2$) and on $\det\Ce(W)$.
\label{P2.7}
\end{pro}

\noindent(The Hilbert structures on $\oplus\kab\Ce\left(X_k,V_k\right)$ and
on $\Ce(W)$ are defined by the orthonormal basis of the basic cochains.)

\noindent{\bf Proof.} Let $\ronce\colon\Ce(W)\to C^j(X_\nu,V)$ be defined
by (\ref{LL8}). Then $\renc\ronce=\id$ and $\ronce$ is an isometry onto
${\Image}(\ronce)$ (relative to the Hilbert structures, defined above).
The subspace $\Image(\ronce)$ is the orthogonal complement to
${\Image}\,j\left(\oplus\kab\Ce\left(X_k,W\cup V_k\right)\right)$
in $C^j\left(X_\nu,V\right)$ and $j$ is an isometry onto $\Image j$.
(Here, $\renc$ and $j$ are the same as in the bottom row of
(\ref{LL1})).%
\footnote{This proposition is essentially equivalent to Proposition~\ref{P5},
Section~1.}.

Thus Lemma \ref{L2.5}\ is proved.\ \ \ $\Box$

\subsubsection{Uniform Sobolev inequalities for $\nu$-transmission interior
boundary conditions. Proof of Proposition \ref{PA89}}

Let $I\times N\subset M$ (where $I=[-1,1]$) be a neighborhood of
$N=0\times N\subset M$ and let $g_{M}$ be a direct product metric
on $I\times N$.
Proposition \ref{PA89} is a consequence of the assertions as follows.

\begin{pro}
The inequality (\ref{S1094}) holds uniformly with respect to $\nu\in\rbo$
for all $\omega\in\DR(M_\nu)$ of the class (\ref{S10941}) and such that
$\supp\,\omega\subset [-4/5,4/5]\times N\subset M$.
\label{PA150}
\end{pro}

\begin{pro}
The inequality (\ref{S1094}) holds for all $\omega\in\DR(M,Z)$
such that $\supp\,\omega\subset M\setminus\left([-1/3,1/3]\times N\right)$
and such that%
\footnote{For $\omega$ with $\supp\,\omega\subset M\setminus N$
the conditions (\ref{S10941}) and (\ref{S1200}) are equivalent.
The domain $D\left(\Delta_{M,Z}\right)$ of $\Delta_{M,Z}$ consists
of smooth forms on $M$ with the Dirichlet boundary conditions on $Z$
and the Neumann ones on $\df M\setminus Z$.}
\begin{equation}
\omega\in D\left(\Delta_{M,Z}\right), \Delta\omega\in D\left(\Delta_{M,Z}
\right),\ldots,\Delta^m\omega\in D\left(\Delta_{M,Z}\right).
\label{S1200}
\end{equation}
\label{PA151}
\end{pro}

The last assertion is well known (\cite{Ch}, Section 5).

Let $f$ be a smooth function on $M$, $0\le f\le 1$, $f\equiv 1$ on
$[-1/2, 1/2]\times N$ and $f\equiv 0$ on
$M\setminus\left([-3/4,3/4]\times N\right)$.
The $2m$-Sobolev norm on the right in (\ref{S1094}), defined as
\begin{equation*}
\left\|\omega\right\|^2_{(2m)}:=\sum^m_{k=0}\left\|\Delta^k_\nu\omega\right\|
^2_{2},
\end{equation*}
is equivalent uniformly in $\nu\in\rbo$ (i.e., with constants $c_{3},c_{4}>0$
independent of $\nu$ and of $\omega$) to the norm%
\footnote{The lower estimate with $c_{3}$ in (\ref{S1222}) is obvious.
Note that $\supp\left((1-f)\omega\right)\subset M\setminus\left([-1/3,1/3]
\times N\right)$. Then the upper estimate with a constant $c'_{4}$ for
$\left\|(1-f)\omega\right\|_{(2m)}$ by $\left\|\omega\right\|_{(2m)}$
is well known (\cite{Ho}, Appendix B and Proposition 20.1.11).}:
$\left\|\omega\right\|^2_{(2m),f}:=\sum^m_{k=0}\left(\left\|\den^k
(f\omega)\right\|^2_{2}+\left\|\Delta^k\left((1-f)\omega\right)\right\|
^2_{2}\right)$:
\begin{equation}
c_{3}\left\|\omega\right\|^2_{(2m)}<\left\|\omega\right\|^2_{(2m),f}\le
c_{4}\left\|\omega\right\|^2_{(2m)} .
\label{S1222}
\end{equation}

It is enough to verify the upper estimate (with $c_{4}$ independent of $\nu$)
for $f\omega$. It is true for $m=1$, since the estimate holds:
$$
\left\|\Delta_\nu(f\omega)\right\|^2_{2}\le C_{1}\cdot\left(\left\|\Delta_\nu
\omega\right\|^2_{2}+\left\|\omega\right\|^2_{2}+\left\|d\omega\right\|^2
_{2}+\left\|\delta\omega\right\|^2_{2}\right)\le 3/2 C_{1}\left(\left\|
\Delta_\nu\omega\right\|^2_{2}+\left\|\omega\right\|^2_{2}\right),
$$
where $C_{1}$ depends on $f$ but it is {\em independent of} $\nu$.
Hence the following estimate holds for $\omega\in D\left(\Delta^k_\nu\right)$
(with $C_{2}$ {\em independent of $\nu$ and of $\omega$}):
$$
\left\|\Delta^k_\nu(f\omega)\right\|^2_{2}\le C_{2}\cdot\left(\left\|
\Delta^k_\nu\omega\right\|^2_{2}+\left\|\Delta^{k-1}_\nu\omega\right\|^2_{2}
+\ldots+\left\|\omega\right\|^2_{2}\right).
$$
The upper estimate is done. Thus Proposition~\ref{PA89} follows from
(\ref{S1222}) and from Propositions~\ref{PA150} and \ref{PA151}.

\noindent{\bf Proof of Proposition \ref{PA150}.} The form $\omega$
on $I\times N$ is the sum $\omega_{0}+\oma$, where $\omega_i$ is an $i$-form
in the direction of $I$ (where $I=[-1,1]$). It is enough to prove
the inequality (\ref{S1094}) separately for $\omega_{0}$ and for $\oma$.
Let us prove it for $\omega_{0}$. For $\nu=(\alpha,\beta)\in\rbo$
the Green function $G(\nu)$ for the Laplacian $\Delta_{\nu,I}$ on
functions on $I$ with the $\nu$-transmission boundary condition
at $0\in I$ and the Dirichlet boundary conditions on $\df I=\{-1,1\}$
is given by the kernel
\begin{align}
\left(G_{I}(\nu)\right)_{x_{1},x_{2}} & =
g_{x_{1},x_{2}}+{\beta^2-\alpha^2\over \alpha^2+\beta^2}\,g_{-x_{1},x_{2}}
\text{ for } x_{1}, x_{2}\in Q_{1}=[-1,0], \notag \\
\left(G_{I}(\nu)\right)_{x_{1},x_{2}} & =
g_{x_{1},x_{2}}+{\alpha^2-\beta^2\over \alpha^2+\beta^2}\,g_{-x_{1},x_{2}}
\text{ for } x_{1}, x_{2}\in Q_{2}=[0,1], \label{S1240}\\
\left(G_{I}(\nu)\right)_{x_{1},x_{2}} & =
{2\alpha\beta\over\alpha^2+\beta^2}\,g_{x_{1},x_{2}}
\quad\qquad\;\text{ for } x_{1}, x_{2} \text{ from different } Q_k. \notag
\end{align}
Here, $g_{x_{1},x_{2}}$ is the Green function for the Laplacian
on functions on $I$ with the Dirichlet boundary conditions on $\df I$:
\begin{equation}
g_{x_{1},x_{2}}=
\begin{cases}
c\cdot(x_{2}+1)(1-x_{1}), & -1\le x_{2}\le x_{1}\le 1, \\
c\cdot(x_{1}+1)(1-x_{2}), & -1\le x_{1}\le x_{2}\le 1,
\end{cases}
\label{S1241}
\end{equation}
where $c\ne 0$ is a constant.

It follows from (\ref{S1240}) and (\ref{S1241}) that $G_{I}(\nu)$ has
a continuous kernel on $\Qq_{r_{1}}\times\Qq_{r_{2}}$ and that
it is estimated uniformly with respect to $\nu\in\rbo$ and to $\xa,\xb$:
\begin{equation}
\sup_{x_{1},x_{2},\nu}\left|\left(G_{I}(\nu)\right)_{x_{1},x_{2}}\right|
< c_{2}.
\label{S1242}
\end{equation}

Since $\supp\omega_{0}\subset(I\setminus\df I)\times N$ and since
the Laplacian $\Delta_{\nu,I}$ has the zero kernel on functions with the
Dirichlet boundary conditions on $\df I$, we have
\begin{equation}
\omega_{0}=\left(\id_{I}\otimes\Pii_{0}(N)\right)\omega_{0}+G_{I}(\nu)
\otimes G_{N}^{m_{2}}\left(\left(\Delta_{\nu,I}\otimes\Delta_{N}^{m_{2}}
\right)\omega_{0}\right),
\label{S1245}
\end{equation}
where $G_{N}$ is the Green function for $\doo_{N}$ and where
$\Pii_{0}(N)$ is the orthogonal projection operator
in $\left(\DR(N)\right)_{2}$ onto $\Ker\doo_{N}$. The operator
$G_{N}^{m_{2}}$ on a closed Riemannian manifold $\left(N,g_N\right)$ has
a square-integrable kernel (relative to the second argument)
for $m_{2}>(n-1)/4$ (where $n-1=\dim N$) and it has a continuous
on $N\times N$ kernel for $m_{2}>(n-1)/2$.

The following estimate holds uniformly with respect to $\nu\in\rbo$
for any $m_2\in\wZ_+$, $m_{2}>(n-1)/4$. From (\ref{S1245}),
(\ref{S1242}), and from the Cauchy inequality we have
\begin{equation}
\left|\omega_{0}(x)\right|^2\le c_{5}\left(\left\|\omega_{0}\right\|^2_{2}+
\left\|\left(\Delta_{\nu,I}\otimes\id\right)\omega_{0}\right\|^2_{2}+
\left\|\left(\Delta_{\nu,I}\otimes\Delta_{N}^{m_{2}}\right)\omega_{0}
\right\|^2_{2}\right).
\label{S1246}
\end{equation}
Indeed, the following two Banach norms on the finite-dimensional
space $\Ker\doo_{N}$
$$
\left\|h\right\|^2_{B}:=\max_{x\in N}\left|h(x)\right|^2\quad\text{and}\quad
\left\|h\right\|^2_{2,N}
$$
are equivalent. So we get (where $x=(x_{1},x_{N})\in I\times N$
and $I=[-1,1]$):
\begin{gather}
\left|\left(\left(\id_{I}\otimes\Pii_{0}(N)\right)\omega_{0}\right)(x)
\right|^2\le\left\|\left(\Pii_{0}(N)\omega_{0}(x_{1},*)\right)\right\|^2_{B}
\le c_{6}\left\|\omega_{0}(x_{1},*)\right\|^2_{2,N}, \label{S1500}\\
\left\|\omega_{0}(x_{1},*)\right\|^2_{2,N}\!\le\!2\sup_{x_{2}}\left|
\left(G_{I}(\nu)\right)_{x_{1},x_{2}}\right|^2\!\cdot\!\left\|\left(\Delta
_{\nu,I}\!\otimes\!\id\right)\omega_{0}\right\|^2_{2,M}\!\le\!2c^2_{2}\!\left\|
\left(\Delta_{\nu,I}\!\otimes\!\id\right)\!\omega_{0}\right\|^2_{2}.
\label{S1501}
\end{gather}

The following estimate is obtained by the similar method:
\begin{multline}
\left|\left(\left(G_{I}(\nu)\otimes G_{N}^{m_{2}}\right)\left(\Delta_{\nu,I}
\otimes\Delta_{N}^{m_{2}}\right)\omega_{0}\right)(x_{1},x_{N})\right|^2\le \\
\le 2c^2_{2}\sup_{y_{1}}\left\|\left(G_{N}^{m_{2}}\right)_{y_{1},*}
\right\|^2_{2,N}\cdot\left\|\left(\Delta_{\nu,I}\otimes\Delta_{N}^{m_{2}}
\right)\omega_{0}\right\|^2_{2,M}.
\label{S1502}
\end{multline}

Hence the estimate (\ref{S1246}) holds for $\omega_{0}$ (even without
the first term on the right in (\ref{S1246})), as follows from (\ref{S1245}),
(\ref{S1500}), (\ref{S1501}), and (\ref{S1502}).

Since $\den=\id_{I}\otimes\Delta_{N}+\den(I)\otimes\id_{N}$ and
since $\Delta_{N}$ and $\Delta_{\nu,I}$ are nonnegative self-adjoint
operators, we have for $m_{2}\in\wZ_+$:
\begin{equation}
\left\|\left(\Delta_{\nu,I}\otimes\Delta_{N}^{m_{2}}\right)\omega_{0}
\right\|_{2}\le \left\|\den^{m_{2}+1}\omega_{0}\right\|_{2}.
\label{S1247}
\end{equation}

The inequality (\ref{S1094}) for $\omega_{0}$ follows from (\ref{S1246})
and (\ref{S1247}).
For $\omega_{1}$ the analogous to (\ref{S1245}) equality holds:
\begin{multline}
\omega_{1}=\left(\Pi^1_{0}(I_\nu)\otimes\left(\id_{N}-\Pi^{\bullet-1}_{0}
(N)\right)\right)\omega_{1}+
\left(\left(\id_{I}-\Pi^1_{0}(I_\nu)\right)\otimes\Pi^{\bullet-1}_{0}(N)
\right)\omega_{1}+\\
+\left(\Pi^1_{0}(I_\nu)\otimes\Pi^{\bullet-1}_{0}(N)\right)
\omega_{1}+
\left(G_{I}(\nu)\otimes G^{m_{2}}_{N}\right)\left(\Delta_{\nu,I}\otimes
\Delta^{m_{2}}_{N}\omega_{1}\right),
\label{S1503}
\end{multline}
where $\Pi^1_{0}(I_\nu)$ is the projection operator
of $\left(DR^1(I)\right)_{2}$ onto the one-dimensional space
$c\cdot dx_{1}$ and $G_{I}(\nu)_{1}$ is the Green function
for the Laplacian $\Delta_{\nu,I}$ on $DR^1(I_\nu)$ (with the Dirichlet
boundary conditions on $\df I=\{-1,1\}$ and with the $\nu$-transmission
boundary conditions at $0$). The kernel $G_{I}(\nu)_{1}$ is continuous
on $\Qq_{r_{1}}\times\Qq_{r_{2}}$ because it can be written in a form
similar to (\ref{S1240}). It is written through the Green function $g_{1}$
of $\Delta_{I}$ on $DR^1(I)$ with the Dirichlet boundary conditions
on $\df I$ where the kernel $(g_{1})_{\xa,\xb}$ of $g_1$ is
continuous on $I\times I$. Hence the second term on the right
in (\ref{S1503}) is estimated similarly to (\ref{S1500}) and to (\ref{S1501}).
The kernel $\Pi^1_{0}(I_\nu)_{\xa,\xb}$ is expressed
in a form analogous to (\ref{S1240}) through the kernel
$2^{-1}d\xa\otimes d\xb$ on $I\times I$ (corresponding
to $\Pi^1_{0}(I_{1,1})$). So the kernel of $\Pi^1_0\left(I_{\nu}\right)$
is continuous on $\Qq_{r_{1}}\times\Qq_{r_{2}}$,
and it satisfies the estimate (\ref{S1242}) (with the upper bound $c$).
The first and the third terms in (\ref{S1503}) are estimated as follows:
\begin{align*}
\left|\left(\Pi^1(I_\nu)\otimes\left(\id_{N}-\Pi^{\bullet-1}_{0}(N)\right)
\right)\oma(x_{1},x_{N})\right|^2 & \!\le
\!2c^2\sup_{y_{1}}\!\left\|\left(G_{N}^{m_{2}}\right)_{y_{1},*}\right\|^2
_{2,N}\!\left\|\left(\id\otimes\Delta_{N}^{m_{2}}\right)\oma\right\|^2_{2,M},\\
\left|\left(\Pi^1(I_\nu)\otimes\Pi^{\bullet-1}_{0}(N)\right)\oma(x)\right|^2 &
\le 2c^2 c_{6}\left\|\oma\right\|^2_{2,M}.
\end{align*}

Hence the estimate (\ref{S1246}) holds uniformly with respect to $\nu\in\rbo$
for any $m\in\wZ_+$, $m\ge m_{0}:=1+\min\{k\in\wZ_+, 4k\ge n\}$.
Thus Proposition \ref{PA150} is proved.\ \ \ $\Box$

\subsection{Actions of the homomorphisms of identifications on
the determinant. Proof of Lemma 2.3} The most simple method to compute
the action of $g_{\nu *}$ on ${\Det}(M,N,Z)$%
\footnote{This action is multiplying by a nonzero factor.}
is to obtain the expression for the action of $v^c_{\nu *}$
on the determinant line $\Det\Ce(X,V,W)$ (\ref{K71}), induced
by the identifications of the corresponding cochain complexes
$v^c_\nu=v^c_{\nu\nu_{0}}\colon\Ce(X_{\nu_{0}},V)\to \Ce(X_\nu,V)$ (where
$v^c_\nu(c_{1},c_{2}):=\left(c_{1},(k_{\nu}/k_{\nu_{0}})c_{2}\right)$
for $\nu,\nu_0\in U$ (\ref{ST3})), and then to use Proposition~\ref{P10}
below. The action of $v^c_{\nu *}$ is defined
by identifications $\psi_\nu$ and $\psi_{\nu_{0}}$, where
$\psi_\nu\colon\det\Ce(X_\nu,V)\rs$\\${\Det}\Ce(X,V,W)$ are defined
by the exact sequence in the bottom row of the diagram (\ref{LL1}).
The following diagram of the identifications is commutative:
\begin{gather*}
\begin{CD}
\det\Ce(X_{\nu_{0}},V) @>{v^c_\nu}>\lrs>  \det\Ce(X_\nu,V)\\
{\sst{\psi_{\nu_{0}}}}\,\big\downarrow\,\wr @.{\sst{\psi_\nu}}\,\big\downarrow
\,\wr @. \\
{\Det}\Ce(X,V,W) @>{v^c_{\nu *}}>\lrs> {\Det}\Ce(X,V,W) \ \ .
\end{CD}
\end{gather*}
\begin{pro}
Under the conditions of Lemma~\ref{L2.3}, the equality holds:
\begin{equation}
g_{\nu *}=v^c_{\nu *}.
\label{S401}
\end{equation}
\label{P10}
\end{pro}

The proof of the equality (\ref{S401}) is done just after the end
of the proof of Lemma~\ref{L2.3}. The expression for the action
of $v^c_{\nu *}$ on the determinant line can be obtained as follows.
Let $\nu\in U$ and let $j$ be the natural inclusion
$j\colon\oplus\kab\Ce(X_k,W\oplus V_k)\to\Ce(X_\nu,V)$.
Then $v^c_\nu$ acts on $\Ce(X_{1},W\cup V_{1})$ as the identity operator
and it acts on $\Ce(X_{2},W\cup V_{2})$ as the operator $(\kn/\kno)\id$.
Proposition~\ref{P2.7} claims that the identification $\psi_\nu$
is an isometry between the combinatorial norm on $\det\Ce(X_\nu,V)$
and the triple tensor product of the combinatorial norms on the components
of ${\Det}\Ce(X,W,V)$. It is enough to compute the action of $\vcna$
on the component $\det\Ce(W)$ of the tensor product ${\Det}\Ce(X,W,V)$.
The inclusion $\ronce\colon\Ce(W)\hookrightarrow\Ce(X_\nu,V)$ (defined
by (\ref{LL8})) is an isometry onto orthogonal complement to $\Image j$
and $\renc\ronce=\id$ on $\Ce(W)$. So the action of
$v^c_\nu$ on this orthogonal complement $\left(\Image\,j\right)^{\perp}$
(identified with $\Ce(W)$ by $r_{\nu_0,c}$ and by $\renc$) can be expressed
as the composition
\begin{multline}
m\in\Ce(W)@>>{\rho_{\nu_0,c}}>
\left((\bo,\ao)\big/\sqrt{\ao^2+\bo^2}
\right)m @>>{v^c_\nu}> \left(\left(\bo,\ao(\kn/\kno)\right)\big/
{\sqrt {\ao^2+\bo^2}} \right)m=\\
=\left((1,\kn)/\sqrt{1+\kno^2}\right)\,m @>>{\renc}>
\left(\sqrt{1+\kn^2}\big/ \sqrt{1+\kno^2}\right)\,m\in\Ce(W).\ \
\label{S410}
\end{multline}
(Here, the signs are written for positive $\beta$ and $\bo$.%
\footnote{The signs are not important for the transformations of the norm
on the determinant line under the actions of $\vcna$.})
The expression for $\vcna$ follows from (\ref{S410}) and from the assertion
that $v^c_\nu$ acts on $\Ce\left(X_2,W\cup V_2\right)$ as
$\left(\kn/\kno\right)\id$. Namely
\begin{equation}
\vcna l=(\kn/\kno)^{-\chi(M_{2},N\cup Z_{2})}\left(\left(1+\kn^2\right)/
\left(1+\kno^2\right)\right)^{-\chi(N)/2}l.
\label{S411}
\end{equation}
for $l\in{\Det}\Ce(X,V,W)$.
It follows from (\ref{S411}) that the equality holds (for $l\ne 0$):
\begin{equation}
\df_{\gamma}\log\|\vcna l\|^2\!=\!-2\chi(M_{2},N\!\cup\! Z_{2})\dfg\!\log(\kn)
\!-\!2\chi(N)\!\left(1+\kn^{-2}\right)^{-1}\!\dfg\!\log(\kn).
\label{S412}
\end{equation}
Proposition \ref{P10}\ claims that the same identity holds also
for the action of $g_{\nu *}$ on ${\Det}(M,N,Z)$.

The right side in the formula (\ref{ST34}) (i.e., in the assertion
of Lemma~\ref{L2.3}) is defined in analytic terms while the right side
in (\ref{S412}) is defined in topological terms.
Each $b_{2,j}\left(\mno,Z\right)$
on the right in (\ref{ST34}) is the sum of integrals over $M_{2}$ and
over $N$ of the locally defined densities, according to Theorem~\ref{TAT706}.
So it is enough to compute (in topological terms) the expression
on the right in (\ref{ST34}) in the case of a mirror-symmetric
$\Me=M_{2}\cup_{N}M_{2}$ with a mirror-symmetric metric $g_{\Me}$
(which is a direct product metric near $\df\Me$ and near $N$) and with
mirror-symmetric boundary conditions on the connected components of $\df\Me$.
In this case, the expression in (\ref{ST34}) is the same as for
a general $M$ (if the piece $M_{2}$ of $M$, $g_{M}|_{TM_{2}}$, and
the boundary conditions on $\df M\cap M_{2}$ are the same as
in the mirror-symmetric case on each piece $M_2$ of $\Me$). It is supposed
from now on in the proof of Lemma~\ref{L2.3} that $M$ and all the data
on $M$ are mirror-symmetric relative to $N$. In this case the kernel
$\eotxy(\nu)$ ($\nu\in\rbo$) of the operator $\exp\left(-t\don\right)$
with the Dirichlet boundary conditions on $Z=Z_2\cup Z_2\subset\df M$ and
with the Neumann conditions on $\df M\setminus N$ is expressed through
the fundamental solution $\eotxy$ for $\df_t+\Delta^{\bullet}$ on $\DR(M,Z)$
(with the same boundary conditions on $\df M$)%
\footnote{It is proved in Proposition~\ref{P1} that $\Delta^{\bullet}$
on $\DR(M,Z)$ for $M$ obtained by gluing two pieces, $M=M_1\cup_NM_2$,
has the same eigenvalues (including their multiplicities)
and eigenforms as $\doaa$ in $\DR(M_{1,1},Z)$.
The analogous assertion is true for the operators $\exp(-t\doaa)$ and
$\exp(-t\Delta^{\bullet})$ in  $\dromb$ and for their kernels.}
as follows%
\footnote{These formulas are analogous to (\ref{S1240}).}%
:
\begin{gather}
\eotxy(\nu)\!=\!\eotxy\!+\!\left(\left(\alpha^2-\beta^2\right)\!/\!\left(
\alpha^2+\beta^2\right)\right)\!(\sigma^*_{1}E^{\bullet})_{t,x,y}\text{ for }
x\in M_{2}, y\in M_{2},
\label{S414}\\
\eotxy(\nu)=\left(2\alpha\beta/\alpha^2+\beta^2\right)\eotxy\text{ for }
x\in M_{1}, y\in M_{2}.
\label{S415}
\end{gather}

Note that the kernel $(E_t+\sigma^*_{1}E_t)_{x,y}=:E^{Neu}_{t,x,y}$
for $x,\,y\in M_{2}$ is the fundamental solution for $\df_t+\domb$,
where $\domb$ is the Laplacian on $\DR(M_{2},Z_{2})$ with the Neumann
boundary conditions on $N$ and the kernel
$(E_t-\sigma^*_{1}E_t)_{x,y}=:E^{Dir}_{t,x,y}$ is the kernel
of $\exp(-t\dombn)$, where $\dombn$ is the Laplacian
on $\DR(M_{2},N\cup Z_{2})$ i.e., with the Dirichlet boundary conditions
on $N$. It follows from (\ref{S414}) that the alternating sum
of zero-order terms (in the asymptotic expansions of the traces
of the heat equation operator relative to $t\to +0$) on the right
in (\ref{ST34}) can be represented in the following form (where
$m_{\nu_{0}}:=2^{-1}\left(1-\kno^{-2}\right)\big/\left(1+\kno^{-2}\right)$):

\begin{multline}
\sum (-1)^j b_{2,j}\left(\mno,Z\right)=
\sum (-1)^j\int_{M_{2}}\tr\left(E^j_{t,x,x}(\nu_{0})\right)^0=\\
=2^{-1}\sum (-1)^j\int_{M}\tr\left(E^j_{t,x,x}\right)^0+
m_{\nu_{0}}\sum(-1)^j\int_{M_{2}}\left(\tr\left(\Ejn\right)^0-
\tr\left(\Ejd\right)^0\right)=\\
=\!2^{-1}\chi(M,Z)+m_{\nu_{0}}\!\left(\chi(M_{2},Z_{2})\!-
\!\chi(M_{2},Z_{2}\cup N)\right)\!=\!\chi(M_{2},Z_{2}\cup N)+\left(1\!+
\!\kno^{-2}\right)^{-1}\!\chi(N).
\label{S416}
\end{multline}

Hence the expression on the right in (\ref{S412}) is equal to the right side
of (\ref{ST34}), and the assertion of Lemma~\ref{L2.3} follows from
Proposition \ref{P10}. The zero superscripts in (\ref{S416}) denote
the densities on $M_j$, $N$, $\df M$, corresponding
to the constant terms (i.e., the $t^0$-coefficients) in the asymptotic
expansions as $t\to +0$ for $\Tr\left(p_j\exp\left(-t\doo\right)\right)$,
where $\doo$ is the Laplacian with appropriate boundary conditions.
In (\ref{S416}) $\int_{M_j}\tr(\centerdot)^0$ denotes the sum
of the integrals over $M_j$, $N$, and over $\df M_j\setminus N$
of the corresponding densities. We use the following equalities
to produce (\ref{S416}):
\begin{align}
\sum (-1)^j\int_{M}\tr\left(\;E^j_{t,x,x}\;\right)^0 & =\chi(M,Z),
\label{S417}\\
\sum  (-1)^j\int_{M_{2}}\tr\left(\Ejn\right)^0 & =\chi(M_{2},Z_{2}),\notag\\
\sum (-1)^j\int_{M_{2}}\tr\left(\,\Ejd\,\right)^0 &=\chi(M_{2},N\cup Z
_{2}). \notag
\end{align}
These equalities are consequences of the analogous equalities without
the zero superscripts and of the existence of asymptotic expansions
in powers of $t$ for the corresponding traces as $t\to +0$
(\cite{Se2}, Theorem~3, or Theorem~\ref{TAT706} below).\ \ \ $\Box$

\noindent{\bf Proof of Proposition \ref{P10}.} The identifications
\begin{equation}
\Det\Ce(X,V,W)\overset{d_c}{@>>\lrs>}\Det(X,V,W)\overset{R}{@<<\lrs<}
\Det(M,N,Z)
\label{S420}
\end{equation}
do~not depend on $\nu$ (the determinant lines in (\ref{S420})
are defined in (\ref{K71})). So the actions of $v^c_{\nu *}$
on $\Det(X,V,W)$ and on $\Det\Ce(X,V,W)$ are the same (i.e., they
multiply by the same number). To prove (\ref{S401}) it is enough
to show that the corresponding operators on $\Det(M,Z,N)$ are the same
(i.e., that $g_{\nu *}=\vcna$ on $\det(M,N,Z)$).

The proof of Proposition \ref{P10}\ uses the following assertion.
\begin{pro}
Let $\phi\colon(\Fe_{0},d_{F_{0}})\to(\Fe_{1},d_{F_{1}})$
be an isomorphism of finite complexes
of finite-dimensional linear spaces. Then the diagram is commutative:
\begin{gather}
\begin{CD}
\det\Fe_{0}     @>{\phi}>\lrs>               \det\Fe_{1}\\
{\sst{d}}\,\big|\,\wr @.              {\sst{d}}\,\big|\,\wr @. \\
\det\Hh(F_{0})  @>{\phi_*}>\lrs> \det\Hh(F_{1})
\end{CD}
\label{S501}
\end{gather}
\label{P11}
\end{pro}

\noindent{\bf Proof.}
The identifications $\det\Fe_j\stackrel{d}{\eqsim}\det\Hh(F_j)$
are defined with the help of differentials $d=d_{F_j}$.
Hence the commutativity of (\ref{S501}) holds.\ \ \ $\Box$

The commutativity of the following diagram of the identifications
(for $\nu$ sufficiently close to $\nu_0$ such that $v_\nu$
is an isomorphism) follows from (\ref{S501}):
\begin{gather}
\begin{array}{ccccc}
\det\Wano & @>{v_\nu}>\lrs> & \det v_\nu\left(\Wano\right)
& @>{\Pi_a}>\lrs> & \det\Wano \\
{\sst{d}}\,\big|\,\wr & & {\sst{d}}\,\big\downarrow\,\wr & & \big|\!\,\wr \,
\sst{d}  \\
\det\Hh(\mno,Z) & @>>{\vna}> & \det\Hh(M_\nu,Z) & = & \det\Hh(M_\nu,Z)
\end{array}
\label{S502}
\end{gather}
where the identification $j_*\colon\Hh\left(v_\nu\left(\Wano\right)\right)
\to\Hh\left(DR(M_\nu,Z)\right)=\Hh(M_\nu,Z)$ is defined by the natural
inclusion $j\colon v_\nu\left(\Wano\right)\hookrightarrow\DR(M_\nu,Z)$
of a quasi-isomorphic subcomplex.
The commutativity of the left square in (\ref{S502}) follows from
(\ref{S501}). The commutativity of the right square in (\ref{S502})
also follows from (\ref{S501}) because the operator induced
by the projection operator $\Pi_a$ on $\Hh\left(DR(M_\nu,Z)\right)$
is the identity operator.

The commutativity of the following diagram is a consequence
of the commutativity of the diagram (\ref{S502}):
\begin{gather}
\begin{array}{ccccc}
\det\Wano & = & \det\Wano & @>{v_\nu}>\lrs> & \det v_\nu\left(\Wano\right) \\
 & & {\sst{d}}\,\big|\,\wr & & {\sst{d}}\,\big|\,\wr \\
{\sst{\phi_{\nu_{0}}(a)}}\,\Bigg|\,\wr & & \det\Hh(\mno,Z) & @>>\vna>
& \det\Hh(M_\nu,Z)\\
 & & {\sst{\fano}}\,\big|\,\wr & & {\sst{\fan}}\,\big|\,\wr \\
\Det(M,N,Z) & = & \Det(M,N,Z) & @>{g_{\nu *}}>\lrs> & \Det(M,N,Z)
\end{array}
\label{S503}
\end{gather}

The action of $\vna\colon\Hh\left(\mno,Z\right)\to\Hh\left(M_\nu,Z\right)$
coincides with the combinatorial action
$\vcna\colon\Hh\left(X_{\nu_{0}},V\right)\to\Hh\left(X_\nu,V\right)$
under the identification of the cohomology
$R\colon\Hh\left(M_\nu,Z\right)\to\Hh\left(X_\nu,V\right)$ induced
by the integration $R$ of closed differential forms over
the simplexes of $X$. Hence the commutativity of (\ref{S502}) involves
also the commutativity of the diagram:
\begin{gather}
\begin{array}{ccccc}
 & & \det\Wano & @>{v_\nu}>\lrs> & \det v_\nu\left(\Wano\right) \\
 & & {\sst{d}}\,\big|\,\wr & & {\sst{d}}\,\big|\,\wr \\
\det\Hh(\mno,Z) & = & \det\Hh(\mno,Z) & @>>\vna> & \det\Hh(M_\nu,Z)\\
 & & {\sst{d_c}}\,\big|\,\wr & & {\sst{d_c}}\,\big|\,\wr \\
{\sst{\phi^c_{\nu_{0}}=\fano}}\,\Bigg{\downarrow}\,\wr &
& \det\Ce(X_{\nu_{0}},V) & @>>{v^c_\nu\ }> & \det\Ce\left(X_\nu,V\right) \\
 & & {\sst{\psi_{\nu_{0}}}}\,\big|\,\wr & & {\sst{\psi_\nu}}\,\big|\,\wr \\
 & & \Det\Ce(X,V,W) & @>>\vcna> & \Det\Ce(X,V,W) \\
 & & {\sst{(\ref{S420})}}\,\big|\,\wr & & {\sst{(\ref{S420})}}\,\big|\,\wr \\
\Det(M,N,Z)\ & = & \Det(M,N,Z) & @>>\vcna> & \Det(M,N,Z)
\end{array}
\label{S504}
\end{gather}

The equality (\ref{S401}) follows immediately from the commutativity
of the right bottom square in (\ref{S503}) and from the commutativity
of (\ref{S504}). Proposition \ref{P10}\ is proved.\ \ \ $\Box$

\subsection{Analytic torsion norm on the cone of a morphism of
complexes. Proof of Lemma~2.4}
Lemma \ref{L2.4} is a particular case of the following assertion.
Let $f$ be a morphism (\ref{S303}) of finite complexes of finite-dimensional
Hilbert spaces.%
\footnote{The morphism $f$ {\em does~not supposed} to be
a quasi-isomorphism.}
Then $\Co f$ is defined by (\ref{X100}).
The exact sequence of complexes:%
\footnote{$\Aa[1]$ is a complex with components $A[1]^j=A^{j+1}$
and with $d_{A[1]}=-d_{A}$. There are the canonical identifications:
$\det\Aa[1]=(\det\Aa)^{-1}$ and
$\det\Hh(A[1])=\left(\det\Hh(A)\right)^{-1}$.}
\begin{equation}
0\to \Ve\to\Co f @>>p>\Aa[1]\to 0,
\label{X101}
\end{equation}
(where the left arrow maps $y\in\Ve$ into $(0,y)\in\Co f$ and
$p(x,y)=x$ for $(x,y)\in A^{j+1}\oplus V^j)$ defines the identification
of the determinants of its cohomology:
\begin{equation}
\phi^H_{\Co f}\colon\det\Hh(\Cone f)\rs\det\Hh(V)\otimes
\left(\det\Hh(A)\right)^{-1}.
\label{X102}
\end{equation}

Let the Hilbert spaces $\Cone^j f$ be the direct sums $A^j\oplus V^{j+1}$
of the Hilbert  spaces.
\begin{lem}
The analytic torsion norm on the determinant of the cohomology
of $\Co f$ is isometric under the identification (\ref{X102})
to the tensor product of the analytic torsions norms%
\footnote{The analytic torsion norm on $\det\Hh(A[1])=\det\Hh(A)^{-1}$
is the dual norm $T_{0}(\Aa)^{-1}$.}:
\begin{equation}
T_{0}(\Co f)=T_{0}(\Ve)\otimes T_{0}(\Aa)^{-1} .
\label{X103}
\end{equation}
\label{L2.6}
\end{lem}

\begin{rem}
Let $h\in\det\Hh(\Cone f)$ be identified by (\ref{X102})
with $\ha\otimes\hb^{-1}$ for $\ha\in\det\Hh(V)$ and $\hb\in\det\Hh(A)$.
Namely
$\phi_{\Co f}h=\ha\otimes\hb^{-1}$,
where $\hb^{-1}$ is an element of the dual one-dimensional space
$\det\Hh(A)^{-1}$ such that $\hb^{-1}(\hb)=1$. In this case, the equality
(\ref{X103}) claims that
\begin{equation}
\left\|h\right\|^2_{T_{0}(\Co f)}=\left\|\ha\right\|^2_{T_{0}(\Ve)}\big/
\left\|\hb\right\|^2_{T_{0}(\Aa)}.
\label{X105}
\end{equation}
\end{rem}

\begin{rem}
The identity (\ref{X103}) (Lemma~\ref{L2.6}) and the equality
(\ref{X105}) also hold under weaker assumptions.
Let the Hilbert structures on $\Aa$, $\Ve$, and on $\Co f$ be such
that the identification
\begin{equation}
\phi_{\Co f}\colon\det\Co f\rs\det\Ve\otimes(\det\Aa)^{-1}
\label{X108}
\end{equation}
(induced by the exact sequence (\ref{X101})) is an isometry.
Then the equality (\ref{X105}) holds.
\end{rem}

\begin{cor}
Let $f:\Aa\to\Ve$ be a quasi-isomorphism. Then $\Hh(\Cone f)=0$. Hence
$\det\Hh(\Cone f)$ is canonically $\wC$ and $1\in{\wC}=\det\Hh(\Cone f)$
is identified by (\ref{X102}) with $\ha\otimes\hb^{-1}$. Here, $f_*\hb=\ha$
under the identification induced by $f$:
\begin{equation}
f_*\colon\det\Hh(A)\rs\det\Hh(V).
\label{X106}
\end{equation}
In this case, the equality (\ref{X105}) claims that
\begin{equation}
\left\|1\right\|^2_{T_{0}(\Cone f)}=
\left\|\ha\right\|^2_{T_{0}(\Ve)}\big/\left\|\hb\right\|^2_{T_{0}(\Aa)}=
T_{0}(\Ve)/T_{0}(\Aa),
\label{X107}
\end{equation}
where $T_{0}(\Ve)/T_{0}(\Aa)$ is the ratio of the two norms
on the same determinant line (since $\det\Hh(V)$ and $\det\Hh(A)$
are identified by (\ref{X106})).
\label{C2.19}
\end{cor}
The equality (\ref{X107}) is the assertion of Lemma~\ref{L2.4}.

\noindent{\bf Proof of Lemma~\ref{L2.6}.} The identification (\ref{X108})
is an isometry of norms on the determinant lines. Let $u$ be a nonzero
element of $\det\Co f$ and let
\begin{equation}
\phi_{\Co f}u=\ua\otimes\ub^{-1},
\label{X109}
\end{equation}
where $\ua\in\det\Ve$ and $\ub\in\det\Aa$. Let $h, \ha, \hb$ be the images
of $u$, $\ua$, $\ub$ under the identifications (defined by the differentials
of the corresponding complexes):
\begin{gather*}
\begin{split}
\det\Co f & \stackrel{d_{\Co f}}{\es}\det\Hh(\Cone f),\\
\det\Aa     \stackrel{d_{A}}{_{\es}}\det\Hh(A), \quad &
\det\Ve     \stackrel{d_{V}}{_{\es}}\det\Hh(V).
\end{split}
\end{gather*}
Then by the definition of ${\phi^H}_{\Co f}$ we have
${\phi^H}_{\Co f}h=\ha\otimes\hb^{-1}$.

The analytic torsion norm on the determinant of the cohomology
of a finite-dimensional complex is the norm, corresponding
to the $L_{2}$-norm on the determinant of this complex defined
by the Hilbert structures on its components. Hence the equalities hold:
\begin{gather}
\begin{split}
\left\|h\right\|^2_{T_{0}(\Co f)} & = \left\|u\right\|^2_{\det(\Co f)},\\
\left\|\ha\right\|^2_{T_{0}(\Ve)}   = \left\|\ua\right\|^2_{\det(\Ve)},
 & \quad
\left\|\hb\right\|^2_{T_{0}(\Aa)}   = \left\|\ub\right\|^2_{\det(\Aa)}.
\end{split}
\label{X112}
\end{gather}

Since the identification (\ref{X108}) is an isometry, we see that
the equality
\begin{equation}
\left\|h\right\|^2_{T_{0}(\Co f)}=\left\|\ha\right\|^2_{T_{0}(\Ve)}
\big/\left\|\hb\right\|^2_{T_{0}(\Aa)}
\label{X114}
\end{equation}
follows from (\ref{X112}) and from (\ref{X109}).\ \ \ $\Box$

The equality (\ref{S304}) in Lemma~\ref{L2.4} is a particular case
of (\ref{X114}) (by Corollary~\ref{C2.19}) corresponding to the case
of a quasi-isomorphism $f$. Lemma~\ref{L2.4} is proved.\ \ \ $\Box$

\subsection{Variation formula for norms of morphisms of identifications.
Proof of Lemma 2.1}

The assertion (\ref{ST23}) of Lemma~\ref{L2.2} can be deduced
from the definition of $g_\nu$ and from the following proposition.
\begin{pro}
There is a neighborhood $\uno$, $\nu_{0}\in\uno\subset\rbo$, such that
a family of finite rank projection operators $\Pi^j_a(\nu):= \Pi_j(\nu;a)$
in $\left(DR^j(M)\right)_{2}$ is smooth on $\uno\ni\nu$.

\label{P19}
\end{pro}

Its proof follows just after the proof of Lemma \ref{L2.2}.
Let $l\in\det\Wano$, $l\ne 0$, and let
$l=\otimes_j l_j^{(-1)^{j+1}}$,
where $l_j\in\det W^j_a(\nu_{0})$, $l_j\ne 0$.
Then we have
\begin{equation}
\log\|g_{\nu}l\|^2_{\dew}=\sum (-1)^{j+1}\log\|g_{\nu}l_j\|^2_{\dewj}.
\label{X506}
\end{equation}

\begin{pro}
For every $j$ the following equality holds (under the conditions
of Lemma~\ref{L2.2}):
\begin{equation}
\dfg\log\left\|g_{\nu}l_j\right\|^2_{\dewj}\bgo=2\,\dfg
\log(\kn)\bgo\Tr \left(p_{2}\Pi^j_a(\nu_{0})\right).
\label{X505}
\end{equation}
\label{P20}
\end{pro}

Thus the assertion (\ref{ST23}) of Lemma~\ref{L2.2} is a consequence
of (\ref{X505}) and of (\ref{X506}).

\noindent{\bf Proof of Proposition \ref{P20}.} It is enough to prove
the equality (\ref{X505}) in the case when $\left\|l_j\right\|^2_{\dewo}=1$,
i.e., when $l=e_{1}\wedge\ldots\wedge e_m$, where $\{e_i\}$ is an orthonormal
basis in $\Wano$ ($\nu_{0}=\nu(0)$). In this case, we have
\begin{equation}
\dfg\log\|g_{\nu}l\|^2_{\dewj}\bgo=\tr\left(\dfg
\left(a_{ij}(\nu)\right)\right)\bgo,
\label{X508}
\end{equation}
where $a(\nu)=\left(a_{ij}(\nu)\right)$ is a matrix of scalar products in
$\left(DR^j\left(\Mm_{1}\right)\oplus DR^j\left(\Mm_{2}\right),
g_{M}\right)$ of the images of the basis elements
$$
a_{ij}(\nu):=<g_\nu e_i,g_\nu e_j>.
$$
The formula (\ref{X508}) is deduced as follows. Since
$\left\|g_\nu l\right\|^2_{\dewj}=\det\,a(\nu)$, we have
\begin{equation*}
\dfg\log\det\,a(\nu)\bgo=\tr\left(\dfg a\cdot a^{-1}(\nu)\right)\bgo=
\tr(\dfg a)\bgo\ .
\end{equation*}

The family of the operators $\Pi^j_a(\nu)$ is smooth in $\nu$
for $\nu\in\uno$ (Proposition~\ref{P19}). Hence the operator
$\df_{\gamma}\Pi^j_a(\nu)$ exists.
Since $\Pi^j_a(\nu)$ are projection operators, we have
\begin{equation*}
\Pi^2_a=\Pi_a,\qquad \dfg\Pi_a\cdot\Pi_a=(id-\Pi_a)\dfg\Pi_a.
\end{equation*}
So $e_i$ are orthogonal to $\dfg\Pi_a\bgo e_j$ and
we get
\begin{multline}
\dfg a_{ij}(\nu)\bgo=\dfg<\Pi_a g_{\nu}e_i,\Pi_a g_{\nu}e_j>\bgo =
<\dfg(\Pi_a g_\nu)e_i,e_j>\bgo+\\
+<e_i,\dfg(\Pi_a g_\nu)e_j>\bgo=\dfg(\log\kn)\bgo(<p_{2}e_i,e_j>+
<e_i,p_{2}e_j>).
\label{X511}
\end{multline}

Since $\{e_i\}$ is an orthonormal basis in $\Wano$, we have
\begin{equation}
\tr\left(<p_{2}e_i,e_j>\right)=\tr\left(p_{2}\Pi^j_a(\nu_{0})\right)
\label{X512}
\end{equation}
Hence (\ref{X505}) follows from (\ref{X508}), (\ref{X511}), and (\ref{X512}).
Thus Proposition~\ref{P20} and Lemma~\ref{L2.2} are proved.\ \ \ $\Box$

\noindent{\bf Proof of Proposition~\ref{P19}.} Let $\uno\subset\rbo$
be the set of $\nu\in\rbo$ such that the Laplacians $\djn$
on $DR^j(M_\nu,Z)$ (for all $j$) have no eigenvalues $\lambda$
from $(a-\epsilon,a+2\epsilon)\subset{\wR}_+$ (where $\eps>0$
is small enough). Then for $\nu\in\uno$ the projection operator
$\Pi^j_a(\nu)$ from $\left(DR^j(M)\right)_2$ onto a linear space
$W^j_a(\nu)$ (spanned by the eigenforms of $\djn$
with the eigenvalues $\lambda$ from $[0,a]$) is equal
to the contour integral
\begin{equation*}
\Pi^j_a(\nu)=i/{2\pi}\int_{\Gamma_{a+\eps}}\gjln d\lambda,
\end{equation*}
where $\gjln=(\djn-\lambda)^{-1}$ is the resolvent for the Laplacian
$\djn$ and $\Gamma_{a+\eps}$ is a circle
$\Gamma_{a+\eps}=\{\lambda: |\lambda|=a+\eps\}$ oriented opposite
to the clockwise. For $\lambda\in\Gamma_{a+\eps}$ and $\nu\in\uno$
the operators $\gjln$ form a smooth in $(\lambda,\nu)$ family
of bounded operators in $\left(DR^j(M)\right)_{2}$.
(It is an immediate consequence of Proposition~\ref{PA3002} below.
Indeed, $\Spec(\djn)$ is discrete and it is a subset of $\wR_+\cup 0$,
according to Theorem~\ref{TAT705}.
Thus if $(a-\eps,a+2\eps)\cap\Spec(\djn)=\emptyset$ then
$\Gamma_{a+\eps}\cap\Spec(\djn)=\emptyset$ and $G^j_\lambda(\nu)$ form
a smooth in $\nu\in\uno$ and in $\lambda\in\Gamma_{a+\eps}$ family
of bounded in $\left(DR^j(M)\right)_{2}$ operators
by Proposition~\ref{PA3002}.)
Proposition~\ref{P20}\ is proved.\ \ \ $\Box$

\subsection{Variation formula for the scalar analytic torsion.
Proof of Lemma 2.2}

First the lemma is proved  in the case when $a>0$ is less
than $4^{-1}\lambda_{1}(\nu_{0})$ (where $\lambda_{1}(\nu)$ is
the minimal positive eigenvalue of the Laplacian $\den$
on $\left(\oplus_j DR^j(M_\nu,Z),g_{M}\right)$). Let $U_{1}(a)$
be a neighborhood of $\nu_{0}$, $\nu_{0}\in U_{1}(a)\subset U$
(\ref{ST3}), such that for $\nu\in U_{1}(a)$ we have
$a<2^{-1}\lambda_{1}(\nu)$.
(Such a neighborhood exists according to Theorem~\ref{TAT705}
and to Proposition~\ref{PA3002}.)

Let $\nu(\gamma)$ be a smooth local map
$(\wR^1,0)\to\left(U(a),\nu_{0}\right)$.
\begin{pro}
For $t>0$ the following variation formula holds%
\footnote{The operator $\exp\left(-t\djn\right)$ acts from
$\left(DR^j(M)\right)_{2}$ into the domain $D\left(\djn\right)$
of $\djn$ defined by (\ref{Y4}). The operators $\exp\left(-t\djn\right)$
and $p_1\exp\left(-t\djn\right)$ are of trace class. Their traces
are equal to the integrals over the diagonal of the traces of their kernels
restricted to the diagonals (by Proposition~\ref{PA75}).}:
\begin{multline}
\dfg\sum(-1)^j j\Tr\exp\left(-t\djn\right)\bgo=\\
=2\dfg\log(\kn)\bgo\sum(-1)^j\left(-t\,\df/\df t\right)
\Tr\left(p_{1}\exp\left(-t\djon\right)\right),
\label{S607}
\end{multline}
where $\kn:=\alpha/\beta$ for $\nu=(\alpha,\beta)\in U$ (i.e.,
for $\alpha\beta\ne 0$).
\label{P25}
\end{pro}

\begin{pro}
Let $\Ree\,s>(\dim M)/2$ and let $0<a<\lambda_{1}(\nu_{0})$.
Then the following equalities hold:
\begin{multline}
\dfg\left[\Gamma(s)^{-1}\int^{\infty}_0 t^{s-1}\sum(-1)^j j\left(\Tr\,
\exp\left(-t\djn\right)-\dim\Ker\djn\right)dt\right]\big|_{\gamma=0}=\\
=\Gamma(s)^{-1}\int^{\infty}_0 t^s\sum(-1)^j(-\df_t)\Tr\left(p_{1}\left
(\exp\left(-t\djon\right)-\Pi^j_a(\nu_{0})\right)\right)dt=\\
=2\,\dfg\log(\kn)\bgo\left(s\big/\Gamma(s)\right)\sum(-1)^j\int^{\infty}_0
t^{s-1}\Tr\left(p_{1}\left(\exp\left(-t\djon\right)-\Pi^j_a(\nu_{0})\right)
\right)dt.
\label{S630}
\end{multline}
\label{PA200}
\end{pro}

\noindent{\bf Proof of Proposition~\ref{PA200}.}
To produce the second row of (\ref{S630}) from (\ref{S607}), it is enough
to prove (\cite{Bo}, II.26, Proposition~7) that for $\Ree\,s>(\dim M)/2$
the integral
\begin{equation}
\int_0^{\infty}t^s(-\df_t)\Tr\left(p_{1}\exp(-t\djn)\right)dt=\int_0^{\infty}
t^s\Tr\left(p_{1}\djn\exp(-t\djn)\right)dt
\label{S950}
\end{equation}
converges uniformly in $\nu$ for $\nu$ from a small neighborhood $\uno$
of $\nu_{0}$ and to prove the convergence of the integral
\begin{equation}
\int_0^{\infty}t^{s-1}\Tr\left(\exp(-t\djon)-\dim\Ker\djon\right)dt
\label{S951}
\end{equation}
together with the uniform convergence in $\nu$ for an arbitrary
$\nu_{1}\in\uno$ (as $\nu\to\nu_{1}$) of the function
\begin{equation}
t^{s+1}\!\sum(-1)^j\!\Tr\!\left(p_{1}\djn\!\exp(-t\dtn)\right)\!\to\! t^{s+1}
\!\sum(-1)^j\!\Tr\!\left(p_{1}\djna\!\exp(-t\dtna)\!\right)
\label{S952}
\end{equation}
for $t$ from any closed finite interval $t\in I\subset (0,+\infty)$.
(To apply the theorem from \cite{Bo}, quoted above, it is useful first
to do the transformation $\wR_+\ni t\to h=\log t\in\wR$, $dt\to tdh$.)

The following estimates are satisfied
\begin{multline}
\left|\Tr\left(p_{1}\djn\exp(-t\djn)\right)\right|\le\Tr\left(\djn
\exp(-t\djn)\right)\le \\
\le\left\|\djn\exp(-t\djn/2)\right\|_{2}\left(\Tr\left(\exp(-t\djn/2)-
\dim\Ker\djn\right)\right),
\label{S957}
\end{multline}
\begin{equation}
\left\|\djn\exp(-t\djn/2)\right\|_{2}\le\max_{\lambda\ge 0}\left(\lambda
\exp(-t\lambda/2)\right)=2/(te),
\label{S953}
\end{equation}
(where $\left\|\cdot\right\|_2$ are the operator norms
in $\left(DR^j(M)\right)_{2}$).
The first estimate in (\ref{S957}) follows from the Mercer theorem.
(The applying of this theorem here is similar to its applying in the proofs
of Propositions~\ref{PA75} and \ref{PA80} below.)

Let $t_0$ be any positive number. Then for $t\ge 2t_{0}$ and for $\nu$
sufficiently close to $\nu_{0}$ we have the following uniform with respect
to $\nu$ estimate
\begin{equation}
\Tr\exp\left(-t\djn/2\right)-\dim W_a^j(\nu)<C\exp(-c_{1}t),
\label{S954}
\end{equation}
where $C$ and $c_{1}$ are positive constants. Indeed, according
to Theorem~\ref{TAT706}, for any $t_{0}> 0$ there is a constant $L> 0$
(depending on $t_{0}$) such that the inequality
\begin{equation*}
\Tr\exp\left(-t_{0}\djn/2\right)\le L.
\end{equation*}
holds uniformly with respect to $\nu\in\rbo$.
So for all $\nu\in U$ sufficiently close to $\nu_{0}$  and such
that $\lambda_{1}(\nu)> 2a$, the estimate (\ref{S954}) is true
for $t\ge 2t_{0}$ with $C=L$ and $c_{1}=a/4$:
\begin{equation}
\Tr\exp\left(-t\djn/2\right)-\dim W_a^j(\nu)\le L\exp(-ta/4).
\label{S955}
\end{equation}

The uniform convergence (with respect to $\nu$) of the integral (\ref{S950})
for $\Ree\,s>(\dim M)/2$ follows from the asymptotic expansion (\ref{S16})
in powers of $t$ as $t\to +0$ and from the estimates (\ref{S957}),
(\ref{S953}), and (\ref{S954}).
The convergence of the integral (\ref{S951}) for $\Ree\,s>(\dim M)/2$
follows from (\ref{S954}) and from the existence of the asymptotic
expansion of the trace as $t\to +0$ (by Theorem~\ref{TAT706}):
\begin{multline}
\Tr\exp(-t\djon)=f_{-\dim M}t^{-(\dim M)/2}+f_{1-\dim M}t^{(1-\dim M)/2}+
\dots +\\
+f_k t^{k/2}+O\left(t^{(k+1)/2}\right),
\label{S631}
\end{multline}
where $k\in\wZ_+$ and $f_i:=f_i(\nu;j)$ are smooth in $\nu\in\rbo$.
This asymptotic expansion (\ref{S631}) is differentiable with respect
to $\nu$, according to Proposition~\ref{PE303}.

The uniform convergence of (\ref{S952}) for $t\in I\subset(0,\infty)$
(if $s$ is fixed and $\Ree\,s>(\dim M)/2$) follows from
Proposition~\ref{PA75} and from the uniform convergence (with respect
to $\nu$) of the functions of $t$
$$
t^{s+1}\df_t\int_{M_{1}}\Tr E^j_{t,x,x}(\nu)\to t^{s+1}\df_t
\int_{M_{1}}\tr E^j_{t,x,x}(\nu_{1})
$$
for $t\in I$. (The latter assertion follows from Proposition~\ref{PE303}
and from Theorem~\ref{TAT706}.)

The last equality in (\ref{S630}) is true for $\Ree\,s>(\dim M)/2$
according to the asymptotic expansion (\ref{S631}), to the estimate
(\ref{S955}), to the absolute convergence of the integral (\ref{S950}),
and to the following estimates (where $0< a<\lambda_{1}(\nu_{0})$):
\begin{equation}
0<\Tr\left(p_{1}\left(\exp\left(-t\djon\right)-\Pi^j_a(\nu_{0})\right)\right)
\le \Tr\left(\exp\left(-t\djon\right)-\Pi^j_a(\nu_{0})\right).
\label{S958}
\end{equation}
(These estimates are deduced from the Mercer theorem the same way
as in the proofs of Propositions~\ref{PA75} and \ref{PA80}.)
Thus Proposition~\ref{PA200} is proved.\ \ \ $\Box$

\begin{pro}
For $0<a <\lambda_{1}(\nu_{0})$ the assertion of Lemma \ref{L2.2}
is true, i.e., it holds:
\begin{equation}
\dfg\!\log\! T\!\left(M_\nu,Z\right)\!\bgo\!=\!2\dfg\!\log(\kn)\bgo\!\sum
(-1)^j\!\Tr\!\left(p_{1}\!\left(\exp\!\left(-t\djon\right)\!-\!\Pi^j_a
(\nu_{0})\right)\!\right)^0.
\label{S690}
\end{equation}
\label{PA201}
\end{pro}

\noindent(The zero superscript denotes the constant coefficient
in the asymptotic expansion as $t\to+0$ for the operator trace.)

\noindent{\bf Proof.} The equality (\ref{S630}) claims that
for $\Ree\,s>(\dim M)/2$ and for $0< a<\lambda_{1}(\nu_{0})/4$
we have
\begin{multline}
\dfg\left(\sum(-1)^j j\znj(s)\right)\bgo= \\
\!=\!2\dfg\!\log(\kn)\bgo\!\left(s/\Gamma(s)\right)\!\sum(-1)^j\!\int_0
^{\infty}\!t^{s-1}\!\Tr\!\left(p_{1}\!\left(\exp\!\left(\!-t\djon\!\right)
\!-\!\Pi^j_a(\nu_{0})\right)\!\right)dt.
\label{S960}
\end{multline}

The final expressions in (\ref{S960}) and (\ref{S630}) are analytic
functions of $s$ for $\Ree\,s>(\dim M)/2$ according to (\ref{S958}),
(\ref{S955}), and to (\ref{S16}).
The meromorphic continuation to the whole complex plane $\wC\ni s$
of this function of $s$ can be produced with the help of the asymptotic
expansion (\ref{S16}) (or of the expansion (\ref{X1100}) below
for $\Tr\left(p_{1}\left(\exp\left(-t\djon\right)-\Pi^j_a(\nu_{0})\right)
\right)$) using the estimates (\ref{S958}) and (\ref{S955}).
Let the asymptotic expansion as $t\to +0$ for the trace
of the operator below be as follows ($n:=\dim M$):
\begin{equation}
\!\Tr\!\left(p_{1}\!\left(\exp\left(\!-t\djon\!\right)\!-\!\Pi^j_a(\nu_{0})
\!\right)\!\right)\!=\!q_{-n}t^{-n/2}\!+\!\ldots\!+\!q_{0}t^0\!+\!\hat{q}_{1}
t^{1/2}\!+\!\ldots\!+\!\hat{q}_k t^{k/2}\!+\!r_{k,j}(t),
\label{X1100}
\end{equation}
where $r_{k,j}(t)$ is $O\left(t^{(k+1)/2}\right)$ as $t\to +0$ and
$r_{k,j}(t)$ is a $C^{[k/2]}$-smooth function of $t\in[0,1]$.%
\footnote{This asymptotic expansion exists and is differentiable
with respect to $t$ by Proposition~\ref{PE303}.}
Then the analytic
continuation to $\Ree\,s>-(k+1)/2$ of the integral on the right
in (\ref{S960}) is given by the expression
\begin{multline}
\int_0^{\infty}t^{s-1}\Tr\left(p_{1}\left(\exp\left(-t\djon\right)-
\Pi^j_a(\nu_{0})\right)\right)dt= \\
=q_{-n}\big/(-n/2+s)+q_{-n+1}\big/(-n/2+1/2-s)+\dots +q_{0}\big/s+\hat{q}_{1}
\big/(1/2+s)+\dots \\
+\hat{q}_k\big/(k/2+s)+\int_0^1\! t^{s-1}r_{k,j}(t)dt+\int_1^{\infty}\!t^{s-1}
\Tr\!\left(p_{1}\!\left(\exp\left(-t\djon\right)\!-\!\Pi^j_a(\nu_{0})\right)
\right)dt.
\label{S961}
\end{multline}
The latter integral in (\ref{S961}) is an analytic function
on the whole complex place $\wC\in s$
(according to (\ref{S958}) and (\ref{S955})). The integral
of $r_{k,j}$ in (\ref{S961}) is an analytic function of $s$
for $\Ree\,s>-(k+1)/2$. The asymptotic expansions (\ref{S631})
for $\Tr\exp\left(-t\djn\right)$ can be differentiated with respect
to $\nu$ according to Proposition~\ref{PE303}. They provide us with
the analytic continuation of $\sum(-1)^j j\znj(s)$ to $\Ree\,s>-(k+1)/2$
as follows:
\begin{multline}
\sum(-1)^j j\znj(s)=\Gamma(s)^{-1}\Bigl(F_{-n}\big/(-n/2+s)+\dots +F_k\big/
(k/2+s)+ \\
+\int_0^1\! t^{s-1}m_{k,\nu}(t)dt+\int_1^{\infty}\!t^{s-1}\sum(-1)^j j\Tr\!
\left(\exp\left(-t\djn\right)\!-\!\dim\Ker\djn\right)dt\Bigr),
\label{S962}
\end{multline}
where $F_k:=\sum_j(-1)^j j\left(f_k(\nu,j)-\delta_{0,k}\dim\Ker\djn\right)$
and the functions $m_{k,\nu}(t)$ are $C^{[k/2]}$-smooth in $t\in[0,1]$ and
in $\nu$ (for $\nu\in\rbo$) and are such that
$m_{k,\nu}=O\left(t^{(k+1)/2}\right)$ uniformly with respect to $\nu$
as $t\to +0$.

The latter integral in (\ref{S962}) is an analytic function
of $s\in\wC$. We obtain its derivative with respect to $\gamma$ taking into
account (\ref{S607}), (\ref{S955}), and (\ref{S958}):
\begin{multline*}
\dfg\int_1^{\infty}t^{s-1}\sum_j(-1)^j j\Tr\left(\exp\left(-t\djn\right)-
\dim\Ker\djn\right)dt\bgo= \\
=2\,\dfg\log(\kn)\bgo\Bigl\{s\int_1^{\infty}t^{s-1}\sum_j(-1)^j\Tr\left(p_{1}
\left(\exp\left(-t\djon\right)-\Pi^j_a(\nu_{0})\right)\right)dt+ \\
+\sum_j(-1)^j\Tr\left(p_{1}\left(\exp\left(-\djon\right)-\Pi^j_a(\nu_{0})
\right)\right)\Bigr\},
\end{multline*}
where $0<a<\lambda_{1}(\nu_{0})$.%
\footnote{For such $a$ the operator $\Pi^j_a(\nu_0)$ is the projection
operator from $\left(DR^j(M)\right)_2$ onto the space of harmonic forms
$\Ker\left(\djon\right)$.}
Using (\ref{S631}), (\ref{S962}), and (\ref{S961}), we get the equalities
(where $q_k(j,\nu_{0}):=q_k$ for $k\le 0$, $q_k(j,\nu_{0}):=\hat{q}_k$
for $k> 0$ and $q_k$, $\hat{q}_k$ are from (\ref{X1100})):
\begin{gather}
\begin{split}
\dfg F_l\bgo & =2\,\dfg\log(\kn)\bgo(-l/2)\sum(-1)^j q_l(j,\nu_{0}), \\
\dfg m_{k,\nu}(t)\bgo & =2\,\dfg\log(\kn)\bgo(-t)\df_t\sum_j(-1)^j r_{k,j}(t).
\end{split}
\label{S964}
\end{gather}
Hence the equality (\ref{S960}) holds on the whole complex plane $\wC\ni s$.
In particular, we obtain
$$
\dfg\log T\left(M_\nu,Z\right)\bgo=2\,\dfg\log(\kn)\bgo\sum(-1)^j q_{0}(j,
\nu_{0}).
$$
Thus Proposition~\ref{PA201} is proved.\ \ \ $\Box$

\begin{rem}
A consequence of (\ref{S964}) is as follows. For any smooth local map
$\nu(\gamma):(\wR^1,0)\to (U,0)$ (where $U$ is defined by (\ref{ST3}))
the identity $\dfg F_{0}\equiv 0$ holds according to (\ref{S964}).
Hence the function on $U$
\begin{equation*}
F_{0}(\nu)=\sum(-1)^j j\left(f_{0}(\nu,j)-\dim\Ker\djn\right)
\end{equation*}
is independent of $\nu$.
The dimension of $\Ker\djn$ is independent of $\nu$ for $\nu\in U$ as
it follows from the cohomology exact sequence in the top row
of (\ref{X2040}). Indeed, for $\nu\in U$ the dimension of $\Image\df_{D}$
(where $\df_{D}\colon H^i(N,\wC)\to\oplus\kab H^{i+1}(M_k,Z_k\cup N;\wC)$
is a differential in this exact sequence) is independent of $\nu$
because for different $\nu=\nu_{0}$ and $\nu_{1}$ from $U$
the maps
$\df_{D,k}(\nu)\colon H^i(N,\wC)\to H^{i+1}(M_k,N\cup Z_k;\wC)$ for fixed
$k=1,2$ (and for a fixed $i$) differ by the nonzero scalar constant
factor (depending on $\nu_0$ and $\nu_1$). Hence
$\sum(-1)^j jf_{0}(\nu,j)$
is a constant function on $U$.
\end{rem}

 \noindent{\bf Proof of Proposition \ref{P25}.} Let $E^j_{t,x,y}(\nu)$
(where $t>0$) be the heat kernel of the operator $\exp\left(-t\djn\right)$.
By the Duhamel principle, a variation in $\nu$ of $\eotxy(\nu)$ can be
written as follows. Let
$\left(\eotaxaa(\nu_{1}),\eotbay(\nu)\right)_{M}
=:\int_{M}\eotaxaz(\nu_{1})\wedge *_z\eotbzxb(\nu)$
be a scalar product (\ref{Y2}) (with respect to the variable $z$).
We have
\begin{multline}
\eotxy(\nu)-\eotxy(\nu_{0})=
\lim_{\eps\to +0}\int^{t-\eps}_\eps d\tau\df/
d\tau\left(\eotuxa(\nu),\eottay(\nu_{0})\right)=\\
=\lim_{\eps\to +0}\int^{t-\eps}_\eps d\tau\left[\left(-\Delta_*\eotuxa(\nu),
\eottay(\nu_{0})\right)+
\left(\eotuxa(\nu),\Delta_*\eottay(\nu_{0})\right)\right].
\label{S601}
\end{multline}
Stokes' formula claims that for any two smooth forms
$\oma,\omb\in DR^j(\Mm_{1})$ on a manifold $M_1$ with boundary
we have
\begin{equation}
(d\oma,\omb)_{M_{1}}-(\oma,\delta\omb)_{M_{1}}=(r\oma,A\omb)_{\df M_{1}}:=
\int_{\df M_1}r\oma\wedge *_{\df M_1}A\omb,
\label{S602}
\end{equation}
where the density $r\oma\wedge*_{\df M}A\omb$ on $\df M_1=:N$ and
the operators $r$ and $A$ are defined as follows.
Let any local orientations be chosen on $TM|_{N}$ and on
$TN$. Then the following forms on $N$ are locally defined:
\begin{equation}
r\oma:=\left[N:M_{1}\right]\ianma\oma,\quad A\omb:=*^{-1}_{N}
\ianma *_{M_{1}}\omb,
\label{S603}
\end{equation}
where $\ianma:\DR(\Mm_{1})\to\DR(N)$ is the geometrical restriction to $N$
(and $[N\colon M_{1}]=1$ if $N$ is locally oriented as $\df M_1$ and
$[N\colon M_{1}]=-1$ in the opposite case). The locally defined form
$r\oma\wedge *_{\df M_1}A\omb$ is a globally defined density on $N=\df M$%
\footnote{It does~not depend on a local orientation on $TM$ and if a local
orientation on $TN$ changes to the opposite then this local form changes
its sign.}.
So the equality (\ref{S601}) can be written in the form
\begin{multline}
\eotxy(\nu)-\eotxy(\nu_{0})
=\lim_{\eps\to +0}\int^{t-\eps}_\eps d\tau\sum\kab\Bigl(\Bigl[
\left(-(r\delta)_*\eotuxa(\nu),A_*\eottay(\nu_{0})\right)
_{\df M_k}+\\
+\left(A_*\eotuxa(\nu),(r\delta)_*\eottay(\nu_{0})\right)_{\df M_k}\Bigr]+
\Bigl[\left((Ad)_*\eotuxa(\nu),r_*\eottay(\nu_{0})\right)_{\df M_k}-\\
-\left(r_*\eotuxa(\nu),(Ad)_*\eottay(\nu_{0})\right)_{\df M_k}\Bigr]\Bigr),
\label{S608}
\end{multline}
where $r=r_k$, $A=A_k$ are the corresponding operators for pairs
$\left(M_k,\df M_k\right)$.

Let any local orientations be chosen on $TM|_{N}$ and on $TN$. Then
the conditions (\ref{Y4}) for the domain $D\left(\don\right)$ claim that
for the kernels
$\eo(\nu)$ and $\eo(\nu_{1})$ (where $\nu=(\alpha,\beta)$ and $\nu_{1}=
(\alpha_{1},\beta_{1})$ are from $\rbo$) the following equalities hold
on $N$:
\begin{gather}
\begin{split}
\alpha\,r_z(M_{1},N)\circ\eo_{t,x,z}(\nu) &=-\beta\,r_z(M_{2},N)\circ
\eo_{t,x,z}(\nu),\\
\beta\,A_z(M_{1},N)\circ\eo_{t,x,z}(\nu) &=\alpha\,A_z(M_{2},N)\circ\eo_{t,x,z}
(\nu),\\
\alpha\,r_z(M_{1},N)\circ\delta_z\eo_{t,x,z}(\nu) & =-\beta\,r_z(M_{2},N)\circ
\delta_z\eo_{t,x,z}(\nu),\\
\beta\,A_z(M_{1},N)\circ d_z\eo_{t,x,z}(\nu) & =\alpha\,A_z(M_{2},N)\circ d_z
\eo_{t,x,z}(\nu),
\end{split}
\label{S610}
\end{gather}
The analogous equalities hold for $\eo_t(\nu_{1})$ (where $(\alpha,\beta)$
are replaced by $(\alpha_1,\beta_1)$). Hence the equality (\ref{S608})
for $\nu\in U(a)\subset U$ can be written in the form
\begin{multline}
\eotxy(\nu)-\eotxy(\nu_{0})=\\
=\lim_{\eps\to +0}\int^{t-\tau}_\eps d\tau\Bigl[-\left(
1-(\kn/\kno)\right)\left((r\delta)_*\eotuxa(\nu),A_*\eottay(\nu_{0})\right)
_{\df M_{1}}+\qquad\\
+\left(1-(\kno/\kn)\right)\left(A_*\eotuxa(\nu),
(r\delta)_*\eottay(\nu_{0})\right)_{\df M_{1}}+\quad\\
+\left(1-(\kno/\kn)\right)\left(A_*d_*\eotuxa(\nu),r_*\eottay(\nu_{0})\right)
_{\df M_{1}}-\\
-\left(1-(\kn/\kno)\right)\left(r_*\eotuxa(\nu),A_*d_*\eottay
(\nu_{0})\right)_{\df M_{1}}\Bigr].
\label{S611}
\end{multline}

The kernel $\eotxy(\nu)$ is smooth in $\nu\in\rbo$ as a smooth double form
on $\Mm_{r_{1}}\times\Mm_{r_{2}}$ (where the limit values
on $N\subset\df\Mm_r$ are taken from the side of $M_r$)
according to Proposition~\ref{PE303}.
So we can conclude from (\ref{S611}) that
\begin{multline}
\dfg\eotxy(\nu)\bgo=\dfg\log(\kn)\bgo\lim_{\eps\to+0}\int^{t-\eps}_\eps d\tau
\Bigl[\left((r\delta)_*\eotuxa(\nu_{0}),A_*\eottay(\nu_{0})\right)
_{\df M_{1}}+\\
+\left(A_*\eotuxa(\nu_{0}),(r\delta)_*\eottay(\nu_{0})\right)_{\df M_{1}
}+\left(A_*d_*\eotuxa(\nu_{0}),r_*\eottay(\nu_{0})\right)_{\df M_{1}}+\\
+\left(r_*\eotuxa(\nu_{0}),A_*d_*\eottay(\nu_{0})\right)_{\df M_{1}}
\Bigr],
\label{S612}
\end{multline}
where the limits of the kernels are taken from the side of $M_{1}$.

By Proposition~\ref{PA75}, we have the equality
\begin{equation}
\Tr\exp\left(-t\djon\right)=\sum_{r=1,2}\int_{\Mm_r}i^*_{M_r}\etxaxb^j
(\nu_{0}),
\label{S614}
\end{equation}
where the exterior product of double forms on the diagonals $i_{M_r}:\Mm_r
\hookrightarrow\Mm_r\times\Mm_r$ are implied.

We deduce from the semigroup property for the kernels
of the operators $\exp\left(-t\djon\right)$ that
\begin{multline}
\left(\qaza\etaxza^j(\nu_{0}),\qbzb\etbzbx^j(\nu_{0})\right)_{M}:=\\
=\int_{M}\qaza\etaxza^j(\nu_{0})\wedge_x*_x\qbzb\etbzbx^j(\nu_{0})=
\qaza\qbzb\etzba^j(\nu_{0}),
\label{S615}
\end{multline}
where $Q_{i,z_i}$ are differential operators acting on differential
forms, $t_i>0$ and the integration is with respect to the variable $x$.
We deduce the following variation formula from (\ref{S612})
using (\ref{S614}) and (\ref{S615}):
\begin{multline}
\dfg\Tr\exp\left(-t\djn\right)\bgo=
2\,\dfg\log(\kn)\bgo t\Bigl[\left(\aza(M_{1},N)\dza\rzb(M_{1},N)\etzazb
^j(\nu_{0})\right)_{N}\!+\\
+\left(\aza(M_{1},N)\rzb(M_{1},N)\dezb\etzazb^j(\nu_{0})\right)_{N}\Bigr],
\label{S620}
\end{multline}
where the summands are the integrals of the densities
on $N$ (as it is defined in (\ref{S602}) and (\ref{S603})). For instance,%
\footnote{We imply (but do~not write) the restriction
to the diagonal $N\hookrightarrow N\times N$ in (\ref{S621}).}
\begin{equation}
\bigl(\aza\dza\rzb\etzazb^j(\nu_{0})\bigr)_{N}
:=\int\left(\aza\dza\wedge*_{z_{2},N}\rzb\etzazb^j(\nu_{0})\right).
\label{S621}
\end{equation}
The local forms
$F_{1}=\left(\iama*_{z_{2}}\dza\dzb\eotzazb(\nu_{0})\right)$ and
$F_{2}=\left(\iama*_{z_{2}}\deza\dza\etzazb^{\bullet+1}(\nu_{0})\right)$
are smooth on the diagonal
$i_{M_1}\colon\Mm_{1}\hookrightarrow\Mm_{1}\times\Mm_{1}$.
(The limit values on $N\subset\df M_{1}$ of these double forms
are taken from the side of $M_{1}$ and the exterior product
of these double forms is implied.)
The local form $i^*_{N}*_{N,z_{2}}\aza\dza\rzb\eotzazb(\nu_{0})$
is a smooth density on the diagonal $N\subset\df M_{1}$,
$i_{N}:N\hookrightarrow N\times N$. Hence we can apply Stokes' formula
(\ref{S602}) to the expression on the right in (\ref{S620}).
The result transformed with the help of the equality
$d_x\eotxy(\nu)=\delta_y E^{\bullet +1}_{t,x,y}(\nu)$ is as follows:
\begin{multline}
\dfg\Tr\exp\left(-t\djn\right)\bgo=\\
=2\dfg\log(\kn)\bgo t\Bigl[-\left(\deza\dza\etzazb^j(\nu_0)\right)_{M_{1}}
+\left(\dza\deza\etzazb^{j+1}(\nu_0)\right)_{M_{1}}+\\
+\left(\dzb\dezb\etzazb^j(\nu_0)\right)_{M_{1}}-\left(\dezb\dzb\etzazb^{j-1}
(\nu_0)\right)_{M_{1}}\Bigr],
\label{S622}
\end{multline}
where the summands in (\ref{S622}) are the integrals of the densities
of the type:
\begin{equation}
\left(\deza\dza\eotzazb(\nu_0)\right)_{M_{1}}:=\int_{M_{1}}\iama*_{z_{2}}
\deza\dza\eotzazb(\nu_0)
\label{S623}
\end{equation}
(and the exterior product of double forms is implied in (\ref{S623})
and (\ref{S622})).

The formula (\ref{S607}) is an immediate consequence of (\ref{S622})
if we take into account the equalities
$$
\df_t\eotxy=-\Delta_x\eotxy=-\Delta_y\eotxy,\quad
\eotxy=E^{\bullet}_{t,y,x}
$$
and use the formula (\ref{S1001}) in Proposition~\ref{PA75}.
Thus Proposition~\ref{P25} is proved. Hence Lemma~\ref{L2.1} is proved
in the case of $a<4^{-1}\lambda_{1}(\nu_{0})$.\ \ \ $\Box$

Let now $a>0$ be an arbitrary number such that
$a\notin\cup_j\Spec\left(\djon\right)$.
Then for any nonzero element $l\in\det\Wano$ we have by Lemma~\ref{L2.2} that
\begin{equation*}
\dfg\log\left\|g_{\nu}l\right\|^2_{\dewo}\bgo=-2\,\dfg\log(\kn)\bgo\sum(-1)^j
\Tr\left(p_{2}\Pi^j_a(\nu_{0})\right),
\end{equation*}
where $g_\nu=\Pii_a v_\nu$ is defined by (\ref{ST22}). We know that
under the identifications (\ref{S5}) and (\ref{S7}), the analytic
torsion norm $T_0\left(M_\nu,Z;a\right)$ on $\dew$ transforms into
the analytic torsion norm $T_0\left(M_\nu,Z\right)$ on $\det\Hh(M_\nu,Z)$,
$\left\|g_\nu l\right\|^2_{T_{0}(M_\nu,Z;a)}=\left\|(g_\nu l)_{H}\right\|
^2_{T_{0}(M_\nu,Z)}$.
(Here, $(g_\nu l)_{H}\in\det\Hh(M_\nu,Z)$ corresponds to $g_\nu l$
under these identifications.) Let $\left(g_\nu l\right)_H$ be fixed.
Then the analytic torsion norm of $g_\nu l\in\dew$
\begin{equation}
\left\|g_\nu l\right\|^2_{T_{0}(M_\nu,Z;a)}:=\left\|g_\nu l\right\|^2
_{\dew}\cdot T(M_\nu,Z;a)
\label{S702}
\end{equation}
is independent of $a>0$. Let $\nu_{0}=(\ao,\bo)\in U$ (i.e., $\ao\bo\ne0$).
Then using (\ref{S702}), (\ref{S690}) we obtain ($\mu:=\dfg\log(\kn)\bgo$):
\begin{multline}
\dfg\log T(M_\nu,Z;a)\bgo=
\lim_{\eps\to +0}\dfg\log T(M_\nu,Z;\eps)\bgo+ \\
+2\mu\left(-\lim_{\eps\to +0}\sum(-1)^j\Tr\left(p_{2}
\Pi^j_\eps(\nu_{0})\right)+
\sum(-1)^j\Tr\left(p_{2}\Pi^j_a(\nu_{0})\right)\right)=\\
=2\mu\Bigl[\sum(-1)^j\left(\Tr p_{1}\exp(-t\djon)\right)^0-
\sum(-1)^j\Tr\Pi^j_\eps(\nu_{0})+\sum(-1)^j\Tr\left(p_{2}\Pi^j_a(\nu_{0})
\right)\Bigr].
\label{S703}
\end{multline}
Note that for arbitrary $c>0$, $a>0$ we have
$$
\sum(-1)^j\Tr\Pi^j_c(\nu_{0})=\chi(\mno;Z)=\sum(-1)^j\Tr\left((p_{1}+p_{2})
\Pi^j_a(\nu_{0})\right).
$$
So the final expression in (\ref{S703}) is equal to
$$
2\,\dfg\log(\kn)\bgo\sum(-1)^j\Tr\left(p_{1}\left(\exp(-t\djon)-
\Pi^j_a(\nu_{0})\right)\right)
$$
Thus Lemma~\ref{L2.1} is proved for an arbitrary $a>0$ such that
$a\notin\cup_j\Spec\left(\djon\right)$.\ \ \ $\Box$

\subsection{Continuity of the truncated scalar analytic torsion. Proof of
Proposition \ref{PA300}}
Taking into account the definition (\ref{S14}) of the truncated scalar
analytic torsion $T(M_\nu,Z;a)$, we see that Proposition~\ref{PA300}
is a consequence of the assertion as follows. For $\Ree\,s>(\dim M)/2$
the truncated $\zeta$-function (\ref{S9}) for $\djn$ is defined
by the integral%
\footnote{The analytic continuation of this integral from
$\Ree\,s>(\dim M)/2$ to the whole complex plane coincides with
the meromorphic continuation of $\znj(s;a)$.}
\begin{equation}
\znj(s;a)=\Gamma(s)^{-1}\int_0^{\infty}t^{s-1}\Tr\left(\exp\left(-t\djn\right)
\left(1-\Pi^j_a(\nu)\right)\right)dt
\label{S1599}
\end{equation}

\begin{pro}
For $a> 0$ the truncated determinant for the $\nu$-transmission interior
boundary conditions
\begin{equation*}
\det\left(\don;a\right):=\exp\left(-\df/\df s\znb(s;a)\big|_{s=0}\right)
\end{equation*}
is a continuous function of $\nu$ for $\nu\in\rbo$ such
that $a\notin\Spec\left(\don\right)$.
\label{PA301}
\end{pro}

\noindent{\bf Proof.} Let $\eotxy(\nu)$, $t> 0$, be the kernel of
$\exp\left(-t\doo_\nu\right)$. According to Proposition~\ref{PA75}
we have
\begin{equation}
\Tr\exp\left(-t\don\right)=\sum\kab\int_{\Mm_k}\tr\left(*_{x_{2}}
i^*_k\eotxaxb\right),
\label{S1601}
\end{equation}
where $i\colon\Mm_k\hookrightarrow\Mm_k\times\Mm_k$ are the diagonal
immersions. (The exterior product of the restriction to the diagonal
of double forms is implied in (\ref{S1601}).) Set $I=[-1,1]$.
Let $I\times N\subset M$ be the inclusion of the neighborhood
of $N=0\times N$ into $M$ and let $g_{M}$ be a direct product metric
on $I\times N$. Let $\doo_{\nu;0}$ be the Laplacian on $\DR(I\times N)$
with the $\nu$-transmission boundary conditions on $N=0\times N$ and
with the Dirichlet boundary conditions on $\df I\times N$.
Let $\eotxy(\nu;0)$, $t> 0$, be the kernel of $\exp\left(-t\dono\right)$.
The equality analogous to (\ref{S1601}) holds also
for $\Tr\exp\left(-t\dono\right)$ (where $\Mm_k$ is replaced
by $Q_k\times N$, $Q_{1}:=[-1,0]$, $Q_{2}:=[0,1]$).

Let $\nu_{0}\ne(0,0)$ and let $a\notin\Spec\left(\donoo\right)$
be a fixed positive number.
Then from Theorem~\ref{TAT706}, Proposition~\ref{PA3002}, and from
the estimate (\ref{S955}) it follows that for an arbitrary $\eps> 0$
there are a neighborhood $U_{0}(\eps)$ of $\nu_{0}$ and $T> 0$
such that for $\nu\in U_{0}(\eps)$ and for $t\ge T$ the estimate holds
\begin{equation}
\left|\Tr\left(\exp\left(-t\don\right)\left(1-\Pii_a(\nu)\right)
\right)\right|\le\eps\exp(-at/2).
\label{S1602}
\end{equation}

To prove the continuity of $\det\left(\don\right)$ in $\nu$ at $\nu=\nu_{0}$,
it is enough to obtain the following estimate.%
\footnote{The integrals (\ref{S1599}) for the values $\nu_{0}$ and $\nu$
of the transmission parameter have the analytic continuations from
$\Ree\,s>(\dim M)/2$ to the whole complex plane. It follows from
the estimates (\ref{S1603}) and (\ref{S1602}) that the difference
of these integrals multiplied by $\Gamma(s)$ is an absolutely convergent
integral for $\Ree\,s>-1$. Hence this difference is an analytic function
of $s$ for such $s$ and it is equal to the difference
of the analytic continuations
of the integrals (\ref{S1599}) for $\nu$ and $\nu_{0}$.}
For a given $b$, $1>b>0$, and for an arbitrary $\eps>0$ there exists
a~neighborhood $U_\eps\ni\nu_{0}$ such that the estimate holds
for $\nu\in U_\eps$, $-(1-b)<s<1$:
\begin{equation}
\int_0^T\!\left|\Tr\!\left(\exp\!\left(-t\don\right)\!\left(1\!-\!\Pii_a(\nu)
\right)\right)\!-\!\Tr\!\left(\exp\!\left(\!-t\donoo\right)\!\left(1\!-
\!\Pii_a(\nu_{0})\right)\!\right)\right|\!t^{s-1}\!dt\!<\!\eps.
\label{S1603}
\end{equation}
The spectrum $\Spec\left(\don\right)$ is discrete and it depends
continuously on $\nu$
(by Proposition~\ref{PA3002}). Since $a\notin\Spec\left(\don\right)$
we see that $\Tr\Pii_a(\nu)=\Tr\Pii_a(\nu_{0})$ ($=\rk\Pii_a$)
in a neighborhood of $\nu_{0}$ and that the following estimate
is satisfied
uniformly with respect to $s$, $-1/2<s<1$ and to $\nu$ for $\nu$
sufficiently close to $\nu_{0}$:
\begin{equation}
\int_0^T\left|\Tr\left(\exp\left(-t\don\right)\Pii_a(\nu)\right)-\Tr\left(\exp
\left(-t\donoo\right)\Pii_a(\nu_0)\right)\right|t^{s-1}dt<\eps/2
\label{S1604}
\end{equation}
The inequality
\begin{equation}
\int_0^T\left|\Tr\exp\left(-t\don\right)-\Tr\exp\left(-t\donoo\right)\right|
t^{s-1}dt<\eps/2
\label{S1605}
\end{equation}
for $\nu$ sufficiently close to $\nu_{0}$ and for $-1/2<s<1$ is obtained
as follows. According to Proposition~\ref{PA3002},
$\Tr\exp\left(-t\don\right)$ is equal to the integral over $\cup\Mm_j$
of the density defined by the restriction to the diagonal
of the corresponding kernel. So it is enough to estimate in (\ref{S1605})
the integrals of the difference between the densities defined
by $\exp\left(-\don\right)$ and by $\exp\left(-t\donoo\right)$ separately
over a fixed neighborhood $U$ of $N=0\times N\hookrightarrow M$ and
over $M\setminus U$. The estimate of the integral over $U\supset N$
is obtained with the help of the kernel $\eotxy(\nu;0)$
of $\exp\left(-t\dono\right)$.
Set $\Eotxy(\nu):=\eotxy(\nu)-\eotxy(\nu;0)$ for $x,y\in I\times N$.
\begin{pro}
For an arbitrary $m\in\wZ_+$ there is a neighborhood of $\nu_{0}$ such that
for all $x,y\in\mab:=[-1/2,0]\times N\cup [0,1/2]\times N\hookrightarrow
M_{1}\cup M_{2}$ and for $t\in(0,1]$ the estimate is satisfied uniformly
with respect to $\nu\in\rbo$
\begin{equation}
\left|\eo_{t,x,y}(\nu)\right|\le c_m t^m,
\label{S1606}
\end{equation}
(where $c_m$ is independent of $t$ and of $\nu$).
\label{PA302}
\end{pro}
\begin{pro}
The following estimate holds uniformly with respect to $s$ for $-(1-b)<s<1$
and to $\nu$ for $\nu$ sufficiently close to $\nu_{0}$
\begin{equation}
\int_0^T t^{s-1}dt\left|\int_{\mab}\tr\left(*i^*_k\Eotxaxb(\nu)\right)-\tr
\left(*i^*_k\Eotxaxb(\nu_{0})\right)\right|<\eps/4.
\label{S16071}
\end{equation}
\label{PA303}
\end{pro}

\begin{rem}
For $x,y\in[-1,1]\times N$ the equalities hold
(analogous to (\ref{S414}), (\ref{S415}), (\ref{S1240})):
\begin{equation*}
\eo_{t,x,y}(\nu;0)=
\begin{cases}
\eo_{t,x,y}+\left(\beta^2-\alpha^2\big/\beta^2+\alpha^2\right)\left(\sigma^*
_{1}\eo\right)_{t,x,y} & \text{for } x,y\in Q_{1}\times N, \\
\eo_{t,x,y}+\left(\alpha^2-\beta^2\big/\beta^2+\alpha^2\right)\left(\sigma^*
_{1}\eo\right)_{t,x,y} & \text{for } x,y\in Q_{2}\times N, \\
\left(2\alpha\beta\big/\alpha^2+\beta^2\right)\eotxy & \text{for $x,y$ from
different } Q_k\!\times\!N.
\end{cases}
\end{equation*}
\begin{equation}
\label{S1607}
\end{equation}
Here, $\eotxy$ is the kernel of $\exp\left(-t\doo\right)$ on $I\times N$
with the Dirichlet boundary conditions on $\df I\times N$
and $\sigma_{1}$ is the mirror symmetry with respect to $N=0\times N$
acting on the variable $x$ of the kernel. So we get
\begin{equation}
\!\int_{\mab}\!\tr\!\left(*_{x_{2}}i^*_k\Eotxaxb(\nu)\!\right)\!=\!\int_{\mab}
\!\tr\!\left(*_{x_{2}}i^*_k\!\left(\eotxaxb(\nu)\!-\!\eotxaxb\right)\!\right).
\label{S1608}
\end{equation}
{}From this equality and from the estimate (\ref{S16071}) it follows
that the integral over $\mab$ of the difference between the kernels
for $\nu$ and for $\nu_{0}$ gives the term in (\ref{S1605}) which is less
than $\eps/4$.
\end{rem}
\begin{pro}
The following estimate holds uniformly with respect to $s$, $-(1-b)<s<1$,
and to $\nu$ for $\nu$ sufficiently close to $\nu_{0}$:
\begin{equation}
\int_0^T\left|\int_{M\setminus\mab}\tr\left(*_{x_{2}}i^*\eotxaxb(\nu)\right)-
\tr\left(*_{x_{2}}i^*\eotxaxb(\nu_{0})\right)\right|dt<\eps/4.
\label{S1609}
\end{equation}
\label{PA305}
\end{pro}
The estimate (\ref{S1605}) is a consequence of (\ref{S16071}),
(\ref{S1608}), and (\ref{S1609}). The estimate (\ref{S1603}) follows
from (\ref{S1604}), (\ref{S1605}). The estimate (\ref{S1605}) together
with (\ref{S1602}) gives us the continuity of $\doo\left(\nu;a\right)$
in $\nu$ at $\nu_{0}$. Thus we get the proofs of Propositions~\ref{PA301}
and \ref{PA300}.\ \ \ $\Box$

\noindent{\bf Proof of Proposition \ref{PA302}.} The following equality
is obtained similarly to (\ref{S608}) by using of (\ref{S610}):
\begin{multline}
e_{t,x,y}(\nu)=-\lim_{\eps\to+0}\int^{t-\eps}_\eps d\tau\df/\df\tau\int
_{\df I\times N}\Bigl[\left((r\delta)_*\eotuxa(\nu),A_*\eottay(\nu;0)
\right)_{\df I\times N}+\\
+\left(r_*\eotuxa(\nu),(Ad)_*\eottay(\nu;0)\right)_{\df I\times N}\Bigr],
\label{S1610}
\end{multline}
where the operators $r$ and $A$ correspond to the pair
$\left(I\times N,\df I\times N\right)$.
So the estimate (\ref{S1606}) follows from the analogous estimates
for the kernels
\begin{equation}
r_z\eo_{t,x,z}(\nu),\quad (r\delta)_z\eo_{t,x,z}(\nu),\quad A_z\eo_{t,z,y}
(\nu;0),\quad  (Ad)_z\eo_{t,z,y}(\nu;0),
\label{X1000}
\end{equation}
where $x,y\in\mab$ and $z\in\df I\times N=\{-1,1\}\times N\hookrightarrow M$.
Such estimates are derived with the help of Proposition~\ref{PA89}
for $\don$ and $\doo_{\nu;0}$ as follows.

Let $m\in\wZ_+$ be taken large enough. Then there is an approximate
fundamental solution $\pom(\nu)$ for $\left(\df_t+\doo_{\nu,x}\right)$
which is the sum of an interior term $\pom_{int}$ and of terms, defined
near the boundaries $\df M$ and $N$. The kernel $\pom(\nu)$ is
a good approximation for $\eotxy(\nu)$ for small $t>0$. Its interior
term is defined as follows.
For any closed Riemannian $(M,g_{M})$ there is a locally defined
{\em parametrix} $p^{\bullet(m)}_{t,x,y}$ (i.e., an approximate fundamental
solution for $\left(\df_t+\doo_{M}\right)$) such that the difference
$\left(p^{\bullet(m)}-\eo\right)_{t,x,y}$ (where $\eotxy$ is the fundamental
solution for $\left(\df_t+\doo_{M}\right)$) is a $C^\infty$-double
form for $t>0$ and such that the following estimates hold uniformly
with respect to $(x,y)\in M\times M$ and to $t\in(0,T]$:
\begin{align*}
\left|p^{\bullet(m)}_{t,x,y}-\eotxy\right| & \le c_mt^{-n/2+m+1},\\
\left|\left(\doo_x\right)^k\left(p^{\bullet(m)}_{t,x,y}-\eotxy\right)\right|
& \le c_{m,k}t^{-n/2+m+1-k}
\end{align*}
\begin{gather}
\begin{split}
\left|\left(\df_t+\doo_{M}\right)p^{\bullet(m)}_{t,x,y}\right| & <c_{(m)}
t^{-n/2+m}, \\
\left|\left(\doo_x\right)^k\left(\df_t+\doo_{M}\right)p^{\bullet(m)}_{t,x,y}
\right| & <c_{(m,k)}t^{-n/2+m-k}
\end{split}
\label{E1965}
\end{gather}
(\cite{RS}, Proposition~5.3, \cite{BGV}, Theorems~2.20, 2.23, 2.26, 2.30).
Such a parametrix can be represented in the following form ($n:=\dim M$):
\begin{equation}
p^{\bullet(m)}_{t,x,y}=\left(4\pi t\right)^{-n/2}\exp\left(-d(x,y)^2/4t\right)
f\left(d(x,y)\right)\sum^m_{i=0}t^i\Phi_i(x,y),
\label{E1968}
\end{equation}
where $d(x,y)$ is the geodesic distance between $x$ and $y$,
$f\in C^\infty_{0}(\wR_+)$, $f(\tau)\equiv 1$ for $0\le\tau\le\eps$ and
$f\equiv 0$ for $\tau>2\eps$. The injectivity radius $i\left(M,g_M\right)$
is supposed to be greater than $2\eps$, i.e., the exponential map
$\exp_xB_{2\eps}$ is a diffeomorphism on its image for any $x\in M$
(where $B_{2\eps}:=\left\{\xi\in T_xM,|\xi|\le 2\eps\right\}$).
The coefficients $\Phi_i(x,y)$ in (\ref{E1968}) are smooth double forms
on $M\times M$ whose germs on the diagonal $M\hookrightarrow M\times M$
are unique. The principal term $\Phi_{0}(x,y)$ is the kernel
of the parallel transport in $\web TM$ along the geodesic line
$\exp_y\xi=x$ from $y$ into $x$ (and it is defined
for $d(x,y)<i\left(M,g_M\right)$).
Each $\Phi_i(x,y)$ is determined through $\Phi_{i-1}(x,y)$ in differential
geometry terms and it is well-defined for $d(x,y)<i\left(M,g_M\right)$
(\cite{RS}, Sect.~5, \cite{BGV}, Theorem~2.26, Lemma~2.49).

Let $0\times N\hookrightarrow I\times N\hookrightarrow M$
be a neighborhood of $N=0\times N\hookrightarrow M$, where
the metric $g_{M}$ is a direct product.
The fundamental solution for $\left(\df_t+\doo_\nu\right)$ on $I\times N$
(with the Dirichlet boundary conditions on $\df I\times N$) is
\begin{equation}
\cPe(\nu)=\sum_{i=0,1}E^i_{I,t}(\nu)\otimes E^{\bullet-i}_{N,t},
\label{E1969}
\end{equation}
where $\eo_{I,t}(\nu)$ is defined by the formulas completely analogous
to (\ref{S1240}) and (\ref{S1607}). (Here, $\eo_{N,t}$ is the fundamental
solution for $\left(\df_t+\doo_{N}\right)$. The operator corresponding
to the kernel $E^i_{I,t}(\nu)\otimes E^{\bullet-i}_{N,t}$
acts on $DR^i(I,\df I)\otimes DR^{\bullet-i}(N)$.
The kernel $E^i_{I,t}(\nu)$ corresponds to the Laplacian on $I$
with the Dirichlet boundary conditions on $\df I$ and with
the $\nu$-transmission boundary conditions at $0\in I$.
The term in (\ref{E1969}) with $i=1$ is equal to zero for $\bullet=0$.)

The parametrix $\pom_{t,x,y}(\nu)$ for
$\left(\df_t+\doo_\nu\right)$ on $M$ is defined by
\begin{equation}
\pom_{t,x,y}(\nu)=\psi\cPe_{t,x,y}(\nu)\phi+\psi_{1}p^{\bullet(m)}
_{t,x,y}(1-\phi).
\label{E1970}
\end{equation}
Here, $\phi=\phi(\ya)$, $\psi=\psi(\xa)$ (in the coordinates
$(\xa,x')=x$ and $(\ya,y')=y$ of points in $I\times N$),
$\phi,\psi\in C^\infty_{0}(I,\df I)$;  $\phi,\psi\ge 0$,
$\phi(\ya)\equiv 1$ for $|\ya|\le2^{-1}+\eps$, $\phi(\ya)\equiv 0$
for $|\ya|\ge 5/8$, $\psi\equiv 1$ in a neighborhood of $\supp\phi$,
and $\psi\equiv 0$ for $|\xa|\ge 3/4$,
$\psi_{1}\in C^\infty_{0}(M\setminus N)$, $\psi_{1}\equiv 1$
in a neighborhood of $\supp(1-\phi)\subset M\setminus\mab$.
Hence the parametrix $\pom_{t,x,y}$ is equal to zero for $y\in\mab$,
$x\in M\setminus\left([-3/4,3/4]\times N\right)$.
The term $\psi_{1}p^{\bullet(m)}(1-\phi)$ in (\ref{E1970}) is defined
from now on as $\pom_{int}$. (In the case of $\df M\ne\emptyset$ the terms,
completely analogous to $\cPe(\nu)$ for $\nu\in\{(0,1),(1,0)\}$,
have to be added to $\pom$. Their supports are in
$\left([0,1]\times\df M\right)^2\hookrightarrow M\times M$ and $g_{M}$
is a direct product metric on $[0,1]\times\df M$.)

\begin{pro}
1. The boundary condition for $\pom_{t,x,y}(\nu)$ on $N$ and on $\df M$
and the boundary condition for $\left(\doo_x\right)^k\pom_{t,x,y}(\nu)$
($k\in\wZ_+$) are the same as for $\eotxy(\nu)$
and for $\left(\doo_{\nu,x}\right)^k\eotxy(\nu)$. Namely
$\pom_{t,x,y}(\nu)$ is a smooth in $t>0$ and
in $(x,y)\in\Mm_{j_{1}}\times\Mm_{j_{2}}$ kernel,
$\Delta^k_{\nu,x}P^{(m)}_{t,x,y}(\nu)\subset D(\doo_{\nu,x})$ for fixed
$y$, $t>0$, and for any $k\in\wZ_+\cup 0$. Here
$D\left(\doo_{\nu,x}\right)\subset\DR\left(M_\nu,Z\right)$ is the domain
of definition of $\doo_{\nu,x}$ on pairs $(\oma,\omb)$ of smooth forms
on $\Mm_j$. It is defined by (\ref{Y4})).\\

2. The following estimates (analogous to (\ref{E1965})) hold for $t\in(0,T]$
uniformly with respect to $\nu\in\rbo$ and
to $(x,y)\in\Mm_{j_{1}}\times\Mm_{j_{2}}$
(with $C_m$, $C_{m,k}$ independent of $t$):
\begin{align}
\left|\left(\df_t+\doo_{\nu,x}\right)\pom_{t,x,y}(\nu)\right| & < C_m
t^{-n/2+m},
\label{S1611}\\
\left|\left(\doo_{\nu,x}\right)^k\left(\df_t+\doo_{\nu,x}\right)\pom_{t,x,y}
(\nu)\right| & < C_{m,k}t^{-n/2+m-k}.
\label{S16111}
\end{align}

The kernel
$\romm_{t,x,y}(\nu):=\left(\df_t+\doo_{\nu,x}\right)\pom_{t,x,y}(\nu)$
is smooth in $(x,y)\in M\times\Mm_j$ and its $C^{2l}$-norm on $M\times\Mm_j$
is estimated through $c_{m,l}t^{-n/2+m-l}$. For any linear differential
operator $F$ of order $d=d(F)$, acting on double forms on $M\times\Mm_j$,
and for any $T>0$ the kernel $F\circ\romm_{t,x,y}(\nu)$ satisfies
the estimate as follows when $t\in(0,T]$. It hold uniformly with respect
to $(x,y)\in M\times\Mm_j$ and to $\nu\in\rbo$
\begin{equation}
\left|F\circ\romm_{t,x,y}(\nu)\right|<c(F)t^{-(n+d)/2+m}.
\label{S16100}
\end{equation}
\\

3. For $0\le k\le[-n/2]+m-1$ the following condition is satisfied
uniformly with respect to $\nu\in\rbo$ and
to $(x,y)\in\Mm_{j_{1}}\times\Mm_{j_{2}}$:
\begin{equation}
\lim_{t\to+0}\left(\doo_{\nu,x}\right)^k\left(\eo-\pom\right)_{t,x,y}(\nu)=0.
\label{E1971}
\end{equation}
\label{PAT310}
\end{pro}

\begin{cor}
1. For $k\in\wZ_+$, $0\le k\le [-n/2]+m-1$ the following estimates
of $L_{2}$-norms (with respect to the variable $x$) of the kernel
$\left(\pom(\nu)-\eo(\nu)\right)_{t,x,y}$ hold for $t\in(0,T]$
uniformly in $y\in\cup_{j=1,2}\Mm_j$ and in $\nu\in\rbo$, where
$C_m$ and $C_{m,k}$ are the same as in (\ref{S1611}) and (\ref{S16111}):
\begin{align}
\left\|\left(\eo(\nu)-\pom(\nu)\right)_{t,*,y}\right\|_{2} & \!\le
\!C_m(-n/2+m+1)^{-1}t^{-n/2+m+1},
\label{S1615}\\
\!\left\|\Delta^k_{\nu,*}\!\left(\eo(\nu)\!-\!\pom\!(\nu)\!\right)_{t,*,y}
\right\|_{2} & \!\le\! C_{m,k}\left(\!-n/2\!+\!m\!+\!1\!-\!k\right)^{-1}
\!t^{-n/2+m+1-k}.
\label{S1617}
\end{align}
\\

2. The following estimate for $\eotxy(\nu)$ holds for $t\in(0,T]$ and
for an arbitrary  $q\in\wZ_+$ uniformly with respect to $\nu\in\rbo$,
to $x\in M\setminus\left([-3/4,3/4]\times N\right)$, and to $y\in\mab$:
\begin{equation}
\left|\eotxy(\nu)\right|\le C_q t^{-n/2+q}.
\label{E1972}
\end{equation}
It holds according to (\ref{S1615}), (\ref{S1617}), and (\ref{S1094}).
(Indeed, for such $x,y$ and for an arbitrary $m\in\wZ_+$ we have
$\eotxy(\nu)=\left(\eo(\nu)-\pom(\nu)\right)_{t,x,y}$.
Hence $m$ can be chosen large enough to get (\ref{E1972}).)
\label{CAT311}
\end{cor}

The estimate (\ref{S1615}) is a consequence of (\ref{S1611}) and
of the following equality
\begin{multline}
\left(\eo-\pom\right)_{t,x,y}(\nu)=
\lim_{\eps\to+0}\int^{t-\eps}_\eps d\tau\df_\tau\left(\eottxa(\nu),\left(\eo-
\pom\right)_{\tau,*,y}(\nu)\right)_{M}=\\
=\lim_{\eps\to+0}\int^{t-\eps}_\eps d\tau\left(\eottxa
(\nu),\left(\df_\tau+\doo_{\nu,*}\right)\left(\eo-\pom\right)_{\tau,*,y}(\nu)
\right)_{M}.
\label{S1612}
\end{multline}
This equality follows from the assertions that
$\left(\eo-\pom\right)_{t,x,y}(\nu)\to 0$ as $t\to+0$ and that
$\left(\eo-\pom\right)_{\tau,x,y}(\nu)\in D\left(\doo_{\nu,x}\right)$
with respect to the variable $x$.

The estimate (\ref{S1615}) is a consequence of (\ref{S1612}), (\ref{S1611}),
and (\ref{S1614}), since the operator $\exp\left(-t\don\right)$ for $t>0$
is bounded in $\dromb$ and its operator norm is less or equal to one:
\begin{equation}
\left\|\exp\left(-t\don\right)\right\|_{2}\le 1.
\label{S1614}
\end{equation}
(This inequality follows from Theorems~\ref{TAT705} and \ref{TAT706}.
They claim that $\don$ is a nonnegative self-adjoint unbounded operator
in $\dromb$ and that $\exp\left(-t\don\right)$ is a trace class operator.)

The inequality (\ref{S1617}) for $1\le k\le[-n/2]+m-1$ is a consequence
of (\ref{S1611}), (\ref{S1614}), and of the following equality
(which is a generalization of (\ref{S1612})):
\begin{multline*}
\Delta^k_{\nu,x}\left(\eo-\pom\right)_{t,x,y}(\nu)=\\
=\lim_{\eps\to+0}\int^{t-\eps}_\eps
\left(\eottxa(\nu),\left(\doo_{\nu,*}\right)^k\left(\df_\tau+\Delta_{\nu,*}
\right)\left(\eo-\pom\right)_{\tau,*,y}(\nu)\right)_{M}.
\end{multline*}
This equality holds since
$\Delta_{\nu,x}^k\left(\eo-\pom\right)_{t,x,y}(\nu)\to0$
as $t\to+0$ (for $0\le k\le[-n/2]+m-1$) and since
$\left(\doo_{\nu,x}\right)^k\left(\eo-\pom\right)_{t,x,y}(\nu)\in D
\left(\doo_{\nu,x}\right)$ for fixed $y$ and $t>0$.

The proof of Proposition~\ref{PAT310} is preceded by the proof
of Proposition~\ref{PA302}.

The estimates analogous to (\ref{E1972}) (with $t\in(0,T]$ and $q\in\wZ_+$)
hold also for the kernel $\eotxy(\nu;0)=\cPe_{t,x,y}$
of $\exp\left(-t\dono\right)$,
where $x\in\left(I\setminus[-3/4,3/4]\right)\times N$, $y\in\mab$.
Indeed, such estimates are true ($n$ is replaced by $1$)
for $\left(E^i_{I,t}(\nu)\right)_{\xa,\ya}$
with $\xa\in I\setminus[-3/4,3/4]$ and  $\ya\in[-1/2,1/2]$, and
the kernel $\left(E^{\bullet-i}_{N}\right)_{x',y'}$
is $O\left(t^{-(n-1)/2}\right)$ for $t\in(0,T]$.

The desired estimates for $(r\delta)\eo(\nu)$ and $Ad\eo(\nu;0)$ are obtained
from the estimates (\ref{S1615}), (\ref{S1617}), and from the generalization
of the inequality (\ref{S1094}) as follows. Let $K$ be an arbitrary
first order differential operator acting on $\DR\left(\Mm_j\right)$.
Let $\omega\in\DR\left(M_\nu,Z\right)$ obeys the conditions
(\ref{S10941}) with $m_{1}=1+\min\{l: 4l\ge\dim M+1\}$. Then the inequality
is satisfied uniformly with respect to $\omega$ and to $x\in\Mm_1\cup\Mm_2$:
\begin{equation}
\left|K\omega(x)\right|^2<C_{K}\sum^{m_{1}}_{i=0}\left\|\Delta^i_\nu\omega
\right\|^2_{2},
\label{S1620}
\end{equation}
where $C_{K}>0$ is independent of $\nu\in\rbo$.
The proof of (\ref{S1620}) is exactly the same as the proof of
(\ref{S1094}) given above except the kernel
$\left(G_{I}(\nu)\otimes G^{m_{2}}_{N}\right)_{x,y}$ has to be replaced
by $K_x\left(G_{I}(\nu)\otimes G^{m_{2}}_{N}\right)_{x,y}$.
Thus Proposition~\ref{PA302} is proved.\ \ \ $\Box$

\noindent{\bf Proof of Proposition~\ref{PAT310}.}
1. For $x$ from a neighborhood of $N=0\times N\hookrightarrow M$,
where $\psi_{1}\equiv 0$, the parametrix $\pom_{t,x,*}(\nu)$ is equal
to ${\cP}^{\bullet(m)}_{t,x,*}(\nu)\phi(*)$. So
$\pom_{t,x,*}(\nu)\in D\left(\left(\doo_{\nu,x}\right)^k\right)$
with respect to the variable $x$, since
${\cP}^{\bullet(m)}_{t,x,*}(\nu)\in D\left(\left(\doo_{\nu,x}\right)^k\right)$
for $k\in\wZ_+$. \\

2. The estimates (\ref{S1611}) and (\ref{S16111}) hold for the term
$\psi\cPe(\nu)\phi$ of $\pom(\nu)$ with an arbitrary $m\in\wZ_+$, since
$\left(\df_t+\doo_{\nu,x}\right)\cPe(\nu)=0$
(for $x\in(I\setminus\df I)\times N$) and since
$\min_{\xa\in A}\min_{\ya\in\supp\phi}\left(|\xa-\ya|,|\xa+\ya|\right)>\delta
>0$ (where $A:=\supp\left(\df_{\xa}\psi\right)$ and the number $\delta>0$
is fixed).
For $\xa\in A$ and $\ya\in\supp\phi$ the estimate (\ref{E1972})
($n$ is replaced by $1$)
holds with an arbitrary $q\in\wZ_+$ for the fundamental solution
$\left(\eo_{I,t}\right)_{x,y}$ of $\left(\df_t+\doo_{I}\right)$
with the Dirichlet boundary conditions on $\df I$.
The same estimate for such $\xa,\ya$ holds for the kernels
$\left(\sigma^*_{1}\eo_{I,t}\right)_{\xa,\ya}$ and
$\left(\eo_{I,t}(\nu)\right)_{\xa,\ya}$. So this estimate holds
also for the kernel $\left(\cPe(\nu)\right)_{(\xa,x'),(\ya,y')}$
(defined by (\ref{E1970})), since the kernel $\left(\eo_{N,t}\right)_{x',y'}$
is $O\left(t^{-(n-1)/2}\right)$ uniformly with respect to $(x',y')$
(\cite{RS}, Proposition~5.3, \cite{BGV}, Theorem~2.23).
By the analogous reasons, for such $\xa,\ya$
the estimate (\ref{E1972}) with an arbitrary $q\in\wZ_+$ holds
for the kernel $\left(F\circ\cPe(\nu)\right)_{(\xa,x'),(\ya,y')}$,
where $F$ is a linear differential operator of finite order $d(F)$
on $M\times M$, acting on double forms on $\Mm_{j_{1}}\times\Mm_{j_{2}}$,
and $n$ in (\ref{E1972}) is replaced by $n+d(F)$.

So the estimate (\ref{E1972}) with an arbitrary $q\in\wZ_+$
is satisfied by $\left(\df_t+\doo_{\nu,x}\right)\left(\psi\cPe\phi\right)$
and by $\left(\doo_{\nu,x}\right)^k\left(\df_t+\doo_{\nu,x}\right)\left(
\psi\cPe\phi\right)$ with $k\in\wZ_+$.

The estimates (\ref{S1611}) and (\ref{S16111}) hold for
$\psi_{1}p^{\bullet(m)}(1-\phi)=:\pom_{int}$, since they hold for
$p^{\bullet(m)}_{t,x,y}$ and since the distance on $M$ between
the closure $\Bb$ of $B$ (where $B:=\{x\colon d_x\psi_{1}\ne 0\}$)
and the support $\supp(1-\phi)$ is greater than a positive number $\delta$.
Hence the uniform with respect to $(x,y)\in\Bb\times\supp(1-\phi)$
estimate (\ref{E1972}) (with an arbitrary $q\in\wZ_+$) is satisfied
by $p^{\bullet(m)}_{t,x,y}$.
This estimate holds also for the kernel $Fp^{\bullet(m)}_{t,x,y}$,
where $F$ is a linear differential operator
of finite order $d(F)$, acting on the smooth kernels, defined
on $M\times M$. (For instance, the function $f\left(d(x,y)\right)$
in the definition (\ref{E1968}) of $p^{\bullet(m)}_{t,x,y}$ can be chosen
such that $f(\tau)\equiv 0$ for $\tau\ge\delta$. These estimates
follow also from \cite{RS}, Proposition~5.3, estimates~(5.5), and from
\cite{BGV}, Theorem~2.23(2).)%
\footnote{For the sake of brievity the proof of Proposition~\ref{PAT310}
is written in the case of $\df M=\emptyset$.}
\\

3. The difference $\left(\eo-\pom\right)_{t,x,y}(\nu)$ can be written
as the Volterra series (analogous to \cite{BGV}, 2.4):
\begin{multline}
\left(\eo-\pom\right)_{t,x,y}(\nu)=\\
=\sum_{l\ge 1}(-t)^l\int_{\Delta_l}\int_{(\ya,\dots,y_l)\in\left(\Mm_{1}\cup
\Mm_{2}\right)^l}\pom_{\sio t,\xa,\ya}(\nu)r^{(m)}_{\sia t,\ya,\yb}(\nu)\dots
r^{(m)}_{\sig_l t,y_l,y}(\nu),
\label{E1962}
\end{multline}
where $\Delta_l=\left\{(\sio,\dots,\sig_l)\colon 0\!\le\!\sig_i\!\le\! 1,
\sum\sig_i\!=\!1 \right\}$%
\footnote{Here $\sig_i=t_{i+1}-t_i$ for $1\le i\le k-1$, $\sio=t_1$,
$\sig_k=1-t_k$.
The volume of $\Delta_k$ with respect to the density $dt_1\ldots dt_k$
is equal to $1/k!$.}%
and $r^{(m)}_{t,x,y}(\nu):=\left(\Delta_{\nu,x}\!+\!\df_t\right)\pom_
{t,x,y}(\nu)$ (a scalar product $\tr(\oma\wedge*\omb)$ with its values in
densities on $M$ is implied in (\ref{E1962})). The assertion (\ref{E1971})
follows from the convergence of the series (\ref{E1962}) in the topology
of uniform convergence of smooth kernels on $\Mm_{j_{1}}\times\Mm_{j_{2}}$
together with their partial derivatives of orders $\le 2k$
on $\Mm_{j_{1}}\times\Mm_{j_{2}}$ (i.e., in the $C^{2k}$-topology
on $\Mm_{j_{1}}\times\Mm_{j_{2}}$).

Indeed, the definition of $\pom_{t,x,y}$ (\ref{E1970}) involves that
the kernel $r^{(m)}_{t,x,y}(\nu)$ is equal to zero for $x$ from a fixed
(independent of $t$, $\nu$, and $m$) neighborhood
of $N=0\times N\hookrightarrow M$
in $M$. So $r^{(m)}_{t,x,y}(\nu)$ is a smooth kernel on $M\times\Mm_j$, and
the inequalities (\ref{S16100}) claim that
the $C^{2k}$-norm of $r^{(m)}_{t,x,y}(\nu)$ is $O\left(t^{-n/2+m-k}\right)$
for $t\in(0,T]$ uniformly with respect to $\nu\in\rbo$. It is $O(t)$
for $0\le k\le[-n/2]+m-1$, and the series (\ref{E1962}) is convergent
in the $C^{2k}$-topology for such $k$, since the volume of $\Delta_l$
is $(l!)^{-1}$ and since the following assertion is true.
For any $T>0$ the parametrix $\pom_{t,x,y}(\nu)$ defines a family
of bounded operators from the space of smooth forms $\DR(M)$ (equiped
with a $C^{2k}$-norm) into the space $\oplus_{j=1,2}\DR\left(\Mm_j\right)$
(equiped with a $C^{2k}$-norm on $\DR\left(\Mm_j\right)$). These operators
are bounded uniformly with respect to $t\in(0,T]$ and to $\nu\in\rbo$.
This assertion for the operators, corresponding to $\pom_{int,t}$,
is proved in \cite{BGV}, Lemma~2.49. It is also true for the operators
corresponding to $\psi{\cP}^{\bullet(m)}_t(\nu)\phi$. Indeed, it holds for
the operators $\exp\left(-t\doo_{N}\right)$ in $\DR(N)$ (equiped
with a $C^{2k}$-norm)
and for the operators $\psi\exp\left(-t\doo_{I}\right)\phi$
in $\DR(I)$ equiped with a $C^{2k}$-norm. (Here $\doo_I$ is defined
on forms with the Dirichlet boundary conditions on $\df I$.)
It holds also for the operators with the kernels
$\psi(\xa)\left(\sia^*\eo_{I,t}\right)_{\xa,ya}\phi(\ya)$,
acting from smooth forms on $I$ into smooth forms on $[0,\pm 1]$
(where the Dirichlet boundary conditions are implied on $\df I$
and $\sia$ is the mirror symmetry with respect to $0\in I$).

The $C^{2k}$-norm of the kernel $\left(\pom_{N,t}-\eo_{N,t}\right)$
on $N\times N$ is $O\left(t^{m-(n-1)/2-k}\right)$ for $t\in(0,T]$,
where $\pom_{N,t}$ and $\eo_{N,t}$ are the parametrix of the type
(\ref{E1968}) and the fundamental solution for $\left(\df_t+\doo_{N}\right)$
(\cite{BGV}, Theorem~2.30). The operators in $\DR(N)$ corresponding
to $\pom_{N,t}$ are uniformly bounded for $t\in(0,T]$ with respect
to a $C^{2k}$-norm (\cite{BGV}, Lemma~2.49). So the operators
$\exp\left(-t\doo_{N}\right)$ in $\DR(N)$ are uniformly bounded
for $t\in(0,T]$ with respect to a $C^{2k}$-norm. The convergence
of the series on the right in (\ref{E1962})
with respect to $C^{2k}$-norms ($k\le[-n/2]+m-1$) for the kernels
on $\Mm_{j_1}\times\Mm_{j_2}$ involves also a proof of the equality
(\ref{E1962}) (in the case of $m\ge-[-n/2]+2$).
Indeed, we have%
\footnote{The proof of (\ref{S16101}) is analogous to the one given
in \cite{BGV}, Lemma~2.22. It follows from the formula for
$\df_t\int^t_0 f(x,t)dx$.}
\begin{equation}
\left(\df_t+\doo_{\nu,x}\right)\left(\pom_l\right)_{t,x,y}=\left(\romm_l+
\romm_{l+1}\right)_{t,x,y},
\label{S16101}
\end{equation}
where $(-1)^m\pom_l$ is the term with the number $l$ in the right side of
(\ref{E1962}), $(-1)^m\romm_l$ is the same term in which $\pom_{\sio t}$
is replaced by $\romm_{\sio t}$, $\pom_0:=\pom$. For any fixed $y$ and $t>0$
and for any $k\in\wZ_+\cup 0$ we have
$\left(\Delta^k_{\nu,x}\right)\left(\pom_l\right)_{t,x,y}\subset D\left(
\doo_{\nu,x}\right)$.
The series $\Pe_t=\sum_{l\ge 0}(-1)^m\pom_l$ for $m\ge[-n/2]+2$
is the fundumental solution for $\left(\df_t+\doo_{\nu}\right)$
since $\left(\df_t+\doo_{\nu,x}\right)\Pe_t=0$ for $t>0$ and since
the operator corresponding to $\pom_t$ tends in a weak sense
to the identity operator in $\dromb$ as $t\to+0$ (i.e.,
$\pom_t\omega\to\omega$ as $t\to+0$ for $\omega\in\dromb$).
The latter assertion holds for $\pom_{t,int}(1-\phi)\omega$ and
for ${\cP}^{\bullet(m)}\phi\omega$.
Proposition~\ref{PAT310} is proved.\ \ \ $\Box$

\noindent{\bf Proof of Proposition \ref{PA303}.} Proposition~\ref{PA302}
involves the following conclusion. For any $\eps>0$ there exist
a $\delta>0$ and a~neighborhood $U:=U(\nu_{0},\eps)\subset\rbo$ of $\nu_0$
such that the estimate holds uniformly with respect to $\nu\in U$ and
to $s$, $-1<s<1$:
\begin{equation*}
\left|\int_0^{\delta}t^{s-1}dt\int_{\mab}\sum\kab\tr\left(*i^*_k\Eo_t
(\nu)\right)\right|<\eps/20.
\end{equation*}

So it is enough to prove the existence of a neighborhood $U_{1}$
of $\nu_{0}$ such that for any $\nu\in U_{1}$ the following estimate holds
uniformly with respect to $s$ for $-(1-b)<s<1$ ($b$, $0<b<1$, is fixed):
\begin{equation}
\int^T_{\delta}t^{s-1}dt\left|\int_{\mab}\sum\kab\left(\tr\left(*i^*_k
\Eo_t(\nu)\right)-\tr\left(*i^*_k\Eo_t(\nu_{0})\right)\right)\right|<
\eps/10.
\label{S1631}
\end{equation}

For $e(\nu)$ and $e(\nu_{0})$ the equalities (\ref{S1610}) hold. So
the estimate (\ref{S1631}) takes place for $\nu$ sufficiently close
to $\nu_{0}$ since the convergence
\begin{equation}
\eotxy(\nu)\to\eotxy(\nu_{0})
\label{S1632}
\end{equation}
is uniform with respect to $t\!\in\![\delta_{1},T]$ (where $\delta_{1}\!>\!0$
is fixed), to $x\!\in\! M\setminus\left((\!-3/4,\!3/4)\!\times\! N\!\right)$,
and to $y\in\mab$. The convergence of the kernels
\begin{equation}
d_x\eotxy(\nu)\to d_x\eotxy(\nu_{0}),\quad
\delta_x\eotxy(\nu)\to\delta_x\eotxy(\nu_{0}).
\label{S1633}
\end{equation}
is a uniform one for such $(t,x,y)$. All the double forms in (\ref{S1632})
and (\ref{S1633}) are uniformly bounded on the set of such $(t,x,y)$
and their norms at $(t,x,y)$ satisfy the upper estimate
for $t\in(0,\delta_{1}]$
(obtained in Proposition~\ref{PA302} above) through $c_m t^m$
with an arbitrary $m\in\wZ_+$ and with $c_m$ independent of $\nu$.
The uniform convergence of the kernels in (\ref{S1632}) and (\ref{S1633})
on the compact set of $(t,x,y)$ defined above follows from the continuity
in $t$, $x$, $y$, and $\nu$ for $(x,y)\in\Mm_{j_{1}}\times\Mm_{j_{2}}$
of the corresponding double forms. (See Proposition~\ref{PE303}, where
it is proved that these double forms are $C^{\infty}$-smooth in $t$, $x$,
$y$, and $\nu$ for $\Ree\,t>0$ and $\nu\ne(0,0)$.)\ \ \ $\Box$

\noindent{\bf Proof of Proposition~\ref{PA305}.} If $(\ao,\bo)=\nu_{0}\in U$
(i.e., if $\ao\cdot\bo\ne 0$) then we can suppose that $\nu\in U$ in
(\ref{S1609}). In this case, the identity (\ref{S611}) holds for
the difference
\begin{equation}
\left(\eo(\nu)-\eo(\nu_{0})\right)_{t,x_{1},x_{2}}.
\label{S1701}
\end{equation}

Let $\nu_{0}\in\wR^2\setminus\left(U\cup(0,0)\right)$. For example, let
$\nu_{0}:=(\ao,0)$, $\ao\ne 0$. Then the identities (\ref{S608}) and
(\ref{S610}) claim in the cases of $\nu_{0}$ and of $\nu:=(\alpha,\beta)$
that (\ref{S1701}) is equal to
\begin{multline}
\left({\beta\over\alpha}\right)\lim_{\eps\to+0}\int^{t-\eps}_{\eps}d\tau\Bigl[
-\left(r_{*,\df M_{2}}\delta\eotuxaa(\nu),A_{*,\df M_{1}}\eottaxb(\nu_{0})
\right)_{N}+\\
+\!\left(\!A_{*,\df M_{1}}\!\eotuxaa(\nu),r_{*,\df M_{2}}\delta_*
\eottaxb(\nu_{0})\!\right)\!_{N}\!+
\!\left(\!A_{*,\df M_{1}}d_*\eotuxaa(\nu),r_{*,\df M_{2}}\eottaxb(\nu_{0})
\!\right)\!_{N}\!- \\
-\left(r_{*,\df M_{2}}\eotuxaa(\nu),A_{*,\df M_{1}}d_*\eottaxb(\nu_{0})
\right)_{N}\Bigr].
\label{S1710}
\end{multline}

The factor $\kn^{-1}=\beta/\alpha$ in (\ref{S1710}) tends to zero
as $\nu$ tends to $\nu_{0}$. The factors $(1-\kn/\kno)$ and $(1-\kno/\kn)$
in (\ref{S611})) also tend to zero as $\nu\to\nu_0$ in the case
$\nu_{0},\nu\in U$.
The estimate (\ref{S1609}) follows from (\ref{S1710}) and (\ref{S611}).
Indeed, there are the uniform with respect to $\nu$ upper estimates
(analogous to (\ref{S1606})) for the kernels (\ref{X1000}),
where $x,y\in M\setminus\mab$ and $z\in N=0\times N\subset\df M_j$.
These estimates follow from
(\ref{S1094}) and their proofs are completely analogous to the proof
of (\ref{S1606}). The main step in these proofs is using
of the parametrix $P^{(m)}$%
\footnote{The parametrix $P^{(m),\bullet}(\nu)$ for $\eo(\nu)$
can be chosen such that $P^{(m),\bullet}_{t,x,z}(\nu)=0$
for $x\notin[-1/3,1/3]\times N \hookrightarrow M$ and
$z\in[-1/6,1/6]\times N\hookrightarrow M$.}
and of the estimates (\ref{S1611}), (\ref{S16111}), and (\ref{S1617})
for $t\in(0,T]$.\ \ \ $\Box$

\subsection{Dependence on the phase of a cut of the spectral plane.
The analytic torsions as functions of the phase of a cut. Gluing formula
for the analytic torsions}

The scalar analytic torsion (\ref{S14}) depends not only on
$\left(M,g_M,Z,\nu\right)$ but also on the phase $\theta$
of a cut on the spectral plane $\wC\ni\lambda$. A zeta-function
$\znb(s;\theta)$ is defined for $\Ree\,s>n/2$ ($n:=\dim M$) as the sum
of absolutely convergent series $\sum m(\lambda_j)\lambda^{-s}_{j,\theta}$,
where the sum is over nonzero $\lambda_j\in\Spec\left(\don\right)$
and $m(\lambda_j)$ are the multiplicities of $\lambda_j$. The function
$\lambda^{-s}_{j,\theta}:=\exp\left(-s\log_{(\theta)}\lambda_j\right)$,
$\theta>\Image(\log_{(\theta)}\lambda_j)>\theta-2\pi$, is defined for
$\theta\notin\arg\lambda_j+2\pi\wZ$. (For positive self-adjoint operators
this condition means that $\theta\notin 2\pi\wZ$.) All the results
for the analytic torsion norm are obtained above in the case
of $0<\theta<2\pi$ (for instance, for $\theta=\pi$).

The zeta-function $\znb(s;\theta)$ does~not depend on $\theta\notin2\pi\wZ$,
if $[\theta/2\pi]$ does~not change. However we have
\begin{equation}
\znb(s;\theta+2\pi)=\exp(-2\pi is)\znb(s;\theta) \quad\text{ for }\Ree\,s>n/2.
\label{X2045}
\end{equation}
Since $\znb(s;\pi)$ can be meromorphically continued to the complex
plane $\wC\ni s$ (Theorem~\ref{TAT705} below), we see that $\znb(s;\theta)$
for $\theta\notin 2\pi\wZ$ also can be meromorphically continued
to $\wC$ with at most simple poles at $s_j:=(n-j)/2$. The continuation
of $\znb(s;\theta)$ is regular at $s_j$ for $(-s_j)\in\wZ_+\cup 0$.
So the equality (\ref{X2045}) holds for all
$\theta\notin 2\pi\wZ$, $s\in\wC$. Hence for such $\theta$ we have
\begin{equation}
\znb(s;\theta)=\znb(s)\exp(-2\pi ism),
\label{X2046}
\end{equation}
where $\znb(s):=\znb(s;\pi)$ and $m:=[\theta/2\pi]$, $\theta\notin2\pi\wZ$.
{}From now on this $\zeta$-function will be denoted as $\znb(s;m)$
with $m=[\theta/2\pi]$. The value $\znb(0,m)$ is independent of $m$
(according to (\ref{X2046})).

The dependence of the scalar analytic torsion (\ref{S14}) on $m$ is
given by
\begin{equation}
T\left(M_\nu,Z;m\right)=T(M_\nu,Z)\exp\left(-2\pi imF\left(M_\nu,Z\right),
\right)
\label{X2048}
\end{equation}
where%
\footnote{By the definition, $F\left(M_\nu,Z\right)\in\wC/\wZ$ but it
can be also defined as $\sum\left(-1\right)^jj\znj(0)\in\wC$.}
$F\left(M_\nu,Z\right):=\sum\left(-1\right)^jj\znb(0)(\operatorname{mod}\wZ)$
and
$T\left(M_\nu,Z\right):=T\left(M_\nu,Z;0\right)$, i.e.,
$T\left(M_\nu,Z\right)$ corresponds to $\theta=\pi$ and it is the scalar
analytic torsion defined by (\ref{S14}).
Here $Z$ is the union of the connected components of $\df M$ where
the Dirichlet boundary conditions are given. The Laplacian $\don$ is defined
on $\omega\in\DR(M_\nu,Z)$ with the $\nu$-transmission boundary conditions
(\ref{Y4}) on $N$, with the Dirichlet and the Neumann boundary conditions
on $Z$ and on $\df M\setminus Z$. The equality (\ref{X2048}) is obtained
by using of
\begin{equation*}
\df_s\znb(s;m)\big|_{s=0}=-2\pi im\znb(0)+\df_s\znb(s)\big|_{s=0}.
\end{equation*}

The number $F(M,Z)$ is defined also in the case of a manifold $M$
without an interior boundary $N$. In this case, $\znb(0)$
in the definition of $F(M,Z)$ is replaced by $\zeta_{\bullet}(0)$
for the Laplacians on $\DR(M,Z)$. The dependence of $T(M,Z;m)$ on $m$
is given by (\ref{X2048}) with $F\left(M_\nu,Z\right)$ replaced
by $F(M,Z)$. In particular, $F(M)$ is defined for a closed $M$ and
also in the case $\df M\ne\emptyset$, $Z=\emptyset$.
Let $M$ be obtained by gluing two pieces $M_1$ and $M_2$ along the common
component $N$ of their boundaries, $M=M_1\cup_N M_2$, where $N\subset M$
is closed and of codimension one.
Then $F(M,Z)=F\left(M_{1,1},Z\right)$, according to Proposition~\ref{P1}.

The class of $F\left(M_\nu,Z\right)$ in $\wC/\wZ$ is the same as the class
$(\operatorname{mod}\wZ)$ of the number $F_1\left(M_\nu,Z\right)\in\wC$, where
\begin{equation*}
F_1\left(M_\nu,Z\right):=\sum\left(-1\right)^jj\left(\znj(0)+\dim\Ker
\left(\djn\right)\right).
\end{equation*}

The Laplacian $\djn\left(M_\nu,Z\right)$ with its domain
$\Dom\left(\djn\right)\subset\left(DR^j(M)\right)_2$ is self-adjoint
according to Theorem~\ref{TAT705}. For $\Ree\,s>2^{-1}\dim M$
the zeta-function $\znj(s)$
is defined by the absolutely convergent series
$\sum m\left(\lambda_k\right)\exp\left(-s\log\lambda_k\right)$
($[\theta/2\pi]=0$, $\theta\ne 0$),
where the sum is over $\lambda_k\in\Spec\left(\djn\right)$,
$\lambda_k\ne 0$, and with the branch of logarithm
$-\pi<\Image\log\lambda<\pi$. Because $\log\lambda_k\in\wR$
for $\lambda_k>0$, the function $\znj(s)$ is real for real $s$.
Hence $F_1\left(M_\nu,Z\right)\in\wR$.
It is supposed from now on that a metric $g_M$ on $M=M_1\cup_N M_2$
is a direct product metric near $N$ and near $\df M$.

\begin{pro}
1. For a closed manifold $M$ the number $F_1(M)$ is an integer.

2. Let $M=M_1\cup_N M_2$ be obtained by gluing two its pieces $M_1$
and $M_2$ along the common component $N$ of their boundaries.
Let the $\nu$-transmission boundary
conditions (\ref{Y4}) be given on $N$, the Dirichlet boundary conditions
be given on a union $Z$ of some connected components of $\df M$ and
the Neumann boundary conditions be given on $\df M\setminus Z$.
Let $L$ be a closed Riemannian manifold. Then the following holds:%
\footnote{The Euler characteristic
$\chi\left(M_\nu,Z\right):=\sum(-1)^i\dim H^i\left(M_\nu,Z\right)$
is equal to the Euler characteristic of the complex
$\left(\Ce(X_\nu,X\cap Z),d_c\right)$ (as it follows from
Proposition~\ref{PA87}). Hence it is equal to $\chi(M,Z)$ and is
independent of $\nu\in\rbo$.}
\begin{equation}
F_1\left(M_\nu\times L,Z\times L\right)=\chi(L)F_1\left(M_\nu,Z\right)+
F_1(L)\chi\left(M_\nu,Z\right).
\label{X2056}
\end{equation}

3. Let $K\subset\df M\setminus Z$ be a union of some connected components
of $\df M$. Then the following holds under the conditions on $M$ above:
\begin{equation}
F_1\left(M_\nu,Z\right)=F_1\left(M_\nu,Z\cup K\right)+F_1(K)+2^{-1}
\chi(K).
\label{X2060}
\end{equation}

4. Under the conditions on $M$ above, the number $F_1\left(M_\nu,Z\right)$
obeys a gluing property analogous to the gluing property (\ref{ST1})
for the analytic torsion norms. Namely the following holds:
\begin{equation}
F_1\left(M_\nu,Z\right)=F_1(M_1,Z_1\cup N)+F_1(M_2,Z_2\cup N)+F_1(N)+
2^{-1}\chi(N),
\label{X2057}
\end{equation}
where $Z_k:=Z\cap\df\Mm_k$.
\label{PA418}
\end{pro}

\begin{cor}
1. For a closed $M$ the scalar analytic torsion
$T\left(M,[\theta/2\pi]\right)$  is independent of $\theta\notin2\pi\wZ$.

2. Under the conditions of (\ref{X2056}), (\ref{X2060}), and
(\ref{X2057}), the following holds in $\wR/\wZ$:
\begin{gather*}
F\left(M_\nu\times L,Z\times L\right)=\chi(L)F\left(M_\nu,Z\right). \\
F\left(M_\nu,Z\right)=F\left(M_\nu,Z\cup K\right)+2^{-1}\chi(K). \\
F\left(M_\nu,Z\right)=F\left(M_1,Z_1\cup N\right)+F(M_2,Z_2\cup N)+
2^{-1}\chi(N).
\end{gather*}
\label{CA419}
\end{cor}

\begin{exmp}
The number $F_1(S^1)$ is equal to $-f_{0;1}=-f_{0;0}$.%
\footnote{The coefficients $f_{k,j}:=f_{k,j}(M,Z)$ are the coefficients
in the asymptotic expansion $\sum f_{k,j}t^{k/2}$ ($k\ge-n$)
for $\Tr\exp\left(-t\Delta_j(M,Z)\right)$ as $t\to+0$ for the Laplacian
on $DR^j(M,Z)$ ($n:=\dim M$).}
The latter one is equal to zero because the asymptotic expansion
for $\Tr\exp\left(-t\Delta_0\left(S^1\right)\right)$ as to $t\to+0$
(where $\Delta_0$ is the Laplacian on functions) is
$f_{-1;0}t^{-1/2}+f_{1;0}t^{1/2}+f_{3;0}t^{3/2}+\ldots$.
\label{EXA}
\end{exmp}

\begin{exmp}
The number $F_1(I,\df I)$ is equal to $-f_{0;1}(I,\df I)=-f_{0;0}(I)$.
Since $S^1$ has a mirror symmetry relative to its diameter, we have,
taking into account (\ref{S414}) and (\ref{S1607}),
\begin{equation}
f_{0,0}\left(S^1\right)=f_{0;0}(I)+f_{0;0}(I,\df I)=0.
\label{X2066}
\end{equation}
Since $f_{0;0}(I,\df I)-f_{0;1}(I,\df I)=\chi(I,\df I)=1$ and
(\ref{X2066}) holds, we see that
\begin{equation}
F_1(I,\df I)=-f_{0;1}(I,\df I)=-f_{0;0}(I)=f_{0;0}(I,\df I)=-2^{-1}.
\label{X2067}
\end{equation}
\label{EXB}
\end{exmp}

\begin{rem}
The equality (\ref{X2067}) means that the analytic torsion
$T\left(I,\df I;[\theta/2\pi]\right)$ is multiplied by the factor
$\exp\left(-2\pi i\cdot\left(-2^{-1}\right)\right)=-1$, if
$\theta$ is replaced by $\theta+2\pi$ ($\theta\notin2\pi\wZ$).

It is necessary to note the following. The scalar analytic torsion
$T(I,\df I):=T(I,\df I;\pi)$ is the factor in the analytic torsion
norm. But the latter one is the {\em square of the norm}
on the determinant line $\det H^1(I,\df I)$. So the factor, corresponding
to the {\em norm} itself, is multiplied in the case of $(M,Z)=(I,\df I)$
by the factor
\begin{equation*}
\exp\left(\pi i\zeta_1(I,\df I)|_{s=0}\right)=\exp(-\pi i/2)=-i,
\end{equation*}
if $\theta$ is replaced by $\theta+2\pi$, $\theta\notin 2\pi\wZ$. Indeed,
we have
$$
-2^{-1}=F_1(I,\df I)=-\left(\dim H^1(I,\df I)+\zeta_1(I,\df I)
|_{s=0}\right),
$$
and so $\zeta_1(I,\df I)|_{s=0}=-2^{-1}=-F(I,\df I)$.
For $M=S^1$ it holds that $-F(S^1)=\zeta_1\left(S^1\right)|_{s=0}=-1$,
and so $\exp\left(-\pi iF\left(S^1\right)\right)=-1$.

It follows from Proposition~\ref{PA422} below that $F(M_\nu,Z)$
and $F(M,Z)$ have a form $(1/2)+\wZ$, if the numbers $n:=\dim M$
and $\chi(M,Z)$ are odd. So in this case the factors
$\exp\left(-\pi iF\left(M_\nu,Z\right)\right)$ and
$\exp\left(-\pi iF(M,Z)\right)$ are equal to $\{\pm i\}$.
\label{RA420}
\end{rem}

\noindent{\bf Proof of Proposition~\ref{PA418}.}
1. Theorem~\ref{TAT706} {\em 1.} claims that the number
$\znj(0)+\dim\Ker\left(\djn\right)$ is equal to the constant term
$f_{0;j}\left(M_\nu,Z\right)$ in the asymptotic expansion (\ref{S631})
for $\Tr\exp\left(-t\djn\right)$ as $t\to+0$. So according to
Theorem~\ref{TAT706} {\em 1.} the number $F_1(M_\nu,Z)$ is equal to the sum
of the integrals over $M_1$, $M_2$, $N$, and $\df M$ of the locally defined
densities. Then we have
\begin{equation}
F_1\left(M_\nu,Z\right)=\sum\left(-1\right)^jjf_{0;j}(M_\nu,Z).
\label{X2062}
\end{equation}

If $\left(M,g_M\right)$ is a closed Riemannian manifold then
$f_{0;j}(M)=f_{0;n-j}(M)$. Hence taking into account (\ref{X2062})
and (\ref{S417}), we get (for even $n:=\dim M$)
\begin{equation}
F_1(M)=\sum\left(-1\right)^jjf_{0;j}(M)=(n/2)\sum(-1)^jf_{0;j}(M)=(n/2)\chi(M).
\label{X2063}
\end{equation}

Let $n$ be odd. Then $f_{0;j}(M)$ is equal to zero since the asymptotic
expansion for $\tr\exp\left(-t\Delta_j\left(M,g_M\right)\right)$
(as $t\to+0$) is $t^{-n/2}\sum t^lf_{2l-n;j}\left(M,g_M\right)$, where
the sum is over $l\in\wZ_+\cup 0$ (\cite{Gr}, Theorem~1.6.1;
\cite{BGV}, Theorem~2.30). Hence $F_1(M)=0=(n/2)\chi(M)$ for an odd $n$ also.

This number $(n/2)\chi(M)$ is an integer for any closed $M$. (The assertion
that $F_1(M)$ is an integer follows also from the equality which holds
for any closed even-dimensional Riemannian $(M,g_M)$:
$$
\sum j\left(-1\right)^j\zeta_j(M,s)=\sum j(-1)^j\zeta_{n-j}(M,s)=(n/2)\sum
\left(-1\right)^j\zeta_j(M,s)=0,
$$
because $\zeta_j(M,s)=\zeta_{n-j}(M,s)$.)

2. Let $\lambda\in\Spec\left(\djn(M,Z)\right)$,
$\mu\in\Spec\left(\Delta_i(L)\right)$ and let $m_\lambda(j;M_\nu,Z)$,
$m_\mu(i;L)$ be their multiplicities. If $\lambda\ne 0$ and $\mu\ne 0$
then we have
\begin{equation}
\sum\left(-1\right)^{j+i}(j+i)m_\lambda(j;M_\nu,Z)m_\mu(i;L)=0,
\label{X2065}
\end{equation}
since the subcomplexes $\left(\Ve_\lambda\left(M_\nu,Z\right),d\right)
\hookrightarrow\left(\DR\left(M_\nu,Z\right),d\right)$ and
$\left(\Ve_\mu(L),d\right)\hookrightarrow\left(\DR(L),d\right)$,
corresponding to the $\lambda$-eigenforms for $\dnb(M,Z)$ and
to the $\mu$-eigenforms for $\Delta_{\bullet}(L)$, are acyclic.
If $\lambda\ne 0$ but $\mu=0$ then the right side of (\ref{X2065}) is equal to
\begin{equation*}
\left(\sum\left(-1\right)^jjm_\lambda\left(j;M_\nu,Z\right)\right)\left(\sum
\left(-1\right)^i\dim\Ker\Delta_i(L)\right)=
=\chi(L)\sum\left(-1\right)^jjm_\lambda\left(j;M_\nu,Z\right).
\end{equation*}
So under the conditions of {\em 2}, by using (\ref{V3}), we have
\begin{multline}
\sum\left(-1\right)^jj\znj\left(M_\nu\times L,Z\times L\right)\big|_{s=0}=\\
=\chi(L)\sum\left(-1\right)^jj\znj\left(M_\nu,Z\right)\big|_{s=0}+
\chi\left(M_\nu,Z\right)\sum\left(-1\right)^jj\zeta_j(L)\big|_{s=0},
\label{X2167}
\end{multline}
\begin{multline}
\sum\left(-1\right)^jj\dim\Ker\left(\djn\left(M_\nu\times L,Z\times L\right)
\right)=\\
=\chi(L)\sum\left(-1\right)^jj\dim\Ker\left(\djn\left(M_\nu,Z\right)\right)+
\chi\left(M_\nu,Z\right)\sum\left(-1\right)^jj\dim\Ker\left(\Delta_j(L)
\right).
\label{X2168}
\end{multline}

The equality (\ref{X2056}) follows now from (\ref{X2167}) and (\ref{X2168}).

3. The numbers $F_1\left(M_\nu,Z\right)$ and $F_1(M_\nu,Z\cup K)$
are the sums of the integrals over $M_1$, $M_2$, $N$, and $\df M$
of the locally defined densities (as it follows from (\ref{X2062})
and from Theorem~\ref{TAT706}). The densities, corresponding to the pairs
$(M_\nu,Z)$ and $(M_\nu,Z\cup K)$, differ only on $K$. So the difference
$F_1(M_\nu,Z)-F_1(M_\nu,Z\cup K)$ depends only on $K$ and on $g_M$
near $K$. Thus taking into account that $g_M$ is a direct product metric
near $K$, we get
\begin{equation}
2\left(F_1\left(M_\nu,Z\right)-F_1\left(M_\nu,Z\cup K\right)\right)=F_1
(K\times I)-F_1\left(K\times(I,\df I)\right)
\label{X2070}
\end{equation}
for any fixed metric on $K$ in all the terms of this equality.

According to (\ref{X2056}) we have
\begin{gather}
\begin{split}
F_1(K\times I)                   & =F_1(K)\chi(I)+F_1(I)\chi(K),\\
F_1\left(K\times(I,\df I)\right) & =F_1(K)\chi(I,\df I)+F_1(I,\df I)\chi(K),\\
F_1(K\times I)-F_1\left(K\times(I,\df I)\right) & =2F_1(K)+\chi(K)
\left(F_1(I)-F_1(I,\df I)\right).
\end{split}
\label{X2071}
\end{gather}
Since $\zeta_1(s;I)=\zeta_1(s;I,\df I)$ (on the same $I$) we have
$$
F_1(I)-F_1(I,\df I)=
-\zeta_1(I)\big|_{s=0}+\zeta_1(I,\df I)\big|_{s=0}+\dim H^1(I,\df I)=1.
$$
By (\ref{X2070}) and (\ref{X2071}) we get
$$
F_1(M_\nu,Z)=F_1(M_\nu,Z\cup K)+F_1(K)+2^{-1}\chi(K).
$$

4. The number $F_1(M_\nu,Z)$ is the sum of the integrals over $M_1$,
$M_2$, $N$, and over $\df M$ of the locally defined densities. (It is
a consequence of Theorem~\ref{TAT706}). So the densities on $M_j$, $N$,
and on $\df M\cap\Mm_j$ are the same as for the number
$F_1\left(M^{(2)}_{j,\nu},Z^{(2)}_j\right)$,
where $M^{(2)}_j:=M_j\cup_N M_j$, $Z^{(2)}_j:=Z_j\cup Z_j$, and
$g_{M^{(2)}_j}$
are mirror symmetric with respect to $N$ (the $\nu$-transmission
boundary conditions are given on $N\hookrightarrow M^{(2)}_j$)
and $g_{M_j^{(2)}}|_{M_j}=g_M|_{M_j}$. Thus we have
\begin{equation*}
2F_1(M_\nu,Z)=\sum_{j=1,2}F_1\left(M^{(2)}_{j,\nu},Z^{(2)}_j\right).
\end{equation*}
Since pairs $\left(M^{(2)}_j,Z^{(2)}_j\right)$ are mirror symmetric
with respect to $N$, it follows from (\ref{S414}) and (\ref{S415}) that
\begin{align}
F_1\left(M^{(2)}_{j,\nu},Z^{(2)}_j\right) & =F_1\left(M^{(2)}_j,Z^{(2)}_j
\right)=F_1(M_j,Z_j)+F_1\left(M_j,Z_j\cup N\right),\\
F_1(M_\nu,Z) & =2^{-1}\sum_{j=1,2}\left(F_1\left(M_j,Z_j\right)+F_1\left(M_j,
Z_j\cup N\right)\right).
\label{X2075}
\end{align}

The equality (\ref{X2057}) follows from (\ref{X2060}) with $M=M_j$, $Z=Z_j$,
$K=N$, and $\nu=(1,1)$ and from (\ref{X2075}), because
$F_1(M_j,Z_j)=F_1(M_j,Z_j\cup N)+F_1(N)+2^{-1}\chi(N)$.
Thus Proposition~\ref{PA418} is proved.\ \ \ $\Box$

The analytic torsion $T_0\left(M_\nu,Z;m\right)$ (where $Z\subset\df M$
is a union of some connected components of $\df M$, $m:=[\theta/2\pi]$,
$\theta\notin 2\pi\wZ$) is defined as the {\em product of the norm}
$\left\|\cdot\right\|^2_{\det\Hh(M_\nu,Z)}$
(given by the natural norm on harmonic forms for $\don(M_\nu,Z)$) and
of the {\em scalar analytic torsion} $T(M_\nu,Z;m)$:
\begin{equation}
T_0\left(M_\nu,Z;m\right):=\left\|\cdot\right\|^2_{\det\Hh(M_\nu,Z)}T\left(
M_\nu,Z;m\right).
\label{X2076}
\end{equation}

The analytic torsion $T_0\left(M,Z;m\right)$ is the norm
$\left\|\cdot\right\|^2_{\det\Hh(M,Z)}T\left(M,Z;m\right)$,
where the norm on the determinant line is given by the harmonic forms for
$\Delta_{\bullet}(M,Z)$. (If $N$ is the interior boundary and
if $g_M$ is a direct product metric near $N$ then
$T_0\left(M,Z;m\right)=T_0\left(M_{1,1},Z;m\right)$ according to
Proposition~\ref{P1}.)

\begin{thm}
1. Let $M$ be obtained by gluing two pieces along $N$, $M=M_1\cup_N M_2$,
where $N$ is a closed of codimension one submanifold in $M$ with
a trivial normal bundle $TM|_N/TN$ and $g_M$ is a direct product metric
near $N$ and near $\df M$. Then for $\nu\in\rbo$ the following gluing
formula holds:
\begin{multline}
\fan T_0\left(M_\nu,Z;m\right):=\\
=(-1)^{m\chi(N)}T_0\left(M_1,Z_1\cup N;m\right)\otimes T_0\left(M_2,Z_2\cup N;m
\right)\otimes T_0(N;m),
\label{X2077}
\end{multline}
where the identification $\fan$ of the determinants lines is defined
by the short exact sequence (\ref{A12}) of the de Rham complexes and by
Lemma~\ref{L1}, $Z_k:=Z\cap\df\Mm_k$. The factor $T_0(N;m):=T_0(N)$
is independent of $m$ (according to Proposition~\ref{PA418} 1).

2. Let $K\subset\df M\setminus Z$ be a union of some connected components
of $\df M$. Then the formula holds for gluing $K$ and
$\left(M_\nu,Z\cup K\right)$:
\begin{equation}
\phi_{an}T_0\left(M_\nu,Z;m\right)=\left(-1\right)^{m\chi(K)}T_0\left(M_\nu,
Z\cup K;m\right)\otimes T_0(K;m).
\label{X2079}
\end{equation}
Here the identification $\phi_{an}$ is defined by the short exact sequence
(analogous to (\ref{B2})):
\begin{equation}
0\to\DR\left(M_\nu,Z\cup K\right)\to\DR\left(M_\nu,Z\right)\to\DR(K)\to 0
\label{X2081}
\end{equation}
(the left arrow in (\ref{X2081}) is the natural inclusion and the right
arrow is the geometrical restriction) and by Lemma~\ref{L1}. The factor
$T_0(K;m):=T_0(K)$ is independent of $m$. The analogous formula holds
for gluing $K$ and $(M,Z\cup K)$:
\begin{equation}
\phi_{an}T_0\left(M,Z;m\right)=\left(-1\right)^{m\chi(K)}T_0\left(M,Z\cup K;
m\right)\otimes T_0(K),
\label{X2080}
\end{equation}
where $\phi_{an}$ is defined by the short exact sequence (\ref{X2081})
with $M_\nu$ replaced by $M$.
\label{TA421}
\end{thm}

\noindent{\bf Proof.}
1. For $T_0\left(M_\nu,Z\right):=T_0\left(M_\nu,Z;0\right)$ the following
gluing formula holds (according to (\ref{A10}) and to Lemma~\ref{L2}):
\begin{equation}
\fan T_0\left(M_\nu,Z\right)=T_0\left(M_1,Z_1\cup N\right)\otimes T_0\left(
M_2,Z_2\cup N\right)\otimes T_0(N).
\label{X2082}
\end{equation}
By the definition of $F\left(M_\nu,Z\right)$ we have
$$
T_0\left(M_\nu,Z;m+1\right)=\exp\left(-2\pi iF\left(M_\nu,Z\right)\right)T_0
\left(M_\nu,Z;m\right).
$$
(Analogous equalities are true for $T_0\left(M_j,Z_j\cup N;m\right)$.
The differences $F_1\left(M_\nu,Z\right)-F(M_\nu,Z)$, $F_1(M_j,Z_j\cup N)-
F(M_j,Z_j\cup N)$, and $F_1(N)-F(N)$ are integers and $F_1(N)$ is an integer
(according to Proposition~\ref{PA418} {\em 1}). Hence (\ref{X2077}) is
a consequence of (\ref{X2057}) and (\ref{X2082}).

2. The gluing formula holds for $T_0(M,Z):=T_0(M,Z;0)$ according to
(\ref{X3}) (Theorem~\ref{T2}):
$\phi_{an}T_0(M,Z)=T_0(M,Z\cup K)\otimes T_0(K)$.
So the gluing formula (\ref{X2080}) follows from (\ref{X2060}) since
$T_0(M,Z;m+1)=\exp\left(-2\pi iF(M,Z)\right)T_0(M,Z;m)$
and since the difference $F_1(M,Z)-F(M,Z)$ is an integer.

Let $N\subset M$ be a disjoint union $N_1\cup N_2$ of two closed
codimension one submanifolds of $M$ with trivial normal bundles
and let the $\nu_j$-transmission interior boundary conditions be given
on $N_j$. Let $M=M_1\cup_{N_1}M_2$ and let $N_2\subset M_1$.
Under these conditions, the equality (\ref{A10}) and the assertion
of Lemma~\ref{L2} are also true. Their proofs are similar to the given
above. The resulting formula is
\begin{equation}
\phi^{an}_{\nu_1}T_0\left(M_{\nu_1,\nu_2},Z\right)=T_0\left(M_{1,\nu_2},Z_1
\cup N_1\right)\otimes T_0\left(M_2,Z_2\cup N_1\right)\otimes T_0(N_1).
\label{X2084}
\end{equation}
(Here $Z\subset\df M$ is a union of some connected components of $\df M$,
$Z_k=Z\cap\df\Mm_k$ and $g_M$ is a direct product metric near $N_j$ and
$\df M$.) As a consequence of (\ref{X2084}) (obtained by the same method
as Theorem~\ref{T2} is obtained from (\ref{X1}) and (\ref{X2}))
we get the following equality
\begin{equation}
\phi_{an}T_0\left(M_{1,\nu_2},Z_1\right)=T_0\left(M_{1,\nu_2},Z_1\cup N_1
\right)T_0(N_1).
\label{X2085}
\end{equation}

The equality (\ref{X2079}) follows from (\ref{X2085}) (where $M_{1,\nu_2}$,
$Z_1$, $N_1$ are replaced by $M_\nu$, $Z$, $K$) and from
(\ref{X2060}) since
$T_0\left(M_\nu,Z;m+1\right)=\exp\left(-2\pi iF\left(M_\nu,Z\right)\right)T_0
\left(M_\nu,Z;m\right)$.
Theorem~\ref{TA421} is proved.\ \ \ $\Box$

\begin{pro}
Let $M=M_1\cup_N M_2$ be obtained by gluing along $N$ and
let the $\nu$-transmission boundary conditions ($\nu\in\rbo$) be given
on $N$. Set $n:=\dim M$. Then the number $F_1\left(M_\nu,Z\right)$
for the scalar analytic torsion $T\left(M_\nu,Z;m\right)$ is expressed by
\begin{equation}
F_1\left(M_\nu,Z\right)=2^{-1}n\chi\left(M_\nu,Z\right)=2^{-1}n\chi\left(M,Z
\right).
\label{X2086}
\end{equation}

The number $F\left(M_\nu,Z\right):=\sum\left(-1\right)^jj\znj(0)$
is as follows:
\begin{equation*}
F\left(M_\nu,Z\right)=\sum\left(-1\right)^j\left(-j+2^{-1}n\right)\dim H^j
\left(M_\nu,Z\right).
\end{equation*}
\label{PA422}
\end{pro}

\noindent{\bf Proof.}
1. Proposition~\ref{PA87} claims that $\chi\left(M_\nu,Z\right)$
(i.e., the Euler characteristic
$\sum\left(-1\right)^j\dim H^j\left(DR\left(M_\nu,Z\right)\right)$)
is equal to the Euler characteristic for the finite-dimensional complex
$\left(\Ce\left(X_\nu,Z\cap X\right),d_c\right)$.
Note that $\dim C^j(X,Z\cap X)$ is equal
to $\dim C^j\left(X_\nu,Z\cap X\right)$. Hence
$\chi\left(M_\nu,Z\right)=\sum\left(-1\right)^j\dim C^j\left(X,Z\cap X
\right)$.
This sum is equal to $\chi(M,Z)$ by the de Rham theorem (\cite{RS},
Proposition~4.2). Thus $\chi\left(M_\nu,Z\right)=\chi(M,Z)$.

2. According to (\ref{X2057}), the gluing formula holds:
$$
F_1\left(M_\nu,Z\right)=F_1\left(M_1,Z_1\cup N\right)+F_1\left(M_2,Z_2\cup N
\right)+F_1(N)+2^{-1}\chi(N).
$$

For the Euler characteristics the analogous formula holds:

\begin{equation}
\chi\left(M_\nu,Z\right)=\chi(M,Z)=\chi\left(M_1,Z_1\cup N\right)+
\chi\left(M_2,Z_2\cup N\right)+\chi(N).
\label{X2087}
\end{equation}

The number $F_1(N)$ for a closed manifold $N$ is equal
to $\chi(N)(\dim N)/2$ by (\ref{X2063}). So $F_1(N)+2^{-1}\chi(N)=n\chi(N)/2$.
Let $\left(M,g_M\right)$ be a closed Riemannian manifold, mirror symmetric
with respect to $N=\df M_1$, $M=M_1\cup_N M_1$.
Let $g_M$ be a direct product metric near $N$.
Then the equality $F_1\left(M_1,N\right)=2^{-1}n\chi\left(M_1,N\right)$
follows from (\ref{X2063}), which claims that $F_1(M)=2^{-1}n\chi(M)$,
and from (\ref{X2087}). So (\ref{X2086}) holds for pairs $(M,Z)$,
where $Z=\df M$.
For any union $Z$ of some connected components of $\df M$ the equality
(\ref{X2086}) follows from (\ref{X2060}) for $K=\df M\setminus Z$, as
$F_1(K)+2^{-1}\chi(K)=n\chi(K)/2$, according to (\ref{X2063}). The equality
(\ref{X2086}) for $F_1\left(M_\nu,Z\right)$ follows from its particular
cases for $F_1\left(M_j,Z_j\cup N\right)$ by using (\ref{X2057})
and (\ref{X2087}).\ \ \ $\Box$

\begin{cor}
1. The analytic torsion $T_0\left(M_\nu,Z;m\right)$ (defined
by (\ref{X2076})) is the following function
of $m=[\theta/2\pi]$ ($\theta\notin2\pi\wZ$):
\begin{equation}
T_0\left(M_\nu,Z;m\right)=\left(-1\right)^{mn\chi(M,Z)}T_0(M_\nu,Z).
\label{X2187}
\end{equation}
Here $T_0\left(M_\nu,Z\right):=T_0\left(M_\nu,Z;0\right)$,
$n:=\dim M$.

2. The analytic torsion $T_0(M,Z;m)$ is equal to $T\left(\mno,Z;m\right)$
for $\nu_0=(1,1)$ according to Proposition~\ref{P1}. The formula
(\ref{X2187}) holds also for $T_0(M,Z;m)$, where $T_0\left(M_\nu,Z\right)$
is replaced by $T_0(M,Z):=T_0(M,Z;0)$.
\label{CA423}
\end{cor}

Let $\left(M,g_M\right)$ be obtained by gluing two Riemannian manifolds
$M_1$ and $M_2$ along a common component $N$ of their boundaries,
$M=M_1\cup_N M_2$.
Let $g_M$ be a direct product metric near $N$
and near $\df M$ and let $Z\subset\df M$ be a union of some connected
components of $\df M$. The following main theorem
is an immediate consequence of Theorems~\ref{T4}, \ref{T5} and of
Corollary~\ref{CA423}.

\begin{thm}[Generalized Ray-Singer conjecture]
1. The analytic torsion \\ $T_0\left(M_\nu,Z;m\right)$ is expressed through
the combinatorial torsion norm (\ref{LL6}) as follows:
\begin{equation*}
T_0\left(M_\nu,Z;m\right)=2^{\chi(\df M)+\chi(N)}\left(-1\right)
^{mn\chi(M,Z)/2}\tau_0\left(M_\nu,Z\right)
\end{equation*}
(where $m=[\theta/2\pi]$, $\theta\notin2\pi\wZ$ is the phase of a cut
of the spectral plane $\wC\ni\lambda$ and $n=\dim M$).

2. The analytic torsion $T_0(M,Z;m)$ is expressed through the combinatorial
torsion norm:
\begin{equation*}
T_0(M,Z;m)=2^{\chi(\df M)}\left(-1\right)^{mn\chi(M,Z)/2}\tau_0(M,Z).
\end{equation*}
\label{TA424}
\end{thm}

\begin{rem}
The combinatorial torsion norms $\tau_0(M,Z)$ and $\tau_0\left(\mno,Z\right)$
(where $\nu_0=(1,1)$) on the determinant line
$\det\Hh\left(\mno,Z\right)=\det\Hh(M,Z)$ are different, if $\chi(N)\ne 0$
(by Remarks~\ref{R4} and \ref{R6}). The canonical identifications
$\Hh\left(\mno,Z\right)=\Ker\left(\donoo\right)=\Ker\left(\doo\right)=
\Hh(M,Z)$ are given by Proposition~\ref{P1} and by the de Rham theorem.
\label{RA425}
\end{rem}

\section{Zeta- and theta-functions for the Laplacians with
$\nu$-transmission interior boundary conditions}
\subsection{Properties of zeta- and theta-functions for
$\nu$-transmission boundary conditions}
Let $M$ be a compact manifold with boundary obtained by gluing
manifolds $M_1$ and $M_2$ along a common component $N$ of their boundaries,
$M=M_1\cup_N M_2$ ($N\subset M$ is a closed codimension one submanifold
of $M$ with a trivial normal bundle $TM|_N/TN$).
Let $g_{M}$ be a direct product metric near
$N=0\times N\hookrightarrow I\times N\hookrightarrow M$.
Let the Dirichlet boundary conditions be given on a union $Z$ of some
connected components of $\df M$, the Neumann ones be given
on $\df M\setminus Z$ and the $\nu$-transmission interior boundary
conditions (\ref{Y4}) be given on $N$.

The operator $\don$ is originally defined on the set $D\left(\don\right)$
of all the pairs of smooth forms
$\omega=(\oma,\omb)\in\DR\left(\Mm_{1}\right)\oplus\DR\left(\Mm_{2}\right)$
such that the Dirichlet boundary conditions hold for $\omega$ on $Z$, the
Neumann boundary conditions hold on $\df M\setminus Z$ and the interior
boundary conditions (\ref{Y4}) hold for $\omega$ on $N$.
Let $\Dom\left(\don\right)$ be the closure of the $D\left(\don\right)$
in $\dromb$ in the topology given by the graph norm%
\footnote{The $L_{2}$-completion $\dromb$ of $\DR(M)$ coincides
with the $L_{2}$-completion
of $\DR\left(\Mm_{1}\right)\oplus\DR\left(\Mm_{2}\right)$.}
$\left\|\omega\right\|^2_{2}+\left\|\don\omega\right\|^2_{2}=:\left\|\omega
\right\|^2_{graph}$.
The closure of the operator $\don$ (with respect to the graph norm)
is an operator with the domain of definition $\Dom\left(\don\right)$.
If $\omega_j\to\omega$ in the graph norm topology,
$\omega_j\in D\left(\don\right)$, then $\don(\omega)$ is defined
as $\lim_j\den\omega_j$ in the $L_{2}$-topology in $\dromb$.

\begin{thm}
1. The operator $\don$ with the domain $\Dom\left(\don\right)$
is self-adjoint in $\dromb$.
Its spectrum $\Spec\left(\don\right)\subset\wR_+\cup 0$ is discrete.%
\footnote{A spectrum is discrete if it consists entirely of isolated
eigenvalues with finite multiplicities.} \\
2. Its {\em zeta-function} is defined for $\Ree\,s>(\dim M)/2$ by the
absolutely convergent series (including the multiplicities)
$\znb(s):=\sum_{\lambda_j\in\Spec\left(\don\right)\setminus 0}\lambda_j^{-s}$.
This series converges uniformly for $\Ree\,s\ge(\dim M)/2+\eps$
(for an arbitrary $\eps>0$). The zeta-function $\znb$ can be contunued
to a meromorphic function on the whole complex plane with at most simple
poles at the points $s_j:=(j-\dim M)/2$, $j=0,1,2,\dots$.
It is regular at $s=0,1,2,\dots$.\\
3. The residues $\res_{s=s_j}\zeta_{\nu,\bullet}(s)$ and the values
$\znb(m)+\delta_{m,0}\dim\Ker\left(\don\right)$ are equal
to the sums of the integrals over $M$, $\df M$, and $N$ of the densities
locally defined on these manifolds.
\label{TAT705}
\end{thm}

\begin{pro}
1. Let $\lambda\notin\Spec\left(\don\right)$. Then the resolvent
$\Ge_\lambda(\nu):=\left(\don-\lambda\right)^{-1}$,
$\Ge_\lambda(\nu):\left(\DR(M)\right)_{2}\rs\Dom(\don)\hookrightarrow
\left(\DR(M)\right)_{2}$, is the isomorphism (in algebraic and topological
senses) onto the closure $\Dom(\don)$ of $D(\don)$ with respect
to the graph norm.%
\footnote{The topology on $\dromb$ is given by $\|\omega\|^2_{2}$,
and on $\Dom(\don)$ it is given by $\|\omega\|^2_{graph}$.}
The operators $\Ge_\lambda(\nu)$ for pairs $(\lambda,\nu)$ such that
$\lambda\notin\Spec\left(\don\right)$ form a smooth in $(\lambda,\nu)$
family of bounded operators in $\dromb$. \\

2. The families $d\circ\Ge_\lambda(\nu)$ and $\delta\circ\Ge_\lambda(\nu)$
for $\lambda\notin\Spec\left(\don\right)$ form a smooth in $(\lambda,\nu)$
family of bounded operators
$\dromb\to\left(DR^{\bullet\pm 1}(M)\right)_{2}$.
\label{PA3002}
\end{pro}

\begin{thm}
1. The operator $\exp\left(-t\don\right)$ in $\dromb$ for an arbitrary
$t>0$ is of trace class. For its trace the asymptotic expansion
(\ref{S631}) (relative to $t\to+0$) holds. The coefficients
$f_{-\dim M+j}$ of this expansion are the sums of the integrals over $M$,
$\df M$, and $N$ of the locally defined densities.
If $j\ne\dim M+2m$, $m\in\wZ_+\cup 0$, the densities on $M$, $\df M$,
and on $N$ for $f_{-\dim M+j}$ are the same as for
$$
\Gamma((\dim M-j)/2)\res_{s=s_j}\znb(s).
$$
If $j=\dim M+2m$, $m\in\wZ_+\cup 0$, these densities are the same as for
$$
(m!)^{-1}(-1)^m\left(\znb(m)+\delta_{m,0}\dim\Ker(\don)\right).
$$

2. Let $p_{1}\colon\dromb\to\left(\DR(M_{1})\right)_{2}\hookrightarrow\dromb$
be the composition of the restriction to $M_1$ and of the extension by zero
of $L_{2}$-forms. Then the operator $p_{1}\exp\left(-t\don\right)$
in $\dromb$ for $t>0$ is of trace class. For its trace the asymptotic
expansion relative to $t\to+0$ holds
\begin{equation}
\Tr\left(p_{1}\exp(-t\don)\right)=q_{-n}t^{-n/2}+\dots +q_{0}t^0+
q_{1}t^{1/2}+\dots+q_m t^{m/2}+r_m(t),
\label{S2005}
\end{equation}
where $r_m(t)$ is $O\left(t^{(m+1)/2}\right)$ uniformly with respect
to $\nu$ and it is smooth in $t$ for $t>0$ ($n:=\dim M$).
The coefficients $q_j$ are equal to the sums of the integrals over $M_{1}$
and over $\df M_{1}=N\cup(\df M\cap\Mm_{1})$ of the locally defined
densities. The coefficients $q_j$ in (\ref{S2005}) depend only on
$(j,M_{1},g_{M}|_{TM_{1}},Z\cap\df\Mm_{1},N,\nu)$ and {\em do~not depend}
on $M_{2}$ and $Z\cap\df\Mm_{2}$, $g_M|_{TM_2}$. \\

3. For any $t>0$ the traces of $\exp\left(-t\don\right)$ are bounded
uniformly with respect to $\nu\in\rbo$:
\begin{equation}
\left|\Tr\exp\left(-t\don\right)\right|< C(t).
\label{E98}
\end{equation}

The traces $\Tr\left(p_j\exp\left(-t\don\right)\right)$ are also
bounded uniformly with respect to $\nu$ for any $t>0$.
\label{TAT706}
\end{thm}

\begin{pro}
1. The kernel $\eotxaxb(\nu)$ for $\exp\left(-t\don\right)$ (where $t>0$)
is smooth in $x_j\in\Mm_{r_j}$, $t$, and in $\nu\in\rbo$. \\

2. The asymptotic expansions (\ref{S2005}), (\ref{S631}) are differentiable
with respect to $\nu\in\rbo$.
\label{PE303}
\end{pro}

\subsection{Zeta-functions for the Laplacians with $\nu$-transmission
interior boundary conditions. Proofs of Theorem~\ref{TAT705} and
of Proposition~\ref{PA3002}}

Let $A$ be an elliptic differential operator on a manifold with boundary
$(M,\df M)$. Let the differential elliptic boundary conditions be given
for $A$ on $\df M$ such that $A$ with these boundary conditions satisfies
{\em Agmon's} condition (formulated below) for $\lambda$ from a sector
$\theta_1<\arg\lambda<\theta_2$ in the spectral plane $\wC\ni\lambda$.
Properties of zeta-functions for $A$ with these boundary
conditions can be investigated
with the help of the parametrix for $\left(A-\lambda\right)^{-1}$.
The analogous statement is true also for elliptic interior boundary
conditions.%
\footnote{Theorem~\ref{TAT705} is analogous to the results
of \cite{Se1}, \cite{Se2} with modifications connected with
the $\nu$-transmission interior boundary conditions. In \cite{Sh}, Ch.II,
the theory \cite{Se1}, \cite{Se2} of the zeta-functions is written
in detail in the case of a closed manifold.}
The parametrix $P^m_\lambda$
for $\left(\don-\lambda\right)^{-1}$ is defined locally in coordinate
charts. Namely
\begin{equation}
P^m_\lambda=\sum\psi_j P^m_{\lambda,U_j}\phi_j,
\label{X1195}
\end{equation}
where $\phi_j$ is a partition of unity subordinate to a finite cover
$\left\{U_j\right\}$ of $M$ by coordinate charts, $\psi_j\phi_j=\phi_j$,
$\psi_j\in C^{\infty}_{0}(U_j)$. If $U_j\cap(\df M\cup N)=\emptyset$
then the operator $P^m_{\lambda,U_j}$ is a pseudodifferential operator
(PDO) with parameter $\lambda$ (\cite{Sh}, Chapter~II, \S~9)
and its symbol is equal to
$\theta(\xi,\lambda)s_{(m)}\left(\left(\doo-\lambda\right)^{-1}\right)
(x,\xi,\lambda)$. This symbol is defined as follows.
Let $s(\doo-\lambda)=\left((b_{2}-\lambda)\id+b_{1}\right)(x,\xi)$
be the symbol of $\doo-\lambda$ (where $\doo$ is the Laplacian
on $\DR\left(U_j\right)$) and let
$$
s\left((\doo-\lambda)^{-1}\right):=\sum_{j\in\wZ_+\cup 0}a_{-2-j}
(x,\xi,\lambda)
$$
be the symbol of $\left(\doo-\lambda\right)^{-1}$
as of a PDO with parameter ($a_{-k}$ is positive homogeneous of degree $-k$
in $\left(\xi,\lambda^{1/2}\right)$).
Set
$s_{(m)}\left(\left(\doo-\lambda\right)^{-1}\right):=\sum^m_{j=0}a_{-2-j}
(x,\xi,\lambda)$.
The condition $s(\doo-\lambda)\circ s\left((\doo-\lambda)^{-1}\right)=1$,
where $\circ$ is the composition of symbols with parameter (\cite{Sh},
\S~11.1), is equivalent to the system of equalities
\begin{gather}
a_{-2}(x,\xi,\lambda)=(b_{2}-\lambda)^{-1},\notag\\
a_{-3}=-(b_{2}-\lambda)^{-1}[b_{1}a_{-2}+\sum_i D_{\xi_i}b_{2}\df_
{x_i}a_{-2}], \label{ST750}\\
a_{-2-j}=-(b_{2}-\lambda)^{-1}\sum_{|\gamma|+i+l=j}{1\over{\gamma !}}
D_\xi^{\gamma}b_{2-i}\df_x^\gamma a_{-2-l}.\notag
\end{gather}
The sum in the last equation of (\ref{ST750}) is over $(\gamma,i,l)$
such that $\gamma=(\gamma_{1},\dots,\gamma_n)\in(\wZ_+\cup 0)^n$, $|\gamma|:=
\gamma_{1}+\dots+\gamma_n$, $0\le|\gamma|\le j$ for $b_j$, $|\gamma|+i\ge 1$
$\left(D:=i^{-1}\df\right)$. The function $\theta(\xi,\lambda)$ (in the symbol
of $P^m_\lambda$) is smooth, $\theta(\xi,\lambda)\equiv 1$
for $|\xi|^2+|\lambda|\ge 1$, and $\theta$ is equal to zero
for $|\xi|^2+|\lambda|\le\eps$.

Let $U_j\cap N\ne\emptyset$. Then the term $P^m_{\lambda,U}$
of the parametrix is the sum of the interior term (which is a PDO
with parameter and its symbol is defined with the help of (\ref{ST750}))
and of the correction terms. (Here $U:=U_j$.)
The latter terms correspond to the $\nu$-transmission interior
boundary conditions on $N$ and to the Dirichlet and the Neumann
boundary conditions, given on the connected components of $\df M$.
First of all we'll verify that these $\nu$-transmission boundary
conditions are {\em Agmon's conditions} on any ray $\arg\lambda=\phi$
in the spectral plane not coinciding with $\wR_+$.

Let $(t,y)\in I\times U_{N}$ be the coordinates on $U:=U_j$ near
$N=0\times N\hookrightarrow I\times N\hookrightarrow M$, $I=[-2,2]$,
and let $t>0$ on $M_{1}$.
{}From now on it is supposed that $\phi_j(t,y)=\phi_{j,I}(t)\phi_{j,N}(y)$
and that $\phi_{j,I}(t)\equiv 1$ for $|t|\le 1$. It is supposed also that
$\psi_j(t,y)=\psi_{j,I}(t)\psi_{j,N}(y)$ and that $\phi_{j,I}$, $\psi_{j,I}$
are even functions: $\phi_{j,I}(-t)=\phi_{j,I}(t)$,
$\psi_{j,I}(-t)=\psi_{j,I}(t)$.  The forms $dy^c$ and $dt\,dy^f$
(where $c=(c_{1},\dots,c_{n-1})$, $f=(f_{1},\dots,f_{n-1})$,
$c_i,f_i\in\{0,1\}$) provide us with a trivialization of
$\wedge^{\bullet}TM|_{I\times U_{N}}$. Namely
$\omega_j=\sum_{|c|=\bullet}\omega_{j,c}dy^c+\sum_{|f|+1=\bullet}
\omega_{j,(1,f)}dt\,dy^f $. Let
$\omega=(\oma,\omb)\in D(\don)\subset\DR(M_\nu)$.
Then on $U_N=0\times U_N\hookrightarrow I\times U_{N}$
the conditions $\omega\in D(\don)$ can be written as follows. Set
$|\nu|=(\alpha^2+\beta^2)^{1/2}$. Let $\cL$ be the transformation
($t\ge 0$)
\begin{gather}
\begin{split}
v_{1,c}(t,y) & :=|\nu|^{-1}\left(\alpha\omega_{1,c}(t,y)-\beta\omega_{2,c}
(-t,y)\right), \\
v_{2,c}(t,y) & :=|\nu|^{-1}\left(\beta\omega_{1,c}(t,y)+\alpha\omega_{2,c}
(-t,y)\right), \\
w_{2,(1,f)}(t,y) & :=|\nu|^{-1}\left(\alpha\omega_{1,(1,f)}(t,y)+\beta
\omega_{2,(1,f)}(-t,y)\right), \\
w_{1,(1,f)}(t,y) & :=|\nu|^{-1}\left(-\beta\omega_{1,(1,f)}(t,y)+\alpha
\omega_{2,(1,f)}(-t,y)\right).
\end{split}
\label{ST751}
\end{gather}
Then the conditions $\omega\in D\left(\don\right)$ are equivalent on $U_N$ to
\begin{equation}
\begin{split}
v_{1,c}(0,y)             & =0, \\
\df_t v_{2,c}\big|_{t=0} & =0,
\end{split}
\qquad
\begin{split}
w_{1,(1,f)}(0,y)             & =0, \\
\df_t w_{2,(1,f)}\big|_{t=0} & =0.
\end{split}
\label{ST752}
\end{equation}
The inverse to (\ref{ST751}) transformation $\cL^{-1}$ is
\begin{equation}
\left(\!\begin{array}{c}
\omega_{1,c}(t,y) \\
\omega_{2,c}(-t,y)
\end{array}\!\right)\!=\!L^{-1}\!
\left(\!\begin{array}{c}
v_{1,c}\\
v_{2,c}
\end{array}\!\right)\!(t,y),\;
\left(\!\begin{array}{c}
\omega_{2,(1,f)}(-t,y) \\
\omega_{1,(1,f)}(t,y)
\end{array}\!\right)\!=\!L^{-1}\!
\left(\!\begin{array}{c}
w_{1,(1,f)} \\
w_{2,(1,f)}
\end{array}\!\right)\!(t,y),
\label{ST760}
\end{equation}
$$
L:=|\nu|^{-1}
\left(\begin{array}{cc}
\alpha & -\beta \\
\beta & \alpha
\end{array}\right)\ .
$$

Agmon's conditions on a ray $l:=\{\arg\lambda=\phi\}$ in the case
of $\nu$-transmission boundary conditions claim that
for $(\xi',\lambda)\ne (0,0)$ and $\lambda\in l$ the equation
on $\wR_+\ni t$
\begin{equation}
\left(-\df_t^2+b_{2}(y,\xi')-\lambda\right)v(t)=0,\quad
v(t)\to 0  \text{ for } t\to +\infty
\label{ST753}
\end{equation}
has a unique solution for each of the initial conditions
$v_{t=0}=v_{0}$ or $\df_t v|_{t=0}=v_{1}$.
(Here $\xi'$ are dual to $y$ and $b_{2}(\Delta_{N})=b_{2}(y,\xi')\id$
is the scalar principal symbol of $\doo_{N}$ on $U_{N}$.)
Agmon's conditions for the $\nu$-transmission boundary value problem
are satisfied on each ray $\arg\lambda=\phi$ not coinciding with $\wR_+$
because the equation (\ref{ST753}) with each of the initial conditions
given above has a unique solution for any $\lambda\notin\wR_+$,
$(\xi',\lambda)\ne (0,0)$.

It is convenient to compute the contribution to $P^m_{\lambda,U}$
from $\nu$-transmission boundary conditions in the coordinates
$v_{j,c}(t,y)$, $w_{j,(1,f)}(t,y)$ defined by (\ref{ST751})
with $t\ge0$. (Then the $\nu$-transmission boundary conditions
are transformed into the conditions (\ref{ST752}).)
These contributions are defined with the help of the symbol
$d=\sum_{j\in\wZ_+\cup0}d_{-2-j}(t,y,\tau,\xi',\lambda)$, which
is the solution of the equation%
\footnote{Here $\;b_{2}(y,\xi')\id+b_{1}(y,\xi)$ is the symbol
$s(\doo_{N})$ on $U_N$ of the Laplacian on $\DR(N)$ for the components
$\omega_{tan,N}(t,y)$ and on $DR^{\bullet-1}(N)$ for $\omega_{norm,N}(t,y)$.
The variable $\tau$ is dual to $t$.}
\begin{equation}
\left(-\df_t^2+\left(b_{2}(y,\xi')-\lambda\right)\id+b_{1}(y,\xi')
\right)\circ d(t,y,\tau,\xi',\lambda)=0
\label{ST761}
\end{equation}
(with the composition $\circ$ of symbols of $(y,\xi')$ in it).
The equation (\ref{ST761}) holds for $t\ne 0$. The boundary conditions
for (\ref{ST761}) are: $d_{-k}\to 0$ as $|t|\to\infty$ and
\begin{gather}
\begin{split}
\left(\cL d_{-k}\right)_{1,c}\big|_{t=0}         &\!=\!\left(\cL a_{-k}\right)
_{1,c}\big|_{t=0},\\
\df_t\left(\cL d_{-k}\right)_{2,c}\big|_{t=0}    &\!=\!i\tau\left(\cL a_{-k}
\right)_{2,c}\big|_{t=0},
\end{split}
\;
\begin{split}
\left(\cL d_{-k}\right)_{1,(1,f)}\big|_{t=0}     &\!=\!\left(\cL a_{-k}\right)
_{1,(1,f)}\big|_{t=0}, \\
\df_t\left(\cL d_{-k}\right)_{2,(1,f)}\big|_{t=0}&\!=\!i\tau\left(\cL a_{-k}
\right)_{2,(1,f)}\big|_{t=0}
\end{split}
\label{ST762}
\end{gather}

Here the transformation $\cL$ acts on the columns of the matrix valued
functions $d$, $a$ (depending on $t$ and on $\tau$).%
\footnote{Note that the function $a_{-2-j}(t,y,\tau,\xi',\lambda)$
is continuous in $N$ and nonsingular for $\lambda\notin\wR_+$ and
$(\tau,\xi',\lambda)\ne(0,0,0)$. (It is also independent of $t$
for $|t|$ small enough.)  So the right sides of (\ref{ST762}) can be
simplified for $a_{-k}$. In (\ref{ST761})--(\ref{ST763}) it is used
that $g_{M}$ is a direct product metric on $I\times N\hookrightarrow M$.}

The equation (\ref{ST761}) is the recurrent system
\begin{equation}
-\df_t^2 d_{-k}+(b_{2}-\lambda)d_{-k}+\sum{1\over{\gamma!}}D^{\gamma}
_{\xi'}b_i\df_y^{\gamma}d_{-m}=0,
\label{ST763}
\end{equation}
where the sum is over $m<k$ and $\gamma$ such that $m+|\gamma|+2-i=k$,
$0\le|\gamma|\le i$ for $b_i$.

For $t=0$ the symbol $d_{-k}$ over $M_j\cap U_{N}$ is positive homogeneous
of degree $(-k)$ in $(\tau,\xi',\lambda^{1/2})$. The boundary contribution
to $P^m_{\lambda,U}$ is an operator $\cD_m$ corresponding to%
\footnote{$\theta_{1}(\xi',\lambda)\in C^{\infty}(\wR^{n-1}\times\wC)$,
$\theta_1\equiv0$ for $|\xi'|^2+|\lambda|<\eps$ and $\theta_{1}\equiv 1$
for $|\xi'|^2+|\lambda|\ge 1$; $n:=\dim M$.}
$\theta_{1}(\xi,\lambda)\sum^m_{j=0}d_{-2-j}(t,y,\tau,\xi',\lambda)$.
This operator acts on $f\in\DR_c(\wR_t\times\wR^{n-1}_y)$ such that
$\supp f\cap(0\times\wR_y^{n-1})=\emptyset$ as follows:
\begin{equation}
\!(\cD_m f)(y,t)\!=\!(2\pi)^{-n}\!\iint\!\exp\!\left(i(y,\xi')\right)\!\sum^m
_{0}\!\theta_{1}d_{-2-j}(t,\!y,\!\tau,\!\xi',\!\lambda)(\cF\! f)(\tau,
\!\xi')d\xi'd\tau
\label{ST764}
\end{equation}
(where $(\cF f)(\tau,\xi')=\iint\exp\left(-i\left(t\tau+(x,\xi')\right)
\right)f(t,x)dxdt$ is Fourier transform of $f$). The term
of the parametrix, corresponding to $U$ (if $U\cap N\ne\emptyset$)
is defined by
\begin{equation}
P^m_{\lambda,U}=P^m_{\lambda,int}-\cD_m,
\label{X1196}
\end{equation}
where $P_{\lambda,int}$ is the PDO with the symbol
$s_{(m)}(t,y,\tau,\xi',\lambda)$ defined by (\ref{ST750}) ($x$ is
replaced by $(t,y)$ and $\xi=(\tau,\xi')$).

For $U_j\cap\df M\ne\emptyset$ the boundary term in $P^m_{\lambda,U_j}$
for the Dirichlet or the Neumann boundary conditions on the connected
components of $\df M$ is defined similarly.

The following assertions are true: \\

1. For $m\ge n$ the operator $\left(\doo-\lambda\right)P^m_\lambda-\id$
(where $\left(\doo-\lambda\right)$ acts on the restrictions of forms
to $\Mm_{1}$ and to $\Mm_{2}$) has a continuous
on $\Mm_{j_{1}}\times\Mm_{j_{2}}$ kernel
which is $O\left(\left(1+|\lambda|^{1/2}\right)^{n-m}\right)$ for
$\lambda\in\Lambda_{\eps}:=\{\lambda\ne 0,\eps<\arg\lambda<2\pi-\eps\}$,
where $\pi>\eps>0$ is fixed (\cite{Se1}, Lemma 5, p. 901).
This estimate is satisfied uniformly with respect to $\nu$ since
the families $d_{-2-j}$ are smooth in $\nu\in\rbo$ and since the estimates
for $\cL d_{-2-j}$ by \cite{Se1}, (29), p. 900, are uniform with respect
to $\nu\ne(0,0)$. \\

2. Let $m\ge n$ and $\nu=(\alpha,\beta)\in\rbo$.
Let $A_j:=A\left(M_j,N\right)$ be the same as in (\ref{S603})
and (\ref{S610}) and $R_j$ be the geometrical restrictions
to $N\subset\df M_j$ of forms on $M_j$. Then the operators
\begin{gather*}
\begin{split}
|\nu|^{-1}(\alpha R_{1}       & -\beta R_{2})P^m_{\lambda},\\
|\nu|^{-1}(\alpha R_{1}\delta & -\beta R_{2}\delta)P^m_{\lambda},
\end{split}
\quad\qquad
\begin{split}
|\nu|^{-1}(\beta A_{1}  & -\alpha A_{2})P^m_{\lambda}, \\
|\nu|^{-1}(\beta A_{1}d & -\alpha A_{2}d)P^m_{\lambda}
\end{split}
\end{gather*}
have smooth kernels on $N\times\Mm_j$ which are
$O\left(\left(1+|\lambda|^{1/2}\right)^{n-m}\right)$ for
$\lambda\in\Lambda_\eps$, where $\pi>\eps>0$ and $\eps$ is fixed
(\cite{Se1}, Lemma~6). These estimates are uniform with respect
to $\nu\in\rbo$. \\

3. Set $B_{1,\nu}:=|\nu|^{-1}(\alpha R_{1}-\beta R_{2})\colon\oplus_j
\DR(\Mm_j)\to\DR(N)$. Let $p_{1}\colon[0,1]\times N\to N$,
$p_{2}\colon[-1,0]\times N\to N$ be the natural projections. The operator
\begin{equation}
q_{1,\nu}:\DR(N)\to\oplus\DR(M_j),\quad q_{1,\nu}(\omega_{N}):=|\nu|^{-1}\phi
(t)(\alpha p^*_{1}\omega_{N},-\beta p^*_{2}\omega_{N})
\label{X1992}
\end{equation}
(where $\phi(t)\in C^{\infty}_{0}(I)$, $\phi(t)\equiv 1$
for $t\in[-1/2,1/2]$) is the right inverse to $B_{1,\nu}$
since $B_{1,\nu}q_{1,\nu}=\id$.
The analogous right inverse operators $q_{k,\nu}$ are naturally defined
for $B_{k,\nu}$
\begin{gather}
\begin{split}
B_{2,\nu}:=|\nu|^{-1}(\beta A_{1}-\alpha A_{2}),\quad
B_{3,\nu} & :=|\nu|^{-1}(\alpha R_{1}\delta-\beta R_{2}\delta), \\
B_{4,\nu}:=|\nu|^{-1}(\beta A_{1}d & -\alpha A_{2}d),
\end{split}
\label{X1993}
\end{gather}
\begin{equation}
B_{i,\nu}q_{j,\nu}=\delta_{ij}\cdot\id.
\label{X1994}
\end{equation}
For instance, $q_{3,\nu}\circ\omega_{N}:=\eps|\nu|^{-1}\phi(t)t\left(\alpha
dt\wedge p^*_{1}\omega_{N},-\beta dt\wedge p^*_{2}\omega_{N}\right)$,
$\eps=\pm 1$. Set $B_\nu:=(B_{j,\nu})$, $q_\nu:=(q_{j,\nu})$, and
$q_\nu B_\nu:=\left(q_{j,\nu}B_{j,\nu}\right)$. For $m\ge n$ the operator
is defined%
\footnote{For simplicity we'll suppose from now on here that
$\df M=\emptyset$. For the Dirichlet or the Neumann boundary conditions on
the components of $\df M$ the appropriate terms have to be added
to $R^m_\lambda$.}
\begin{equation}
R^m_\lambda:=P^m_\lambda-q_\nu B_\nu P^m_\lambda.
\label{X1682}
\end{equation}
It maps (according to \cite{Se1}, Lemma~12, p. 912)
$C^{\infty}$-forms $\omega\in\DR(M)$, $\supp\omega\cap N=\emptyset$,
to $C^{\infty}$-forms on $M_j$. Moreover
$R^m_\lambda:\DR_c(M\backslash N)\to D\left(\don\right)$.

4. For $m\ge n$ and $\lambda\notin\wR_+\cup 0$ the operator $R_\lambda$ can be
continued to a bounded operator in $\left(\DR(M)\right)_{2}$,
$R_\lambda:\left(\DR(M)\right)_{2}\to \Dom\left(\don\right)$.

Indeed, for any fixed differential operator $F$ of order $d=d(F)\le 2$
the operator $FR_\lambda$ is bounded in $\dromb$ with its
norm $O\left(\left(1+|\lambda|^{1/2}\right)^{d-2}\right)$ in a sector
$\lambda\in\Lambda_\eps$ ($\pi>\eps>0$ and $\eps$ is fixed) according
to \cite{Se1}, Lemmas~7, 13, 14. This estimate is uniform with respect
to $\nu\in\rbo$.
The continuation of $R_\lambda$ to $\dromb$ is as follows.
If $\omega_j\in\DR_c(M\setminus N)$ and $\omega_j\to\omega$ in
$\dromb$ then $R_\lambda\omega_j$ converges in $\dromb$
and $R_\lambda\omega$ is defined as its limit.
We see that $R_\lambda\omega_j\in D\left(\don\right)$ and
$\left(\don-\lambda\right)R_\lambda\omega_j$
converges in $\dromb$. Hence $R_\lambda\omega\in\Dom\left(\don\right)$. \\

5. The operator $\Ge_\lambda(\nu):=(\don-\lambda)^{-1}\colon\dromb
\to\Dom(\don)$ exists%
\footnote{This means that $\don-\lambda$ maps $\Dom^{\bullet}(\den)$
one-to-one to $\dromb$. It is equivalent to the existence of
$(\don-\lambda)^{-1}\colon\dromb\to\Dom(\don)$,
$(\don-\lambda)\circ(\don-\lambda)^{-1}=\id$ on $\dromb$.}
for $\lambda\in\Lambda_\eps:=\{\lambda\ne 0\colon\eps<\arg\lambda<2\pi-\eps\}$
and $|\lambda|$ sufficiently large. Its operator norm is $O(|\lambda|^{-1})$
for such $\lambda$ uniformly
with respect to $\nu\in\rbo$ (\cite{Se1}, Lemma~15). \\

6. The Laplacian $\don$ is a closed unbounded operator in $\dromb$
with its domain of definition $\Dom\left(\don\right)$.
Actually, if $\{u_i\}\subset\Dom\left(\don\right)$ and if the limits
$\lim_i u_i=:u$ and $\lim_i\left(\left(\don-\lambda\right)u_i\right)=:v$
exist in $\dromb$ then for sufficiently large $\lambda\in\Lambda_\eps$
we have
$u=\lim_i\Ge_\lambda(\nu)\left(\left(\don-\lambda\right)u_i\right)=
G_\lambda(\nu)v\in\Dom\left(\don\right)$. Hence $\left(\don-\lambda\right)u=
\left(\don-\lambda\right)\left(\Ge_\lambda(\nu)v\right)=v$,
i.e., the operators $\don-\lambda\id$ and $\don$ are closed in $\dromb$.
The operator $\don$ is defined on $\Dom\left(\don\right)$.
It is a self-adjoint unbounded operator in $\dromb$. Indeed, the domain
of definition $\Dom\left(\left(\don-\lambda\right)^*\right)$
of the adjoint operator $\left(\don-\lambda\right)^*$ in $\dromb$
is the set of $v\in\dromb$ such that the linear functional
$\left(\left(\don-\lambda\right)\omega,v\right)$ is continuous
on $\Dom\left(\don\right)\ni\omega$ in the $L_2$-topology of $\dromb$.
If $v\in\Dom\left(\don\right)$ then for any $\omega\in\Dom\left(\don\right)$
we have
$\left(\left(\don-\lambda_{0}\right)\omega,v\right)=\left(\omega,\left(\don-
\lambda_{0}\right)v\right)$ for $\lambda_{0}\in\wR_-$. Indeed, for each
$\omega$ and $v$ from $\Dom\left(\don\right)$ there exist sequences
$\left\{\omega_j\right\}$ and $\left\{v_j\right\}$ of elements
$D\left(\don\right)$ whose limits in the graph norm topology
are $\omega$ and $v$. Hence we have
$$
\lim_j\left(\den\omega_j,v\right)_{2}=\lim_j\lim_i\left(\den\omega_j,v_i
\right)_{2}=\lim_j\lim_i\left(\omega_j,\den v_i\right)_{2}=\left(\omega,
\den v\right)_{2}.
$$
So $\left(\left(\don-\lambda_{0}\right)u,v\right)$ is a continuous
linear functional on $\Dom\left(\don\right)\ni u$ with respect
to the $L_2$-topology of $\dromb$ for any $v\in\Dom\left(\don\right)$. Hence
$\Dom\left(\don\right)\subset\Dom\left(\left(\don-\lambda_{0}\right)^*\right)$
and $\left(\den-\lambda_{0}\right)^*v=\left(\den-\lambda_{0}\right)v$
for $v\in\Dom\left(\don\right)$.

Let $\lambda_{0}\in\wR_-$ and $|\lambda_{0}|$ be sufficiently large.
Then for any $w\in\Dom\left(\left(\don-\lambda_{0}\right)^*\right)$
there exists an element $v\in\Dom\left(\don\right)$ such that
$\left(\don-\lambda_{0}\right)^*w=\left(\don-\lambda_{0}\right)v=\left(\don-
\lambda_{0}\right)^*v$ (since $\Image(\don-\lambda_{0})=\dromb$).
So $w-v\in\Ker\left(\left(\don-\lambda_{0}\right)^*\right)$ and
for any $u\in\Dom\left(\don\right)$ we have
$0=\left(u,\left(\don-\lambda_{0}\right)^*(w-v)\right)=\left(\left(\don-
\lambda_{0}\right)u,w-v\right)$.
Then $w-v=0$, as $\Image\left(\don-\lambda_{0}\right)=\dromb$. Hence
$\Dom\left(\left(\don-\lambda_{0}\right)^*\right)=\Dom\left(\don\right)=
\Dom\left(\left(\don\right)^*\right)$, and $\don$ is a self-adjoint
unbounded operator in $\dromb$.

The operator $\don$ is nonnegative, $\left(\don\omega,\omega\right)_{2}\ge0$
for any $\omega\in\Dom\left(\don\right)$, since there exists a sequence
$\left\{\omega_j\right\}$, $\omega_j\in D\left(\don\right)$, such that
its limit in the graph norm topology is $\omega$. So we have
$\lim_j\left(\don\omega_j,\omega_j\right)_{2}=\lim_j(d_\nu\omega_j,d_\nu
\omega_j)_{2}+\lim_j(\delta_\nu\omega_j,\delta_\nu\omega_j)_{2}\ge 0$. \\

7. The spectrum $\Spec\left(\don\right)$ of the operator $\don$
is discrete because the operator
$$
\left(\don-\lambda\right)\left(\don-\lambda_{0}\right)^{-1}=\id+\left(\don-
\lambda_{0}\right)^{-1}\cdot(\lambda_{0}-\lambda)
$$
differs from the identity operator in $\dromb$ by a compact operator.
Here, $\lambda_0\in\Lambda_\eps$ and $|\lambda_0|$ is large enough.
The assertion {\em 5} above claims that $\left(\don-\lambda_0\right)^{-1}$
exists for such $\lambda_0$.
The operator $G_{\lambda_0}(\nu):=\left(\don-\lambda_0\right)^{-1}$
is compact since it is bounded in $\dromb$ and since the operators
$I-\left(\don-\lambda_{0}\right)R^m_{\lambda_{0}}$ (for $m\ge n$)
and $R^m_{\lambda_{0}}$ are compact in $\dromb$ (\cite{Se1}, Lemmas~4, 5,
9~({\em iv})). So the operator
$$
\left(\don-\lambda_{0}\right)^{-1}=R^m_{\lambda_{0}}+\left(\don-\lambda_{0}
\right)^{-1}\left(I-\left(\don-\lambda_{0}\right)R^m_{\lambda_{0}}\right)
$$
is compact in $\dromb$. Since $G_{\lambda_{0}}(\nu)$ is a compact
operator for $\lambda_{0}\in\Lambda_\eps$, $|\lambda_{0}|>>1$, and
since $\don$ is a closed operator in $\dromb$, it follows that $\don$
is an operator in $\dromb$ with compact resolvent.
So (according to \cite{Ka}, Ch.~3, Theorem~6.29) its spectrum
$\Spec\left(\don\right)$ consists of isolated eigenvalues
with finite multiplicities (i.e., $\Spec\left(\don\right)$ is discrete)
and the operator $\Ge_\lambda(\nu)$ is compact in $\dromb$
for $\lambda\in\wC\setminus\Spec(\don)$. The operator $\don$
is nonnegative. Hence $\Spec\left(\don\right)\subset\wR_+\cup 0$.

If $\lambda_{0}\notin\Spec\left(\don\right)$ then
$\Ker\left(\don-\lambda_{0}\right)=0$ and
$\Image\left(\don-\lambda_{0}\right)=\dromb$. Hence
$\ind\left(\don-\lambda_{0}\right)$ is equal to $0$ (as the index
of the operator from
$\left(\Dom\left(\don\right),\left\|\cdot\right\|^2_{graph}\right)$ into
$\left(\dromb,\left\|\cdot\right\|^2_{2}\right)$). The operator
$(\lambda_{0}-\lambda)\id$ from $\Dom\left(\don\right)$
into $\dromb$ is compact (since $\Ge_{\lambda_{0}}(\nu)$ is a compact
operator in $\dromb$ and since it is a topological isomorphism
$G_{\lambda_0}(\nu)\colon\dromb\rs\Dom\left(\don\right)$).
So $\ind\left(\don-\lambda\right)=0$ for an arbitrary $\lambda\in\wC$
(according to \cite{Ka}, Ch.~4, Theorem~5.26, Remarks~1.12, 1.4). \\

8. The operator $\left(\don\right)^{-s}$ for $\Ree\,s>0$ is defined
by the integral
\begin{equation}
{i\over 2\pi}\int_{\Gamma}\lambda^{-s}\left(\don-\lambda\right)^{-1}d\lambda
=:T_{-s}(\nu),
\label{ST808}
\end{equation}
where the contour $\Gamma$ is
$$
\{\lambda=re^{i\pi},\infty>r>\eps\}\cup\{\lambda=
\eps e^{i\phi},\pi>\phi>-\pi\}\cup\{\lambda=re^{-i\pi},\eps<r<\infty\}.
$$
Here the number $\eps>0$ is such that $\Spec\left(\don\right)\cap(0,\eps]=
\emptyset$. The integral (\ref{ST808}) is absolutely convergent
(with respect to the operator norm $\left\|\cdot\right\|_2$ in $\dromb$)
because the estimate
$\left\|\left(\don-\lambda\right)^{-1}\right\|_{2}\le C|\lambda|^{-1}$
is satisfied as $\lambda\to -\infty$ for $\lambda\in\wR_-$.
So $T_{-s}(\nu)$ is a bounded operator in $\dromb$ for $\Ree\,s<0$.

For $-k\ge\Ree\,s>-(k+1)$, $k\in\wZ_+$, the operator $T_{-s}$ is defined
as $\Delta^{k+1}_\nu T_{-(s+k+1)}$. Its domain is
$\Dom(T_{-s})=\left\{\omega\in \left(\DR(M)\right)_{2},\ T_{-(s+k+1)}\omega
\in\Dom\left((\don)^{k+1}\right)\right\}$, where
$$
\Dom\left(\left(\don\right)^{k+1}\right):=\left\{\omega\in\Dom\left(\don
\right),\don\omega\in\Dom\left(\don\right), \dots,\left(\don\right)^k
\omega\in\Dom\left(\don\right)\right\}.
$$
The restriction $T_{-s}^{0}$ of $T_{s}$ to the orthogonal
complement $L_{0}$ of $\Ker\left(\don\right)$ in $\dromb$ is defined
on $\Dom\left(T^{0}_{-s}\right):=D\left(T_{-s}\right)\cap L_{0}$,
$T^{0}_{-s}:=T_{-s}|_{L_{0}}$. Then $T_{-s}$ is the direct sum%
\footnote{If $v\in L_{0}$ and $\Ree\,s>0$ then we have $T_{-s}v\in L_{0}$
since for $h\in\Ker\left(\don\right)$ and $\lambda\in\Gamma$ it holds
$$
0=(v,h)=\left(\left(\don-\lambda\right)\Ge_\lambda(\nu)v,h\right)=-\lambda
\left(\Ge_\lambda(\nu)v,h\right)
$$
and since the integral (\ref{ST808}) is absolutely convergent.
For $h\in\Ker\left(\don\right)$ and for $\Ree\,s>0$ we have
$T_{-s}h=0$ because for such $s$ the integral $\int_{\Gamma}\lambda^{-s-1}
d\lambda$ is absolutely convergent and is equal to zero. Since
$T_{-s}h=\left(\don\right)^{k+1}T_{-(s+k+1)}h=0$ for $-k\ge\Ree\,s>-(k+1)$,
we get $T_{-s} h\equiv 0$ for all $s$.}
of $T^{0}_s$ and of the zero operator on $\Ker\left(\don\right)$.
Theorem~1 in \cite{Se2} claims that the family $T^{0}_{-s}$ of operators
in the Hilbert space $L_{0}$ for $\Ree\,s_{2}>0$ satisfies the equation
$T^{0}_{-s_{1}}T^{0}_{-s_{2}}=T^{0}_{-(s_{1}+s_{2})}$ and that the same
is true for $-s_{1}\in\wZ_+$ and for each $s_{2}$.
This theorem claims also that
$$
T^{0}_{0}=\id \text{ on } L_{0},\quad T^{0}_{-l}=\left(\left(\don\right)
^{-1}|_{L_{0}}\right)^l,\text{ for } l\in\wZ_+,\text{ and } T^{0}_{1}=
\don|_{L_{0}}
$$
(the domain of $T^0_1$ is $\Dom(\don)\cap L_{0}$) and that
$T^{0}_{-s}$ for $\Ree\,s>0$ is a holomorphic function%
\footnote{A function with the values in a Banach space is holomorphic in
a strong sense if it is  weakly holomorphic (\cite{Ka}, Ch.~3, \S~1,
Theorem~1.37, p. 139)}
with its values in a Banach space $B(L_{0})$ of bounded operators
in $L_{0}$ where the Banach norm is the operator norm as the norm.

9. For $\Ree\,s>n/2$ the kernel of $T_{-s}(\nu)$ is continuous on
$\Mm_{j_{1}}\times\Mm_{j_{2}}$ and analytic in $s$ (\cite{Se2}, Theorem~2(i)).
For $\Ree\,s>n/2$ the zeta-function $\zeta_{\nu,\bullet}(s)$ is equal
to the sum of integrals over the diagonals
$\Mm_j\hookrightarrow\Mm_j\times\Mm_j$ ($j=1,2$) of the densities
defined by the restrictions to these diagonals of the kernel $T_{-s}(\nu)$,
according to Proposition~\ref{PA80} below.
So $\zeta_{\nu,\bullet}(s)$ is holomorphic for $\Ree\,s>n/2$.

The operator $\Ge_\lambda(\nu)-P^m_\lambda$ for $m\ge n$
(where $P^m_\lambda$ is the parametrix (\ref{X1195})) is a bounded
in $\dromb$ operator with a continuous on $\Mm_{j_{1}}\times\Mm_{j_{2}}$
kernel $\left(r^m_\lambda\right)_{\xa,\xb}$ which is
\begin{equation}
\left(r^m_\lambda\right)_{\xa,\xb}=O\left(\left(1+\left|\lambda\right|^{1/2}
\right)^{-(2+m)+n}\right)
\label{X1794}
\end{equation}
as $|\lambda|\to+\infty$, $\lambda\in\Lambda_\eps$ (\cite{Se1},
Theorem~1, or also the assertions {\em 5}, {\em 1}, {\em 2}, {\em 7} above).
So the operator
\begin{equation}
{i\over 2\pi}\int_{\Gamma}\lambda^{-s}\left(\Ge_\lambda(\nu)-P^m_\lambda
\right)d\lambda
\label{X1210}
\end{equation}
for $\Ree\,s>(n-m)/2$ is of trace class
and its kernel is continuous on $\Mm_{j_{1}}\times\Mm_{j_{2}}$
and analytic in $s$.

The trace of the operator (\ref{X1210}) is holomorphic in $s$
for $\Ree\,s>(n-m)/2$. Let us denote by $K^{int}_{x,y}(s)$ the kernel
of the operator
$(i/2\pi)\int_{\Gamma}\lambda^{-s}P^m_{\lambda,int}d\lambda$
(where $P^m_{\lambda,int}:=\sum\psi_j P^m_{\lambda,U_j,int}\phi_j$
is a term of (\ref{X1195}) and $P^m_{\lambda,U_j,int}$ is a PDO
with symbol $\theta s_{(m)}$, defined by (\ref{ST750})).
This kernel is continuous on $\Mm_{j_{1}}\times\Mm_{j_{2}}$ for
$\Ree\,s>n/2$. Off the diagonals $\Mm_j\hookrightarrow\Mm_j\times\Mm_j$
it extends to a kernel which is an entire function of $s\in\wC$
equal to zero for $(-s)\in\wZ_+\cup 0$. The density on $\cup_j\Mm_j$
defined
by the restriction of this kernel to the diagonals also can be continued
to a meromorphic in $s\in\wC$ density. This density has at most
simple poles at $s_j=(n-j)/2$ for $(-s_j)\notin\wZ_+\cup0$,
$0\le j\le m$, and it is regular at $s_j$ for $(-s_j)\in\wZ_+\cup0$.

The residue at $s=s_j$ is completely defined by the component
$a_{-2-j}(x,\xi,\lambda)$ of the symbol
$s\left(\left(\doo-\lambda\right)^{-1}\right)$
(\cite{Se2}, Lemma 1 or \cite{Sh}, Theorem 12.1).
These components are given by (\ref{ST750}). The value of this density
at $s=s_j$ for $(-s_j)\in\wZ_+\cup 0$ is completely defined by $a_{-2-j}$
(by the formulas (11), (12) in \cite{Se2} with changing of the sign in (11)
to the opposite one). Here, $j=n+2m$, $m\in\wZ_+\cup 0$.

The kernel $K^{\df}_{x,y}(s)$ of the operator%
\footnote{Here $\cD_{m,\lambda}=\sum\psi_j\cD_{m,U}(\lambda,\nu)\psi_j$.
The operator $\cD_{m,U}$ from (\ref{X1196}) is defined for
$U\cap N\ne\emptyset$ by (\ref{ST764}), (\ref{ST761}), and (\ref{ST762}).}
$(i/2\pi)\int_{\Gamma}\lambda^{-s}\cD_{m,\lambda}d\lambda$
for $\Ree\,s>n/2$ is continuous on $\Mm_{j_{1}}\times\Mm_{j_{2}}$ and
analytic in $s$ (\cite{Se2}, Lemma~4). Let $(x,y)$ be off the diagonals
or let either $x$ or $y$ be not from $\cup_j\df\Mm_j\supset N$.
Then $K^{\df}_{x,y}(s)$ is an entire function of $s\in\wC$
and it is equal to zero at $s$ for $(-s)\in\wZ_+\cup0$ (\cite{Se2}, Lemma~4).
For $\Ree\,s>n/2$ the densities defined by the restriction $K^{\df}_{x,x}(s)$
of $K^{\df}_{x,y}(s)$ to the diagonals $\Mm_j$ are integrable over
the fibers of the natural projections $p_{1}\colon [0,1]\times N\to N$
and $p_{2}\colon[-1,0]\times N\to N$. These integrals are densities
on $N$. They can be continued%
\footnote{Let $\cL\cD_{m,U}(\lambda,\nu)$ be an operator acting
on $\cL\omega$ as $\cL\left(\cD_{m,U}(\lambda,\nu)\omega\right)$
(for any $\omega\in\DR_c(\wR^n)$ such that
$\supp\omega\cap\wR^{n-1}=\emptyset$, where $\wR^{n-1}$ are local
coordinates on $N$). All the assertions about the kernels, analogous
to $K_{x,y}^{\df}(s)$ in the case of $\cL\cD_{m,U}(\lambda,\nu)$,
and about the corresponding densities in this case, are proved
in \cite{Se2}. Thus the transformation (\ref{ST760}) provides us
with all the assertions about the kernel $K^{\df}_{x,y}(s)$ (and about
the corresponding densities) connected with $\cD_{U,\lambda}(\nu)$.}
to meromorphic on $s\in\wC$ densities (on $N$) with at most simple poles
at $s_j=(n-j)/2$, $1\le j\le m$, such that $(-s_j)\notin\wZ_+\cup 0$.
Its residues at $s_j$ for $1\le j\le m+1$ are completely defined
by a term $d_{-2-j+1}$ in $d$ (\cite{Se2}, Theorem~2({\em iv}),
formula~{\em (II)}). The values of these densities at $s_j$
for $(-s_j)\in\wZ_+\cup0$ (where $n\le j\le m+1$) are also completely
defined by $d_{-2-j+1}$ (\cite{Se2}, Lemmas~2, 3, 4, Theorem~2({\em iv}),
formula~{\em (II$'$)} with changing of the sign to the opposite one). \\

10. The kernel of the operator (\ref{X1210}) is the difference
of the kernels
\begin{equation}
(T_{-s})_{x,y}-\left(K^{int}_{x,y}(s)-K^{\df}_{x,y}(s)\right).
\label{X1214}
\end{equation}
For $\Ree\,s>(n-m)/2$ it is holomorphic in $s$ and continuous
on $\Mm_{j_{1}}\times\Mm_{j_{2}}$. The term
$\left(K^{int}-K^{\df}\right)_{x,y}(s)$ is equal to zero for $x\ne y$
and $(-s)\in\wZ_+\cup 0$. The term $(T_{-s})_{x,y}$ is equal to zero
for $x\ne y$ and $(-s)\in\wZ_+$ (according to the assertions of {\em 8}
above, since $\don$ is a differential operator).
We have $(T_{0})_{x,y}=-\cHh_{x,y}$, where $\cHh$ is the kernel
of the orthogonal projection operator in $\dromb$
onto $\Ker\left(\don\right)$ (the assertion~{\em 8}).
The properties of $\znb(s)$%
\footnote{The values of $\zeta_{\nu,\bullet}(s)$ at $(-s)\in\wZ_+$
and the residues of $\zeta_{\nu,\bullet}$ at $s=s_j$ can be also
expressed in terms of noncommutative residues (\cite{Wo} or \cite{Kas}).
The density on $M$ whose integral over $M$ is equal to a volume term
in $\operatorname{Res}_{s=s_j}\znb(s)$ can be written
as $2^{-1}\res\left(x,\Delta^{-s_j}_{\theta=\pi}\right)$. Here $\res$
is a noncommutative residue for the symbol of PDO
$\Delta^{-s_j}_{\theta=\pi}$. This symbol is defined with the help
of the symbol $\sum a_{-2-j}(x,\xi,\lambda)$
of $\left(\doo-\lambda\right)^{-1}$ (\cite{Sh}, 11.2). The boundary
term in $\operatorname{Res}_{s=s_j}\znb(s)$ for $(-s_j)\notin\wZ_+\cup0$
is expressed similarly.}
formulated in Theorem~\ref{TAT705}, follow from the assertions of {\em 9}
and {\em 10}. The theorem is proved.\ \ \ $\Box$

\begin{rem}
The kernel (\ref{X1214}) of the operator (\ref{X1210}) is holomorphic
in $s$ and continuous in $(x,y)\in\Mm_{j_{1}}\times\Mm_{j_{2}}$
for $\Ree\,s>(n-m)/2$. It is equal to zero for $x\ne y$ at $s=-k$,
$k\in\wZ_+$, and to $-\cHh_{x,y}(\nu)$ at $s=0$.
So the analytic continuations of the densities on $\Mm_j$ and on $N$
defined by the kernels $\left(T_{-s}\right)_{x,y}$ and
$\left(K^{int}(s)-K^{\df}(s)\right)_{x,y}$ have the same residues
at $s=s_j$, $0\le j\le m$, $(-s_j)\notin\wZ_+\cup0$, and the same
values at $s=s_j$, $(-s_j)\in\wZ_+$, $n+2\le j\le m$.
They differ at $s=0$ (i.e., for $j=n$) by the densities on $\Mm_j$,
defined by $-\cHh_{x,x}(\nu)$. Hence the densities on $\Mm_j$ and on $N$,
corresponding to the residues and to the values at $s=s_j$, $0\le j\le m-1$,
of $\left(K^{int}(s)-K^{\df}(s)\right)_{x,x}$ are the same
for all the parametrixes $P^m_\lambda$ defined by (\ref{X1195}) (with
different covers $\{U_j\}$, partitions of unity $\{\phi_j\}$ subordinate
to $\{U_j\}$, and $\{\psi_j\}$).

And back, the values and the residues of the analytic continuation
for the integral
$\int_{\Mm_j}\tr\left(i^*_j(T_{-s})\right)+\delta_{s,0}\int_{\Mm_j}\tr
\left(i^*_j\cHh(\nu)\right)$
at $s_j$, $0\le j\le m$, are defined by an arbitrary parametrix $P^m_\lambda$.
\label{RAT300}
\end{rem}

\noindent{\bf Proof of Proposition~\ref{PA3002}.}
Let $m\ge n:=\dim M$, $m\in\wZ_+$, and $\lambda\in\Lambda_\eps$.
Then the parametrix $R^m_{\lambda}$ for $\Ge_\lambda(\nu)$%
\footnote{The statement that
$\Ge_\lambda(\nu)\colon\dromb\to\Dom\left(\don\right)$ is an isomorphism
for $\lambda\notin\Spec\left(\don\right)$ is proved in Theorem~\ref{TAT705}.}
(defined by (\ref{X1682})) is a bounded operator%
\footnote{The terms $P^{(m)}_{\lambda,int}$, $\psi_j\cD_m\phi_j$, and
$q_{\nu}B_{\nu}P^m_{\lambda}$ in $R^m_{\lambda}$ are bounded operators
with the same estimate of their norms in $\dromb$
for $\lambda\in\Lambda_\eps$ (the proof of Theorem~\ref{TAT705}).
For the sake of brevity the proof of Proposition~\ref{PA3002} is given
in the case of $\df M=\emptyset$.}
in $\dromb$ with its norm estimated
by $O\left(\left(1+|\lambda|^{1/2}\right)^{-2}\right)$
for $\lambda\in\Lambda_\eps$. It holds that
$R_{\lambda}\colon\dromb\to\Dom\left(\don\right)$.
For a linear differential operator $F$ of order $d=d(F)\le 2$ the operator
$FR_\lambda$ is defined on smooth forms $\omega\in\DR_c(M\setminus N)$ and
its closure in $\dromb$ is a bounded operator in $\dromb$ with its norm
estimated by $O\left(\left(1+|\lambda|^{1/2}\right)^{d-2}\right)$
for $\lambda\in\Lambda_\eps$. (All these estimates are uniform
with respect to $\nu\in\rbo$.)
The only terms of $R^m_{\lambda}$ depending on $\nu$ are the terms
$\cD-q_\nu B_\nu P^m_\lambda(\nu)$, where%
\footnote{$\cD_m=\cD_{m,U_j}$ is defined by (\ref{ST763}),
(\ref{ST762}), and (\ref{ST764}).}
$\cD=\cD(\nu):=-\sum\psi_j\cD_{m,U_j}\phi_j$
has the kernel with support in the neighborhood $I\times N$ of the
interior boundary $N\hookrightarrow M$.%
\footnote{In the case $\df M\ne\emptyset$ the terms connected with
the Dirichlet and the Neumann boundary conditions are added to $\cD(\nu)$.
Then the corresponding kernel has its support in a neighborhood
of $\df M$ in $M$.}
(The parametrix
$P^m_\lambda(\nu):=P^m_{\lambda,int}-\cD(\nu)$ is defined by (\ref{X1196})).
We need the following assertion now.
\begin{pro}
The operators $\cD(\nu)$ and $q_{\nu}B_{\nu}P^m_\lambda(\nu)$
depend smoothly on $\nu\in\rbo$ as bounded operators in $\dromb$.
For a $C^{\infty}$-map $\nu=\phi(\gamma): [-a,a]\to\rbo$ the operator
$\dfg\cD(\nu)$ is a bounded operator in $\dromb$ whose norm is uniformly
with respect to $\gamma$ estimated by
$O\left(\left(1+|\lambda|^{1/2}\right)^{-2}\right)$
for $\lambda\in\Lambda_{\eps}$. Let $F$ be a linear differential operator
of order $d=d(F)\le 2$ from $\DR(\Mm_j)$ into $DR^{\bullet+k}(\Mm_j)$,
$k\in\wZ$. Then $F\cD(\nu)$, $F\dfg\cD(\nu)$ are bounded operators
from $\dromb$ into $\left(DR^{\bullet+k}(M)\right)_2$ whose norms
are estimated by
$O\left(\left(1+|\lambda|^{1/2}\right)^{d(F)-2}\right)$ for
$\lambda\in\Lambda_{\eps}$. The operator
$\dfg\left(q_\nu B_\nu P^m_\lambda(\nu)\right)$ is uniformly with respect
to $\gamma$ estimated by $O\left(\left(1+|\lambda|^{1/2}\right)^{-1}\right)$
for $\lambda\in\Lambda_\eps$.
\label{PX3003}
\end{pro}

\noindent{\bf Proof.} The kernel of the operator
$\cL\psi_j\cD_{m,U_j}\phi_j\cL^{-1}$
(where $\cL=\cL(\nu)$ and $\cL^{-1}$ are defined by (\ref{ST751}) and
(\ref{ST760})) has a support in $\left((U_j\cap N)\times[0,1]\right)^2$.
The operator $\cD_{m,U_j}$ is defined in $(U_j\cap N)\times\wR_+$
by (\ref{ST763}) and (\ref{ST762}).
The right sides of the boundary conditions (\ref{ST762}) depend on $\nu$
only by their dependence on $L(\nu)$ (where $L$ is the matrix
defined by (\ref{ST760})).
Since $g_{M}$ is a direct product metric near $N$, a mirror symmetry
(relative to $N$) acts as the identity operator on the symbol
$\sum a_{-2-j}(x,\xi)$ of the Laplacian $\doo$ on $M$ for $x=(t,x')$
from the neighborhood $I\times N$ of $N$. The symbol
$\sum a_{-2-j}(t,x',\tau,\xi',\lambda)$ is independent of $t$ for $t\in I$.

So the symbol $\cL\sum a_{-2-j}$ (for $t\in I$) is expressed as $LaL^{-1}$,
where $L$ and $L^{-1}$ act on the components of a matrix valued functions
$a_{-2-j}$ in the coordinates $\omega_{j,c}$ and $\omega_{j,(1,f)}$
as follows (according to (\ref{ST751})):
\begin{gather}
\begin{split}
(La)_{1,c;*}     & =|\nu|^{-1}(\alpha-\beta)(a)_{c;*}\;,\\
(La)_{2,c;*}     & =|\nu|^{-1}(\beta+\alpha)(a)_{c;*}\;,
\end{split}
\;
\begin{split}
(La)_{1,(1,f);*} & =|\nu|^{-1}(-\beta+\alpha)(a)_{(1,f);*}\;,\\
(La)_{2,(1,f);*} & =|\nu|^{-1}(\alpha+\beta)(a)_{(1,f);*}\;.
\end{split}
\label{M150}
\end{gather}

The boundary conditions (\ref{ST762}) (according to (\ref{M150}))
depend on $\nu$ only by the matrix transformation whose coefficients
are independent of $(t,x')$ and smooth in $\nu$. This transformation
acts separately on each homogeneous component $a_{-2-j}$. The right
sides in (\ref{M150}) are nonsingular in $(x',\tau,\xi',\lambda)$
for $b_{2}(x',\tau,\xi')-\lambda\ne 0$ (where $b_{2}$ is the principal
symbol of the Laplacian on $M$ for $(t,x')\in I\times N$).
Hence Lemma~2 in \cite{Se1} holds also for the symbol
$\dfg\left(\cL(\nu)\sum d_{-2-j}\right)$. Thus the desired estimates
for the norm of $\dfg\cD(\nu)$ in $\dromb$ and for the norms of $F\cD(\nu)$
and of $F\dfg\cD(\nu)$ are consequences of \cite{Se1}, Lemma~7.

The operators $q_{j,\nu}$ ($1\le j\le 4$) from (\ref{X1992})
and (\ref{X1994}) (for $\phi(t)$ even on $t$) can be defined such that%
\footnote{The operator $B_\nu P^m_\lambda(\nu)$ has a continuous on
$\Mm_{j_{1}}\times\Mm_{j_{2}}$ kernel which is estimated uniformly
with respect to $(\xa,\xb)\in\left(N\cap\Mm_{j_{1}}\right)\times\Mm_{j_{2}}$
and to $\nu\in\rbo$ by $O\left(\left(1+|\lambda|^{1/2}\right)^{n-m}\right)$
for $\lambda\in\Lambda_\eps$ (\cite{Se1}, Lemma~6). Such an estimate holds
also for the kernel of $q_\nu B_\nu P^m_\lambda(\nu)$.}
$$
\cL q_\nu B_\nu P^m_\lambda(\nu)f(t)\cL^{-1}=\sum q_j B_j\cL P^m_\lambda(\nu)
f(t)\cL^{-1},
$$
where $f\in C^\infty_0(I)$, $f(t)\equiv 1$ for $t\in [0,1/2]$
and $f(t)\equiv 0$ for $t>3/4$. The operators $B_j$ and $q_j$
are independent of $\nu\in\rbo$ and correspond to $B_{j,\nu}$ and $q_{j,\nu}$
from (\ref{X1993}), (\ref{X1994}). (Here $B_{1}$, $B_2$, $B_3$, $B_{4}$
are the operators (\ref{ST752}) acting respectively on $v_{1,c}$,
$w_{2,(1,f)}$, $w_{1,(1,f)}$, $w_{2,c}$.)
These operators are such that $B_i q_j=\delta_{ij}\id$.

The operator $\sum q_jB_j\cL\left(\dfg P^m_\lambda(\nu)\right)f\cL^{-1}$
is equal to $\sum q_jB_j\cL\left(-\dfg\cD(\nu)f\right)\cL$
(since $P^m_{\lambda,int}$ is independent of $\nu$). The operator
$B_j\cL\left(\dfg\cD(\nu)\right)\cL^{-1}$ is defined on smooth forms
$\omega\in\DR_c\left((0,1)\times N\right)$ and its $L_{2}$-norm
is estimated by $O\left(\left(1+|\lambda|^{1/2}\right)^{-1}\right)$
for $\lambda\in\Lambda_\eps$ (\cite{Se1}, Lemma~7). The operators
$\dfg(q_\nu B_\nu)P^m_{\lambda,int}$ are defined on smooth forms
$\omega\in\DR_c(M\setminus N)$ and their operators $L_{2}$-norms
are estimated by $O\left(\left(1+|\lambda|^{1/2}\right)^{-1}\right)$
for $\lambda\in\Lambda_\eps$ uniformly with respect to $\nu$ (according
to \cite{Se1}, Lemma~7). The proposition is proved.\ \ \ $\Box$

\medskip
Let $\lambda\in\Lambda_\eps$ and $|\lambda|$ be large enough.
Then the Green function $\Ge_\lambda(\nu)$ can be represented
by the series
\begin{equation}
\Ge_\lambda(\nu)=R^m_{\lambda}\sum^{\infty}_{i=0}\left(L^m_\lambda\right)^i,
\label{X1986}
\end{equation}
where $\left(L^m_\lambda\right):=\id-\left(\don-\lambda\right)R^m_{\lambda}$
is a bounded operator in $\dromb$ for $\lambda\in\Lambda_\eps$.
The norm of $L^m_{\lambda}$ in $\dromb$ is
$O\left(\left(1+|\lambda|^{1/2}\right)^{n-m+2}\right)$ (where $n:=\dim M$)
because the norm of $\left(\id-(\doo-\lambda)P^m_{\lambda}\right)$ is
$O\left(\left(1+|\lambda|^{1/2}\right)^{n-m}\right)$ and the norm of
$(\doo-\lambda)q_{\nu}B_{\nu}P^m_{\lambda}$ is
$O\left(\left(1+|\lambda|^{1/2}\right)^{n-m+2}\right)$ (according
to the proof of Theorem~\ref{TAT705}). Hence if $m>n+2$ and
if $\lambda\in\Lambda_\eps$ with $|\lambda|$ large enough then
the series (\ref{X1986}) is convergent with respect to the operator norm
in $\dromb$. The operator $L^m_{\lambda}$ depends smoothly
on $\nu\in\rbo$. The norm of $\dfg L^m_{\lambda}$ is estimated
by $O\left(1+|\lambda|^{1/2}\right)$ (according to Proposition~\ref{PX3003}).
Let $\nu\colon[-a,a]\to\rbo$ be a smooth map. Then the series
for $\dfg\Ge_\lambda(\nu)$ is convergent (in the operator norm)
if $\lambda\in\Lambda_\eps$ and if $|\lambda|$ is large enough.
Hence for such $\lambda$ the resolvent
$\Ge_\lambda(\nu):=\left(\don-\lambda\right)^{-1}$ depends smoothly
on $\gamma$. So the family $\Ge_\lambda(\nu)$ of bounded operators
in $\dromb$ is smooth in $(\nu,\lambda)$ for such $\lambda$.
Their operator norms are estimated by $O\left(|\lambda|^{-1}\right)$
uniformly with respect to $\nu\in\rbo$. Let $F$ be a linear differential
operator of degree $d(F)\le 2$. Then the operators $F\Ge_\lambda(\nu)$
for such $\lambda$ are bounded in $\dromb$ with their operator norms
estimated by $O\left(|\lambda|^{(d-2)/2}\right)$ uniformly with respect
to $\nu$. These operators depend smoothly on $\nu$ for such $\lambda$
and we have
\begin{equation}
\dfg F\Ge_\lambda(\nu)=F\dfg\Ge_\lambda(\nu).
\label{X1996}
\end{equation}
Hence for a given $\nu_{0}\in\rbo$ there exists $\lambda_{1}\in\Lambda_\eps$
such that $\Ge_{\lambda_{1}}(\nu)$ depends smoothly on $\nu$ for $\nu$
sufficiently close to $\nu_{0}$.
For $\lambda\in\wC\setminus\Spec\left(\donoo\right)$ the resolvent
$\Ge_\lambda(\nu_{0})$ can be represented as follows:
\begin{equation}
\Ge_\lambda(\nu_{0})=-(\lambda-\lambda_{1})^{-1}-(\lambda-\lambda_{1})^{-2}
R\left((\lambda-\lambda_{1})^{-1},\Ge_{\lambda_{1}}(\nu_{0})\right),
\label{X1985}
\end{equation}
where $R\left(\eta,\Ge_{\lambda_{1}}(\nu_{0})\right):=\left(\Ge_{\lambda_{1}}
(\nu_{0})-\eta\right)^{-1}$ is the resolvent of a bounded operator
$\Ge_{\lambda_{1}}(\nu_{0})$ in $\dromb$ (\cite{Ka}, Ch.~IV, (3.6),
Ch.~III, (6.18)). The bounded operator $R(\eta,B)$ is an analytic function
of a bounded operator $B$ and of $\eta$ for $\eta\notin\Spec B$ (i.e., near
$(\eta_{0},B_{0})$, $\eta_{0}\notin\Spec B_{0}$, it is locally
defined by a convergent double power series in $(\eta-\eta_{0})$ and
$(B-B_{0})$). The operator $\Ge_{\lambda_{1}}(\nu)$ depends
smoothly on $\nu$ for $\nu$ sufficiently close to $\nu_{0}$. Then it follows
from (\ref{X1985}) that $\Ge_\lambda(\nu)$ depends smoothly on $\nu$ for
$\lambda\in\wC\setminus\Spec\left(\don\right)$ and for $\nu$ sufficiently
close to $\nu_{0}$.

Let $F\colon\oplus_j\DR(\Mm_j)\to\oplus_j DR^{\bullet+k}(\Mm_j)$,
$k\in\wZ$, be a linear differential operator of degree $d(F)\le 2$.
Then for $\lambda\in\Lambda_\eps$ and $|\lambda|$ large enough the operators
$F\Ge_\lambda(\nu)\colon\dromb\to\left(DR^{\bullet+k}(M)\right)_{2}$ are
defined, bounded, and smooth in $\nu\in\rbo$. (It is proved above.)
For example, $d\Ge_\lambda(\nu)$ and $\delta\Ge_\lambda(\nu)$ are smooth
in $\nu$. According to (\ref{X1996}) we have
$\dfg\left(d\Ge_\lambda(\nu)\right)=d\dfg\Ge_\lambda(\nu)$,
$\dfg\left(\delta\Ge_\lambda(\nu)\right)=\delta\dfg\Ge_\lambda(\nu)$.

The operators $d\Ge_\lambda(\nu)$ are defined for $\lambda\notin\Spec(\don)$.
{}From (\ref{X1996}) and (\ref{X1985}) we get
\begin{multline*}
\dfg\left(d\Ge_\lambda(\nu)\right)\bgo=
d\left(\dfg\Ge_\lambda(\nu)\bgo\right)=\\
=-(\lambda-\lambda_{1})^{-2}d\left({\df\over\df B}R\left((\lambda-\lambda_{1})
^{-1},B\right)\big|_{B=\Ge_{\lambda_{1}}(\nu_{0})}\dfg\Ge_{\lambda_{1}}
(\nu)\bgo\right)
\end{multline*}
for a $C^\infty$-local map
$\left(\wR^1_\gamma,0\right)\to\left(\rbo,\nu_{0}\right)$, where
$\lambda_{1}\in\Lambda_\eps$ with $|\lambda_{1}|$ large enough and
$\lambda\notin\Spec\left(\donoo\right)$.
Proposition~\ref{PA3002} is proved.\ \ \ $\Box$

\subsection{Theta-functions for the Laplacians with $\nu$-transmission
boundary conditions. Proofs of Theorem~\ref{TAT706} and
of Proposition~\ref{PE303}}

Let $\eps$ be fixed, $0<\eps<\pi/2$.
The operator $\exp\left(-t\don\right)$ is defined for $\Ree\,t>0$,
$\pi/2-\eps>\arg t>-(\pi/2-\eps)$, by the integral
\begin{equation}
\exp\left(-t\don\right)={i\over 2\pi}\int_{\Gamma_{L,\eps}}\exp(-\lambda t)
\Ge_\lambda(\nu)d\lambda,
\label{E100}
\end{equation}
where $\Gamma_{L,\eps}=\Gamma_{L,\eps}^1\cup\Gamma_{L,\eps}^2$,
$\Gamma^1_{L,\eps}=\{\lambda=-L+x\exp(i\eps),+\infty\!>x\!\ge 0\}$,
$\Gamma^2_{L,\eps}=\{\lambda=-L+x\exp(-i\eps),0\!\le x<\!+\infty\}$,
$L>0$.
The integral (\ref{E100}) is absolutely convergent because the operator
norm in $\dromb$ of the operator $\Ge_\lambda(\nu)$ (which is bounded
in $\dromb$) is estimated by $O\left(|\lambda|^{-1}\right)$
for $\lambda\in\Gamma_{L,\eps}$, according to Theorem~\ref{TAT705}.%
\footnote{The constant factor in this estimate depends on $\eps$.}
This integral is independent of $L>0$ and of $\eps$, $\pi/2>\eps>0$,
for $t$ such that $|\arg t|<\pi/2-\eps$, since the spectrum of $\don$
is discrete and since $\Spec\left(\don\right)\subset\wR_+\cup 0$.
With the help of the inverse {\em Mellin transform} $f\to M^{-1}f$
\begin{equation*}
\left(M^{-1}f\right)(t):=\left(2\pi i\right)^{-1}\int_{\Ree\,s=c}\Gamma(s)
t^{-s}f(s)ds
\end{equation*}
it is possible to obtain the results about the asymptotic expansion
for $\Tr\exp\left(-t\don\right)$ as $\Ree\,t\to+0$ (when
$\pi/2-\eps>|\arg t|$) from the results about $\znb(-m)$,
$m\in\wZ_+\cup 0$, and about $\res_{s=s_j}\zeta_{\nu,\bullet}(s)$
obtained in Theorem~\ref{TAT705}. The integral (\ref{E100}) can be
transformed as follows
\begin{equation*}
\exp(-t\don)=\cH^{\bullet}(\nu)+{i\over 2\pi}\int_{\Gamma_{-\delta,
\eps}}\exp(-t\lambda)\Ge_\lambda(\nu)d\lambda,
\end{equation*}
where $\cH^{\bullet}(\nu)$ is the kernel of the orthogonal projection
operator of $\dromb$ onto $\Ker\don$ and where $\delta>0$ and $\rho$,
$\rho\ge\delta$, is such that $\Spec\left(\don\right)\cap(0,\rho]=\emptyset$.
The operator $\exp\left(-t\don\right)$ for $|\arg t|<\pi/2-\eps$
can be represented as follows (where $\Gamma$ is the same
as in (\ref{ST808}) and $c>0$):
\begin{multline}
\exp(-t\don)=\cHh(\nu)+{i\over 2\pi}\int_{\Gamma_{-\delta,\eps}}\exp(-t
\lambda)\Ge_\lambda(\nu)d\lambda=\\
=\cHh(\nu)+{i\over 2\pi}\int_{\Gamma_{-\delta,\eps}}(2\pi i)^{-1}\Ge_
\lambda(\nu)\left(\int_{\Ree\,s=c}(\lambda t)^{-s}\Gamma(s)ds\right)d\lambda=\\
=\cHh(\nu)+(2\pi i)^{-1}\int_{\Ree\,s=c}t^{-s}\Gamma(s)\left({i\over 2\pi}\int
_{\Gamma_{-\delta,\eps}}\lambda^{-s}\Ge_\lambda(\nu)d\lambda\right)ds=\\
=\cHh(\nu)+(2\pi i)^{-1}\int_{\Ree\,s=c}t^{-s}\Gamma(s)\left({i\over 2\pi}\int
_{\Gamma}\lambda^{-s}\Ge_\lambda(\nu)d\lambda\right)ds=\\
=\cHh(\nu)+(2\pi i)^{-1}\int_{\Ree\,s=c}\Gamma(s)t^{-s}T_{-s}(\nu)ds.
\label{E103}
\end{multline}
Here, the integration is over $\Ree\,s=c$ from $c-i\infty$
to $c+i\infty$ (where $c>0$). The operator $T_{-s}(\nu)$ for $\Ree\,s>0$
is defined by the integral (\ref{ST808}). The transformations we apply
in (\ref{E103}) are correct by the Fubini theorem since the estimate
\begin{equation*}
\left\|\Ge_\lambda(\nu)\right\|_{2}<C\cdot\left|\lambda\right|^{-1},\qquad
\lambda\in\Gamma_{-\delta,\eps},
\end{equation*}
is satisfied by the operator norm of $\Ge_\lambda(\nu)$ in $\dromb$
and since for $\Ree\,s>0$ the gamma-function can be estimated as follows.
We have
\begin{equation*}
\Gamma(s)=\int^{\infty}_{0}t^{s-1}\exp(-t)dt=\int^{\infty}_{0}t^{s-1}\exp
(i\phi s)\exp\left(-t\exp(i\phi)\right)dt
\end{equation*}
for $\Ree\,s>0$ and for an arbitrary $\phi\in\wR$ such that $\pi/2>|\phi|$.
So the estimate holds for any $\eps_1$, $0<\eps_{1}\le\pi/2$ and
for $\Ree\,s>0$:
\begin{equation}
\left|\Gamma(s)\right|\le(\sin\eps_{1})^{-\Ree\,s}\Gamma(\Ree\,s)\exp\left(-
\left({\pi\over 2}-\eps_{1}\right)\left|\Image\,s\right|\right).
\label{E106}
\end{equation}

The kernel of $\left(T_{-s}(\nu)\right)_{\xa,\xb}$ is continuous
in $(\xa,\xb)\in\Mm_{j_{1}}\times\Mm_{j_{2}}$ for $\Ree\,s>n/2$
(according to Theorem~\ref{TAT705}). The equality (\ref{E103}) holds
also for $c=\Ree\,s>n/2$. For such $s$ the integral
$\int_{\Ree\,s=c}\Gamma(s)t^{-s}\left(T_{-s}(\nu)\right)_{\xa,\xb}ds$
is absolutely convergent (by Proposition~\ref{PE200} below and
by (\ref{E106})). Hence it defines a continuous
on $\Mm_{j_{1}}\times\Mm_{j_{2}}$ kernel. So the kernel $\eotxaxb(\nu)$
of $\exp(-t\don)$ is continuous on $\Mm_{j_{1}}\times\Mm_{j_{2}}$ because
we have
\begin{equation}
\eotxaxb(\nu)=\cHh(\nu)_{\xa,\xb}+(2\pi i)^{-1}\int_{\Ree\,s=c}\Gamma(s)t^{-s}
\left(T_{-s}(\nu)\right)_{\xa,\xb}dt,
\label{E108}
\end{equation}
where $c>n/2$. (The integral in (\ref{E108}) converges uniformly
with respect to $\xa,\xb$ for any fixed $c>n/2$ by Proposition~\ref{PE200}.)

{}From the functional equation
$\Gamma(s)=s^{-1}(s+1)^{-1}\dots(s+l-1)^{-1}\Gamma(s+l)$
it follows that $|\Gamma(s)|$ for $\Ree\,s>-l$ is also estimated
by $\exp\left(-\left(\pi/2-\eps_{1}\right)|\Image\,s|\right)$ as
$|\Image\,s|\to\infty$ (with any fixed $\eps_1$, $0<\eps_{1}\le\pi/2$).
The operator $\exp\left(-t\don\right)$ for $\Ree\,t>0$ is a trace class
operator.
Namely its kernel is continuous on $\Mm_{j_{1}}\times\Mm_{j_{2}}$
(as it follows from (\ref{E108})). Hence it is a trace class operator
and its trace is equal to the sum of the integrals over the diagonals
$\Mm_j$ of the corresponding densities (according to Proposition~\ref{PA75}
below).

The {\em theta-function} $\tno(t)$ for $\don$ is defined as the trace
of $\exp\left(-t\don\right)$ for $\Ree\,t>0$. The analogous theta-function
$\tno(t;p_j)$ is defined as the trace
$\Tr\left(p_j\exp\left(-t\don\right)\right)$
for $\Ree t>0$ (where
$p_j\colon\dromb\to\left(\DR(M_j)\right)_{2}\hookrightarrow\dromb$
is the composition of the natural restriction and of the prolongation
by zero). Proposition~\ref{PA75} claims that $\tno(t;p_j)$ is equal to
the integral over $\Mm_j$ of the density
$\tr\left(*_{\xb}i^*_{M_j}\eotxaxb(\nu)\right)$.

The zeta-function $\znb(s)$ is defined by (\ref{S9}) for $\Ree\,s>n/2$
($n:=\dim M$). It is equal to $\Tr T_{-s}(\nu)$ for $\Ree\,s>n/2$
(according to Theorem~\ref{TAT705} and to Proposition~\ref{PA80}).
The zeta-function $\zeta(s;p_j):=\Tr\left(p_j T_{-s}(\nu)\right)$
is equal for such $s$ to the integral over the diagonal
$i\left(\Mm_j\right)\hookrightarrow\Mm_j\times\Mm_j$ of the density,
corresponding to the restriction of the kernel $T_{-s}(\nu)$
to $i\left(\Mm_j\right)$. The integral of this density over $\Mm_j$
can be represented as the sum of the integrals of densities on $\Mm_j$
and on $\df M_j$ (they are defined by the parametrix (\ref{X1195})
and can be continued to meromorphic functions on the whole complex
plane $\wC\ni s$)
and of a density on $\Mm_j$, which is holomorphic for $\Ree\,s>(n-m)/2$.
(This assertion follows from the proof of Theorem~\ref{TAT705}.)
The contour of the integration in (\ref{E103}) can be moved
to $\Ree\,s=a$ for an arbitrary $a$ such that $(-2a)\notin\wZ_+\cup 0$
(according to the estimates of $|\Gamma(s)|$ as $|\Image\,s|\to+\infty$
and to Proposition~\ref{PE201} below). Then it follows from (\ref{E103})
that
\begin{multline}
\theta_{\nu,\bullet}(t;p_j)=\sum t^{-(n-k)/2}\res_{s=s_k}\left(\Gamma(s)
\zeta_{\nu,\bullet}(s;p_j)\right)+\\
+(2\pi i)^{-1}\int_{\Ree\,s=a}t^{-s}\Gamma(s)\zeta_{\nu,\bullet}(s;p_j)ds+
\Tr\left(p_j\cHh(\nu)\right),
\label{E110}
\end{multline}
where the sum is over $k$ such that $s_k:=(n-k)/2>a$.
The estimate of the integral over $\Ree\,s=a$ in (\ref{E110})
is obtained with the help of (\ref{E106}) and (\ref{E109}) as follows.
For $\Ree\,t>0$, $|\arg t|<\pi/2-\eps$ ($\eps$, $0<\eps<\pi/2$,
is fixed) and for $\Ree\,s=a$ the estimate is satisfied:
\begin{multline}
\left|t^{-s}\Gamma(s)\zeta_{\nu,\bullet}(s;p_j)\right|<
\left(\sin{\eps\over 4}\right)^{-(a-[a])}\left|\Gamma(a)\right|\left|t\right|
^{-a}\exp\left(-{\eps\over 2}|\Image\,s|\right)C\left(a,{\eps\over 4}\right)
\times\\
\times\left\{c_{\eps/4}^{2(a-1)}\Gamma\left(2(1-a),c_{\eps/4}\rho^{1/2}\right)
+\max\left(\rho^{-a},1\right)\left(1+\sum|s_k-a|^{-1}\right)\right\},
\label{E114}
\end{multline}
where the sum is over $k<n-2a$. The constants $C(a,\eps/4)$, $c_{\eps/4}$
in (\ref{E114}) and the function $\Gamma(u;x)$ are
as in Proposition~\ref{PE201}, (\ref{E109}).
The latter estimate is a consequence of (\ref{E106}) and (\ref{E109})),
where $\eps_1$ and $\eps$ are replaced by $\eps/4$. We see that
\begin{equation}
\left|\int_{\Ree\,s=a}t^{-s}\Gamma(s)\zeta_{\nu,\bullet}(s;p_j)ds\right|<C_{1}
(\eps,a)|t|^{-a},
\label{E115}
\end{equation}
where $\Ree\,t>0$, $|\arg t|<\pi/2-\eps$, $\pi/2>\eps>0$, $a<0$, and
$(-2a)\notin\wZ_+$. The assertions of Theorem~\ref{TAT706} about
the asymptotic expansion (\ref{S2005}) for $\tno(t;p_j)$ (relative
to $t\to+0$ when $|\arg t|<\pi/2-\eps$) follow from the equality
(\ref{E110}) and from the estimate (\ref{E115}).
The estimates analogous to (\ref{E109}) below and to (\ref{E115})
are satisfied also by the analytic continuation to $\wC\ni s$
of the densities (on $\Mm_j$ and on $\df M_j$) defined by the parametrix
$P^m_\lambda(\nu)$ (as in Proposition~\ref{PE200} below).
Thus we see that the equalities between the densities in the integral
representation for the coefficients of the expansion (\ref{S2005})
and the corresponding densities for the residues and the values
of $\znb(s;p_j)$ are satisfied.

The uniform with respect to $\nu\in\rbo$ estimate (\ref{E98})
for the traces of $\exp\left(-t\don\right)$ (for a fixed $t$, $\Ree\,t>0$)
follows
from (\ref{E110}) and (\ref{E114}) because for $a=-m-1/4$, $m\in\wZ_+$,
$m>>1$, the integral over $\Ree\,s=a$ on the right in (\ref{E110})
is absolutely convergent. The estimate%
\footnote{For $a<0$ the function
$\Gamma\left(2(1-a),c_{\eps/4}\rho^{1/2}\right)$ tends to
$\Gamma\left(2(1-a)\right)$ as $\rho\to+0$ and so it is bounded for
$0<\rho<1$.}
(\ref{E114}) and the equality (\ref{E110}) provide us with the uniform
in $\nu$ upper estimate for $\Tr\left(p_j\exp\left(-t\don\right)\right)$,
$\nu\in\rbo$. Indeed, the estimate%
\footnote{It follows from the exact sequence (\ref{A12}) (where
$Z_j:=Z\cap\df M_j$) and from Lemma~\ref{L1} that
$$
\dim\Ker\don=\dim\Hh(M_\nu,Z)\le\sum_j\dim\Hh(M_j,N\cup Z_j)+\dim\Hh(N).
$$}
$\dim\Ker\don<C$ is satisfied uniformly with respect to $\nu$.
The formulas $q_{-n+k}=\res_{s=s_k}\left(\Gamma(s)\znb(s;p_j)\right)+
\delta_{n,k}\Tr\left(p_j\cHh(\nu)\right)$ for the coefficients $q_{-n+k}$
of the asymptotic expansion (\ref{S2005}) are consequences of (\ref{E110})
for $a=-m-1/4$, where $m\in\wZ_+$.
For $a=-m-1/4$ the absolute value of the integral over $\Ree\,s=a$ in
(\ref{E110}) is estimated (with the help of (\ref{E114}))
by $C|t|^{m+1/4}$ uniformly with respect to $\nu\in\rbo$ (where $\Ree\,t>0$,
$|\arg t|<\pi/2-\eps$, $\pi/2>\eps>0$ and $\eps$ is an arbitrary but fixed).
So it holds
\begin{equation}
\tno(t;p_j)=\sum^{n+2m-1}_{k=0}q_{-n+k}t^{-(n-k)/2}+\left\{q_{-n+2m}t^m+
O\left(\left|t\right|^{m+1/4}\right)\right\}.
\label{X1691}
\end{equation}

The latter two terms in (\ref{X1691}) are $O\left(|t|^m\right)$
relative to $t\to+\infty$ uniformly with respect to $\nu\ne(0,0)$
(for $|\arg t|<\pi/2-\eps$). The statements about the structure
of the values and the residues of $\znb(s;p_j)$ (Theorem~\ref{TAT705})
provide us with the desired information about coefficients $q_{-n+k}$
in (\ref{S2005}). These values and residues
(up to $\delta_{n,k}\Tr\left(p_j\cHh(\nu)\right)$) are the sums
of the integrals over $M_j$, $\df M$, and $N$ of the densities
which are defined by the absolutely convergent integrals
of the components $a_{-2-k}$ and $d_{-2-k+1}$ (\cite{Se2}, Theorem~2,
and the proof of Theorem~\ref{TAT705} above). The latter symbols
are defined by (\ref{ST750}), (\ref{ST763}), and (\ref{ST762}).
These integrals are smooth in $\nu\in\rbo$. Hence the coefficients
$q_{-n+k}$ in (\ref{S2005}) are smooth in $\nu\ne(0,0)$ (and are invariant
under $\nu\to c\nu$, $c\ne 0$). Theorem~\ref{TAT706} is proved.\ \ \ $\Box$

\begin{rem}
The coefficients $q_{-n+k}$ of (\ref{S2005}) for $0\le k\le m$ are
completely defined (according to (\ref{X1691}) and to Remark~\ref{RAT300})
by an arbitrary parametrix $P^m_\lambda(\nu)$ (\ref{X1195})
for $\left(\don-\lambda\right)^{-1}$.
\label{RAT301}
\end{rem}

\noindent{\bf Proof of Proposition~\ref{PE303}.} The parametrix%
\footnote{The properties of such a parametrix are summarized
in Proposition~\ref{PAT310}. For the sake of brevity the proof
of Proposition~\ref{PE303} is given in the case of $\df M=\emptyset$.
The terms of $P^{(m)}_t(\nu)$, connected with the Dirichlet and
the Neumann boundary conditions on the components of $\df M$,
are independent of $\nu$ and the proof in the case of $\df M\ne\emptyset$
does~not contain any additional difficulties.}
$\pom_{t,x,y}(\nu)$ for $\eotxy(\nu)$ (defined by (\ref{E1970}))
is such that it is smooth in $(x,y)\in\Mm_{j_{1}}\times\Mm_{j_{2}}$ and
in $\nu\in\rbo$. The $\nu$-transmission boundary conditions (\ref{Y4})
are satisfied for $\left(\doo_{\nu,x}\right)^k\pom_{t,x,y}(\nu)$
(i.e., the image of $\dromb$ under the action of the operator with
the kernel $\pom_{t,x,y}(\nu)$ belongs to $D\left(\left(\don\right)^k\right)$
for an arbitrary $k\in\wZ_+$). The uniform with respect to $\nu\in\rbo$
estimates (\ref{S1611}), (\ref{S16111}) are satisfied and for $x$
from an appropriate neighborhood $U$ of $N\subset M$ we have
$\left(\df_t+\doo_{\nu,x}\right)\pom_{t,x,y}(\nu)\equiv 0$
(where $U$ is independent of $\nu$).

Set $r^{(m)}_{t,x,y}(\nu):=\left(\df_t+\doo_{\nu,x}\right)\pom
_{t,x,y}(\nu)$. Then the estimates are satisfied for any $k\in\wZ_+\cup 0$
\begin{equation}
\left|\Delta^k_{\nu,x}r^{(m)}_{t,x,y}(\nu)\right|<C_{m,k}t^{-n/2+m-k},
\label{X1998}
\end{equation}
where $C_{m,k}$ is independent of $\nu\in\rbo$ and of $t\in(0,T]$
($n:=\dim M$).

The kernel $\eotxy(\nu)$ can be represented as the Volterra series%
\footnote{This series was used in the case of a closed manifold $M$
in \cite{BGV}, 2.4, 2.7. See also the formula (\ref{E1962}) above.}
\begin{multline}
\eotxaxb(\nu)=\\
=\sum_{k\ge 0}(-t)^k\int_{\Delta_k}\int_{(\ya,\dots,y_k)\in\left(\Mm_{1}\cup
\Mm_{2}\right)^k}\pom_{\sio t,\xa,\ya}(\nu)r^{(m)}_{\sia t,\ya,\yb}
(\nu)\dots r^{(m)}_{\sig_k t,y_k,\xb}(\nu),
\label{X2001}
\end{multline}
where $\Delta_k=\{(\sio,\dots,\sig_k)\colon 0\le\sig_i\le 1,\sum\sig_i=1\}$
(and the scalar product $\tr(\oma\wedge*\omb)$ with the values in densities
on $M$ is implied in (\ref{X2001})). The proof of (\ref{X2001}) (or of
(\ref{E1962})) is given in the proof of Proposition~\ref{PAT310}.

Let $\phi\colon I\to\rbo$, $\nu=\phi(\gamma)$, be a $C^\infty$-map
(where $\gamma\in[-a,a]=:I$). Then the only term
in $\pom_{t,(\xa,x'),(\ya,y')}(\nu)$ depending on $\gamma$ is
\begin{equation}
E_{N,t}\otimes\psi(\xa)E_{I,t}(\nu)\phi(\ya)=:\eo_{N,I}(\nu)
\label{X2002}
\end{equation}
but it does~not depend on $m$.
(Here $E_{I,t}$ corresponds to $\Delta_{I}$ with
the Dirichlet boundary conditions on $\df I$,
$\phi,\psi\in C^\infty_{0}(I\setminus\df I)$, $\psi\equiv 1$
in a neighborhood of $\supp\phi\subset I\setminus\df I$, $\phi(\xa)\equiv1$
for $\xa\in[-1/2,1/2]$ and $\phi,\psi$ are even: $\phi(-\xa)=\phi(\xa)$,
$\psi(-\xa)=\psi(\xa)$.)
So, as it follows from the explicit formulas (\ref{S1240}) for
$\left(G_{I}(\nu)\right)_{\xa,\ya}$ (and from the analogous formulas
(\ref{S414}) and (\ref{S415}) for $\left(E_{I,t}(\nu)\right)_{\xa,\ya}$),
the uniform with respect to $\gamma$ estimates are satisfied
for any $k,q\in\wZ_+$
\begin{equation}
\left|\dfg^kr^{(m)}_{t,x,y}(\nu)\right|<c_{(q,k)}t^{-n/2+q}.
\label{X2012}
\end{equation}
because they are true for
$(\df_t+\Delta_{x'}+\Delta_{\xa})\left\{\left(E_{N,t}\right)_{x',y'}
\otimes\psi(\xa)\left(E_{I,t}(\nu)\right)_{\xa,\ya}\phi(\ya)\right\}$
and for
$(\df_t+\Delta_{x'}+\Delta_{\xa})\left\{\left(E_{N,t}\right)_{x',y'}
\otimes\psi(\xa)\left(\sigma^*_{1}E_{I,t}(\nu)\right)_{\xa,\ya}\phi(\ya)
\right\}$ (where $\sigma_{1}$ is the reflection of $I=[-1,1]$ with respect
to $0\in I$ which acts on the variable $\xa$).
The kernels $\dfg^k\eo_{N,I}(\nu)$
are the linear combinations of these two kernels with the coefficients
independent of $x$ and $y$. (These coefficients are smooth in $\gamma$).
The estimate (\ref{X2012}) is satisfied for $t\in(0,T)$ and for an arbitrary
$q\in\wZ_+$ uniformly with respect to $\gamma$, because if
$\df_{\xa}\psi(\xa)\ne 0$ then $\rho(\xa,\supp\phi)>\delta>0$.

Let $\DR_M(l)$ be the space of forms on $M$ of a class $C^l$ (i.e.,
of forms with $l$ continuous derivatives on $M$) equiped with a $C^l$-norm.%
\footnote{This norm corresponds to a smooth partition of unity $\{\phi_i\}$
subordinate to a finite cover $\{U_i\}$ of $\Mm_{1}$ and of $\Mm_{2}$
by coordinate charts (i.e., $\phi_i\in C^\infty_{0}(U_i)$).
For $v\in\DR_{M,N}(l)$ its $C^l$-norm $\|v\|_l$ is equal
to $\sum\sup_{x\in U_i}\sup_{|\alpha|\le l}\left|D^\alpha_x(\phi_iv)\right|$
i.e., to the sum of the suprema of partial derivatives of orders $\le l$.
The $C^l$-norm for an arbitrary smooth finite cover $\{U'_i\}$ and
for a partition of unity $\{\phi'_i\}$ subordinate to $\{U'_i\}$
is equivalent to the one defined by $\left\{U_i\right\}$ and
by $\left\{\phi_i\right\}$.}
Let $\DR_{M,N}(l):=\DR_{(l)}\left(\Mm_{1}\right)\oplus\DR_{(l)}\left(
\Mm_{2}\right)$ be the space of pairs $(\oma,\omb)$ of forms $\omega_j$
of a class $C^l$ on $\Mm_j$ with a $C^l$-norm.
The operators with the kernels $\pom_{t,x,y}(\nu)$ for $\nu\in\rbo$
and the operators corresponding
to $\dfg^k\pom_{t,x,y}\left(\nu(\gamma)\right)$ (for a fixed
$k\in\wZ_+\cup 0$) are families of uniformly (with respect to $\nu$ and
to $t\in(0,T]$) bounded operators acting from $\DR_M(l)$ into $\DR_{M,N}(l)$.
For the operators, corresponding to the interior terms
in $\pom_{t,x,y}(\nu)$ this assertion is proved in \cite{BGV},
Theorem~2.29, Lemma~2.49. This proof uses that this statement is local
in $x\in M$ (for a closed $M$) and it uses also the explicit definition
of $P^{(m)}_{int}$ over a geodesic ball $\exp_x B\subset M$ (where $B$
is a ball $\|v\|\le c$ in $T_x M$ and $\exp_x$ is the exponential map
for $(M,g_{M})$ from $T_x M$).

The kernel (\ref{X2002}) (i.e., the term of $\pom_{t,x,y}(\nu)$
corresponding to the interior boundary $N$) is equal (up to the factor
$\psi(\xa)\phi(\ya)$) to a linear combination given by (\ref{S414})
and (\ref{S415}) of the kernels $\eo_t$ for $N\times I$
and $\sia^*\eo_t$ ($\sia$ is the mirror symmetry with respect
to $N\times 0$).%
\footnote{The operators $\dfg^kP^{(m)}(\nu)=\dfg^kE_{N,I}(\nu)$ for
$k\in\wZ_+$ are expressed similarly.}
Its coefficients depend on $\left(j_1,j_2,\nu\right)$, where
$(x,y)\in\Mm_{j_1}\times\Mm_{j_2}$. These coefficients
and their derivatives of a fixed order on $\gamma$ are uniformly bounded.

For a closed $N$ the operators defined by
$\left(P^{(m)}_{N}\right)_t:=\left(P^{(m)}_{int,N}\right)_t$
are uniformly bounded for $0<t\le T$ with respect to a $C^l$-norm
in the space $\DR_{N}(l)$ of $C^l$-smooth forms on $N$.
The equality (\ref{X2001}) is satisfied by $\eo_{N,t}$, $\pom_{N,t}$
and $r^{(m)}_{N,t}$. Since the estimate (\ref{X2004}) below (as well
as the analogous estimate (\ref{X1998})) is satisfied by $r^{(m)}_{N,t}$
(where $n$ is replaced by $n-1$) we see that the series of operators
on the right in (\ref{X2001}) is convergent for $m\ge(n+l-1)/2$ relative
to a $C^l$-norm in $\DR_{N}(l)$.
Hence the sum of this series defines a family of the operators
$\eo_{N,t}$ in $\DR_{N}(l)$ bounded uniformly with respect to $t\in(0,T]$.
The analogous assertion is also true for a family
of operators defined by the kernels
$\psi(\xa)\left(\eo_{I,t}\right)_{\xa,\ya}\phi(\ya)$ acting
on smooth forms with compact supports on $(I,\df I)$ (with respect
to a $C^l$-norm). So the kernels $\eo_{N,I}(\nu)_{t,(\xa,x'),(\ya,y')}$
and $\dfg^q\eo_{N,I}(\nu(\gamma))_{t,(\xa,x'),(\ya,y')}$ (for a fixed
$q\in\wZ_+$) define collections of uniformly (in $t\in(0,T]$ and $\nu$
or in $t$ and $\gamma$) bounded with respect to a $C^l$-norm operators
from $\DR_M(l)$ into $\DR_{M,N}(l)$.

The kernel $r^{(m)}_{t,x,y}(\nu)$ is smooth on $M\times\Mm_j$ according
to the definition (\ref{E1970}) of $\pom_{t,x,y}(\nu)$.
The $C^l$-norm of $r^{(m)}_{t,x,y}(\nu)$ on each $M\times\Mm_j$
satisfies the estimate (analogous to (\ref{X1998}))
\begin{equation}
\left\|r^{(m)}_{t,x,y}(\nu)\right\|_l\le C_lt^{-n/2+m-l/2}.
\label{X2004}
\end{equation}
uniformly with respect to $\nu\in\rbo$ and to $t\in(0,T]$. The estimates
analogous to (\ref{X2012}) and to (\ref{S16100}) are satisfied uniformly
with respect to $\nu$ and to $t\in(0,T]$ also by the $C^l$-norms
of $\dfg^kr^{(m)}_{t,x,y}(\nu)$ on each $M\times\Mm_j$
for any $k,q\in\wZ_+$:
\begin{equation}
\left\|\dfg^kr^{(m)}_{t,x,y}(\nu)\right\|_l\le c_{(q,k,l)}t^{-n/2+q}.
\label{X2022}
\end{equation}

Leibnitz's rule claims that $P^{(m)}_{t,x,y}(\nu(\gamma))$ is a bounded
operator from the space of $C^p$-maps $\omega\colon[-a,a]\to\DR_{M}(l)$
(equiped with the norm
$\sum^p_{i=0}\sup_{\gamma\in[-a,a]}\left\|\dfg^i\omega\right\|_l$)
into $C^p\left([-a,a],\DR_{M,N}(l)\right)$.
Indeed, the kernel $\dfg^kP^{(m)}_{t,x,y}(\nu(\gamma))$ depends
smoothly on $\gamma\in[-a,a]$ on $\Mm_{j_{1}}\times\Mm_{j_{2}}$
for $k\in\wZ_+\cup0$, since
$P^{(m)}_{t,x,y}(\nu):=\left(P^{(m)}_{int}\right)_{t,x,y}+\eo_{N,I}(\nu)$
and since $\eo_{N,I}(\nu)$ is smooth in $\nu\in\rbo$. For $0<t\le T$
the operators $\left(P^{(m)}_t(\nu(\gamma))\right)$ from
$C^p\left([-a,a],\DR_M(l)\right)$
into $C^p\left([-a,a],\DR_{M,N}(l)\right)$ are uniformly bounded, because
$\dfg^kP^{(m)}_t$ (for a fixed $k\in\wZ_+\cup0$) are the operators
from $\DR_M(l)$ into $\DR_{M,N}(l)$ bounded uniformly
in $\gamma$ and $t$, $0<t\le T$, with respect to a $C^l$-norm .
Hence, according to (\ref{X2004}),
(\ref{X2022}), and to the fact that the volume of $\Delta_k$ is equal
to $\left(k!\right)^{-1}$, the series (\ref{X2001}) for the derivative
$\dfg^p\etxaxb\left(\nu(\gamma)\right)$ is convergent in the $C^l$-norm
on $\bigcup\left(\Mm_{j_{1}}\times\Mm_{j_{2}}\right)$ for $m>(n+l)/2$.
(The number $m$ in the definition $P^{(m)}$ is greater than $(n+l)/2$.)

This proves that $\dfg^k\eotxaxb(\nu(\gamma))$ is $C^{\infty}$-smooth
on $\Mm_{j_{1}}\times\Mm_{j_{2}}$.
(For instance, for $k=0$ this proves that $\eotxaxb(\nu)$ is
$C^{\infty}$-smooth on $\Mm_{j_{1}}\times\Mm_{j_{2}}$.)

So the restrictions $i^*_j\eo_{t,x,x}(\nu(\gamma))$ to the diagonals
$i_j\colon\Mm_j\hookrightarrow\Mm_j\times\Mm_j$ are
$C^{\infty}$-smooth double forms on $\Mm_j$ which are $C^{\infty}$-smooth
in $\gamma$ . Since $r_m(t,\nu)$ in (\ref{S2005}) are
$O\left(t^{(m+1)/2}\right)$ uniformly with respect to $\nu\in\rbo$
and since $q_i$ are $C^{\infty}$-smooth in $\nu$ we see that
the asymptotic series (\ref{S2005}) can be differentiated on $\gamma$.
Actually, the equality (\ref{X2001}) holds for $\eo_t(\nu)$, $\pom_t(\nu)$,
and $r^{(m)}_t(\nu)$. The kernel $r^{(m)}_t(\nu)$ satisfies the estimates
(\ref{X1998}), (\ref{X2012}), and (\ref{X2022}) and the kernel
$\pom_t(\nu(\gamma))$ defines a family of uniformly with respect
to $t\in(0,T]$ and to $\gamma$ bounded operators
from $C^p\left([-a,a],\DR_M(l)\right)$
into $C^p\left([-a,a],\DR_{M,N}(l)\right)$. Hence the power terms
$t^{(-n+j)/2}$, $0\le j\le 2m$, in the asymptotic expansion
of $\int_{M_j}\tr i^*_j\dfg\eo_t(\nu)$ as $t\to+0$ are equal
to the appropriate terms in the asymptotic expansion
of $\int_{M_j}i^*_j\dfg \pom_t(\nu)$.
(The kernel $\dfg\left(\eo_t(\nu)-\pom_t(\nu)\right)_{\xa,\xb}$
is $O\left(t^{-n/2+m+1}\right)$, according to (\ref{X2001}).)
But the coefficients $q_i$, $0\le i\le 2m$, in (\ref{S2005}) are
completely defined by $i^*_j\pom_t(\nu)$, because the kernel
$\left(\eo_t(\nu)-\pom_t(\nu)\right)_{\xa,\xb}$
is $O\left(t^{-n/2+m+1}\right)$ uniformly with respect
to $(\xa,\xb)\in\Mm_{j_{1}}\times\Mm_{j_{2}}$ and to $t\in(0,T]$,
according to (\ref{X2001}). Thus Proposition~\ref{PE303}
is proved.\ \ \ $\Box$

\subsection{Estimates for zeta-functions and for the corresponding kernels
in vertical strips in the complex plane}
\begin{pro}
The meromorphic continuation of the zeta-function
$\zeta_{\nu,\bullet}(s;p_j):=\Tr\left(p_jT_{-s}(\nu)\right)$ for
$\Ree\,s>n/2$ is estimated by $C(\eps)\exp\left(\eps|\Image\,s|\right)$
as $|\Image\,s|\to+\infty$ for any fixed $\eps>0$.
Namely for any $\eps>0$ and for an arbitrary $a\in\wR$
the following estimate is satisfied if $\Ree\,s\ge a$:
\begin{multline}
\left|\zeta_{\nu,\bullet}(s;p_j)\right|\le C(a,\eps)\exp\left(\eps|\Image\,s|
\right)\times \\
\times\left(c_\eps^{2(\Ree\,s-1)}\Gamma\left(2(1-\Ree\,s),c_\eps\rho^{1/2}
\right)+\max\left(\rho^{-\Ree\,s},1\right)\left(1+\sum\left|s-s_j\right|^{-1}
\right)\right),
\label{E109}
\end{multline}
where $\rho>0$ is such that  $\Spec\left(\don\right)\cap(0,\rho]=\emptyset$
and the sum is over $s_j:=(n-j)/2$, $-s_j\notin\wZ_+\cup 0$, $s_j\ge a$.
The constants $C(a,\eps)$ and $c_\eps$ are positive and independent
of $\nu\in\rbo$, and $\Gamma(u,x):=\int_x^{\infty}t^{u-1}\exp(-t)dt$
for $x>0$.
\label{PE201}
\end{pro}

\begin{pro}
For $\Ree\,s>n/2$ ($n:=\dim M$) and for any $\eps>0$ the following
estimate is satisfied (where $\rho>0$ is such that
$\Spec\left(\don\right)\cap(0,\rho]=\emptyset$):
\begin{equation}
\left|\left(T_{-s}(\nu)\right)_{\xa,\xb}\right|\le C_\eps\rho^{-\Ree\,s}\exp
\left(\eps|\Image\,s|\right)\left(\left(\Ree\,s-{n\over 2}\right)^{-1}+1
\right).
\label{E107}
\end{equation}
\label{PE200}
\end{pro}

\noindent{\bf Proof of Proposition~\ref{PE201}.} It is proved
in Theorem~\ref{TAT705} that the operator norm
$\left\|\Ge_\lambda(\nu)\right\|_{2}$ in $\dromb$ of the Green function
$\Ge_\lambda(\nu)$ for the Laplacian $\don$ is estimated
by  $C_\eps|\lambda|^{-1}$ for
$\lambda\in\Lambda_\eps:=\{\lambda\in\wC,\eps\le\arg\lambda\le 2\pi-\eps\}$,
where $\eps$, $0<\eps\le\pi$ is fixed. The spectrum $\Spec\left(\don\right)$
is a discrete subset of $\wR_+\cup0$ by Theorem~\ref{TAT705}.
So the operator $T_{-s}(\nu)$ defined by the integral (\ref{ST808})
is equal to the same integral with the contour $\Gamma$ replaced by
$\Gamma_{(\eps)}:=\Gamma_{1,\eps}\cup\Gamma^\eps_{\rho}\cup\Gamma_{2,\eps}$,
\begin{gather}
\begin{split}
\Gamma_{1,\eps}=\{\lambda=x\exp(i\eps),\infty>x\ge\rho\},\quad &
\Gamma^\eps_\rho=\{\lambda=\rho\exp(i\phi),\eps>\phi>-\eps\}, \\
\Gamma_{2,\eps}=\{\lambda=x\exp(-i\eps),\, & \rho\le x<\infty\}.
\end{split}
\label{X1790}
\end{gather}

There is a constant $c>0$ such that the principal symbol
$\left(b_{2}(x,\xi)-\lambda\right)\id$ of $\doo-\lambda\id$ on $M$
is invertible for
\begin{equation}
\left|\xi\right|^2>c|\lambda|
\label{X1791}
\end{equation}
in the coordinate charts $U_l$ (of the same finite cover
$\left\{U_l\right\}$ of $M$ as in (\ref{X1195})). The integral (\ref{ST808})
over the contour $\Gamma_{(\eps)}$ (the latter one is defined
by (\ref{X1790})) does not depend on $\rho$ for all $\rho>0$ such that
$(0,\rho]\cap\Spec(\don)=\emptyset$. We suppose from now on that
\begin{equation}
0<\rho<(2c+1)^{-1}
\label{X1792}
\end{equation}
and that $(0,\rho]\cap\Spec(\don)=\emptyset$.

The kernel $\left(r^m_\lambda\right)_{\xa,\xb}$ of the operator
$r^m_\lambda:=\Ge_\lambda(\nu)-P^m_\lambda$ for $m\ge n$
is continuous on $\left(\Mm_{j_{1}}\times\Mm_{j_{2}}\right)$
for $\lambda\in\Lambda_\eps\setminus 0$. (The parametrix $P^m_\lambda$
is defined by (\ref{X1195}).) It is estimated for $|\lambda|\ge\rho$,
$\lambda\in\Lambda_\eps$ (according to (\ref{X1794})) uniformly
with respect to $\nu\in\rbo$ by
\begin{equation}
\left|\left(r^m_\lambda\right)_{\xa,\xb}\right|<C_{\eps,\eps_{1}}\left(1+
\left|\lambda\right|^{1/2}\right)^{-(2+m)+n+\eps_{1}}
\label{E125}
\end{equation}
for any $\eps_{1}>0$. Since (\ref{E125}) is satisfied for all
$\lambda\in\Gamma_{(\eps)}$, we have
for $\Ree\,s>2^{-1}\left(-m+n+\eps_{1}\right)$ that
\begin{multline}
\left|{i\over 2\pi}\int_{\Gamma_{(\eps)}}\lambda^{-s}\left(r^m_\lambda
\right)_{\xa,\xb}d\lambda\right|=
\left|{i\over 2\pi}
\left(\sum_j\int_{\Gamma_{j,\eps}}+\int_{\Gamma^\eps_\rho}\right)
\lambda^{-s}\left(r^m_\lambda\right)_{x,x}d\lambda\right|\le \\
\le 2C_{\eps,\eps_{1}}\exp\left(\eps|\Image\,s|\right)\times\\
\times\left\{\int^{\infty}_\rho\left|\lambda\right|^{-\Ree\,s}\left(1+
\left|\lambda\right|^{1/2}\right)^{-(2+m)+n+\eps_{1}}d|\lambda|+
\pi\rho\max\left(\rho^{-\Ree\,s},1\right)\right\}.
\label{E127}
\end{multline}

The estimate (\ref{E127}) claims that for the proof of (\ref{E109}) in the
domain $\Ree\,s\ge a$ it is enough to prove the analogous estimate for the
analytical continuation of the densities on $\Mm_j$, $N$, and on $\df M$,
corresponding (for $\Ree\,s>n/2$) to the kernel of
\begin{equation}
{i\over 2\pi}\int_{\Gamma_{(\eps)}}\lambda^{-s}P^m_\lambda d\lambda,
\label{E400}
\end{equation}
where $m=m(a)\in\wZ_+$ is sufficiently large. (These densities were
introduced in the proof of Theorem~\ref{TAT705}, and the sum of their
integrals is equal to the trace of (\ref{E400}).)

Let $\Ree\,s>n/2$  and $p^{s,\eps}_{int}(x)$ be the density on $M_j$,
corresponding to the restriction to the diagonal
$i_j\colon M_j\hookrightarrow M_j\times M_j$ of the kernel
$P^{s,\eps}_{int}(\xa,\xb)$ of the operator
$\int_{\Gamma_{(\eps)}}\lambda^{-s}\left(p_j\sum_i\psi_i P^m_{\lambda,
int,U_i}\phi_i\right)d\lambda$ (where $P^m_{\lambda,int}$ is defined
by (\ref{X1196}) and by (\ref{ST750})). Then $p^{s,\eps}_{int}(x)$
can be continued to a whole complex plane $\wC\ni s$ as a meromorphic
density (\cite{Se2}, Lemma~1, or \cite{Sh}, Theorem~12.1).

\begin{pro}
The density $p^{s,\eps}_{int}(x)$ satisfies the following estimate for any
$\eps>0$:
\begin{equation}
\left|p^{s,\eps}_{int}(x)\right|<C_\eps\max\left(\rho^{-\Ree\,s},1\right)\exp
\left(\eps|\Image\,s|\right)\sum\left|s-s_k\right|^{-1},
\label{E128}
\end{equation}
where the sum is over $0\le k\le m$ such that $(-s_k)\notin\wZ_+\cup 0$.
\label{PE203}
\end{pro}

\noindent{\bf Proof.} The density $p^{s,\eps}_{int}(x)$ for $\Ree\,s>n/2$
corresponds to the sum of the integrals
\begin{equation}
\left(2\pi\right)^{-n}\sum_j\phi_j(x)\int d\xi{i\over 2\pi}
\int_{\Gamma_{(\eps)}}\lambda^{-s}\theta(\xi,\lambda)\sum^m_{j=0}
a_{-2-j}(x,\xi,\lambda) d\lambda,
\label{E300}
\end{equation}
where $a_{-k}$ is a positive homogeneous of degree $(-k)$
in $\left(\xi,\lambda^{1/2}\right)$ component of the symbol
$s\left(\left(\doo-\lambda\right)^{-1}\right)$ in the coordinate
chart $U_i$ defined by (\ref{ST750}). The integral (\ref{E300})
is the sum
of the integrals $J^s_{0,\eps}(x)+J^s_{1,\eps}(x)+J^s_{2,\eps}(x)$
over the three corresponding domains:
\begin{align}
K_{0} & :=\left\{(\xi,\lambda)\colon\left|\xi\right|^2\le 1-\rho,\ \lambda\in
\Gamma_{(\eps)},\ |\lambda|\le\left(1-\left|\xi\right|^2\right)\right\},
\notag\\
I_{1} & :=\left\{(\xi,\lambda)\colon\left|\xi\right|^2\le 1-\rho,\ \lambda\in
\Gamma_{(\eps)},\ |\lambda|>\left(1-\left|\xi\right|^2\right)\right\},
\label{E301}\\
I_{2} & :=\left\{(\xi,\lambda)\colon\left|\xi\right|^2>1-\rho,\ \lambda\in
\Gamma_{(\eps)}\right\},\notag
\end{align}

Since $K_{0}$ is compact and since
$\lambda^{-s}\theta\sum^m_{0}a_{-2-j}(x,\xi,\lambda)$ is continuous on
$M_j\times K_{0}$, the density $J^s_{0,\eps}(x)$ is holomorphic
in $s\in\wC$ and it is estimated by
\begin{multline}
\left|J^s_{0,\eps}(x)\right|\le \\
\le C_\eps\exp\left(\eps|\Image\,s|\right)\max\left(\rho^{-\Ree\,s},1\right)
\max_{M_j\times K_{0}}\sum_i\left|\phi_i(x)\theta\sum^m_{0}a_{-2-j}
(x,\xi,\lambda)\right|.
\label{E302}
\end{multline}

The latter factor on the right in (\ref{E302}) does~not depend on $s$
and $\rho$. Hence it is estimated by a constant.

Set $J^s_k(x):=J^s_{k,\eps}(x)$ from now on. For $\Ree\,s>n/2$ the
density $J^s_{2}(x)$ does not change if the interior integral
in (\ref{E300}) is replaced by the integral over
\begin{equation}
\Gamma_{(\eps),|\xi|}:=|\xi|^2(1-\rho)^{-1}\Gamma_{(\eps)},
\label{E398}
\end{equation}
because $\theta\sum^m_{0}a_{-2-j}$ is holomorphic in $\lambda$ in the domain
between the contours $\Gamma_{(\eps)}$ and $\Gamma_{(\eps),|\xi|}$.
Indeed, this symbol is holomorphic in $\lambda$ for $(\xi,\lambda)$
such that $|\xi|^2+|\lambda|>1$ and $|\xi|^2>c|\lambda|$
(where $c>0$ is the same as in (\ref{X1791}) and (\ref{X1792})). Since
$0<\rho<(2c+1)^{-1}$, we have for $\lambda$ between $\Gamma_{(\eps)}$
and $\Gamma_{(\eps),|\xi|}$
\begin{gather*}
\begin{split}
\rho\le|\lambda|\le\left|\xi\right|^2\rho\left(1-\rho\right)^{-1},\qquad &
c|\lambda|\le c|\xi|^2\rho(1-\rho)^{-1}<2^{-1}|\xi|^2<|\xi|^2, \\
\text{ and }\qquad |\xi|^2+|\lambda| & >1\qquad\text{ for }\quad (\xi,\lambda)
\in I_{2}.
\end{split}
\end{gather*}

The density $J^s_{2}(x)$ is represented as the sum
$J^s_{2,\rho}(x)+\sum_{j=1,2}J^s_{2,j}(x)$, where $J^s_{2,j}$ and
$J^s_{2,\rho}$ correspond to (\ref{E300}), with the interior integral
replaced by the integral over $\Gamma^\eps_{j,|\xi|}:=\left|\xi\right|^2
\left(1-\rho\right)^{-1}\Gamma_{j,\eps}$ and
over $\Gamma^\eps_{\rho,|\xi|}:=|\xi|^2(1-\rho)^{-1}\Gamma^\eps_\rho$.
The density $J^s_{1}(x)$ is equal to the sum $\sum_{j=1,2}J^s_{1,j}(x)$,
where the interior integral in (\ref{E300}) for the term $J^s_{1,j}$
is over the contour $\Gamma_{j,\eps}\setminus D_{1-|\xi|^2}$
($D_r:=\{\lambda,|\lambda|<r\}$).

Set $\lambda:=\exp\left(i\eps\left(-1\right)^{j+1}\right)t^2$ on
$\Gamma^\eps_{j,|\xi|}$ (for $|\xi|^2>1-\rho$) and on
$\Gamma_{j,\eps}\setminus D_{1-|\xi|^2}$ (for $|\xi|^2\le 1-\rho$),
where $t>0$ is a new variable. Then we have
\begin{multline}
\left(J^s_{1,j}+J^s_{2,j}\right)(x)=\\
=\!2\!\left(2\pi\right)^{-(n+1)}\!i\!\exp\!\left(\!(\!-1\!)^ji\eps(\!s\!-\!1\!
)\!\right)\!\sum_l\!\phi_l(x)
\!\int_{F}\!t^{-2s+1}\!\tr\!\left(\!\sum^m_{0}\!a_{-2-k}\!\left(x,\xi,\lambda
_\eps(t)\!\right)\!\right)\!dt\,d\xi,
\label{E310}
\end{multline}
where $\lambda_\eps(t):=\exp\left((-1)^{j+1}i\eps\right)t^2$, $\Ree\,s>n/2$,
and $F$ is the domain
$$
\left\{(\xi,t)\colon\left|\xi\right|^2+t^2\ge 1,\left|\xi\right|^2\le 1-\rho,
t\ge\rho^{1\over 2}\right\}\cup\left\{(\xi,t)\colon|\xi|^2\ge 1-\rho,t\ge|\xi|
\rho^{1\over 2}\big/(1-\rho)^{1\over 2}\right\}.
$$

Since $a_{-k}(x,\xi,\lambda)$ are nonsingular for $\lambda\in\Gamma_{j,\eps}$
and since $r\cdot(\xi,t)\in F$ for $(\xi,t)\in F$ and $r\ge 1$,
we see that (\ref{E310}) can be written as follows:
\begin{multline}
\left(J^s_{1,j}+J^s_{2,j}\right)(x)=
2\left(2\pi\right)^{-(n+1)}i\exp\left((-1)^ji\eps(s-1)\right)\times\\
\times\sum_l\sum^m_{k=0}\left(2s+k-n\right)^{-1}\phi_l(x)\int_{F_{1}}
t^{-2s+1}\tr a_{-2-k}\left(x,\xi,\lambda_\eps(t)\right)dt\,\omega_{n+1},
\label{E311}
\end{multline}
where $F_{1}=F\cap\{(\xi,t)\colon\xi^2+t^2=1\}$ and $\omega_{n+1}$ is the
volume form on the unit sphere in $\wR^{n+1}_{\xi,t}$. The integral over
the compact $F_{1}$ in (\ref{E311}) is an entire function of $s\in\wC$.
So the right side of (\ref{E311}) realizes the analytic continuation
of the density $\left(J^s_{1,j}+J^s_{2,j}\right)(x)$ to a meromorphic
in $s\in\wC$ density with no more than simple poles at the points
$s_k=(n-k)/2$, $0\le k\le m$.

Since $\lambda=\rho\left(1-\rho\right)^{-1}|\xi|^2\exp(i\phi)$ on
$\Gamma^\eps_{\rho,|\xi|}$ (where $\eps\ge\phi\ge-\eps$), we have
\begin{equation}
J^s_\rho(x)=-i\left(2\pi\right)^{-(n+1)}\sum_l\sum^m_{k=0}\phi_l(x)\int I
^{\eps,\rho}_{l,k}(s,\xi)d\xi,
\label{E320}
\end{equation}
where the integral is over $\{\xi\colon|\xi|^2\ge 1-\rho\}$ and
\begin{gather}
I^{\eps,\rho}_{l,k}(s,\xi):=\!\int^\eps_{-\eps}\!\exp\!\left(-i\phi(s-1)\right)
\!a_{-2-k}\!\left(x,\xi,\lambda(\phi,|\xi|)\!\right)\!d\phi\left|\xi\right|
^{-2(s-1)}\!\rho_+^{-(s-1)}, \label{E321}\\
\lambda(\phi,|\xi|):=|\xi|^2\rho_+\exp(i\phi),\qquad \rho_+:=\rho(1-\rho)
^{-1}. \notag
\end{gather}

The symbol (\ref{E321}) is positive homogeneous of degree $(-2s-k)$
in $\xi$. It is analytic in $s\in\wC$ and nonsingular. So (\ref{E320})
realizes a meromorphic continuation of $J^s_\rho(x)$ to the whole
complex plane $\wC\ni s$. Namely
\begin{equation}
\!J^s_\rho(x)\!=\!2^{-1}\!\left(2\pi\right)^{-(n+1)}\!\sum_l\!\sum^m_{k=0}
\!\phi_l(x)\!\left(\!\int_{|\xi|=1}\!I^{\eps,\rho}_{l,k}(s,\xi)_x\omega_n
\!\right)\!\left(s\!-\!s_k\right)^{-1}\!\left(1\!-\!\rho\right)^{-(s-s_k)},
\label{E322}
\end{equation}
where $\omega_n$ is the volume form on the unit sphere in $\wR^n$.
The formulas (\ref{E311}) and (\ref{E322}) provide us with
a meromorphic continuation of the density $P^{s,\eps}_{int}(x)$.
Together with the estimate (\ref{E302}) they provide us with
the estimate (\ref{E128}). (However, with the sum in it over
all $s_k$, $0\le k\le m$.) The analytic continuation of the density
defined by the sum of the integrals (\ref{E300}) with the interior
integral over the contour $\Gamma_{(\pi)}$ (i.e., with $\eps=\pi$)
is regular in $s=s_k$ for $(-s_k)\in\wZ_+\cup 0$.
(This assertion is obtained in the proof of Theorem~\ref{TAT705}.)
For $|\xi|^2>1-\rho$ the interior integral over $\Gamma_{(\eps)}$
in (\ref{E300}) is equal to the integral over $\Gamma_{(\pi)}$.
So the estimate (\ref{E128}) is satisfied, where the sum over $s_k$,
$0\le k\le m$, such that $(-s_k)\notin\wZ_+\cup 0$.\ \ \ $\Box$

Let $\Ree\,s>n/2$ and let $p^{s,\eps}_{j,\df}(x)$ be the density on $M_j$
corresponding to the term in (\ref{E400}) determined
by the $\nu$-transmission interior boundary conditions.%
\footnote{From now on we'll suppose that $\df M=\emptyset$. Estimates of the
contributions into $\zeta_{\nu,\bullet}(s;p_j)$ from the Dirichlet and the
Neumann boundary conditions on the components of $\df M_j\setminus N$ are
analogous to the estimates for the contributions from the $\nu$-transmission
interior boundary conditions.}
It is defined by the restrictions to the diagonal
$\Mm_j\hookrightarrow\Mm_j\times\Mm_j$ of the kernel for the operator
\begin{equation}
{i\over 2\pi}\int_{\Gamma_{(\eps)}}\lambda^{-s}d\lambda\,p_j\sum_l\psi_l
\cD_{m,U_l}\phi_l,
\label{E415}
\end{equation}
where $\cD_m=\cD_m(\lambda,\nu)$ is defined by (\ref{ST761}),
(\ref{ST762}), and (\ref{ST764}).

The operator $\cD_m$ is defined by the symbols $\sum d_{-2-k}$,
$0\le k\le m$, in a coordinate chart $(x',t)$,
$U_i\subset\wR^{n-1}\times\wR^1$ (where
$N\cap U_i=\left(\wR^{n-1}\times 0\right)\cap U_i$ and the structure
$\wR^{n-1}\times\wR^1$ corresponds to the direct product structure
of the metric $g_{M}$ near $N$). Its action on $f$,
$f\in\DR_c\left(\wR^{n-1}\times\left(\wR_+\setminus 0\right)\right)$,
can be represented for $t>0$, $t_{1}>0$, as follows (\cite{Se1},
(26)--(28)):
\begin{align*}
\sum^m_{k=0}\Op(\theta_{1}d_{-2-k})f(x',t) & :=
\sum^m_{0}(2\pi)^{-n}\iint\exp\left(i(\xi',x')\right)\theta_{1}
d_{-2-k}^- f^-(\xi',t_{1})dt_{1}d\xi', \\
d^-_{-2-k}(x',t,\xi',t_{1},\lambda)        & :=
-\int_{\Gamma_-}\exp(-i\tau t_{1})d_{-2-k}(x',t,\xi',\tau,\lambda),\\
f^-(\xi',t_{1}) & =\int\exp\left(-i(\xi',y)\right)f(y,t_{1})dy, \notag
\end{align*}
where $\Gamma_-=\Gamma_-(\xi',\lambda)$ is a simple contour in the half-plane
$\Image\,\tau<0$ which once goes round (in the direction opposite
to the clockwise) the only zero of the principal symbol
$\left(b_{2}(x',\xi',\tau)-\lambda\right)\id$ of the Laplacian
$\left(\doo-\lambda\id\right)$.%
\footnote{The whole symbol of $\doo$ does~not depend on $t$
in the neighborhood of $N$.}
Lemmas~2 and 3 of \cite{Se2} and Lemma~2 of \cite{Se1} claim that
the integral over $\wR_+$ (where
$U_i\cap\left(\wR^{n-1}\times\wR_+\right)=U_i\cap M_j$)
$\int d^-_{-2-k}(x',t,\xi',t,\lambda)dt$ is a symbol
of $(x',\xi',\lambda)$ positive homogeneous of degree $(-2-k)$
in $\left(\xi',\lambda^{1/2}\right)$. It claims also that the kernel
on $\wR^{n-1}$ defined by the integral over $[T,\infty)\subset\wR_+$
(for an arbitrary $T>0$)
\begin{equation*}
\int\exp\left(i(\xi',x'-y')\right)d\xi'\int_{\Gamma}d\lambda\int^{\infty}
_{T}dt\,\theta_{1}(\xi',\lambda)\lambda^{-s}d_{-2-k}(x',t,\xi',t,\lambda)
\end{equation*}
is an entire function of $s\in\wC$, smooth in $x'$, $y'$, $s$ and
vanishing at $s$ for $(-s)\in\wZ_+\cup 0$. The latter assertion is
an immediate consequence of the estimate (\cite{Se1}, (29))
for $d^-_{-2-k}$:
\begin{multline}
\left|D^{\gamma}_{x'}D^q_{\xi'}t^rt^l_{1}D^m_t D^n_{t_{1}}D^p_{\lambda}d^-
_{-2-k}(x',t,\xi',t_{1},\lambda)\right|\le \\
C_{1}\exp\left(-c_\eps\left(|t|+|t_{1}|\right)\left(|\xi|+\left|\lambda
\right|^{1/2}\right)\right)\left(1+|\xi|+\left|\lambda\right|^{1/2}\right)
^{-1-k-(|q|+2p)-(r+l)+m+n}
\label{E420}
\end{multline}
with positive constants $C_{1}(\eps)$ and $c_\eps$ independent
of $\nu\in\rbo$.

\begin{pro}
The analytic continuation to the whole complex plane $\wC\ni s$
of the integral over $M_j$ of $p^{s,\eps}_{j,\df}(x)$ (which is defined
for $\Ree\,s>n/2$) is estimated by
\begin{multline}
\left|\int_{M_j}p^{s,\eps}_{j,\df}(x)\right|<
c_{1,\eps}\max\left(\rho^{-\Ree\,s},1\right)\exp\left(\eps|\Image\,s|\right)
\sum\left|s-s_k\right|^{-1}+ \\
+c_{2,\eps}\exp\left(\eps|\Image\,s|\right)\left(2c_\eps^{2(\Ree\,s-1)}\Gamma
\left(2(1-\Ree\,s),c_\eps\rho^{1/2}\right)+\pi\rho\max\left(\rho^{-\Ree\,s},1
\right)\right),
\label{E427}
\end{multline}
where the sum is over $1\le k\le m+1$ such that $(-s_k)\notin\wZ_+\cup 0$.
\label{PE205}
\end{pro}

\noindent{\bf Proof.}
The trace of the operator (\ref{E415}) for $\Ree\,s>n/2$ is given by
the integral
\begin{equation*}
\left(2\pi\right)^{-n}\sum_l\int dx'\,dt\phi_l(x',t)\int d\xi'\left({i\over
2\pi}\int_{\Gamma_{(\eps)}}\lambda^{-s}d\lambda\theta_{1}(\xi',\lambda)
\sum^m_{k=0}d^-_{-2-k}(x',t,\xi',t,\lambda)\right)\ \ \ .
\end{equation*}
(We suppose that $\phi_l(x',t)$ are independent of $t$ for
$0\le |t|\le 1$.) It follows from the estimate (\ref{E420}) that
the density on $N$ corresponding to the integral
\begin{equation*}
\int_{\wR_+}dt\left(\phi_l(x',t)-\phi_l(x',0)\right)\int d\xi'\tr\left(
{i\over 2\pi}\int_{\Gamma_{(\eps)}}\lambda^{-s}d\lambda\,\theta_{1}\sum d^-
_{-2-k}\right)
\end{equation*}
has the analytic continuation which is an entire function of $s\in\wC$
and which satisfies the estimate
\begin{equation}
C_{\eps,m,n_{1}}\exp\left(\eps|\Image\,s|\right)h_\rho(c_\eps,\Ree\,s)\int
\left(1+|\xi|\right)^{-n_{1}}\exp(-c_\eps|\xi'|)d\xi',
\label{E423}
\end{equation}
\begin{multline*}
h_\rho(c_\eps,\Ree\,s):=\int^{\infty}_\rho\exp\left(-c_\eps t^{1/2}\right)
t^{-\Ree\,s}dt+\pi\rho\max\left(\rho^{-\Ree\,s},1\right)= \\
=2c_\eps^{2(\Ree\,s-1)}\Gamma\left(2(1-\Ree\,s),c_\eps\rho^{1/2}\right)+
\pi\rho\max\left(\rho^{-\Ree\,s},1\right),
\end{multline*}
where $n_{1}\in\wZ_+$ is sufficiently large.

The density on $N$ is defined by the integral
\begin{equation}
p^{s,\eps}_{l,j,\df}(x'):=\left(2\pi\right)^{-n}\phi_l(x',0)\int_{\wR_+}dt
\int d\xi'\tr\left({i\over 2\pi}\int_{\Gamma_{(\eps)}}\lambda^{-s}d\lambda\,
\theta _{1}\sum^m_{k=0}d^-_{-2-k}\right)
\label{E425}
\end{equation}
which is absolutely convergent for $\Ree\,s>(n-1)/2$.
Hence it is analytic in $s$ for such $s$.
The integral over $t\in\wR_+$ of $d^-_{-2-k}$ is a positive homogeneous of
degree $(-2-k)$ in $(\xi',\lambda^{1/2})$ symbol, which is smooth in
$(x',\xi',\lambda)$ and analytic in $\lambda$ for
$c|\lambda|<\left|\xi'\right|^2$ (where $c$ is the same as in (\ref{X1791}),
and (\ref{X1792})) and in
$\lambda\in\Lambda_{\eps/2}:=\{\lambda\colon\eps/2<\arg\lambda<2\pi-\eps/2\}$
for $(\xi',\lambda)\ne (0,0)$ (\cite{Se1}, Lemma~2, \cite{Se2}, Lemma~2).
So the proof of Proposition~\ref{PE203} is valid also for the density
(\ref{E425}) (where $(x,\xi,n)$ are replaced by $(x',\xi',n-1)$).
We conclude that this density has a meromorphic continuation
$p^{s,\eps}_{l,j,\df}(x')$ with no more than simple poles at the points
$s_{1},\dots,s_{m+1}$. This proof provides us with the estimate
\begin{equation}
\left|p^{s,\eps}_{l,j,\df}(x')\right|<c_{1,\eps}\max\left(\rho^{-\Ree\,s},1
\right)\exp\left(\eps|\Image\,s|\right)\sum\left|s-s_k\right|^{-1},
\label{E426}
\end{equation}
where the sum is over $1\le k\le m+1$.  The analytic continuation of the
density on $N$, defined by the sum (over $l$) of the integrals (\ref{E425})
for the interior integrals over the contour $\Gamma_{(\pi)}$ (i.e., with
$\eps=\pi$) is regular at $s=s_k$ for $(-s_k)\in\wZ_+\cup 0$. (It is proved
in Theorem~\ref{TAT705}.) For $\left|\xi'\right|^2>1-\rho$ the integral over
$\Gamma_{(\eps)}$ in (\ref{E425}) is equal to the integral over
$\Gamma_{(\pi)}$. Hence the estimate (\ref{E426}) is satisfied if the sum
is over $k$, $1\le k\le m+1$, such that $(-s_k)\notin\wZ_+\cup 0$.
The estimate (\ref{E427}) follows from (\ref{E426}),
(\ref{E423}).\ \ \ $\Box$

The estimate (\ref{E109}) follows from Propositions~\ref{PE203}, \ref{PE205},
and from (\ref{E127}). Thus Proposition~\ref{PE201} is proved.\ \ \ $\Box$

\noindent{\bf Proof of Proposition~\ref{PE200}.}
The estimate (\ref{E127}) in the proof of Proposition~\ref{PE201}
is satisfied by the integral of $\left(r^m_\lambda\right)_{\xa,\xb}$.
So it is enough to obtain the estimate (\ref{E107}) for the kernel of
\begin{equation*}
{i\over 2\pi}\int_{\Gamma_{(\eps)}}\lambda^{-s}P^m_\lambda d\lambda,
\end{equation*}
where $\Ree\,s>n/2$. The term $p^{s,\eps}_{int}(\xa,\xb)$ in this kernel
has the same form as in (\ref{E300}) but with the addition factor
$\exp\left(i\xi(\xa-\xb)\right)$ under the integral sign (where $x$
and $\phi_j(x)$ are replaced by $x_1$ and by $\psi_j(\xa)\phi_j(\xb)$).
The integration over the domains (\ref{E301}) in the integral
corresponding to (\ref{E300}) represents this kernel
as the sum $\left(J^s_{0,\eps}+J^s_{1,\eps}+J^s_{2,\eps}\right)(\xa,\xb)$,
where $J^s_{l,\eps}$ corresponds to the integration over
the appropriate domain in (\ref{E301}). The term $J^s_{0,\eps}(\xa,\xb)$
satisfies the estimate (\ref{E302}) if $\Ree\,s>n/2$. (In this estimate
$\max\left(\rho^{-\Ree\,s},1\right)$ can be replaced
by $\rho^{-\Ree\,s}$ since $\Ree\,s>0$ and since $0<\rho<1$.)

The contour $\Gamma_{(\eps)}$ of the interior integral in (\ref{E300})
for $J^s_{2,\eps}$ can be replaced by the contour $\Gamma_{(\eps),|\xi|}$
defined by (\ref{E398}). The sum of the integrals over the straight line
pieces of $\Gamma_{(\eps)}$ and of $\Gamma_{(\eps),|\xi|}$ in the kernel
$\left(J^s_{1,\eps}+J^s_{2,\eps}\right)(\xa,\xb)$ has the same form as
(\ref{E310}) (but with the factor $\exp\left(i(\xa-\xb)\xi\right)$
under the integral sign). The integral over the circle part of
$\Gamma_{(\eps),|\xi|}$ for the kernel $J^s_{2,\eps}(\xa,\xb)$ is also
completely analogous to (\ref{E320}). This provides us with
the estimate for the kernel $p^{s,\eps}_{int}(\xa,\xb)$
(where $\Ree\,s>n/2$):
\begin{equation}
\left|p^{s,\eps}_{int}(\xa,\xb)\right|<C_\eps\exp\left(\eps|\Image\,s|\right)
\rho^{-\Ree\,s}\left(\Ree\,s-n/2\right)^{-1}.
\label{E340}
\end{equation}

The proof of (\ref{E107}) for an arbitrary closed manifold $M$
follows from the estimate (\ref{E340}) together with the estimate
(\ref{E127}) for $(r^m_\lambda)_{\xa,\xb}$.
(They also give us the proof
of (\ref{E107}) for a part $p^{s,\eps}_{int}(\xa,\xb)$ of the kernel
$\left(T_{-s}\right)_{\xa,\xb}$ defined by a local parametrix
$\sum_j\psi_jP^m_{\lambda,int}\phi_j$.)
If $(M,g_{M})$ is mirror symmetric with respect to $(N,g_{N})$
(and the $\nu$-transmission interior boundary conditions are given on $N$)
then the kernel $\left(T_{-s}\right)_{\xa,\xb}$ for $\Ree\,s>-n/2$
can be represented by the formulas
(analogous to (\ref{S414}), (\ref{S415}), and to (\ref{S1607})),
where $\nu=(\alpha,\beta)\ne(0,0)$ and $T^M_{-s}$ corresponds to a closed
manifold $(M,g_{M})$ (or to $\nu_{0}=(1,1)$ that is the equivalent
according to Proposition~\ref{P1}):
\begin{align}
\left(T_{-s}\right)_{\xa,\xb} & =\left(T^M_{-s}\right)_{\xa,\xb}+{\beta^2-
\alpha^2\over\alpha^2+\beta^2}\left(\sigma^*_{1}T^M_{-s}\right)_{\xa,\xb}
\quad\text{ for } \xa,\xb\in M_{1}, \notag\\
\left(T_{-s}\right)_{\xa,\xb} & =\left(T^M_{-s}\right)_{\xa,\xb}+{\alpha^2-
\beta^2\over\alpha^2+\beta^2}\left(\sigma^*_{1}T^M_{-s}\right)_{\xa,\xb}
\quad\text{ for } \xa,\xb\in M_{2}, \label{X1911}\\
\left(T_{-s}\right)_{\xa,\xb} & ={2\alpha\beta\over\alpha^2+\beta^2}
\left(T^M_{-s}\right)_{\xa,\xb} \quad\text{ for } \xa,\xb
\text{ from different } M_k, \notag
\end{align}
where $\sigma_{1}$ is the mirror symmetry on $M$ with respect to $N$,
acting on the variable $x_{1}$.
The kernel $\left(T^M_{-s}\right)_{\xa,\xb}$ can be analytically
(meromorphically) continued to the whole complex plane $\wC\ni s$
(separately on the diagonal $\xa=\xb$ and off the diagonal). It follows
from \cite{Se2}, Theorem~1 or from the proof of Theorem~\ref{TAT705}.
Hence (\ref{X1911}) is true for all $s\in\wC$. So the estimate (\ref{E107})
is satisfied also in the case of the $\nu$-transmission interior
boundary conditions on $N$ if $(M,g_{M})$ is mirror-symmetric with respect
to $N$.

The boundary term%
\footnote{The operator $\cD_m=\cD_{m,\lambda}(\nu)$ is defined
in the coordinate chart $\wR^{n-1}\times\wR^1\ni(x',t)$ by (\ref{ST764}),
(\ref{ST763}), (\ref{ST762}).}
$\sum\psi_j\cD_{m,U}\phi_j$ of the parametrix $P^m$ can be identified
with the same term in the mirror-symmetric case (as it is defined in
a neighborhood $N\times I$ of $N=N\times 0$, $I=[-2,2]$). The estimate
(\ref{E127}) for the integral over $\Gamma_{(\eps)}$ of
$\left(r^m_\lambda\right)_{\xa,\xb}$ is satisfied for
the mirror-symmetric case also. So the estimate (\ref{E107})
is satisfied by the kernel
$p^{\df,\eps}_{\xa,\xb}(s)$ of the operator $\sum\psi_j\cD_{m,U}\phi_j$.
This estimate for $p^{\df,\eps}_{\xa,\xb}(s)$ together with the estimates
(\ref{E340}), (\ref{E127}) for $p^{s,\eps}_{int}(\xa,\xb)$, and with
the estimate (\ref{E127}) of the integral
of $\left(r^m_\lambda\right)_{\xa,\xb}$over $\Gamma_{(\eps)}$ provides us
with the estimate (\ref{E107}).\ \ \ $\Box$

\subsection{Appendix. Trace class operators and their traces}
A bounded linear operator $A$ acting in a separable Hilbert space $H$
is a {\em trace class} operator if the series of its singular
numbers (i.e., of the arithmetic square roots of the eigenvalues
for the self-adjoint operator $A^*A$) is absolutely convergent.
If $A$ is a trace class operator then its matrix trace exists
for any orthonormal basis $(e_i)$ in $H$:
$$
\sum(Ae_i,e_i)=:\Spp A
$$
and this sum is independent of the orthonormal basis (\cite{Kr}).
It is called the matrix trace of $A$. The Lidskii theorem (\cite{Li})
claims that if $A$ is a trace class operator then the series
of its eigenvalues is absolutely convergent: $\sum|\lambda_j(A)|<\infty$
and its trace $\Tr A:=\sum\lambda_j(A)$ is equal to its matrix trace:
$\Tr A=\Spp A$. (Here the sums are over all the eigenvalues $\lambda_j(A)$
of $A$ including their {\em algebraic} multiplicities, \cite{Ka}, Ch.~1,
\S~5.4.)
\begin{pro}
For $t> 0$ the operators $\exp\left(-t\djon\right)$ and
$p_{1}\exp\left(-t\djon\right)$ are trace class operators
in the $L_{2}$-completion $\left(DR^j(M)\right)_{2}$ of $DR^j(M)$%
\footnote{$DR^j\left(\Mm_{1}\right)\oplus DR^j\left(\Mm_{2}\right)\subset
\left(DR^j(M)\right)_{2}$}
and their traces are equal to the integrals of the densities defined
by the restrictions to the diagonals of their kernels:
\begin{align}
\Tr\exp(-t\djon) & =\sum_{r=1,2}\int_{\Mm_r}\tr\left(*_{x_{2}}
i^*_{M_r}\etxaxb^j(\nu_{0})\right), \label{S1000}\\
\Tr\left(p_{1}\exp(-t\djon)\right) & =\int_{\Mm_{1}}\tr\left(*_{x_{2}}
i^*_{M_{1}}\etxaxb^j(\nu_{0})\right).
\label{S1001}
\end{align}
Here $p_k\colon\left(DR^j(M)\right)_2\to\left(DR^j\left(M_k\right)\right)_2
\hookrightarrow\left(DR^j(M)\right)_2$ is the composition
of the restriction to $M_k$ of differential forms and of their
prolongation to $M$ by zero on another piece of $M$,
$i_{M_r}:\Mm_r\hookrightarrow \Mm_r\times \Mm_r$ is an immersion
of the diagonal, and the exterior product of the double forms
(restricted to the diagonal) is implied.
\label{PA75}
\end{pro}

\noindent{\bf Proof.} The operator $A_t:=\exp(-t\djon)$
is positive definite on $\left(DR^j(M)\right)_{2}$ and for an arbitrary
$f\in\left(DR^j(M)\right)_{2}$, $f\ne 0$, it holds
$\left(A_t f,f\right)> 0$
(where the scalar product on $\left(DR^j(M)\right)_{2}$ corresponds
to (\ref{Y2})).

The operator $B_t:=p_{1}\exp\left(-t\djon\right)$ is
positive definite on the subspace $\left(DR^j(M_{1})\right)_{2}$
of $\left(DR^j(M)\right)_{2}$. Namely $\left(B_t m,m\right)> 0$
for $m\in\left(DR^j(M_{1})\right)_{2}$, $m\ne 0$, and it is a nonnegative
operator on $\dromb\colon\left(B_t f,f\right)\ge 0$
for $f\in\left(DR^j(M)\right)_{2}$.

The operator $\exp(-t\djon)$ is self-adjoint on $\left(DR^j(M)\right)_{2}$
by Theorem~\ref{TAT706}. Its kernel $A(\xa,\xb)$ is smooth
on $\Mm_{r_{1}}\times\Mm_{r_{2}}$ (as it is proved
in Proposition~\ref{TAT706}) and its trace is equal to
\begin{equation}
\!\Tr\!\exp\!\left(-t\djon\right)\!=\!\sum\kab\!\Tr\!\left(p_k\!\exp\!\left(
-t\djon\right)\!\right)\!=\!\sum\kab\!\Tr\!\left(p_k\!\exp\!\left(-t\djon
\right)\!p_k\right).
\label{S1005}
\end{equation}
(The matrix trace for $\exp\left(-t\djon\right)$
in $\left(DR^j(M)\right)_{2}=:H$ can be computed with the help
of an orthonormal basis $\left(e_i(1)\right),\left(e_i(2)\right)$
in $H$, where $\left(e_i(k)\right)$ is an orthonormal basis
in $H_k:=\left(DR^j\left(M_k\right)\right)_{2}$.)

The operator $A_k:=p_k\exp\left(-t\djon\right)p_k$ acting
in $H=H_{1}\oplus H_{2}$ has a continuous kernel
$A_k(\xa,\xb)=\etxaxb^j(\nu_{0})$ on $\Mm_k\times\Mm_k$
(and it has the zero kernel on $\Mm_{k_{1}}\times\Mm_{k_{2}}$
for $k_{1}\ne k_{2}$). The operator $A_k$ is a self-adjoint operator
acting in the Hilbert space  $H_k$ and it is positive definite,
$(A_k f,f)> 0$ for $f\in H_k$, $f\ne 0$. So according to the Mercer
theorem \cite{GG}, IV.3, \cite{RiN}, \S~98, the Fourier series
for the kernel of $A_k$ by the eigenforms of $A_k$
\begin{equation}
A_k(\xa,\xb)=\sum \mu_j\omega_j(\xa)\otimes\omega_j(\xb),
\label{S1006}
\end{equation}
(where $\mu_j>0$ are the eigenvalues of $A_k$) converges absolutely
and uniformly with respect to $\Mm_k\times\Mm_k$. Hence integrating
this series over the diagonals in $\Mm_k\times\Mm_k$ (for $k=1,2$)
we obtain the equality (\ref{S1000}):
\begin{equation*}
\Tr\exp\left(-t\djon\right)=\sum\kab\Tr A_k=\sum\kab\int_{\Mm_k}\tr\left(
i^*_{M_k}A_k(\xa,\xb)\right).
\end{equation*}

The equality (\ref{S1001}) is obtained similarly
\begin{equation*}
\Tr\left(p_{1}\exp\left(-t\djot\right)\right)=\Tr\left(p_{1}\exp\left(-t\djot
\right)p_{1}\right)=
\int_{\Mm_{1}}\tr\left(i^*_{M_{1}}A_{1}(\xa,\xb)\right).
\end{equation*}
The proposition is proved.\ \ \ $\Box$

\begin{pro}
For $\Ree\,s>n/2$ the operator $T_{-s}$ defined by the integral%
\footnote{The operator $T_{-s}$ for such $s$ is defined on $\dromb$ and
it is equal to the direct sum of the operator $\left(\don\right)^{-s}$
on the orthogonal complement to $\Ker\left(\don\right)$ and of the zero
operator on $\Ker\left(\don\right)$, by Theorem~\ref{TAT705}.}
(\ref{ST808}) and the operators $p_j\left(\don\right)^{-s}$ are trace
class operators ($n:=\dim M$). The traces of these operators
for $\Ree\,s>n/2$ are equal to the integrals over the diagonals
of the denisities, defined by the restrictions of their kernels
to these diagonals.
\begin{align}
\Tr\left(\left(\don\right)^{-s}\right) & =\sum_{r=1,2}\int_{\Mm_r}\tr\left(
*_{\xb}i^*_{M_r}T_{-s}(\xa,\xb)\right), \label{E362}\\
\Tr\left(p_j\left(\don\right)^{-s}\right) & =\int_{\Mm_j}\tr\left(*_{\xb}
i^*_{M_j}T_{-s}(\xa,\xb)\right).
\label{E360}
\end{align}
\label{PA80}
\end{pro}

\noindent{\bf Proof.} The kernel $T_{-s}(\xa,\xb)$ for $\Ree\,s>n/2$ is
continuous on $\Mm_{j_{1}}\times\Mm_{j_{2}}$ (Theorem~\ref{TAT705}).
The operator $T_{-s}$ for such $s$ is nonnegative,
$\left(T_{-s}f,f\right)\ge 0$, and self-adjoint. It is a trace class
operator (Theorem~\ref{TAT705}). For $\Ree\,s>n/2$ the equality holds
(analogous to (\ref{S1005}):
\begin{equation*}
\Tr T_{-s}=\sum_{j=1,2}\Tr(p_j T_{-s}p_j).
\end{equation*}
The operator $p_j T_{-s}p_j$ is self-adjoint in the Hilbert subspace
$\left(\!\DR\!\left(M_j\!\right)\right)_{2}$ of $\dromb$ and its kernel
coincides with the kernel $K_{-s,j}$ of $T_{-s}$ on $\Mm_j\times\Mm_j$.
So for $\Ree\,s>n/2$ its kernel is continuous and, according to the Mercer
theorem, the series on $\Mm_j\times\Mm_j$ for $K_{-s,j}$
by the eigenforms of $p_jT_{-s}p_j$ (analogous to (\ref{S1006}))
is absolutely and uniformly convergent on $\Mm_j\times\Mm_j$.
Hence for such $s$ the integral over the diagonal of the density,
defined by the restriction of the kernel $K_{-s,j}$,
is equal to $\Tr\left(p_jT_{-s}p_j\right)$.
Thus the equalities (\ref{E360}) and (\ref{E362}) are proved.\ \ \ $\Box$

\vspace{3mm}
{\em Acknowledgements.} I express my gratitude to R.~Bott,
W.~M\"uller, C.~Taubes, M.~Wodzicki, and S.-T.~Yau for stimulating
discussions. I am indebted to J.~Bernstein, A.~Beilinson, D.~Kazhdan,
and I.M.~Singer for a series of conversations on the theme of this paper.
I am very grateful to the Department of Mathematics of Harvard University
and to the Max-Planck-Institut f\"ur Mathematik for their hospitality and
for financial support, which made it possible for me to work
in the excellent intellectual atmosphere of these mathematical centers.
The results and the methods of this paper were reported
in the seminars of R.~Bott, C.~Taubes, and S.-T.~Yau in Harvard during
the Fall of 1992, in Oberseminar Max-Planck-Institut f\"ur Mathematik
in May of 1993, and to H.~McKean and L.~Nirenberg in the Courant Institute
in December of 1992. This paper was written in the Max-Planck-Institut
f\"ur Mathematik.

\end{document}